\documentclass[11pt]{article}
\pdfoutput=1

\usepackage{cite}
\usepackage{booktabs}
\usepackage[english]{babel}
\usepackage{amsmath,amssymb,amsbsy,amstext, amsthm, simplewick}
\usepackage{hyperref}
\usepackage{graphicx}
\usepackage{amsfonts}
\usepackage{amssymb}
\usepackage[small]{caption}
\usepackage{upgreek}
\usepackage[svgnames,dvipsnames,x11names,table]{xcolor}
\usepackage{multirow}
\usepackage{geometry}
\usepackage[hang,flushmargin]{footmisc}
\usepackage{bm}
\usepackage{braket}
\usepackage{subcaption}
\usepackage{mathtools}
\usepackage{setspace}
\usepackage{cleveref}
\usepackage{comment}
\usepackage{scalerel}
\usepackage[normalem]{ulem}
\usepackage{slashed}
\usepackage{tensor}
\usepackage{enumitem}
\usepackage{ifthen}
\usepackage{appendix}
\usepackage{floatrow}
\usepackage{siunitx}
\usepackage{framed}
\usepackage{tikz}
\usetikzlibrary{decorations.markings}
\usetikzlibrary{shapes.misc}
\usetikzlibrary{arrows}

\makeatletter
\g@addto@macro\bfseries{\boldmath}
\makeatother

\numberwithin{equation}{section} 

\newcolumntype{M}[1]{>{\centering\arraybackslash}m{#1}}
\newcommand\crossmark[1][]{%
  \tikz[scale=0.4,#1]{
    \fill(0,0)--(0.1,0) .. controls (0.5,0.4) .. (1,0.7)--(0.9,0.7) ..  controls (0.5,0.5) ..(0,0.1) --cycle;
    \fill(1,0.1)--(0.9,0.1) .. controls (0.5,0.3) .. (0,0.7)--(0.1,0.7) .. controls (0.5,0.4) ..(1,0.2) --cycle;
  }%
}

\definecolor{pyblue}{RGB}{31, 119, 180}
\definecolor{pyorange}{RGB}{255, 127, 14}
\definecolor{pygreen}{RGB}{44, 160, 44}
\definecolor{pyred}{RGB}{214, 39, 40}

\definecolor{light-gray}{gray}{0.95}
\definecolor{lightgray}{gray}{0.9}

\hypersetup{
    colorlinks=true,
    linkcolor={Purple!80!black},
    citecolor={pyblue},
    urlcolor={Purple!80!black}
}

\usepackage{colortbl}

\makeatletter
\newlength{\apb@width}
\newcommand{\autoparbox}[2][c]{\settowidth{\apb@width}{#2}\parbox[#1]{\apb@width}{#2}}

\makeatother

\usepackage[framemethod=default]{mdframed}
\newmdenv[skipabove=7pt,
skipbelow=7pt,
rightline=false,
leftline=false,
topline=false,
bottomline=false,
backgroundcolor=gray!10,
linecolor=gray,
innerleftmargin=5pt,
innerrightmargin=5pt,
innertopmargin=5pt,
innerbottommargin=5pt,
leftmargin=0cm,
rightmargin=0cm,
linewidth=4pt]{eBox}

\usepackage[most]{tcolorbox}
\tcbset{colback=white, colframe=black,
        highlight math style= {enhanced, 
            colframe=red,colback=red!10!white,boxsep=0pt}
        }

\crefname{table}{Table}{Tables}
\crefname{equation}{Eq.}{Eqs.}
\crefname{appendix}{App.}{Apps.}
\crefname{section}{Section}{Secs.}
\crefname{figure}{Fig.}{Figs.}

\def \Re {\mathcal{R}}
\def \Im {\mathcal{I}}
\def \d {\mathrm{d}}
\def \ks {k_{\text{s}}}
\def \kl {k_{\ell}}
\def \Nl {N_{\ell}}

\def\fnl{f_{\rm NL}}
\def\Mp{M_{\rm pl}}

\begin{document}

\begin{titlepage}
\setcounter{page}{1} \baselineskip=15.5pt
\thispagestyle{empty}
$\quad$
\vskip 0 pt

\vspace*{-1cm}

\begin{center}
{\fontsize{20}{18} \bf The Cosmological Flow:}\\[14pt] 
{\fontsize{15.3}{18} \bf  A Systematic Approach to Primordial Correlators}
\end{center}

\vskip 20pt
\begin{center}
\noindent
{\fontsize{12}{18}\selectfont Lucas Pinol,$^{1, 2}$ Sébastien Renaux-Petel,$^{2}$ and Denis Werth$^2$}
\end{center}

\begin{center}
\textit{$^1$ Laboratoire de Physique de l’École Normale Supérieure, ENS, CNRS, Université PSL,\\ Sorbonne Université, Université Paris Cité, F-75005, Paris, France} 

\vskip 5pt
\textit{$^2$ Sorbonne Université, CNRS, UMR 7095, Institut d’Astrophysique de Paris,\\ 98 bis bd Arago, 75014 Paris, France}
\end{center}

\vspace{0.4cm}
\begin{center}{\bf Abstract}
\end{center}

\noindent The time evolution of primordial fluctuations conceals a wealth of insights into the high-energy physics at play during the earliest moments of our Universe, which is ultimately encoded in late-time spatial correlation functions. However, the conventional procedure to compute them is technically challenging, and a complete dictionary mapping the landscape of inflationary theories and the corresponding observable signatures is not yet available.
In this paper, we develop a framework to compute tree-level cosmological correlators based on following their time evolution from their origin as quantum zero-point fluctuations to the end of inflation.
From first principles, the structure of the bulk time evolution imposes a set of universal differential equations in time satisfied by equal-time correlators.
We automatise the process of systematically solving these equations. This allows us to accurately capture all physical effects and obtain exact results in theories formulated at the level of inflationary fluctuations that include any number of degrees of freedom with arbitrary dispersion relations and masses, coupled through any time-dependent interactions.
We then illustrate the power of this formalism by exploring the phenomenology of cosmological correlators emerging from the interaction with a massive scalar field.
After an extensive analysis of the quadratic theory and classifying perturbativity bounds, we study both the size and the shape dependence of non-Gaussianities in the entire parameter space, including the strong mixing regime. We present novel characteristics of cosmological collider signals in (would be) single-, double-, and triple-exchange three-point correlators. In the presence of primordial features, after subtracting gauge artefacts unavoidably generated by a breaking of scale-invariance, we show that soft limits of cosmological correlators offer a new possibility to probe the inflationary landscape. Finally, we provide templates to search for in future cosmological surveys.

\end{titlepage}

\setcounter{page}{2}

\restoregeometry

\begin{spacing}{1.2}
\newpage
\setcounter{tocdepth}{3}
\tableofcontents
\end{spacing}

\setstretch{1.1}
\newpage

\section{Introduction}

Time plays a crucial role in cosmology. It orchestrates the gravitational collapse of cosmological structures that has shaped our Universe from the commencement to its present-day stage.
While we only have access to late-time observables, such as statistical properties of cosmological structures, our challenge is to understand the cosmic history and explain the correlations we see. In this sense, cosmology is about tracing the evolution of the Universe back in time or, instead going forward in time, tracking down the ``cosmological flow" of the observed correlations.

\vskip 4pt
According to the current paradigm, the early Universe underwent an inflationary phase~\cite{Starobinsky:1979ty,Starobinsky:1980te,Guth1981inflation,Guth1982inflation-perturbations,Linde:1981mu,Linde:1983gd,Linde:1986fc,Linde:1986fd}. Primordial quantum fluctuations were stretched to cosmological scales and were ultimately imprinted as spatial correlations that reside on the future boundary of inflationary spacetime, marking the beginning of the subsequent hot Big Bang evolution~\cite{Mukhanov:1981xt,Hawking:1982cz,Starobinsky:1982ee,Guth:1982ec,Linde:1982uu,Vilenkin:1982wt,Mukhanov:1985rz,Sasaki:1986hm}. These equal-time correlators of primordial density fluctuations and tensor modes encode valuable information about inflation, and---for connected $n$-point correlators with $n\geq3$---are conventionally called non-Gaussianities \cite{Maldacena:2002vr, Creminelli:2003iq, Chen:2010xka, Renaux-Petel:2015bja, Meerburg:2019qqi}. Recent years have shown increasing interests in computing these correlators as they ultimately hold the key to high-energy physics.

\vskip 4pt
On observational grounds, it is expected that substantial progress is to be made in the coming years. Compared to future CMB experiments that would lead to interesting yet modest improvements on bounds for non-Gaussianities \cite{Abazajian:2019eic}, the qualitative leap ahead of us lies in the constraining power of three-dimensional large-scale structure surveys \cite{Alvarez:2014vva, Achucarro:2022qrl, Karkare:2022bai, Ferraro:2022cmj}. Indeed, a host of galaxy surveys ambition to improve current constraints by one order of magnitude in the near future and reach an important threshold marking the border between a weakly or strongly coupled description of the inflationary background dynamics ($\fnl \sim 1$)~\cite{Creminelli:2003iq, Baumann:2011su}.
In the longer term, 21cm surveys will have the raw statistical power to improve limits by several orders of magnitude \cite{CosmicVisions21cm:2018rfq, Munoz:2015eqa, Chen:2016zuu, Liu:2022iyy}, reaching the gravitational floor ($\fnl \sim 10^{-2}$) \cite{Maldacena:2002vr}. 

\vskip 4pt
The phenomenological interest in primordial correlators lies in the fact that the energy scale of inflation---the Hubble scale---can be as high as $10^{14}\,$GeV~\cite{Planck:2018jri, BICEP:2021xfz}.
This energy makes the study of inflationary correlators a unique opportunity to learn about potential new physics at the highest reachable energies, much higher than those achievable in ground-based experiments. This one-of-a-kind window to fundamental physics has opened a cosmological collider physics program~\cite{Chen:2009we,Chen:2009zp, Noumi:2012vr,Arkani-Hamed:2015bza, Chen:2015lza, Chen:2016nrs, Lee:2016vti, Chen:2016uwp, Chen:2017ryl, An:2017hlx, Iyer:2017qzw, Chen:2018xck, Lu:2019tjj, Wang:2019gbi, Alexander:2019vtb, Hook:2019vcn, Liu:2019fag, Wang:2020ioa, Wang:2021qez, Maru:2021ezc, Cui:2021iie, Pinol:2021aun, Lu:2021wxu, Tong:2021wai, Tong:2022cdz, Qin:2022lva, Jazayeri:2022kjy, Pimentel:2022fsc, Werth:2023pfl,  Tong:2023krn, Jazayeri:2023xcj, Jazayeri:2023kji}. Essentially, even very heavy particles have been spontaneously produced during inflation, whose subsequent decays lead to imprints in inflationary correlators. These specific signatures---whose prospects for detection are bright~\cite{Sefusatti:2012ye, Norena:2012yi, Meerburg:2016zdz, MoradinezhadDizgah:2017szk, Bartolo:2017sbu, MoradinezhadDizgah:2018ssw, Kogai:2020vzz} although the experimental challenge to observe them is enormous---could inform us about the inflationary field content, including the mass spectra of particles, their spins and sound speeds. 

\vskip 4pt
On the theory side, the study of inflationary correlators has triggered recent developments both in deriving analytical closed-form solutions for correlators, and revealing formal aspects such as their analytic structure. These advancements mirror our understanding of flat-space scattering amplitudes or that of boundary correlators in anti-de Sitter (AdS) space, see e.g.~\cite{Kundu:2014gxa, Anninos:2014lwa, Kundu:2015xta, Shukla:2016bnu, Benincasa:2022omn, Benincasa:2022gtd, Benincasa:2021qcb, Benincasa:2020aoj, Benincasa:2018ssx, 
Arkani-Hamed:2017fdk, Pajer:2020wxk, Bonifacio:2021azc, Pajer:2020wnj, Maldacena:2011nz, Goodhew:2020hob, Melville:2021lst, Goodhew:2021oqg, Jazayeri:2021fvk, DiPietro:2021sjt, Meltzer:2021zin, Hogervorst:2021uvp, Baumann:2021fxj, Heckelbacher:2022hbq, Bonifacio:2022vwa, Salcedo:2022aal, Agui-Salcedo:2023wlq, DuasoPueyo:2023viy}. The motivations for such progress lie in the fact that, on cosmological scales probed in the CMB and LSS, inflation is reliably described by weakly-coupled fluctuations on top of a fixed quasi-de Sitter (dS) space. New techniques, such as solving boundary equations in external momenta~\cite{Arkani-Hamed:2018kmz} (see also the review~\cite{Baumann:2022jpr} and the references therein) and AdS-inspired techniques~\cite{Sleight:2019mgd, Sleight:2019hfp, Sleight:2020obc} have led to analytical results for the tree-level dS-invariant four-point exchange correlator mediated by a massive field with integer spin. These closed-form expressions have also been reproduced using partial Mellin-Barnes representations~\cite{Qin:2022lva, Qin:2022fbv} and extended to boost-breaking interactions and reduced sound speeds~\cite{Jazayeri:2022kjy, Pimentel:2022fsc}. Such methods opened up new analytical results for more complex diagrams, e.g.~at loop-level or including a chemical potential~\cite{Xianyu:2022jwk, Qin:2023ejc, Qin:2023bjk, Qin:2023nhv, Stefanyszyn:2023qov, Xianyu:2023ytd}.

\vskip 4pt
A priori, we do have a well-defined method for evaluating equal-time correlators \textit{without approximation}: the acclaimed in-in formalism \cite{Weinberg:2005vy}.
From first principles, calculations can be carried to arbitrary orders in perturbation theory, including loop contributions.
Having a complete formalism at hand, the question is then the following: how to efficiently and systematically compute inflationary correlators given \textit{any} inflationary theory?

\vskip 4pt
The standard procedure comes with its own set of challenges. The background time dependence leads to complicated (or even unknown) mode functions that are rapidly oscillating in the infinite past, and perturbative calculations require us to compute difficult time integrals over the entire inflationary evolution. As a consequence, primarily due to technical considerations, the vast landscape of inflationary theories currently remains beyond the scope of existing predictions, potentially leading to a biased interpretation of forthcoming data. A more pragmatic attitude, evading the intricacy of Feynman-like diagram computations, is to directly focus on correlators as they are the observables today. Assuming that primordial correlators emerge from a universal initial vacuum, they only depend on the specific theory active in the bulk of spacetime, or equivalently, only depend on how they ``flow" towards the end of inflation. Our aim is to \textit{automate} this picture.

\begin{figure}[b!]
\centering
    \includegraphics[scale=1]{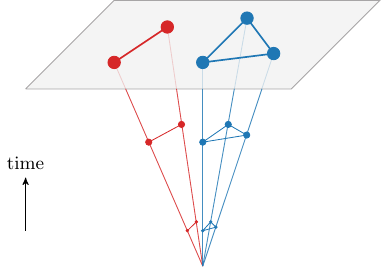}
    \caption{Schematic diagram of the cosmological flow. Inflationary correlators are tracked from initial quantum fluctuations in the asymptotic past to the end of inflation where they become spatial correlations on the reheating surface. The systematic resolution of their bulk time evolution is achieved by solving differential equations in time satisfied by the correlators.} 
    \label{fig:CosmoFlow}
\end{figure}

\vskip 4pt
In this paper, which follows the shorter letter~\cite{Werth:2023pfl}, we present a systematic approach to compute primordial correlators, tracing their cosmological flow, see Figure~\ref{fig:CosmoFlow}. This method is based on computing inflationary equal-time correlators by solving differential equations in \textit{time} governing their time evolution throughout the entirety of spacetime during inflation, from their origin as quantum fluctuations in the deep past to the end of inflation. It takes into account all physical effects at tree-level without approximation, in any theory conveniently formulated at the level of inflationary fluctuations, which can include an arbitrary number of degrees of freedom with any dispersion relations and masses, coupled through any time-dependent interactions. Systematically solving the cosmological flow of inflationary correlators does not require far-past UV regulators as correlators are evolved on the real-time axis, eliminates the need for mode functions, and bypasses the computation of intricate bulk time integrals.

\vskip 4pt
Being able to have direct access to the bulk time evolution of correlators is particularly useful in conjunction with analytical works and to extend the present reach of inflationary correlator phenomenology. It enables one to easily shed light on characteristic time scales at play and give new insights on identifying various physical regimes in complex scenarios that cannot be reached analytically. As an application, we show that the frequency of the cosmological collider signal may not be solely dictated by the mass of an additional field. Specifically, at strong mixing or in the presence of features---emerging from time-dependent couplings---, we derive new cosmological collider signals that break the direct link relating frequency and mass. For the numerous possibilities that our approach provides, we have made our numerical code \textsf{\href{https://github.com/deniswerth/CosmoFlow}{CosmoFlow}}\footnote{\href{https://github.com/deniswerth/CosmoFlow}{github.com/deniswerth/CosmoFlow}} available for the community. With this tool as a first building-block, we pave the way for a far-reaching program of exploring the rich physics of inflation. Although we focus on tree-level three-point correlators of scalars in this work, the cosmological flow allows for straight extensions, for example to compute higher-order correlators or to include spinning fields. 

\vskip 4pt 
Finally, similar to what automated tools brought to the particle physics community, we believe that the cosmological flow will offer new possibilities for the study of inflation. This shift in perspective enables us to move beyond the technical achievement of computing inflationary correlators within a limited subset of the landscape of theories and instead embark on a journey to explore the physics of inflation in full generality.

\paragraph{Outline.} In Section \ref{sec:time_evolution}, we first review how primordial correlators are computed from a bulk perspective using the standard in-in formalism. Then, we derive already-known differential equations in time satisfied by the two- and three-point correlators. To do so, we introduce convenient diagrammatic representations of these equations that enable us to derive these flow equations for any $n$-point correlators. In Section \ref{sec:cosmoflow_formalism}, we concretely apply this approach to compute the two- and three-point correlators in generic theories of inflationary fluctuations. Specifically, we show how the theory dependence is readily encoded in the flow equations and derive initial conditions for the correlators. The rest of the paper is dedicated to applications.
In Section \ref{sec:Application_EFT_multifield}, we construct a general theory of inflationary fluctuations where the Goldstone boson of broken time translations is coupled to a massive scalar field and provide a comprehensive study of the quadratic theory in the full parameter space. In Section \ref{sec:strong_coupling_scales}, we derive bounds on couplings for this theory to remain weakly coupled and discuss naturalness. In Sections \ref{sec:NG_pheno} and \ref{sec:CC_phycis}, for the first time, we provide exact results for the size and shape of non-Gaussianities and a complete understanding of the physics at play in the entire parameter space, including the strong mixing regime.
In Section \ref{sec:features}, we showcase that our method can systematically account for time-dependent couplings. We study both features that cannot be described by an effective single-field theory, and the corresponding cosmological collider signals. We show, with minimal assumptions, that the link between the frequency of the cosmological collider signal and the mass of an additional particle can be explicitly broken, leading to striking new observable signatures. Our conclusions are presented in Section \ref{sec:Conclusion}.
For the enthusiastic readers, appendices contain additional details. The two parts of the paper can be read independently. In particular, readers who want to skip the details of the cosmological flow derivation and only want to see it in action can skip Sections \ref{sec:time_evolution} and \ref{sec:cosmoflow_formalism} and go straight to the applications.

\paragraph{Relation to previous works.} The possibility of using differential equations in time to follow primordial correlators on super-horizon scales has been first recognised in \cite{Mulryne:2009kh, Mulryne:2010rp, Dias:2011xy, Seery:2012vj, Anderson:2012em, Elliston:2012ab}. This method has been extended to sub-horizon scales in \cite{Mulryne:2013uka} where its equivalence with the in-in formalism has been demonstrated. The modern transport approach has been implemented and publicly released as freely-available software packages in~\cite{Dias:2015rca, Dias:2016rjq,Mulryne:2016mzv, Seery:2016lko, Ronayne:2017qzn}. As such, the cosmological flow is built upon the pre-established transport approach. Notably, out of courtesy to the authors, most of the cosmological flow implementation and notations have been borrowed from \cite{Dias:2016rjq}. Other numerical schemes, see e.g.~\cite{Chen:2006xjb, Chen:2008wn, Arroja:2011yu, Hazra:2012yn, Funakoshi:2012ms, Adshead:2013zfa, Assassi:2013gxa, Sreenath:2014nca, Clarke:2020znk,Tran:2022euk} rely on directly computing the in-in integrals with numerical mode functions. All the existing approaches are either tied to non-linear sigma models, governing both the inflationary background and the fluctuations, or not automatic as the form of the bulk time integrals is strongly theory dependent. For instance, even the single-field effective field theory of inflation at leading order in derivatives is not encapsulated in existing automated tools, as some operators simply do not exist in non-linear sigma models. Conversely, as it is implemented directly at the level of fluctuations, the cosmological flow is applicable to any theories in a systematic manner, including of course the subset mentioned above, but offering a much richer and direct access to the physics of inflation.

\paragraph{Notation and conventions.}
We will use natural units in which the speed of light $c$ and Planck's constant $\hbar$ are set to unity $c = \hbar = 1$, with reduced Planck mass $\Mp = 1/\sqrt{8\pi G} = 2.4\times 10^{18}$ GeV, where $G$ denotes Newton's constant. Our metric signature is $(-++\,+)$. Spatial three-dimensional vectors are written in boldface $\bm{k}$. We will use Greek letters ($\mu, \nu, \ldots$) for spacetime indices, and Latin letters ($i, j, \ldots$) for spatial indices as usual. Overdots and primes will denote derivatives with respect to cosmic (physical) time $t$ and conformal time $\tau$ defined by $\mathrm{d}\tau = \mathrm{d}t/a$, respectively. Our Fourier convention is
\begin{equation}
f_{\bm{k}} = \int \mathrm{d}^3x f(x) e^{i \bm{k}.\bm{x}}\,, \hspace*{1cm} f(x) = \int \frac{\mathrm{d}^3k}{(2\pi)^3} f_{\bm{k}} \,e^{-i\bm{k}.\bm{x}}\,.
\end{equation}
A prime on a correlator is defined to mean that we drop the momentum conserving delta function
\begin{equation}
    \braket{\mathcal{O}_{\bm{k}_1} \ldots \mathcal{O}_{\bm{k}_4}} = (2\pi)^{3}\delta^{(3)}(\bm{k}_1 + \ldots + \bm{k}_4)\braket{\mathcal{O}_{\bm{k}_1} \ldots \mathcal{O}_{\bm{k}_4}}' \,.
\end{equation} 
The extended Fourier summation---used in the formalism presented in this paper---will be introduced throughout the main text when needed. Details are given in Appendix \ref{app:Fourier_summation}.

\newpage
\section{Following the Time Evolution of Correlators}
\label{sec:time_evolution}

We will begin our study with a general discussion of inflationary correlators in light of the cosmological flow. Our focus will be to derive differential equations in \textit{time} satisfied by cosmological correlators. Specifically, starting from the in-in formalism in \ref{subsec:cosmo_correlators}, we will derive the so-called \textit{flow equations} for the two-point correlators in \ref{subsec:2pt_flow} and three-point correlators in \ref{subsec:3pt_flow}. We then show how the flow equations for the $n$-point correlators can be found using a diagrammatic representation in \ref{subsec:npt_flow}.

\subsection{Equal-time Primordial Correlators}
\label{subsec:cosmo_correlators}

We will derive flow equations for cosmological correlators starting from the in-in formalism, which we briefly review here, setting up relevant notations and definitions.

\paragraph{Heisenberg-picture operators.} Our aim is to compute the vacuum expectation value of operators $\mathcal{O}(t)$ made up of products of various fields at a given time $t$, $\bra{\Omega} \mathcal{O}(t) \ket{\Omega}$, where $\ket{\Omega}$ is the vacuum of the full interacting theory. 
We will consider a generic theory of $N$ degrees of freedom $\bm{\varphi}^{\alpha}$  (with Greek letters running from $1$ to $N$) and remain voluntarily agnostic throughout this section on the precise form of the theory. It may contain arbitrary interactions and time-dependent coupling constants. The theory is described by a Hamiltonian $H(\bm{\varphi}^\alpha, \bm{p}^\beta)$ which is a functional of the phase-space coordinates i.e. the field-space coordinates $\bm{\varphi}^\alpha$ and their conjugate momenta $\bm{p}^\beta$.

\paragraph{Interaction-picture operators.} Following the standard procedure, we go from the Heisenberg to the interaction picture, therefore splitting the Hamiltonian into a \textit{free} and an \textit{interacting} part (we will describe later the implications of various choices of splitting)
\begin{equation}
\label{eq:Hamiltonian_split}
    H(\bm{\varphi}^\alpha, \bm{p}^\alpha) = H_0(\bm{\varphi}^\alpha, \bm{p}^\alpha) + H_{\text{I}}(\bm{\varphi}^\alpha, \bm{p}^\alpha)\,.
\end{equation}
Let us also introduce the interaction-picture operators $\varphi^\alpha$ and $p^\beta$ defined in terms of the Heisenberg-picture operators $\bm{\varphi}^\alpha$ and $\bm{p}^\beta$ by
\begin{equation}
    \varphi^\alpha \equiv \mathcal{U}^\dagger \bm{\varphi}^\alpha\, \mathcal{U}\,, \hspace*{0.5cm} p^\alpha \equiv \mathcal{U}^\dagger \bm{p}^\alpha\, \mathcal{U}\,,
\end{equation}
where $\mathcal{U}$ is a unitary operator that will be explicitly given below.
Inserting the identity operator inside the expectation value of the Heisenberg-picture operator of interest $\mathcal{O}(\bm{\varphi}^\alpha, \bm{p}^\beta)$, one obtains
\begin{equation}
\label{eq:correlators_interaction_picture}
    \bra{\Omega} \mathcal{O}(\bm{\varphi}^\alpha, \bm{p}^\beta)\ket{\Omega} = \bra{\Omega} \mathcal{U}\,\mathcal{O}(\varphi^\alpha, p^\beta)\,\mathcal{U}^\dagger\ket{\Omega}\,.
\end{equation}
For practical purposes, we gather all fields $\bm{\varphi}^\alpha$ and conjugate momenta $\bm{p}^\beta$ in a phase-space vector\footnote{The Latin index $a$ runs over all phase-space coordinates from $1$ to $2N$. We will also assume that such indices are organised so that a block of field labels is followed by a block of momentum labels, in the same order.} $\bm{X}^a\equiv (\bm{\varphi}^\alpha, \bm{p}^\beta)$, and we do the same for the interaction-picture operators $X^a\equiv (\varphi^\alpha,p^\beta)$. Now, by choosing the unitary operator $\mathcal{U}$ to satisfy the evolution equation
\begin{equation}
\label{eq:EOM_unitary_operator}
    \frac{\mathrm{d}\, \mathcal{U}}{\mathrm{d}t} = i H_{\text{I}}(\bm{X}^a)\, \mathcal{U} = i\, \mathcal{U} H_{\text{I}}(X^a)\,,
\end{equation}
one can verify that the interacting-picture fields evolve with the free Hamiltonian
\begin{equation}
    \frac{\mathrm{d} X^a}{\mathrm{d}t} =  i[H_0(X^b), X^a]\,. 
\end{equation}
The solution to Eq.~(\ref{eq:EOM_unitary_operator}) approaching identity in the asymptotic past is well known and is formally given by Dyson's formula
\begin{equation}
\label{eq:Dyson_formula}
    \mathcal{U} = \bar{\text{T}} \exp\left(i\int_{-\infty^+}^t H_{\text{I}}(t')\,\mathrm{d}t'\right)\,,
\end{equation}
where we have denoted $H_{\text{I}}(t') = H_{\text{I}}[X^a(t')]$. The interacting Hamiltonians evaluated at $t'$ in the power series expansion of the exponential need to be written in time increasing order, with $\bar{\text{T}}$ being the anti-time ordering operator. For the Hermitian conjugate operator $\mathcal{U}^\dagger$, the time ordered operator $\bar{\text{T}}^\dagger=\text{T}$ rewrites the interacting Hamiltonians in decreasing time order. The lower limit $-\infty^+$ accounts for deforming the integration contour above the real axis at early times, with the fields appearing in the integral defined by analytical continuation $t\rightarrow t(1+i\epsilon)$. Inside correlators, this $i\epsilon$ prescription enables us to evaluate the expectation value in the interacting vacuum by adiabatically switching off the interactions in the distant past. This procedure makes it possible to match the \textit{free} theory vacuum $\ket{0}$ onto the full theory vacuum $\ket{\Omega}$ in the infinite past. Collecting Eqs.~(\ref{eq:correlators_interaction_picture}) and (\ref{eq:Dyson_formula}), we end up with the well-known in-in formula \cite{Weinberg:2005vy}
\begin{equation}
\label{eq:in-in}
    \bra{\Omega} \mathcal{O}(\bm{X}^a) \ket{\Omega} = \bra{0} \left[\bar{\text{T}}\, e^{i\int_{-\infty^+}^t H_{\text{I}}(t')\mathrm{d}t'}\right] \mathcal{O}(X^a) \left[\text{T}\, e^{-i\int_{-\infty^-}^t H_{\text{I}}(t')\mathrm{d}t'}\right]\ket{0}\,.
\end{equation}

\paragraph{Equations of motion.} In order to go further, we need to specify a certain form for the Hamiltonian. Since we are going to essentially work in Fourier space, we will adopt the extended Fourier summation convention that consists in using \textsf{sans serif} indices to denote a sum including integrals over Fourier modes, making the following expressions compact and manageable. Hence, in what follows, an index contraction reads
\begin{equation}
    A_{\sf{a}}B^{\sf{a}} = \sum_a \int \frac{\mathrm{d}^3k_a}{(2\pi)^3}\, A_a(\bm{k}_a) B^a(\bm{k}_a)\,.
\end{equation}
This summation convention has been used and detailed in \cite{Dias:2016rjq}. For completeness, we give more details in Appendix \ref{app:Fourier_summation}. Using this notation, we take the Hamiltonian to be
\begin{equation}
\label{eq:full_general_Hamiltonian}
    H = \frac{1}{2!}H_{\sf{ab}} \bm{X}^{\sf{a}} \bm{X}^{\sf{b}} + \frac{1}{3!}H_{\sf{abc}} \bm{X}^{\sf{a}} \bm{X}^{\sf{b}} \bm{X}^{\sf{c}} + \frac{1}{4!}H_{\sf{abcd}} \bm{X}^{\sf{a}} \bm{X}^{\sf{b}} \bm{X}^{\sf{c}} \bm{X}^{\sf{d}} + \ldots\,.
\end{equation}
Without loss of generality, we take $H_{\sf{ab}}, H_{\sf{abc}}, \dots$ to be symmetric under the exchange of any indices.\footnote{In a quantum theory, operator ordering of the Hamiltonian a priori matters. Canonical quantisation consisting in promoting the phase-space variables of the classical Hamiltonian to quantum operators is ambiguous, as an infinite number of quantum theories that reduce to the classical one in the limit $\hbar \rightarrow 0$ may be defined.
In the context of quantum field theory though, ambiguities in the operator ordering eventually show up as different types of UV divergences to be renormalised. Since we will not perform renormalisation explicitly---which amounts to assume we directly work with the renormalised theory---we will overlook the operator ordering ambiguity in this work. We can therefore safely and without loss of generality symmetrise the classical Hamiltonian and then promote the fields and momenta to quantum operators. We wish to especially thank David Mulryne and David Seery for helpful discussions on this matter.}
These tensors encode the various interactions in the theory, and conventionally encompass functions of time and of the various momenta for derivative interactions.
It is worth emphasising that this form of the Hamiltonian is completely general and captures \textit{all} known effective field theories for scalar degrees of freedom at the level of inflationary fluctuations.\footnote{Higher-order time derivative interactions can lead to equations of motion with more than two time derivatives. As this situation signals the presence of an additional dynamical degree of freedom in the theory, one can introduce it in Eq.~(\ref{eq:full_general_Hamiltonian}) so that the Hamiltonian always leads to equations of motion with no more than two time derivatives.}
The fully non-linear equations of motion then read
\begin{equation}
\label{eq:Heisenberg_equations}
\begin{aligned}
    \frac{\mathrm{d} \bm{X}^{\sf{a}}}{\mathrm{d}t} &= i\,[H, \bm{X}^{\sf{a}}] \\
    &= \epsilon^{\sf{ac}}H_{\sf{cb}}\bm{X}^{\sf{b}} + \frac{1}{2!}\epsilon^{\sf{ad}}H_{\sf{dbc}}\bm{X}^{\sf{b}}\bm{X}^{\sf{c}} + \frac{1}{3!}\epsilon^{\sf{ae}}H_{\sf{ebcd}}\bm{X}^{\sf{b}}\bm{X}^{\sf{c}}\bm{X}^{\sf{d}} + \dots \\
    &= \tensor{u}{^{\sf{a}}}{_{\sf{b}}}\bm{X}^{\sf{b}} + \frac{1}{2!} \tensor{u}{^{\sf{a}}}{_{\sf{bc}}}\bm{X}^{\sf{b}}\bm{X}^{\sf{c}} + \frac{1}{3!} \tensor{u}{^{\sf{a}}}{_{\sf{bcd}}}\bm{X}^{\sf{b}}\bm{X}^{\sf{c}}\bm{X}^{\sf{d}}+ \ldots \,,
\end{aligned}
\end{equation}
where in the second line we have introduced the tensor $\epsilon^{\sf{ab}}$ that is defined so that the commutation relation is written in the following compact form\footnote{More details on this commutator in Fourier space is presented in Appendix \ref{app:Fourier_summation}.}
\begin{equation}
\label{eq:commutation_relation}
    [\bm{X}^{\sf{a}}, \bm{X}^{\sf{b}}] = i \epsilon^{\sf{ab}}\,,
\end{equation}
where $\epsilon^{\sf{a}\sf{b}} \equiv (2\pi)^3\delta^{(3)}(\bm{k}_a + \bm{k}_b)\,\epsilon^{ab}$ and the $2N\times2N$ matrix can be written in block form
\begin{equation}
\epsilon^{ab} = 
\begin{pmatrix}
\bf{0} & \bf{1} \\
\bf{-1} & \bf{0}
\end{pmatrix}\,.
\end{equation}
The third line of~(\ref{eq:Heisenberg_equations}) should be seen as a definition for the tensors $\tensor{u}{^{\sf{a}}}{_{\sf{b}}}, \tensor{u}{^{\sf{a}}}{_{\sf{bc}}}, \tensor{u}{^{\sf{a}}}{_{\sf{bcd}}}, \dots\,$. We also have used the fact that the Hamiltonian has been taken to be fully symmetric\footnote{Note that this implies that the tensors $\tensor{u}{^{\sf{a}}}{_{\sf{bc}}}, \tensor{u}{^{\sf{a}}}{_{\sf{bcd}}}, \ldots\,$ are symmetric under the exchange of any lower indices.} and $\epsilon^{\sf{ab}}$ is antisymmetric. Written in this form, it is clear that the equations of motion (\ref{eq:Heisenberg_equations}) encode both the full evolution of $\bm{X}^{\sf{a}}$ \textit{and} the commutation relations. In practice though, they are hard---if not impossible---to solve. One must therefore choose a simpler ``free" Hamiltonian $H_0$ to evolve the interaction-picture operators, thus resorting to a perturbative description of the interactions encoded in $H_\text{I}=H-H_0$, as described by the in-in formula in Eq.~\eqref{eq:in-in}.

\paragraph{Tree-level cosmological correlators.} We choose the free Hamiltonian to be the \textit{full quadratic Hamiltonian}, $H_0 = \frac{1}{2!}H_{\sf{ab}}\bm{X}^{\sf{a}}\bm{X}^{\sf{b}}$, with the interacting part of the Hamiltonian $H_{\text{I}}$ being given by the cubic and higher orders. Doing so, the equations of motion verified by the interaction-picture fields and momenta are linear. Still, all quadratic mixings between the various fields and momenta are taken into account in a non-perturbative manner, as we will show more clearly in \ref{subsec:resumming_quadratic_mixings}.
With this splitting choice, the equations of motion for the interaction-picture operators read
\begin{equation}
\label{eq:EOM_interaction_picture}
    \frac{\mathrm{d}X^{\sf{a}}}{\mathrm{d}t} = \tensor{u}{^{\sf{a}}}{_{\sf{b}}} X^{\sf{b}}\,, 
\end{equation}
which makes it explicit that $X^{\sf{a}}$ evolves with the full quadratic Hamiltonian. By now, we have derived all the necessary fundamental elements to write all possible $n$-point correlation functions. Up to three-point correlators, expanding the exponentials in the in-in formula (\ref{eq:in-in}) and working at tree-level yields
\begin{eBox}
\begin{equation}
\label{eq:correlators_inin}
    \begin{aligned}
    \bra{\Omega}\bm{X}^{\sf{a}} \bm{X}^{\sf{b}} \ket{\Omega} &= \bra{0}X^{\sf{a}}X^{\sf{b}}\ket{0}\,,\\
    \bra{\Omega}\bm{X}^{\sf{a}} \bm{X}^{\sf{b}}\bm{X}^{\sf{c}} \ket{\Omega} &= \bra{0} \frac{i}{3!}\int_{-\infty}^t\mathrm{d}t' H_{\mathsf{def}}\left[X^{\mathsf{d}}X^{\mathsf{e}}X^{\mathsf{f}}, X^{\mathsf{a}}X^{\mathsf{b}}X^{\mathsf{c}} \right]  \ket{0}\,.
    \end{aligned}
\end{equation}
\end{eBox}
From now on, we will not write $\ket{\Omega}$ nor $\ket{0}$ to make the notations less cluttered, assuming it is clear from the context whether the operators are in the Heisenberg ($\bm{X}^{\sf{a}}$) or interaction picture ($X^{\sf{a}}$). We retain the commutator form of the correlators because it will be more convenient in the following development. Note that no mention of mode functions is needed in this language.

\subsection{Two-point Correlators}
\label{subsec:2pt_flow}

We start by deriving the flow equations for the two-point functions $\langle \bm{X}^{\sf{a}} \bm{X}^{\sf{b}} \rangle$.
This object is the equal-time tree-level two-point correlator, taking into account the full quadratic Hamiltonian. In what follows, it will be more convenient to adopt a diagrammatic representation of such objects. We use a white dot to denote an external field insertion at the time $t$
\begin{equation}
\raisebox{0pt}{
\begin{tikzpicture}[line width=1. pt, scale=2]
\draw[fill=white] (0, 0) circle (.05cm) node[above=0.5mm] {\scriptsize$\sf{a}$};
\end{tikzpicture} 
}
= \bm{X}^{\sf{a}}(t)\,,
\end{equation}
and a black dot to denote a single vertex insertion in the bulk at a time $t'<t$ which needs to be integrated over. The two-point correlator can be represented as
\begin{equation}
\label{eq:diagram_resummed_2pt}
\begin{aligned}
\langle \bm{X}^{\sf{a}} \bm{X}^{\sf{b}} \rangle =\raisebox{0pt}{
\begin{tikzpicture}[line width=1. pt, scale=2]
\draw[fill=white] (0, 0) circle (.05cm) node[above=0.5mm] {\scriptsize$\sf{a}$};
\draw[black, double] (0.05, 0) -- (1, 0);
\draw[fill=white] (1, 0) circle (.05cm) node[above=0.5mm] {\scriptsize$\sf{b}$};
\end{tikzpicture} 
}\,.
\end{aligned}
\end{equation}
We will discuss various choices of interaction schemes and the resummation of quadratic mixings in \ref{subsec:resumming_quadratic_mixings}, which will make the double-line notation clear. Since the states are time independent, one can formally differentiate
(\ref{eq:diagram_resummed_2pt}) with respect to time by acting on the
interaction-picture operators $X^{\sf{a}}$ inside the correlator in (\ref{eq:correlators_inin}). We then use the equation of motion (\ref{eq:EOM_interaction_picture}) to find the following closed system 
\begin{equation}
\begin{aligned}
\textcolor{pyblue}{\frac{\mathrm{d}}{\mathrm{d}t}}
\raisebox{0pt}{
\begin{tikzpicture}[line width=1. pt, scale=2]
\draw[fill=white] (0, 0) circle (.05cm) node[above=0.5mm] {\scriptsize$\sf{a}$};
\draw[black, double] (0.05, 0) -- (1, 0);
\draw[fill=white] (1, 0) circle (.05cm) node[above=0.5mm] {\scriptsize$\sf{b}$};
\end{tikzpicture} 
}
&=
\raisebox{0pt}{
\begin{tikzpicture}[line width=1. pt, scale=2]
\draw[color=pyblue, fill=white] (0, 0) circle (.05cm) node[above=0.5mm] {\scriptsize$\sf{a}$};
\draw[pyblue, double] (0.05, 0) -- (0.5, 0);
\draw[black, double] (0.5, 0) -- (1, 0);
\draw[fill=white] (1, 0) circle (.05cm) node[above=0.5mm] {\scriptsize$\sf{b}$};
\end{tikzpicture} 
}
+ 
\raisebox{0pt}{
\begin{tikzpicture}[line width=1. pt, scale=2]
\draw[fill=white] (0, 0) circle (.05cm) node[above=0.5mm] {\scriptsize$\sf{a}$};
\draw[black, double] (0.05, 0) -- (0.5, 0);
\draw[pyblue, double] (0.5, 0) -- (1, 0);
\draw[color=pyblue, fill=white] (1, 0) circle (.05cm) node[above=0.5mm] {\scriptsize$\sf{b}$};
\end{tikzpicture} 
}\\
&= \tensor{u}{^{\sf{a}}}{_{\sf{c}}}
\raisebox{0pt}{
\begin{tikzpicture}[line width=1. pt, scale=2]
\draw[fill=white] (0, 0) circle (.05cm) node[above=0.5mm] {\scriptsize$\sf{c}$};
\draw[black, double] (0.05, 0) -- (1, 0);
\draw[fill=white] (1, 0) circle (.05cm) node[above=0.5mm] {\scriptsize$\sf{b}$};
\end{tikzpicture} 
}
+ \hspace*{0.2cm}
\tensor{u}{^{\sf{b}}}{_{\sf{c}}}
\raisebox{0pt}{
\begin{tikzpicture}[line width=1. pt, scale=2]
\draw[fill=white] (0, 0) circle (.05cm) node[above=0.5mm] {\scriptsize$\sf{a}$};
\draw[black, double] (0.05, 0) -- (1, 0);
\draw[fill=white] (1, 0) circle (.05cm) node[above=0.5mm] {\scriptsize$\sf{c}$};
\end{tikzpicture} 
}\,,
\end{aligned}
\end{equation}
where we have colored in \textcolor{pyblue}{blue} the parts of the correlator that have been differentiated with respect to time. Note that the introduced diagrammatic representation follows the usual Leibniz product rule of differentiation. The sum over repeated indices makes it explicit that---as usual in quantum physics---we sum over all possible diagrams. The flow equations for the two-point correlators are then 
\begin{eBox}
\begin{equation}
\label{eq:flow_equation_2pt}
\frac{\mathrm{d}}{\mathrm{d}t} \langle \bm{X}^{\sf{a}} \bm{X}^{\sf{b}} \rangle = \tensor{u}{^{\sf{a}}}{_{\sf{c}}} \langle \bm{X}^{\sf{c}} \bm{X}^{\sf{b}} \rangle + \tensor{u}{^{\sf{b}}}{_{\sf{c}}} \langle \bm{X}^{\sf{a}} \bm{X}^{\sf{c}} \rangle\,.
\end{equation}
\end{eBox}
Of course, these equations are completely equivalent---in the sense that they encode the same physics---to the linear equations of motion supplemented by the quantisation condition. Consequently, they correctly capture all physical effects arising from quadratic operators in the theory.
Note also that (\ref{eq:flow_equation_2pt}) couples all two-point correlators through the tensor $\tensor{u}{^{\sf{a}}}{_{\sf{b}}}$, including mixed propagators and correlators which contain conjugate momenta.

\subsection{Three-point Correlators}
\label{subsec:3pt_flow}

Let us now derive the flow equations for the three-point correlators. From (\ref{eq:correlators_inin}), we see that such correlators carry two time dependencies: (i) the external operators, and (ii) the upper limit of the integral over a bulk vertex, in the following denoted by a black dot. Diagrammatically, differentiating the three-point correlators with respect to time is then represented by
\begin{equation}
\textcolor{pyblue}{\frac{\mathrm{d}}{\mathrm{d}t}}
\vcenter{\hbox{
\begin{tikzpicture}[line width=1. pt, scale=2]
\draw[fill=white] (-0.3, -0.3) circle (.05cm) node[below=0.5mm] {\scriptsize$\sf{b}$};
\draw[fill=white] (0.3, -0.3) circle (.05cm) node[below=1.2mm] {\scriptsize$\sf{c}$};
\draw[fill=white] (0, 0.4) circle (.05cm) node[above=0.5mm] {\scriptsize$\sf{a}$};
\draw[fill=black] (0, 0) circle (.05cm);
\draw[black, double] (-0.04, -0.04) -- (-0.26, -0.26);
\draw[black, double] (0.04, -0.04) -- (0.26, -0.26);
\draw[black, double] (0, 0.055) -- (0, 0.345);
\end{tikzpicture}}}
= 
\vcenter{\hbox{
\begin{tikzpicture}[line width=1. pt, scale=2]
\draw[fill=white] (-0.3, -0.3) circle (.05cm) node[below=0.5mm] {\scriptsize$\sf{b}$};
\draw[fill=white] (0.3, -0.3) circle (.05cm) node[below=1.2mm] {\scriptsize$\sf{c}$};
\draw[color=pyblue, fill=white] (0, 0.4) circle (.05cm) node[above=0.5mm] {\scriptsize$\sf{a}$};
\draw[fill=black] (0, 0) circle (.05cm);
\draw[black, double] (-0.04, -0.04) -- (-0.26, -0.26);
\draw[black, double] (0.04, -0.04) -- (0.26, -0.26);
\draw[black, double] (0, 0.055) -- (0, 0.2);
\draw[pyblue, double] (0, 0.2) -- (0, 0.345);
\end{tikzpicture}}}
+
\vcenter{\hbox{
\begin{tikzpicture}[line width=1. pt, scale=2]
\draw[color=pyblue, fill=white] (-0.3, -0.3) circle (.05cm) node[below=0.5mm] {\scriptsize$\sf{b}$};
\draw[fill=white] (0.3, -0.3) circle (.05cm) node[below=1.2mm] {\scriptsize$\sf{c}$};
\draw[fill=white] (0, 0.4) circle (.05cm) node[above=0.5mm] {\scriptsize$\sf{a}$};
\draw[fill=black] (0, 0) circle (.05cm);
\draw[black, double] (-0.04, -0.04) -- (-0.15, -0.15);
\draw[pyblue, double] (-0.15, -0.15) -- (-0.26, -0.26);
\draw[black, double] (0.04, -0.04) -- (0.26, -0.26);
\draw[black, double] (0, 0.055) -- (0, 0.345);
\end{tikzpicture}}}
+
\vcenter{\hbox{
\begin{tikzpicture}[line width=1. pt, scale=2]
\draw[fill=white] (-0.3, -0.3) circle (.05cm) node[below=0.5mm] {\scriptsize$\sf{b}$};
\draw[color=pyblue, fill=white] (0.3, -0.3) circle (.05cm) node[below=1.2mm] {\scriptsize$\sf{c}$};
\draw[fill=white] (0, 0.4) circle (.05cm) node[above=0.5mm] {\scriptsize$\sf{a}$};
\draw[fill=black] (0, 0) circle (.05cm);
\draw[black, double] (-0.04, -0.04) -- (-0.26, -0.26);
\draw[black, double] (0.04, -0.04) -- (0.15, -0.15);
\draw[pyblue, double] (0.15, -0.15) -- (0.26, -0.26);
\draw[black, double] (0, 0.055) -- (0, 0.345);
\end{tikzpicture}}}
+
\vcenter{\hbox{
\begin{tikzpicture}[line width=1. pt, scale=2]
\draw[fill=white] (-0.3, -0.3) circle (.05cm) node[below=0.5mm] {\scriptsize$\sf{b}$};
\draw[fill=white] (0.3, -0.3) circle (.05cm) node[below=1.2mm] {\scriptsize$\sf{c}$};
\draw[fill=white] (0, 0.4) circle (.05cm) node[above=0.5mm] {\scriptsize$\sf{a}$};
\draw[color=pyblue, fill=pyblue] (0, 0) circle (.05cm);
\draw[pyblue, double] (-0.04, -0.04) -- (-0.15, -0.15);
\draw[black, double] (-0.15, -0.15) -- (-0.26, -0.26);
\draw[pyblue, double] (0.04, -0.04) -- (0.15, -0.15);
\draw[black, double] (0.15, -0.15) -- (0.26, -0.26);
\draw[pyblue, double] (0, 0.055) -- (0, 0.2);
\draw[black, double] (0, 0.2) -- (0, 0.345);
\end{tikzpicture}}}\,.
\end{equation}
Similar to the case of two-point correlators in the previous section, using the equation of motion (\ref{eq:EOM_interaction_picture}), differentiating an external operator leads to 
\begin{equation}
\vcenter{\hbox{
\begin{tikzpicture}[line width=1. pt, scale=2]
\draw[fill=white] (-0.3, -0.3) circle (.05cm) node[below=0.5mm] {\scriptsize$\sf{b}$};
\draw[fill=white] (0.3, -0.3) circle (.05cm) node[below=1.2mm] {\scriptsize$\sf{c}$};
\draw[color=pyblue, fill=white] (0, 0.4) circle (.05cm) node[above=0.5mm] {\scriptsize$\sf{a}$};
\draw[fill=black] (0, 0) circle (.05cm);
\draw[black, double] (-0.04, -0.04) -- (-0.26, -0.26);
\draw[black, double] (0.04, -0.04) -- (0.26, -0.26);
\draw[black, double] (0, 0.055) -- (0, 0.2);
\draw[pyblue, double] (0, 0.2) -- (0, 0.345);
\end{tikzpicture}}}
= \tensor{u}{^{\sf{a}}}{_{\sf{d}}} \vcenter{\hbox{
\begin{tikzpicture}[line width=1. pt, scale=2]
\draw[fill=white] (-0.3, -0.3) circle (.05cm) node[below=0.5mm] {\scriptsize$\sf{b}$};
\draw[fill=white] (0.3, -0.3) circle (.05cm) node[below=1.2mm] {\scriptsize$\sf{c}$};
\draw[fill=white] (0, 0.4) circle (.05cm) node[above=0.5mm] {\scriptsize$\sf{d}$};
\draw[fill=black] (0, 0) circle (.05cm);
\draw[black, double] (-0.04, -0.04) -- (-0.26, -0.26);
\draw[black, double] (0.04, -0.04) -- (0.26, -0.26);
\draw[black, double] (0, 0.055) -- (0, 0.345);
\end{tikzpicture}}}\,.
\end{equation}
The other term---coming from deriving the cubic bulk vertex---requires a few manipulations. We give a formal derivation of this contribution in the insert below. In terms of diagrams, one needs to differentiate all combinations of a bulk vertex together with an external operator insertion, resulting in cutting the diagram in two-point correlators. The resulting two-point correlators should then be contracted with a $\tensor{u}{^{\sf{a}}}{_{\sf{bc}}}$ tensor. This leads to
\begin{equation}
\label{eq:diagram_cubic_bulk_vertex}
\begin{aligned}
\vcenter{\hbox{
\begin{tikzpicture}[line width=1. pt, scale=2]
\draw[fill=white] (-0.3, -0.3) circle (.05cm) node[below=0.5mm] {\scriptsize$\sf{b}$};
\draw[fill=white] (0.3, -0.3) circle (.05cm) node[below=1.2mm] {\scriptsize$\sf{c}$};
\draw[fill=white] (0, 0.4) circle (.05cm) node[above=0.5mm] {\scriptsize$\sf{a}$};
\draw[color=pyblue, fill=pyblue] (0, 0) circle (.05cm);
\draw[pyblue, double] (-0.04, -0.04) -- (-0.15, -0.15);
\draw[black, double] (-0.15, -0.15) -- (-0.26, -0.26);
\draw[pyblue, double] (0.04, -0.04) -- (0.15, -0.15);
\draw[black, double] (0.15, -0.15) -- (0.26, -0.26);
\draw[pyblue, double] (0, 0.055) -- (0, 0.2);
\draw[black, double] (0, 0.2) -- (0, 0.345);
\end{tikzpicture}}}
&=
\vcenter{\hbox{
\begin{tikzpicture}[line width=1. pt, scale=2]
\draw[fill=white] (-0.3, -0.3) circle (.05cm) node[below=0.5mm] {\scriptsize$\sf{b}$};
\draw[fill=white] (0.3, -0.3) circle (.05cm) node[below=1.2mm] {\scriptsize$\sf{c}$};
\draw[color=pyblue, fill=white] (0, 0.4) circle (.05cm) node[above=0.5mm] {\scriptsize$\sf{a}$};
\draw[color=pyblue, fill=pyblue] (0, 0) circle (.05cm);
\draw[black, double] (-0.04, -0.04) -- (-0.26, -0.26);
\draw[black, double] (0.04, -0.04) -- (0.26, -0.26);
\draw[pyblue, double] (0, 0.055) -- (0, 0.345);
\end{tikzpicture}}}
+
\vcenter{\hbox{
\begin{tikzpicture}[line width=1. pt, scale=2]
\draw[color=pyblue, fill=white] (-0.3, -0.3) circle (.05cm) node[below=0.5mm] {\scriptsize$\sf{b}$};
\draw[fill=white] (0.3, -0.3) circle (.05cm) node[below=1.2mm] {\scriptsize$\sf{c}$};
\draw[fill=white] (0, 0.4) circle (.05cm) node[above=0.5mm] {\scriptsize$\sf{a}$};
\draw[color=pyblue, fill=pyblue] (0, 0) circle (.05cm);
\draw[pyblue, double] (-0.04, -0.04) -- (-0.26, -0.26);
\draw[black, double] (0.04, -0.04) -- (0.26, -0.26);
\draw[black, double] (0, 0.055) -- (0, 0.345);
\end{tikzpicture}}}
+
\vcenter{\hbox{
\begin{tikzpicture}[line width=1. pt, scale=2]
\draw[fill=white] (-0.3, -0.3) circle (.05cm) node[below=0.5mm] {\scriptsize$\sf{b}$};
\draw[color=pyblue, fill=white] (0.3, -0.3) circle (.05cm) node[below=1.2mm] {\scriptsize$\sf{c}$};
\draw[fill=white] (0, 0.4) circle (.05cm) node[above=0.5mm] {\scriptsize$\sf{a}$};
\draw[color=pyblue, fill=pyblue] (0, 0) circle (.05cm);
\draw[black, double] (-0.04, -0.04) -- (-0.26, -0.26);
\draw[pyblue, double] (0.04, -0.04) -- (0.26, -0.26);
\draw[black, double] (0, 0.055) -- (0, 0.345);
\end{tikzpicture}}}\\
&= \tensor{u}{^{\sf{a}}}{_{\sf{de}}}
\vcenter{\hbox{
\begin{tikzpicture}[line width=1. pt, scale=2]
\draw[fill=white] (0, 0.4) circle (.05cm) node[above=0.5mm] {\scriptsize$\sf{d}$};
\draw[fill=white] (0, -0.3) circle (.05cm) node[below=0.5mm] {\scriptsize$\sf{b}$};
\draw[fill=white] (0.4, 0.4) circle (.05cm) node[above=0.5mm] {\scriptsize$\sf{e}$};
\draw[fill=white] (0.4, -0.3) circle (.05cm) node[below=1.2mm] {\scriptsize$\sf{c}$};
\draw[black, double] (0, 0.35) -- (0, -0.25);
\draw[black, double] (0.4, 0.35) -- (0.4, -0.25);
\end{tikzpicture}}}
\hspace*{0.2cm}+\hspace*{0.2cm}
\tensor{u}{^{\sf{b}}}{_{\sf{de}}}
\vcenter{\hbox{
\begin{tikzpicture}[line width=1. pt, scale=2]
\draw[fill=white] (0, 0.4) circle (.05cm) node[above=0.5mm] {\scriptsize$\sf{d}$};
\draw[fill=white] (0, -0.3) circle (.05cm) node[below=0.5mm] {\scriptsize$\sf{a}$};
\draw[fill=white] (0.4, 0.4) circle (.05cm) node[above=0.5mm] {\scriptsize$\sf{e}$};
\draw[fill=white] (0.4, -0.3) circle (.05cm) node[below=0.5mm] {\scriptsize$\sf{c}$};
\draw[black, double] (0, 0.35) -- (0, -0.25);
\draw[black, double] (0.4, 0.35) -- (0.4, -0.25);
\end{tikzpicture}}}
\hspace*{0.2cm}+\hspace*{0.2cm}
\tensor{u}{^{\sf{c}}}{_{\sf{de}}}
\vcenter{\hbox{
\begin{tikzpicture}[line width=1. pt, scale=2]
\draw[fill=white] (0, 0.4) circle (.05cm) node[above=0.5mm] {\scriptsize$\sf{d}$};
\draw[fill=white] (0, -0.3) circle (.05cm) node[below=1.2mm] {\scriptsize$\sf{a}$};
\draw[fill=white] (0.4, 0.4) circle (.05cm) node[above=0.5mm] {\scriptsize$\sf{e}$};
\draw[fill=white] (0.4, -0.3) circle (.05cm) node[below=0.5mm] {\scriptsize$\sf{b}$};
\draw[black, double] (0, 0.35) -- (0, -0.25);
\draw[black, double] (0.4, 0.35) -- (0.4, -0.25);
\end{tikzpicture}}}
\,,
\end{aligned}
\end{equation}
which should be regarded as the diagrammatic rule corresponding to differentiating a cubic bulk vertex. Collecting the various terms,\footnote{The diagrammatic derivation of the flow equations we present is by all means completely equivalent to the formal derivation that can be found in \cite{Dias:2016rjq}.} the flow equations for the three-point correlators are
\begin{eBox}
\begin{equation}
\label{eq:flow_equation_3pt}
\frac{\mathrm{d}}{\mathrm{d}t} \langle \bm{X}^{\sf{a}} \bm{X}^{\sf{b}}\bm{X}^{\sf{c}} \rangle = \tensor{u}{^{\sf{a}}}{_{\sf{d}}} \langle \bm{X}^{\sf{d}} \bm{X}^{\sf{b}}\bm{X}^{\sf{c}} \rangle
+ \tensor{u}{^{\sf{a}}}{_{\sf{de}}}\langle \bm{X}^{\sf{b}} \bm{X}^{\sf{d}} \rangle\langle \bm{X}^{\sf{c}} \bm{X}^{\sf{e}} \rangle + (2\text{ perms})\,.
\end{equation}
\end{eBox}
Similar to Eq.~(\ref{eq:flow_equation_2pt}), these equations couple all correlators, including mixed correlators and those involving conjugate momenta. A few comments are in order.

\paragraph{Cosmological flow.} These equations are quantum in nature, as they evolve correlators in time including the effects of quantum interactions. Therefore, they provide an alternative way of computing correlators, following the ``cosmological flow": instead of directly computing the in-in integrals, the correlators are transported from a finite time in the past to any time of interest on the \textit{real time axis}. Importantly, the use of the $i\epsilon$ prescription is transferred to the derivation of initial conditions (we will come back to this point in \ref{sec:Initial_Conditions}).

\begin{framed}
{\small \noindent {\it Cubic bulk vertex.}---From (\ref{eq:correlators_inin}), the term coming from differentiating the cubic bulk vertex formally reads
\begin{equation}
\vcenter{\hbox{
\begin{tikzpicture}[line width=1. pt, scale=2]
\draw[fill=white] (-0.3, -0.3) circle (.05cm) node[below=0.5mm] {\scriptsize$\sf{b}$};
\draw[fill=white] (0.3, -0.3) circle (.05cm) node[below=1.2mm] {\scriptsize$\sf{c}$};
\draw[fill=white] (0, 0.4) circle (.05cm) node[above=0.5mm] {\scriptsize$\sf{a}$};
\draw[color=pyblue, fill=pyblue] (0, 0) circle (.05cm);
\draw[pyblue, double] (-0.04, -0.04) -- (-0.15, -0.15);
\draw[black, double] (-0.15, -0.15) -- (-0.26, -0.26);
\draw[pyblue, double] (0.04, -0.04) -- (0.15, -0.15);
\draw[black, double] (0.15, -0.15) -- (0.26, -0.26);
\draw[pyblue, double] (0, 0.055) -- (0, 0.2);
\draw[black, double] (0, 0.2) -- (0, 0.345);
\end{tikzpicture}}}
= \frac{i}{3!} \langle  H_{\sf{def}}\left[X^{\sf{d}}X^{\sf{e}}X^{\sf{f}}, X^{\sf{a}}X^{\sf{b}}X^{\sf{c}}\right]\rangle\,.
\end{equation}
Rearranging the commutator, introducing the commutation relation, and using the definition of the $\tensor{u}{^{\sf{a}}}{_{\sf{bc}}}$ tensor, this term can be rewritten
\begin{equation}
     \frac{i}{3!}\langle H_{\sf{def}}\left[X^{\sf{d}}X^{\sf{e}}X^{\sf{f}}, X^{\sf{a}}X^{\sf{b}}X^{\sf{c}}\right]\rangle = \frac{1}{2!} \tensor{u}{^{\sf{a}}}{_{\sf{de}}}\langle X^{\sf{b}}X^{\sf{c}}X^{\sf{d}}X^{\sf{e}}\rangle + (2\text{ perms})\,.
\end{equation}
The right-hand side involves products of interaction-picture operators obeying Gaussian statistics. Consequently, one can use Wick's theorem to contract the various operators. Retaining only connected three-point functions, one obtains
\begin{equation}
\frac{i}{3!} \langle  H_{\sf{def}}\left[X^{\sf{d}}X^{\sf{e}}X^{\sf{f}}, X^{\sf{a}}X^{\sf{b}}X^{\sf{c}}\right]\rangle = \tensor{u}{^{\sf{a}}}{_{\sf{de}}}\langle \bm{X}^{\sf{b}}\bm{X}^{\sf{d}}\rangle \langle \bm{X}^{\sf{c}}\bm{X}^{\sf{e}}\rangle + (2\text{ perms})\,,
\end{equation}
which, in the end, is given by the diagrammatic representation in (\ref{eq:diagram_cubic_bulk_vertex}).
 }
\end{framed}

\paragraph{Linearity.} These equations are linear. This specific structure allows the flow of each kinematic configuration to be tracked independently. However, the source terms of these equations are non-linear, in the sense that the three-point correlators are sourced by a product of two two-point correlators. As a result, Eq.~(\ref{eq:flow_equation_3pt}) does not form a closed system and one needs to solve for both the two- and three-point correlators. Intuitively, solving the two-point correlators is equivalent to solving the mode functions that are needed to compute higher-point correlators. The non-linear structure of the source is directly related to the fact that fields are quantum operators. Indeed, going from a three-point correlator---that is composed of six interaction-picture operators---to a product of two-point correlators---each composed of two operators---, we have used the commutation relation to reduce the number of operator insertions. We also remind the reader that working at tree-level means that non-linearities of the three-point correlators on the two-point correlators are neglected.

\subsection{Higher-order Correlators}
\label{subsec:npt_flow}

We now turn to higher-order correlators. We will see that the diagrammatic rules we introduced enable us to easily find the flow equations for any tree-level $n$-point correlator.

\paragraph{Four-point correlators.} We first illustrate the machinery with four-point correlators. In what follows, we will only consider the connected parts of higher-point correlators, hence defining them by subtracting the disconnected contribution. In full generality, a four-point correlator can be contact-like---with quartic interactions encoded in $H_{\sf{abcd}}$---or exchange-like---with two cubic interactions from $H_{\sf{abc}}$. Both contributions are formally written
\begin{gather}
\label{eq:4pt_in-in_formal}
\begin{aligned}
\langle\bm{X}^{\sf{a}} \bm{X}^{\sf{b}}\bm{X}^{\sf{c}}\bm{X}^{\sf{d}} \rangle &= \frac{i}{4!} \langle  \int_{-\infty}^{t}\mathrm{d}t' H_{\sf{efgh}}(t^\prime)\left[X^{\sf{e}}X^{\sf{f}}X^{\sf{g}}X^{\sf{h}}(t^\prime), X^{\sf{a}}X^{\sf{b}}X^{\sf{c}}X^{\sf{d}}\right]\rangle \\
&\hspace*{-2.5cm}- \frac{1}{(3!)^2}\langle  \int_{-\infty}^t \int_{-\infty}^{t'}\mathrm{d}t'\mathrm{d}t''H_{\sf{efg}} (t'')H_{\sf{hij}}(t')\left[X^{\sf{e}}X^{\sf{f}}X^{\sf{g}}(t''),\left[X^{\sf{h}}X^{\sf{i}}X^{\sf{j}}(t'), X^{\sf{a}}X^{\sf{b}}X^{\sf{c}}X^{\sf{d}}\right]\right]\rangle\,.
\end{aligned}
\raisetag{60pt}
\end{gather}
We first treat the case of contact-like four-point correlators. As for the three-point correlators in \ref{subsec:3pt_flow}, such correlators carry a time dependence in external operator insertions, and in the upper limit of the integral over the bulk vertex. Following the Leibniz product rule of differentiation, the diagrammatic representation of differentiating a contact-like four-point correlator is 
\begin{equation}
\label{eq:diagram_trispectrum_1}
\textcolor{pyblue}{\frac{\mathrm{d}}{\mathrm{d}t}}
\vcenter{\hbox{
\begin{tikzpicture}[line width=1. pt, scale=2]
\draw[fill=white] (-0.3, -0.3) circle (.05cm) node[below=1.2mm] {\scriptsize$\sf{c}$};
\draw[fill=white] (0.3, -0.3) circle (.05cm) node[below=0.5mm] {\scriptsize$\sf{d}$};
\draw[fill=white] (-0.3, 0.3) circle (.05cm) node[above=0.5mm] {\scriptsize$\sf{a}$};
\draw[fill=white] (0.3, 0.3) circle (.05cm) node[above=0.5mm] {\scriptsize$\sf{b}$};
\draw[fill=black] (0, 0) circle (.05cm);
\draw[black, double] (-0.04, -0.04) -- (-0.26, -0.26);
\draw[black, double] (0.04, -0.04) -- (0.26, -0.26);
\draw[black, double] (-0.04, 0.04) -- (-0.26, 0.26);
\draw[black, double] (0.04, 0.04) -- (0.26, 0.26);
\end{tikzpicture}}}
= 
\vcenter{\hbox{
\begin{tikzpicture}[line width=1. pt, scale=2]
\draw[fill=white] (-0.3, -0.3) circle (.05cm) node[below=1.2mm] {\scriptsize$\sf{c}$};
\draw[fill=white] (0.3, -0.3) circle (.05cm) node[below=0.5mm] {\scriptsize$\sf{d}$};
\draw[color=pyblue, fill=white] (-0.3, 0.3) circle (.05cm) node[above=0.5mm] {\scriptsize$\sf{a}$};
\draw[fill=white] (0.3, 0.3) circle (.05cm) node[above=0.5mm] {\scriptsize$\sf{b}$};
\draw[fill=black] (0, 0) circle (.05cm);
\draw[black, double] (-0.04, -0.04) -- (-0.26, -0.26);
\draw[black, double] (0.04, -0.04) -- (0.26, -0.26);
\draw[black, double] (-0.04, 0.04) -- (-0.15, 0.15);
\draw[pyblue, double] (-0.15, 0.15) -- (-0.26, 0.26);
\draw[black, double] (0.04, 0.04) -- (0.26, 0.26);
\end{tikzpicture}}}
\hspace*{0.2cm}+\hspace*{0.2cm}(3\text{ perms})\hspace*{0.2cm}+
\vcenter{\hbox{
\begin{tikzpicture}[line width=1. pt, scale=2]
\draw[fill=white] (-0.3, -0.3) circle (.05cm) node[below=1.2mm] {\scriptsize$\sf{c}$};
\draw[fill=white] (0.3, -0.3) circle (.05cm) node[below=0.5mm] {\scriptsize$\sf{d}$};
\draw[fill=white] (-0.3, 0.3) circle (.05cm) node[above=0.5mm] {\scriptsize$\sf{a}$};
\draw[fill=white] (0.3, 0.3) circle (.05cm) node[above=0.5mm] {\scriptsize$\sf{b}$};
\draw[color=pyblue, fill=pyblue] (0, 0) circle (.05cm);
\draw[pyblue, double] (-0.04, -0.04) -- (-0.15, -0.15);
\draw[black, double] (-0.15, -0.15) -- (-0.26, -0.26);
\draw[pyblue, double] (0.04, -0.04) -- (0.15, -0.15);
\draw[black, double] (0.15, -0.15) -- (0.26, -0.26);
\draw[pyblue, double] (-0.04, 0.04) -- (-0.15, 0.15);
\draw[black, double] (-0.15, 0.15) -- (-0.26, 0.26);
\draw[pyblue, double] (0.04, 0.04) -- (0.15, 0.15);
\draw[black, double] (0.15, 0.15) -- (0.26, 0.26);
\end{tikzpicture}}}\,.
\end{equation}
As usual, differentiating an external operator makes use of the equations of motions (\ref{eq:EOM_interaction_picture})
\begin{equation}
\vcenter{\hbox{
\begin{tikzpicture}[line width=1. pt, scale=2]
\draw[fill=white] (-0.3, -0.3) circle (.05cm) node[below=1.2mm] {\scriptsize$\sf{c}$};
\draw[fill=white] (0.3, -0.3) circle (.05cm) node[below=0.5mm] {\scriptsize$\sf{d}$};
\draw[color=pyblue, fill=white] (-0.3, 0.3) circle (.05cm) node[above=0.5mm] {\scriptsize$\sf{a}$};
\draw[fill=white] (0.3, 0.3) circle (.05cm) node[above=0.5mm] {\scriptsize$\sf{b}$};
\draw[fill=black] (0, 0) circle (.05cm);
\draw[black, double] (-0.04, -0.04) -- (-0.26, -0.26);
\draw[black, double] (0.04, -0.04) -- (0.26, -0.26);
\draw[black, double] (-0.04, 0.04) -- (-0.15, 0.15);
\draw[pyblue, double] (-0.15, 0.15) -- (-0.26, 0.26);
\draw[black, double] (0.04, 0.04) -- (0.26, 0.26);
\end{tikzpicture}}}
=
\tensor{u}{^{\sf{a}}}{_{\sf{e}}}
\vcenter{\hbox{
\begin{tikzpicture}[line width=1. pt, scale=2]
\draw[fill=white] (-0.3, -0.3) circle (.05cm) node[below=1.2mm] {\scriptsize$\sf{c}$};
\draw[fill=white] (0.3, -0.3) circle (.05cm) node[below=0.5mm] {\scriptsize$\sf{d}$};
\draw[fill=white] (-0.3, 0.3) circle (.05cm) node[above=0.5mm] {\scriptsize$\sf{e}$};
\draw[fill=white] (0.3, 0.3) circle (.05cm) node[above=0.5mm] {\scriptsize$\sf{b}$};
\draw[fill=black] (0, 0) circle (.05cm);
\draw[black, double] (-0.04, -0.04) -- (-0.26, -0.26);
\draw[black, double] (0.04, -0.04) -- (0.26, -0.26);
\draw[black, double] (-0.04, 0.04) -- (-0.26, 0.26);
\draw[black, double] (0.04, 0.04) -- (0.26, 0.26);
\end{tikzpicture}}}\,,
\end{equation}
and differentiating the bulk vertex, as previously done for the three-point correlators, results in cutting the diagram in two-point correlators contracted with the tensor $\tensor{u}{^{\sf{a}}}{_{\sf{bcd}}}$
\begin{gather}
\label{eq:diagram_quartic_bulk_vertex}
\begin{aligned}
\vcenter{\hbox{
\begin{tikzpicture}[line width=1. pt, scale=2]
\draw[fill=white] (-0.3, -0.3) circle (.05cm) node[below=1.2mm] {\scriptsize$\sf{c}$};
\draw[fill=white] (0.3, -0.3) circle (.05cm) node[below=0.5mm] {\scriptsize$\sf{d}$};
\draw[fill=white] (-0.3, 0.3) circle (.05cm) node[above=0.5mm] {\scriptsize$\sf{a}$};
\draw[fill=white] (0.3, 0.3) circle (.05cm) node[above=0.5mm] {\scriptsize$\sf{b}$};
\draw[color=pyblue, fill=pyblue] (0, 0) circle (.05cm);
\draw[pyblue, double] (-0.04, -0.04) -- (-0.15, -0.15);
\draw[black, double] (-0.15, -0.15) -- (-0.26, -0.26);
\draw[pyblue, double] (0.04, -0.04) -- (0.15, -0.15);
\draw[black, double] (0.15, -0.15) -- (0.26, -0.26);
\draw[pyblue, double] (-0.04, 0.04) -- (-0.15, 0.15);
\draw[black, double] (-0.15, 0.15) -- (-0.26, 0.26);
\draw[pyblue, double] (0.04, 0.04) -- (0.15, 0.15);
\draw[black, double] (0.15, 0.15) -- (0.26, 0.26);
\end{tikzpicture}}}
&=
\vcenter{\hbox{
\begin{tikzpicture}[line width=1. pt, scale=2]
\draw[fill=white] (-0.3, -0.3) circle (.05cm) node[below=1.2mm] {\scriptsize$\sf{c}$};
\draw[fill=white] (0.3, -0.3) circle (.05cm) node[below=0.5mm] {\scriptsize$\sf{d}$};
\draw[color=pyblue, fill=white] (-0.3, 0.3) circle (.05cm) node[above=0.5mm] {\scriptsize$\sf{a}$};
\draw[fill=white] (0.3, 0.3) circle (.05cm) node[above=0.5mm] {\scriptsize$\sf{b}$};
\draw[color=pyblue, fill=pyblue] (0, 0) circle (.05cm);
\draw[black, double] (-0.04, -0.04) -- (-0.26, -0.26);
\draw[black, double] (0.04, -0.04) -- (0.26, -0.26);
\draw[pyblue, double] (-0.04, 0.04) -- (-0.26, 0.26);
\draw[black, double] (0.04, 0.04) -- (0.26, 0.26);
\end{tikzpicture}}}
+
\vcenter{\hbox{
\begin{tikzpicture}[line width=1. pt, scale=2]
\draw[fill=white] (-0.3, -0.3) circle (.05cm) node[below=1.2mm] {\scriptsize$\sf{c}$};
\draw[fill=white] (0.3, -0.3) circle (.05cm) node[below=0.5mm] {\scriptsize$\sf{d}$};
\draw[fill=white] (-0.3, 0.3) circle (.05cm) node[above=0.5mm] {\scriptsize$\sf{a}$};
\draw[color=pyblue, fill=white] (0.3, 0.3) circle (.05cm) node[above=0.5mm] {\scriptsize$\sf{b}$};
\draw[color=pyblue, fill=pyblue] (0, 0) circle (.05cm);
\draw[black, double] (-0.04, -0.04) -- (-0.26, -0.26);
\draw[black, double] (0.04, -0.04) -- (0.26, -0.26);
\draw[black, double] (-0.04, 0.04) -- (-0.26, 0.26);
\draw[pyblue, double] (0.04, 0.04) -- (0.26, 0.26);
\end{tikzpicture}}}
+
\vcenter{\hbox{
\begin{tikzpicture}[line width=1. pt, scale=2]
\draw[color=pyblue, fill=white] (-0.3, -0.3) circle (.05cm) node[below=1.2mm] {\scriptsize$\sf{c}$};
\draw[fill=white] (0.3, -0.3) circle (.05cm) node[below=0.5mm] {\scriptsize$\sf{d}$};
\draw[fill=white] (-0.3, 0.3) circle (.05cm) node[above=0.5mm] {\scriptsize$\sf{a}$};
\draw[fill=white] (0.3, 0.3) circle (.05cm) node[above=0.5mm] {\scriptsize$\sf{b}$};
\draw[color=pyblue, fill=pyblue] (0, 0) circle (.05cm);
\draw[pyblue, double] (-0.04, -0.04) -- (-0.26, -0.26);
\draw[black, double] (0.04, -0.04) -- (0.26, -0.26);
\draw[black, double] (-0.04, 0.04) -- (-0.26, 0.26);
\draw[black, double] (0.04, 0.04) -- (0.26, 0.26);
\end{tikzpicture}}}
+
\vcenter{\hbox{
\begin{tikzpicture}[line width=1. pt, scale=2]
\draw[fill=white] (-0.3, -0.3) circle (.05cm) node[below=1.2mm] {\scriptsize$\sf{c}$};
\draw[color=pyblue, fill=white] (0.3, -0.3) circle (.05cm) node[below=0.5mm] {\scriptsize$\sf{d}$};
\draw[fill=white] (-0.3, 0.3) circle (.05cm) node[above=0.5mm] {\scriptsize$\sf{a}$};
\draw[fill=white] (0.3, 0.3) circle (.05cm) node[above=0.5mm] {\scriptsize$\sf{b}$};
\draw[color=pyblue, fill=pyblue] (0, 0) circle (.05cm);
\draw[black, double] (-0.04, -0.04) -- (-0.26, -0.26);
\draw[pyblue, double] (0.04, -0.04) -- (0.26, -0.26);
\draw[black, double] (-0.04, 0.04) -- (-0.26, 0.26);
\draw[black, double] (0.04, 0.04) -- (0.26, 0.26);
\end{tikzpicture}}}\\
&=
\tensor{u}{^{\sf{a}}}{_{\sf{efg}}}
\vcenter{\hbox{
\begin{tikzpicture}[line width=1. pt, scale=2]
\draw[fill=white] (0, 0.4) circle (.05cm) node[above=0.5mm] {\scriptsize$\sf{e}$};
\draw[fill=white] (0, -0.3) circle (.05cm) node[below=1.2mm] {\scriptsize$\sf{b}$};
\draw[fill=white] (0.3, 0.4) circle (.05cm) node[above=0.5mm] {\scriptsize$\sf{f}$};
\draw[fill=white] (0.3, -0.3) circle (.05cm) node[below=1.5mm] {\scriptsize$\sf{c}$};
\draw[fill=white] (0.6, 0.4) circle (.05cm) node[above=0.5mm] {\scriptsize$\sf{g}$};
\draw[fill=white] (0.6, -0.3) circle (.05cm) node[below=1.2mm] {\scriptsize$\sf{d}$};
\draw[black, double] (0, 0.35) -- (0, -0.25);
\draw[black, double] (0.3, 0.35) -- (0.3, -0.25);
\draw[black, double] (0.6, 0.35) -- (0.6, -0.25);
\end{tikzpicture}}}
+\hspace*{0.2cm}
\tensor{u}{^{\sf{b}}}{_{\sf{efg}}}
\vcenter{\hbox{
\begin{tikzpicture}[line width=1. pt, scale=2]
\draw[fill=white] (0, 0.4) circle (.05cm) node[above=0.5mm] {\scriptsize$\sf{e}$};
\draw[fill=white] (0, -0.3) circle (.05cm) node[below=1.5mm] {\scriptsize$\sf{a}$};
\draw[fill=white] (0.3, 0.4) circle (.05cm) node[above=0.5mm] {\scriptsize$\sf{f}$};
\draw[fill=white] (0.3, -0.3) circle (.05cm) node[below=1.5mm] {\scriptsize$\sf{c}$};
\draw[fill=white] (0.6, 0.4) circle (.05cm) node[above=0.5mm] {\scriptsize$\sf{g}$};
\draw[fill=white] (0.6, -0.3) circle (.05cm) node[below=1.2mm] {\scriptsize$\sf{d}$};
\draw[black, double] (0, 0.35) -- (0, -0.25);
\draw[black, double] (0.3, 0.35) -- (0.3, -0.25);
\draw[black, double] (0.6, 0.35) -- (0.6, -0.25);
\end{tikzpicture}}}
+\hspace*{0.2cm}
\tensor{u}{^{\sf{c}}}{_{\sf{efg}}}
\vcenter{\hbox{
\begin{tikzpicture}[line width=1. pt, scale=2]
\draw[fill=white] (0, 0.4) circle (.05cm) node[above=0.5mm] {\scriptsize$\sf{e}$};
\draw[fill=white] (0, -0.3) circle (.05cm) node[below=1.5mm] {\scriptsize$\sf{a}$};
\draw[fill=white] (0.3, 0.4) circle (.05cm) node[above=0.5mm] {\scriptsize$\sf{f}$};
\draw[fill=white] (0.3, -0.3) circle (.05cm) node[below=1.2mm] {\scriptsize$\sf{b}$};
\draw[fill=white] (0.6, 0.4) circle (.05cm) node[above=0.5mm] {\scriptsize$\sf{g}$};
\draw[fill=white] (0.6, -0.3) circle (.05cm) node[below=1.2mm] {\scriptsize$\sf{d}$};
\draw[black, double] (0, 0.35) -- (0, -0.25);
\draw[black, double] (0.3, 0.35) -- (0.3, -0.25);
\draw[black, double] (0.6, 0.35) -- (0.6, -0.25);
\end{tikzpicture}}}
+\hspace*{0.2cm}
\tensor{u}{^{\sf{d}}}{_{\sf{efg}}}
\vcenter{\hbox{
\begin{tikzpicture}[line width=1. pt, scale=2]
\draw[fill=white] (0, 0.4) circle (.05cm) node[above=0.5mm] {\scriptsize$\sf{e}$};
\draw[fill=white] (0, -0.3) circle (.05cm) node[below=1.5mm] {\scriptsize$\sf{a}$};
\draw[fill=white] (0.3, 0.4) circle (.05cm) node[above=0.5mm] {\scriptsize$\sf{f}$};
\draw[fill=white] (0.3, -0.3) circle (.05cm) node[below=1.2mm] {\scriptsize$\sf{b}$};
\draw[fill=white] (0.6, 0.4) circle (.05cm) node[above=0.5mm] {\scriptsize$\sf{g}$};
\draw[fill=white] (0.6, -0.3) circle (.05cm) node[below=1.5mm] {\scriptsize$\sf{c}$};
\draw[black, double] (0, 0.35) -- (0, -0.25);
\draw[black, double] (0.3, 0.35) -- (0.3, -0.25);
\draw[black, double] (0.6, 0.35) -- (0.6, -0.25);
\end{tikzpicture}}}\,.
\end{aligned}
\raisetag{112pt}
\end{gather}
We give a formal derivation of this diagrammatic rule in the following insert.

\begin{framed}
{\small \noindent {\it Quartic bulk vertex.}---From (\ref{eq:4pt_in-in_formal}), the term coming from differentiating the quartic bulk vertex formally reads
\begin{equation}
\vcenter{\hbox{
\begin{tikzpicture}[line width=1. pt, scale=2]
\draw[fill=white] (-0.3, -0.3) circle (.05cm) node[below=1.2mm] {\scriptsize$\sf{c}$};
\draw[fill=white] (0.3, -0.3) circle (.05cm) node[below=0.5mm] {\scriptsize$\sf{d}$};
\draw[fill=white] (-0.3, 0.3) circle (.05cm) node[above=0.5mm] {\scriptsize$\sf{a}$};
\draw[fill=white] (0.3, 0.3) circle (.05cm) node[above=0.5mm] {\scriptsize$\sf{b}$};
\draw[color=pyblue, fill=pyblue] (0, 0) circle (.05cm);
\draw[pyblue, double] (-0.04, -0.04) -- (-0.15, -0.15);
\draw[black, double] (-0.15, -0.15) -- (-0.26, -0.26);
\draw[pyblue, double] (0.04, -0.04) -- (0.15, -0.15);
\draw[black, double] (0.15, -0.15) -- (0.26, -0.26);
\draw[pyblue, double] (-0.04, 0.04) -- (-0.15, 0.15);
\draw[black, double] (-0.15, 0.15) -- (-0.26, 0.26);
\draw[pyblue, double] (0.04, 0.04) -- (0.15, 0.15);
\draw[black, double] (0.15, 0.15) -- (0.26, 0.26);
\end{tikzpicture}}}
= \frac{i}{4!} \langle  H_{\sf{efgh}}\left[X^{\sf{e}}X^{\sf{f}}X^{\sf{g}}X^{\sf{h}}, X^{\sf{a}}X^{\sf{b}}X^{\sf{c}}X^{\sf{d}}\right]\rangle\,.
\end{equation}
After a tedious rearrangement of the commutator, we use the commutation relation $[X^{\sf{a}}, X^{\sf{b}}] = i\epsilon^{\sf{ab}}$. This enables us to introduce the $\tensor{u}{^{\sf{a}}}{_{\sf{bcd}}}$ tensor. These manipulations lead to
\begin{equation}
     \frac{i}{4!} \langle  H_{\sf{efgh}}\left[X^{\sf{e}}X^{\sf{f}}X^{\sf{g}}X^{\sf{h}}, X^{\sf{a}}X^{\sf{b}}X^{\sf{c}}X^{\sf{d}}\right]\rangle = \frac{1}{3!} \tensor{u}{^{\sf{a}}}{_{\sf{efg}}}\langle X^{\sf{b}}X^{\sf{c}}X^{\sf{d}}X^{\sf{e}}X^{\sf{f}}X^{\sf{g}}\rangle + (3\text{ perms})\,.
\end{equation}
Only retaining connected four-point functions (there are $3!$ ways to contract the fields) gives
\begin{equation}
\frac{i}{4!} \langle  H_{\sf{efgh}}\left[X^{\sf{e}}X^{\sf{f}}X^{\sf{g}}X^{\sf{h}}, X^{\sf{a}}X^{\sf{b}}X^{\sf{c}}X^{\sf{d}}\right]\rangle = \tensor{u}{^{\sf{a}}}{_{\sf{efg}}}\langle \bm{X}^{\sf{b}}\bm{X}^{\sf{e}}\rangle \langle \bm{X}^{\sf{c}}\bm{X}^{\sf{f}}\rangle \langle \bm{X}^{\sf{d}}\bm{X}^{\sf{g}}\rangle+ (3\text{ perms})\,,
\end{equation}
which, in the end, is given by the diagrammatic representation in (\ref{eq:diagram_quartic_bulk_vertex}).
 }
\end{framed}

\noindent The case of exchange-like four-point correlators follows the same line. The diagrammatic representation of differentiating an exchange-like four point correlator is
\begin{equation}
\label{eq:diagram_trispectrum_2}
\textcolor{pyblue}{\frac{\mathrm{d}}{\mathrm{d}t}}
\vcenter{\hbox{
\begin{tikzpicture}[line width=1. pt, scale=2]
\draw[fill=white] (-0.3, -0.3) circle (.05cm) node[below=1.2mm] {\scriptsize$\sf{c}$};
\draw[fill=white] (-0.3, 0.3) circle (.05cm) node[above=0.5mm] {\scriptsize$\sf{a}$};
\draw[fill=black] (0, 0) circle (.05cm);
\draw[black, double] (-0.04, -0.04) -- (-0.26, -0.26);
\draw[black, double] (-0.04, 0.04) -- (-0.26, 0.26);
\draw[black, double] (0.05, 0) -- (0.26, 0);
\draw[fill=white] (0.6, -0.3) circle (.05cm) node[below=0.5mm] {\scriptsize$\sf{d}$};
\draw[fill=white] (0.6, 0.3) circle (.05cm) node[above=0.5mm] {\scriptsize$\sf{b}$};
\draw[fill=black] (0.3, 0) circle (.05cm);
\draw[black, double] (0.34, -0.04) -- (0.56, -0.26);
\draw[black, double] (0.34, 0.04) -- (0.56, 0.26);
\end{tikzpicture}}}
=
\vcenter{\hbox{
\begin{tikzpicture}[line width=1. pt, scale=2]
\draw[fill=white] (-0.3, -0.3) circle (.05cm) node[below=1.2mm] {\scriptsize$\sf{c}$};
\draw[color=pyblue, fill=white] (-0.3, 0.3) circle (.05cm) node[above=0.5mm] {\scriptsize$\sf{a}$};
\draw[fill=black] (0, 0) circle (.05cm);
\draw[black, double] (-0.04, -0.04) -- (-0.26, -0.26);
\draw[black, double] (-0.04, 0.04) -- (-0.15, 0.15);
\draw[pyblue, double] (-0.15, 0.15) -- (-0.26, 0.26);
\draw[black, double] (0.05, 0) -- (0.26, 0);
\draw[fill=white] (0.6, -0.3) circle (.05cm) node[below=0.5mm] {\scriptsize$\sf{d}$};
\draw[fill=white] (0.6, 0.3) circle (.05cm) node[above=0.5mm] {\scriptsize$\sf{b}$};
\draw[fill=black] (0.3, 0) circle (.05cm);
\draw[black, double] (0.34, -0.04) -- (0.56, -0.26);
\draw[black, double] (0.34, 0.04) -- (0.56, 0.26);
\end{tikzpicture}}}
\hspace*{0.2cm}+\hspace*{0.2cm} (3\text{ perms}) \hspace*{0.2cm}+
\vcenter{\hbox{
\begin{tikzpicture}[line width=1. pt, scale=2]
\draw[fill=white] (-0.3, -0.3) circle (.05cm) node[below=1.2mm] {\scriptsize$\sf{c}$};
\draw[fill=white] (-0.3, 0.3) circle (.05cm) node[above=0.5mm] {\scriptsize$\sf{a}$};
\draw[color=pyblue, fill=pyblue] (0, 0) circle (.05cm);
\draw[pyblue, double] (-0.04, -0.04) -- (-0.15, -0.15);
\draw[black, double] (-0.15, -0.15) -- (-0.26, -0.26);
\draw[pyblue, double] (-0.04, 0.04) -- (-0.15, 0.15);
\draw[black, double] (-0.15, 0.15) -- (-0.26, 0.26);
\draw[pyblue, double] (0.05, 0) -- (0.26, 0);
\draw[fill=white] (0.6, -0.3) circle (.05cm) node[below=0.5mm] {\scriptsize$\sf{d}$};
\draw[fill=white] (0.6, 0.3) circle (.05cm) node[above=0.5mm] {\scriptsize$\sf{b}$};
\draw[color=pyblue, fill=pyblue] (0.3, 0) circle (.05cm);
\draw[pyblue, double] (0.34, -0.04) -- (0.45, -0.15);
\draw[black, double] (0.45, -0.15) -- (0.56, -0.26);
\draw[pyblue, double] (0.34, 0.04) -- (0.45, 0.15);
\draw[black, double] (0.45, 0.15) -- (0.56, 0.26);
\end{tikzpicture}}}\,.
\end{equation}
Differentiating with respect to external operator insertions introduces the usual $\tensor{u}{^{\sf{a}}}{_{\sf{b}}}$ tensor. Differentiating the bulk vertices is given by contracting a three-point correlator with a two-point correlator using the tensor $\tensor{u}{^{\sf{a}}}{_{\sf{bc}}}$ in the following way
\begin{equation}
\label{eq:diagram_trispectrum_3}
\begin{aligned}
\vcenter{\hbox{
\begin{tikzpicture}[line width=1. pt, scale=2]
\draw[fill=white] (-0.3, -0.3) circle (.05cm) node[below=1.2mm] {\scriptsize$\sf{c}$};
\draw[fill=white] (-0.3, 0.3) circle (.05cm) node[above=0.5mm] {\scriptsize$\sf{a}$};
\draw[color=pyblue, fill=pyblue] (0, 0) circle (.05cm);
\draw[pyblue, double] (-0.04, -0.04) -- (-0.15, -0.15);
\draw[black, double] (-0.15, -0.15) -- (-0.26, -0.26);
\draw[pyblue, double] (-0.04, 0.04) -- (-0.15, 0.15);
\draw[black, double] (-0.15, 0.15) -- (-0.26, 0.26);
\draw[pyblue, double] (0.05, 0) -- (0.26, 0);
\draw[fill=white] (0.6, -0.3) circle (.05cm) node[below=0.5mm] {\scriptsize$\sf{b}$};
\draw[fill=white] (0.6, 0.3) circle (.05cm) node[above=0.5mm] {\scriptsize$\sf{d}$};
\draw[color=pyblue, fill=pyblue] (0.3, 0) circle (.05cm);
\draw[pyblue, double] (0.34, -0.04) -- (0.45, -0.15);
\draw[black, double] (0.45, -0.15) -- (0.56, -0.26);
\draw[pyblue, double] (0.34, 0.04) -- (0.45, 0.15);
\draw[black, double] (0.45, 0.15) -- (0.56, 0.26);
\end{tikzpicture}}}
&=
\vcenter{\hbox{
\begin{tikzpicture}[line width=1. pt, scale=2]
\draw[fill=white] (-0.3, -0.3) circle (.05cm) node[below=1.2mm] {\scriptsize$\sf{c}$};
\draw[color=pyblue, fill=white] (-0.3, 0.3) circle (.05cm) node[above=0.5mm] {\scriptsize$\sf{a}$};
\draw[color=pyblue, fill=pyblue] (0, 0) circle (.05cm);
\draw[black, double] (-0.04, -0.04) -- (-0.26, -0.26);
\draw[pyblue, double] (-0.04, 0.04) -- (-0.26, 0.26);
\draw[black, double] (0.05, 0) -- (0.26, 0);
\draw[fill=white] (0.6, -0.3) circle (.05cm) node[below=0.5mm] {\scriptsize$\sf{d}$};
\draw[fill=white] (0.6, 0.3) circle (.05cm) node[above=0.5mm] {\scriptsize$\sf{b}$};
\draw[fill=black] (0.3, 0) circle (.05cm);
\draw[black, double] (0.34, -0.04) -- (0.56, -0.26);
\draw[black, double] (0.34, 0.04) -- (0.56, 0.26);
\end{tikzpicture}}}
+
\vcenter{\hbox{
\begin{tikzpicture}[line width=1. pt, scale=2]
\draw[fill=white] (-0.3, -0.3) circle (.05cm) node[below=1.2mm] {\scriptsize$\sf{c}$};
\draw[fill=white] (-0.3, 0.3) circle (.05cm) node[above=0.5mm] {\scriptsize$\sf{a}$};
\draw[fill=black] (0, 0) circle (.05cm);
\draw[black, double] (-0.04, -0.04) -- (-0.26, -0.26);
\draw[black, double] (-0.04, 0.04) -- (-0.26, 0.26);
\draw[black, double] (0.05, 0) -- (0.26, 0);
\draw[fill=white] (0.6, -0.3) circle (.05cm) node[below=0.5mm] {\scriptsize$\sf{d}$};
\draw[color=pyblue, fill=white] (0.6, 0.3) circle (.05cm) node[above=0.5mm] {\scriptsize$\sf{b}$};
\draw[color=pyblue, fill=pyblue] (0.3, 0) circle (.05cm);
\draw[black, double] (0.34, -0.04) -- (0.56, -0.26);
\draw[pyblue, double] (0.34, 0.04) -- (0.56, 0.26);
\end{tikzpicture}}}
+
\vcenter{\hbox{
\begin{tikzpicture}[line width=1. pt, scale=2]
\draw[color=pyblue, fill=white] (-0.3, -0.3) circle (.05cm) node[below=1.2mm] {\scriptsize$\sf{c}$};
\draw[fill=white] (-0.3, 0.3) circle (.05cm) node[above=0.5mm] {\scriptsize$\sf{a}$};
\draw[color=pyblue, fill=pyblue] (0, 0) circle (.05cm);
\draw[pyblue, double] (-0.04, -0.04) -- (-0.26, -0.26);
\draw[black, double] (-0.04, 0.04) -- (-0.26, 0.26);
\draw[black, double] (0.05, 0) -- (0.26, 0);
\draw[fill=white] (0.6, -0.3) circle (.05cm) node[below=0.5mm] {\scriptsize$\sf{d}$};
\draw[fill=white] (0.6, 0.3) circle (.05cm) node[above=0.5mm] {\scriptsize$\sf{b}$};
\draw[fill=black] (0.3, 0) circle (.05cm);
\draw[black, double] (0.34, -0.04) -- (0.56, -0.26);
\draw[black, double] (0.34, 0.04) -- (0.56, 0.26);
\end{tikzpicture}}}
+
\vcenter{\hbox{
\begin{tikzpicture}[line width=1. pt, scale=2]
\draw[fill=white] (-0.3, -0.3) circle (.05cm) node[below=1.2mm] {\scriptsize$\sf{c}$};
\draw[fill=white] (-0.3, 0.3) circle (.05cm) node[above=0.5mm] {\scriptsize$\sf{a}$};
\draw[fill=black] (0, 0) circle (.05cm);
\draw[black, double] (-0.04, -0.04) -- (-0.26, -0.26);
\draw[black, double] (-0.04, 0.04) -- (-0.26, 0.26);
\draw[black, double] (0.05, 0) -- (0.26, 0);
\draw[color=pyblue, fill=white] (0.6, -0.3) circle (.05cm) node[below=0.5mm] {\scriptsize$\sf{d}$};
\draw[fill=white] (0.6, 0.3) circle (.05cm) node[above=0.5mm] {\scriptsize$\sf{b}$};
\draw[color=pyblue, fill=pyblue] (0.3, 0) circle (.05cm);
\draw[pyblue, double] (0.34, -0.04) -- (0.56, -0.26);
\draw[black, double] (0.34, 0.04) -- (0.56, 0.26);
\end{tikzpicture}}}\\
&=\left(\tensor{u}{^{\sf{a}}}{_{\sf{ef}}}
\vcenter{\hbox{
\begin{tikzpicture}[line width=1. pt, scale=2]
\draw[fill=white] (-0.3, 0.3) circle (.05cm) node[above=0.5mm] {\scriptsize$\sf{e}$};
\draw[fill=white] (-0.3, -0.3) circle (.05cm) node[below=0.5mm] {\scriptsize$\sf{c}$};
\draw[black, double] (-0.3, -0.25) -- (-0.3, 0.25);
\draw[fill=white] (0, 0) circle (.05cm) node[below=0.5mm] {\scriptsize$\sf{f}$};
\draw[fill=black] (0.3, 0) circle (.05cm);
\draw[fill=white] (0.6, 0.3) circle (.05cm) node[above=0.5mm] {\scriptsize$\sf{b}$};
\draw[fill=white] (0.6, -0.3) circle (.05cm) node[below=0.5mm] {\scriptsize$\sf{d}$};
\draw[black, double] (0.05, 0) -- (0.25, 0);
\draw[black, double] (0.34, 0.04) -- (0.56, 0.26);
\draw[black, double] (0.34, -0.04) -- (0.56, -0.26);
\end{tikzpicture}}}
+ \hspace*{0.2cm} (2\text{ perms})\right)\hspace*{0.2cm}  + \hspace*{0.2cm} (3\text{ perms})\,,
\end{aligned}
\end{equation}
where the $2$ permutations are the number of ways to choose a pair of indices where one is summed over and the other is not summed over (exchanging the two dummy indices is already taken into account in the counting), and the $3$ permutations are the usual cyclic permutations $(\sf{a}\rightarrow\sf{b}\rightarrow\sf{c}\rightarrow\sf{d})$. The second line then contains $12$ terms. This diagrammatic rule is proven in the following insert.

\begin{framed}
{\small \noindent {\it Cubic exchange vertex.}---From (\ref{eq:4pt_in-in_formal}), the term coming from deriving the cubic exchange vertex reads
\begin{equation}
\vcenter{\hbox{
\begin{tikzpicture}[line width=1. pt, scale=2]
\draw[fill=white] (-0.3, -0.3) circle (.05cm) node[below=1.2mm] {\scriptsize$\sf{c}$};
\draw[fill=white] (-0.3, 0.3) circle (.05cm) node[above=0.5mm] {\scriptsize$\sf{a}$};
\draw[color=pyblue, fill=pyblue] (0, 0) circle (.05cm);
\draw[pyblue, double] (-0.04, -0.04) -- (-0.15, -0.15);
\draw[black, double] (-0.15, -0.15) -- (-0.26, -0.26);
\draw[pyblue, double] (-0.04, 0.04) -- (-0.15, 0.15);
\draw[black, double] (-0.15, 0.15) -- (-0.26, 0.26);
\draw[pyblue, double] (0.05, 0) -- (0.26, 0);
\draw[fill=white] (0.6, -0.3) circle (.05cm) node[below=0.5mm] {\scriptsize$\sf{d}$};
\draw[fill=white] (0.6, 0.3) circle (.05cm) node[above=0.5mm] {\scriptsize$\sf{b}$};
\draw[color=pyblue, fill=pyblue] (0.3, 0) circle (.05cm);
\draw[pyblue, double] (0.34, -0.04) -- (0.45, -0.15);
\draw[black, double] (0.45, -0.15) -- (0.56, -0.26);
\draw[pyblue, double] (0.34, 0.04) -- (0.45, 0.15);
\draw[black, double] (0.45, 0.15) -- (0.56, 0.26);
\end{tikzpicture}}}
=
- \frac{1}{(3!)^2}\langle  \int_{-\infty}^{t}\mathrm{d}t'H_{\sf{efg}}(t')H_{\sf{hij}}\left[X^{\sf{e}}X^{\sf{f}}X^{\sf{g}}(t'),\left[X^{\sf{h}}X^{\sf{i}}X^{\sf{j}}, X^{\sf{a}}X^{\sf{b}}X^{\sf{c}}X^{\sf{d}}\right]\right]\rangle\,.
\end{equation}
We first start by examining the inner commutator where the operators $X^{\sf{a}}$ are evaluated at $t$. After rearranging it and introducing the tensor $\tensor{u}{^{\sf{a}}}{_{\sf{bc}}}$, we find
\begin{equation}
    -\frac{1}{3!}H_{\sf{hij}}\left[X^{\sf{h}}X^{\sf{i}}X^{\sf{j}}, X^{\sf{a}}X^{\sf{b}}X^{\sf{c}}X^{\sf{d}}\right] = \frac{i}{2!}\tensor{u}{^{\sf{a}}}{_{\sf{hi}}}X^{\sf{h}}X^{\sf{i}}X^{\sf{b}}X^{\sf{c}}X^{\sf{d}} + (3\text{ perms})\,.
\end{equation}
Inside the expectation value, we now Wick contract a pair of external operators where one is the operator whose index is summed over. Among the six such possibilities, half of them lead to the same contribution after relabelling the indices. This cancels the factor $2!$. In the end, after identifying the three-point correlators, we end up with 
\begin{equation}
\begin{aligned}
- \frac{1}{(3!)^2}\langle  \int_{-\infty}^{t}\mathrm{d}t'H_{\sf{efg}}(t')H_{\sf{hij}}&\left[X^{\sf{e}}X^{\sf{f}}X^{\sf{g}}(t'),\left[X^{\sf{h}}X^{\sf{i}}X^{\sf{j}}, X^{\sf{a}}X^{\sf{b}}X^{\sf{c}}X^{\sf{d}}\right]\right]\rangle = \\ &\tensor{u}{^{\sf{a}}}{_{\sf{ef}}}\langle \bm{X}^{\sf{c}}\bm{X}^{\sf{e}}\rangle \langle \bm{X}^{\sf{b}}\bm{X}^{\sf{d}}\bm{X}^{\sf{f}}\rangle + (2\text{ perms}) + (3\text{ perms})\,,
\end{aligned}
\end{equation}
where the $2$ permutations account for choosing a pair among $(\sf{e, b, c, d})$ with one index being $\sf{e}$, and the $3$ permutations are the usual cyclic permutations $(\sf{a}\rightarrow\sf{b}\rightarrow\sf{c}\rightarrow\sf{d})$.
 }
\end{framed}

\noindent Collecting the various terms in Eqs.~(\ref{eq:diagram_trispectrum_1}, \ref{eq:diagram_quartic_bulk_vertex}, \ref{eq:diagram_trispectrum_2}, \ref{eq:diagram_trispectrum_3}), the flow equations for the connected four-point correlator are
\begin{gather}
\begin{aligned}
\label{eq: flow 4-pt}
    \frac{\mathrm{d}}{\mathrm{d}t} \langle\bm{X}^{\sf{a}} \bm{X}^{\sf{b}}\bm{X}^{\sf{c}}\bm{X}^{\sf{d}} \rangle &= \tensor{u}{^{\sf{a}}}{_{\sf{e}}}\langle\bm{X}^{\sf{e}} \bm{X}^{\sf{b}}\bm{X}^{\sf{c}}\bm{X}^{\sf{d}} \rangle + \tensor{u}{^{\sf{a}}}{_{\sf{efg}}} \langle\bm{X}^{\sf{b}} \bm{X}^{\sf{e}}\rangle \langle\bm{X}^{\sf{c}} \bm{X}^{\sf{f}}\rangle \langle\bm{X}^{\sf{d}} \bm{X}^{\sf{g}}\rangle + (3\text{ perms})\\ &+\tensor{u}{^{\sf{a}}}{_{\sf{ef}}}\langle\bm{X}^{\sf{b}} \bm{X}^{\sf{e}}\rangle \langle\bm{X}^{\sf{c}} \bm{X}^{\sf{d}} \bm{X}^{\sf{f}}\rangle + (11\text{ perms})\,.
\end{aligned}
\raisetag{18pt}
\end{gather}
Similar equations were found in \cite{Anderson:2012em} for super-horizon interactions. Here they are derived in a generic context for the first time. It is interesting to note that contact-like four-point diagrams are only sourced by products of two-point functions. Removing the last term on the right-hand side---that come from deriving the bulk vertices of an exchange-like diagram---, the above equations form a closed system along with the flow equations for the two-point correlators only. Contrariwise, exchange-like four-point diagrams are sourced by three-point correlators, reflecting their intrinsic complexity. One would then need to solve all flow equations for the two-, three-, and four-point correlators, which still form a closed triangular system.

\paragraph{Diagrammatic rules for $n$-point correlators.} Having treated the simplest exchange-like diagram appearing in four-point correlators, we now state---without formal proof---general diagrammatic rules to derive the flow equations for any tree-level $n$-point correlators. These rules are the following:

\begin{itemize}
    \item Draw a tree-level graph assigning an operator $\bm{X}^{\sf{a}}(t)$ to each of the $n$ external field, denoted with a white dot. Bulk vertices, denoted with black dots, should be at least of cubic order because quadratic interactions are fully resummed. Hence, the external operators are connected to bulk vertices with doubled lines. Time dependence appears in external operators and in the upper limit of the integral over bulk vertices, considered as a whole.
    \item Following the Leibniz product rule of differentiation, taking the time derivative of the drawn diagram equals the sum of initial-like diagrams where time dependence has been isolated, in our notations coloured in blue.
    \item Differentiating an external operator $\bm{X}^{\sf{a}}$ introduces the tensor $\tensor{u}{^{\sf{a}}}{_{\sf{b}}}$ which contracts -- $\sf{b}$ being a dummy index here -- with the external operator
\begin{equation}
    \vcenter{\hbox{
\begin{tikzpicture}[line width=1. pt, scale=2]
\draw[color=pyblue, fill=white] (0, 0) circle (.05cm) node[left=0.5mm] {\scriptsize$\sf{a}$};
\draw[pyblue, double] (0.05, 0) -- (0.2, 0);
\draw[black, double] (0.2, 0) -- (0.4, 0);
\draw[fill=black] (0.4, 0) circle (.05cm);
\draw[black, double] (0.45, 0) -- (0.6, 0)node[right=0.1mm] {..};
\draw[black, double] (0.4, 0.05) -- (0.4, 0.2) node[above=0.1mm] {:};
\draw[black, double] (0.4, -0.05) -- (0.4, -0.2) node[below=0.1mm] {:};
\end{tikzpicture}}}
=
\tensor{u}{^{\sf{a}}}{_{\sf{b}}}
\vcenter{\hbox{
\begin{tikzpicture}[line width=1. pt, scale=2]
\draw[fill=white] (0, 0) circle (.05cm) node[left=0.5mm] {\scriptsize$\sf{b}$};
\draw[black, double] (0.05, 0) -- (0.4, 0);
\draw[fill=black] (0.4, 0) circle (.05cm);
\draw[black, double] (0.45, 0) -- (0.6, 0)node[right=0.1mm] {..};
\draw[black, double] (0.4, 0.05) -- (0.4, 0.2) node[above=0.1mm] {:};
\draw[black, double] (0.4, -0.05) -- (0.4, -0.2) node[below=0.1mm] {:};
\end{tikzpicture}}}\,.
\end{equation}

    \item Differentiating a contact-like bulk vertex of order $n$---hence exclusively connected to $n$ external operators---leads to the product of $n-1$ two-point correlators contracted with the tensor $\tensor{u}{^{\sf{a}}}{_{\sf{b_1b_2\dots b_{n-1}}}}$, where $\sf{b_1, b_2, \ldots, b_{n-1}}$ are dummy indices 
\begin{equation}
    \vcenter{\hbox{
\begin{tikzpicture}[line width=1. pt, scale=2]
\draw[color=pyblue, fill=white] (0, 0) circle (.05cm) node[left=0.5mm] {\scriptsize$\sf{a}$};
\draw[pyblue, double] (0.05, 0) -- (0.4, 0);
\draw[color=pyblue, fill=pyblue] (0.4, 0) circle (.05cm);
\draw[black, double] (0.45, 0) -- (0.6, 0);
\draw[black, double] (0.4, 0.05) -- (0.4, 0.2);
\draw[black, double] (0.4, -0.05) -- (0.4, -0.2);
\draw[fill=white] (0.65, 0) circle (.05cm);
\draw[fill=white] (0.4, 0.25) circle (.05cm);
\draw[fill=white] (0.4, -0.25) circle (.05cm);
\end{tikzpicture}}}
=
\tensor{u}{^{\sf{a}}}{_{\sf{bcd}}}
\vcenter{\hbox{
\begin{tikzpicture}[line width=1. pt, scale=2]
\draw[fill=white] (0, 0.2) circle (.05cm) node[above=0.5mm] {\scriptsize$\sf{b}$};
\draw[fill=white] (0, -0.2) circle (.05cm);
node[below=0.5mm] { };
\draw[fill=white] (0.2, 0.2) circle (.05cm) node[above=0.5mm] {\scriptsize$\sf{c}$};
\draw[fill=white] (0.2, -0.2) circle (.05cm)node[below=0.5mm] { };
\draw[fill=white] (0.4, 0.2) circle (.05cm) node[above=0.5mm] {\scriptsize$\sf{d}$};
\draw[fill=white] (0.4, -0.2) circle (.05cm)node[below=0.5mm] { };
\draw[black, double] (0, -0.15) -- (0, 0.15);
\draw[black, double] (0.2, -0.15) -- (0.2, 0.15);
\draw[black, double] (0.4, -0.15) -- (0.4, 0.15);
\end{tikzpicture}}}\,.
\end{equation}
    
    \item Differentiating an exchange-like bulk vertex of order $m$, attached to an external field, results in cutting the diagram in $m-1$ subdiagrams contracted with the tensor $\tensor{u}{^{\sf{a}}}{_{\sf{b_1b_2\ldots b_{m-1}}}}$. One should consider all possible contractions. An example is
\begin{equation}
    \vcenter{\hbox{
\vspace*{-0.1cm}
\begin{tikzpicture}[line width=1. pt, scale=2]
\draw[fill=white] (0.1, 0) circle (.05cm);
\draw[black, double] (0.15, 0) -- (0.4, 0);
\draw[fill=black] (0.4, 0) circle (.05cm);
\draw[black, double] (0.45, 0) -- (0.65, 0);
\draw[black, double] (0.4, 0.05) -- (0.4, 0.2);
\draw[black, double] (0.4, -0.05) -- (0.4, -0.2);
\draw[fill=white] (0.4, 0.25) circle (.05cm);
\draw[fill=white] (0.4, -0.25) circle (.05cm);
\draw[color=pyblue, fill=pyblue] (0.7, 0) circle (.05cm);
\draw[color=pyblue, fill=white] (0.7, 0.25) circle (.05cm)node[right=0.5mm] {\scriptsize$\sf{a}$};
\draw[fill=white] (0.7, -0.25) circle (.05cm)node[below=0.5mm] { };
\draw[black, double] (0.7, -0.05) -- (0.7, -0.2);
\draw[pyblue, double] (0.7, 0.05) -- (0.7, 0.2);
\draw[black, double] (0.75, 0) -- (0.95, 0);
\draw[fill=white] (1, 0) circle (.05cm);
\end{tikzpicture}}}
=
\tensor{u}{^{\sf{a}}}{_{\sf{bcd}}}
\vcenter{\hbox{
\vspace*{0.4cm}
\begin{tikzpicture}[line width=1. pt, scale=2]
\draw[fill=white] (0.1, 0) circle (.05cm);
\draw[black, double] (0.15, 0) -- (0.4, 0);
\draw[fill=black] (0.4, 0) circle (.05cm);
\draw[black, double] (0.45, 0) -- (0.65, 0);
\draw[black, double] (0.4, 0.05) -- (0.4, 0.2);
\draw[black, double] (0.4, -0.05) -- (0.4, -0.2);
\draw[fill=white] (0.4, 0.25) circle (.05cm);
\draw[fill=white] (0.4, -0.25) circle (.05cm);
\draw[fill=white] (0.7, 0) circle (.05cm)node[right=0.5mm] {\scriptsize$\sf{b}$};
\draw[fill=white] (1, 0.25) circle (.05cm)node[above=0.5mm] {\scriptsize$\sf{c}$};
\draw[fill=white] (1, -0.25) circle (.05cm);
\draw[black, double] (1, 0.2) -- (1, -0.2);
\draw[fill=white] (1.2, 0.25) circle (.05cm)node[above=0.5mm] {\scriptsize$\sf{d}$};
\draw[fill=white] (1.2, -0.25) circle (.05cm);
\draw[black, double] (1.2, 0.2) -- (1.2, -0.2);
\end{tikzpicture}}}
\hspace*{0.2cm}+
\hspace*{0.2cm}(19\text{ perms})\,.
\end{equation}
    
\end{itemize}

\noindent In general, the flow equations for an exchange-like diagram, together with those for the two-point correlators, do not form a closed system. One then needs to also find the flow equations for lower-order diagrams in order to close the system. 

\vskip 4pt
Our aim is to solve the flow equations (\ref{eq:flow_equation_2pt}), (\ref{eq:flow_equation_3pt}), (\ref{eq: flow 4-pt}), etc. 
Yet, two ingredients are missing: (i) initial conditions for the correlators, and (ii) explicit forms for the tensors $\tensor{u}{^{\sf{a}}}{_{\sf{b}}}$, $\tensor{u}{^{\sf{a}}}{_{\sf{bc}}}, \ldots\,$. In the rest of the paper, we will deal with these points and apply the formalism to study inflationary two- and three-point correlators, after a brief digression that we dedicate to comments on the role of quadratic interactions in this approach.

\subsection{Resumming Quadratic Mixings}
\label{subsec:resumming_quadratic_mixings}

The flow equations previously derived encode an exact treatment of possible quadratic interactions appearing in $H_{\sf{ab}}\bm{X}^{\sf{a}}\bm{X}^{\sf{b}}$. We argue in this section that this is equivalent to dressing the two-point correlators with an infinite number of quadratic mixing insertions, when using an interaction scheme where quadratic mixings are treated perturbatively.

\vskip 4pt
Instead of choosing the ``free" Hamiltonian to be the full quadratic one, a commonly used choice of interaction scheme is to split the quadratic Hamiltonian into two contributions
\begin{equation}
    \frac{1}{2!}H_{\sf{ab}}\bm{X}^{\sf{a}}\bm{X}^{\sf{b}} = \frac{1}{2!}H^{\text{diag}}_{\sf{ab}}\bm{X}^{\sf{a}}\bm{X}^{\sf{b}} + \frac{1}{2!}H^{\text{mix}}_{\sf{ab}}\bm{X}^{\sf{a}}\bm{X}^{\sf{b}}\,,
\end{equation}
with $H^{\text{diag}}_{\sf{ab}}$ being diagonal. One can then define new interaction-picture operators $\bar{X}^{\sf{a}}$ that evolve with $H^{\text{diag}}_{\sf{ab}}$ and derive approximate analytical forms for them. Assuming that quadratic mixings can be treated perturbatively, one can derive analytical expressions for correlators in this regime. The \textit{fully dressed} two-point correlator can be formally written as an infinite sum of nested integrals
\begin{equation}
\begin{aligned}
\langle\bm{X}^{\sf{a}} \bm{X}^{\sf{b}} \rangle &= 
\sum_{n=0}^{\infty} \frac{i^n}{(2!)^n}\int_{-\infty}^t\mathrm{d}t' \int_{-\infty}^{t'}\mathrm{d}t'' \ldots \int_{-\infty}^{t^{(n-1)}}\mathrm{d}t^{(n)}\\
&\hspace*{2cm}\left\langle\left[H^{\text{mix}}_{\sf{cd}}\bar{X}^{\sf{c}}\bar{X}^{\sf{d}},\left[H^{\text{mix}}_{\sf{ef}}\bar{X}^{\sf{e}}\bar{X}^{\sf{f}}, \ldots, \left[H^{\text{mix}}_{\sf{gh}}\bar{X}^{\sf{g}}\bar{X}^{\sf{h}},\bar{X}^{\sf{a}}\bar{X}^{\sf{b}} \right]\ldots\right]\right]\right\rangle\,, \\
\raisebox{0pt}{
\begin{tikzpicture}[line width=1. pt, scale=2]
\draw[fill=white] (0.3, 0) circle (.05cm) node[above=0.5mm] {\scriptsize$\sf{a}$};
\draw[black, double] (0.35, 0) -- (1, 0);
\draw[fill=white] (1, 0) circle (.05cm) node[above=0.5mm] {\scriptsize$\sf{b}$};
\end{tikzpicture} 
}
&= 
\raisebox{0pt}{
\begin{tikzpicture}[line width=1. pt, scale=2]
\draw[fill=white] (0, 0) circle (.05cm) node[above=0.5mm] {\scriptsize$\sf{a}$};
\draw[black] (0.05, 0) -- (0.5, 0);
\draw[fill=white] (0.5, 0) circle (.05cm) node[above=0.5mm] {\scriptsize$\sf{b}$};
\end{tikzpicture} 
+
\begin{tikzpicture}[line width=1. pt, scale=2]
\draw[fill=white] (0, 0) circle (.05cm) node[above=0.5mm] {\scriptsize$\sf{a}$};
\draw[black] (0.05, 0) -- (0.5, 0);
\draw[fill=black] (0.5, 0) circle (.05cm);
\draw[black] (0.5, 0) -- (1, 0);
\draw[fill=white] (1, 0) circle (.05cm) node[above=0.5mm] {\scriptsize$\sf{b}$};
\end{tikzpicture}
+ 
\hspace*{0.1cm}\ldots\hspace*{0.1cm} 
+ 
\begin{tikzpicture}[line width=1. pt, scale=2]
\draw[fill=white] (0, 0) circle (.05cm) node[above=0.5mm] {\scriptsize$\sf{a}$};
\draw[black] (0.05, 0) -- (0.5, 0);
\draw[fill=black] (0.5, 0) circle (.05cm);
\draw[black] (0.5, 0) -- (0.7, 0);
\draw[black, dotted] (0.7, 0) -- (1, 0);
\draw[black] (1, 0) -- (1.2, 0);
\draw[fill=black] (1.2, 0) circle (.05cm);
\draw[black] (1.2, 0) -- (1.7, 0);
\draw[fill=white] (1.7, 0) circle (.05cm) node[above=0.5mm] {\scriptsize$\sf{b}$};
\end{tikzpicture} 
}\,,
\end{aligned}
\end{equation}
where black dots denote insertions of quadratic mixings and single lines stand for propagators of the fields $\bar{X}^{\sf{a}}$ in this interaction picture. In contrast, the formalism developed in this work intrinsically contains such resummation. We have converted the problem of computing complicated nested integrals to solving a set of coupled ordinary differential equations.

\section{Cosmological Flow}
\label{sec:cosmoflow_formalism}

We will now concretely apply the cosmological flow formalism to compute the two- and three-point correlation functions in generic theories of inflationary fluctuations.

\subsection{Landscape of Theories}

We embrace the point of view of the EFT of inflationary fluctuations, without relying on a specific background mechanism. Fluctuations are the fundamental degrees of freedom and the background determines the time-dependent parameters in the Lagrangian.

\paragraph{Action.} Up to cubic order, the most general action in Fourier space for the fluctuations is
\begin{equation}
\label{eq:action}
\begin{aligned}
    S = \frac{1}{2} \int \mathrm{d}t\, a^3 &\left(\bar{\Delta}_{\upalpha\upbeta}\dot{\bm{\varphi}}^{\upalpha}\dot{\bm{\varphi}}^{\upbeta} + \bar{M}_{\upalpha\upbeta}\bm{\varphi}^{\upalpha}\bm{\varphi}^{\upbeta} + 2\bar{I}_{\upalpha\upbeta}\bm{\varphi}^{\upalpha}\dot{\bm{\varphi}}^{\upbeta} \right.\\
    &\left. + \bar{A}_{\upalpha\upbeta\upgamma}\bm{\varphi}^{\upalpha}\bm{\varphi}^{\upbeta}\bm{\varphi}^{\upgamma} + \bar{B}_{\upalpha\upbeta\upgamma}\bm{\varphi}^{\upalpha}\bm{\varphi}^{\upbeta}\dot{\bm{\varphi}}^{\upgamma} +
    \bar{C}_{\upalpha\upbeta\upgamma}\dot{\bm{\varphi}}^{\upalpha}\dot{\bm{\varphi}}^{\upbeta}\bm{\varphi}^{\upgamma}+
    \bar{D}_{\upalpha\upbeta\upgamma}\dot{\bm{\varphi}}^{\upalpha}\dot{\bm{\varphi}}^{\upbeta}\dot{\bm{\varphi}}^{\upgamma}  \right)\,,
\end{aligned}
\end{equation}
where $a$ is the scale factor whose time evolution is arbitrary.\footnote{The formalism presented in this paper is valid in all FLRW spacetimes, including de Sitter, slow-roll inflation and all power-law cosmologies.}
We have used the extended Fourier summation convention introduced in Section \ref{subsec:cosmo_correlators} and recalled in Appendix \ref{app:Fourier_summation}, meaning that repeated \textsf{sans serif} indices $\upalpha, \upbeta$---running from $1$ to the number of fields $N$---denote a sum that includes an integral over Fourier modes. All the introduced tensors carry hidden arbitrary momentum and time dependencies, as we will see later.
The tensor $\bar{\Delta}_{\upalpha\upbeta}$ is real and symmetric and therefore can be considered diagonal without loss of generality.\footnote{We also consider that $\bar{\Delta}_{\upalpha\upbeta}$ is positive definite to avoid ghosts that would render the theory unstable.}
As such, this tensor only accounts for non-canonically normalised fields.
The tensor $\bar{M}_{\upalpha\upbeta}$ is also taken to be symmetric $\bar{M}_{\upalpha\upbeta}=\bar{M}_{(\upalpha\upbeta)}$. It contains the kinetic gradient terms and the mass matrix. The non-symmetric $\bar{I}_{\upalpha\upbeta}$ tensor captures the remaining possible linear mixings among the fields. For the cubic part of the action, the tensors $\bar{A}_{\upalpha\upbeta\upgamma} = \bar{A}_{(\upalpha\upbeta\upgamma)}$ and $\bar{D}_{\upalpha\upbeta\upgamma} = \bar{D}_{(\upalpha\upbeta\upgamma)}$ should be symmetrised over all indices, and $\bar{B}_{\upalpha\upbeta\upgamma} = \bar{B}_{(\upalpha\upbeta)\upgamma}$ and $\bar{C}_{\upalpha\upbeta\upgamma} = \bar{C}_{(\upalpha\upbeta)\upgamma}$ over the first two indices, with corresponding exchange of momenta. No further assumptions than the ones stated above are made.\footnote{Note that the cubic interaction parameterised by the tensor $\bar{D}_{\upalpha\upbeta\upgamma}$ has not been considered in \cite{Dias:2016rjq} as it does not appear in the action for the fluctuations in non-linear sigma models. It does appear in generalised $P(X^{IJ},\phi^K)$ models though, where $X^{IJ}=-\tfrac{1}{2}g^{\mu\nu}\partial_\mu \phi^I \partial_\nu \phi^J$ are mixed kinetic terms~\cite{Seery:2005wm, Chen:2006nt, Langlois:2008qf}, and is also present in the effective field theory of inflationary fluctuations~\cite{Creminelli:2006xe, Cheung:2007st, Senatore:2010wk, Noumi:2012vr}.
The generic approach that we are adopting therefore encompasses all these classes of models directly at the level of fluctuations.}

\paragraph{Conjugate momenta.} Given the action (\ref{eq:action}), the conjugate momenta to the fields are defined by the following functional derivative
\begin{equation}
    \bm{p}_{\upalpha}(t) = \frac{\delta S}{\delta \dot{\bm{\varphi}}^{\upalpha}(t)}\,, \hspace*{0.5cm} \text{with} \hspace*{0.5cm} \frac{\delta \bm{\varphi}^{\upalpha}(t)}{\delta \bm{\varphi}^{\upbeta}(t')} = \delta^{\upalpha}_{\upbeta}\, \delta(t - t')\,.
\end{equation}
At cubic order, it is enough to restrict ourselves to linear order to derive the conjugate momenta and use $\mathcal{H}^{(3)} = - \mathcal{L}^{(3)}$, where $\dot{\bm{\varphi}}^\upalpha$ should be expressed in terms of the linear momenta. Indeed, higher-order terms would give quartic terms in the Hamiltonian density, not relevant to compute three-point correlators.
However, for the sake of completeness, we derive the conjugate momenta at second order (which would be necessary, e.g., for finding the quartic Hamiltonian),
\begin{equation}
\label{eq:conjugate_momentum}
\bm{p}_{\upalpha} = a^3\left(\bar{\Delta}_{\bar{\upalpha}\upbeta}\dot{\bm{\varphi}}^{\upbeta} + \bar{I}_{\bar{\upalpha}\upbeta}\bm{\varphi}^{\upbeta} + \frac{1}{2}\bar{B}_{\upbeta\upgamma\bar{\upalpha}}\bm{\varphi}^{\upbeta}\bm{\varphi}^{\upgamma} + \bar{C}_{\bar{\upalpha}\upbeta\upgamma}\dot{\bm{\varphi}}^{\upbeta}\bm{\varphi}^{\upgamma} + \frac{3}{2}\bar{D}_{\bar{\upalpha}\upbeta\upgamma} \dot{\bm{\varphi}}^{\upbeta}\dot{\bm{\varphi}}^{\upgamma} \right)\,,
\end{equation}
where bar indices indicate that the sign of the corresponding momentum $\bm{k}$ has been reversed i.e. $T^{\bar{\upalpha}} = T^{\upalpha}(-\bm{k}_{\alpha})$ where $T$ can be a tensor or a field. This follows from the fact that Fourier-space expressions can produce the delta function $\delta_{\upalpha\upbeta} = (2\pi)^3\delta_{\alpha\beta}\,\delta^{(3)}(\bm{k}_\alpha + \bm{k}_\beta)$. Consistently, a contraction with $\delta^{\upalpha}_{\upbeta}$ bars an index $\delta^{\upalpha}_{\upbeta}\,T^{\upbeta} = T^{\bar{\upalpha}}$. More details can be found in Appendix \ref{app:Fourier_summation}.

\paragraph{Hamiltonian.} The Hamiltonian---viewed as a functional of $(\bm{\varphi}^{\upalpha}, \bm{p}_{\upbeta})$---is defined from the Lagrangian by a Legendre transform
\begin{equation}
    H = \int \mathrm{d}t \left(\bm{p}_{\upalpha}\dot{\bm{\varphi}}^{\bar{\upalpha}} - \mathcal{L} \right)\,.
\end{equation}
From now on, we rescale the momenta $\bm{p}_{\upalpha} \rightarrow a^3 \bm{p}_{\upalpha}$ to avoid an extra overall factor of $a^3$ coming from $\mathcal{L}$ once the field time derivatives have been replaced by their conjugate momenta. Inverting Eq.~(\ref{eq:conjugate_momentum}) perturbatively gives
\begin{equation}
\begin{aligned}
    \dot{\bm{\varphi}}^{\upalpha} = \bar{\nabla}^{\upalpha\bar{\updelta}}&\left[\bm{p}_{\updelta} - \bar{I}_{\bar{\updelta}\upbeta}\bm{\varphi}^{\upbeta} - \left(\frac{1}{2}\bar{B}_{\upbeta\upgamma\bar{\updelta}} - \bar{C}_{\bar{\updelta}\upsigma\upgamma}\bar{\nabla}^{\upsigma\uptau}\bar{I}_{\bar{\uptau}\upbeta} + \frac{3}{2}\bar{D}_{\bar{\updelta}\uprho\upmu}\bar{\nabla}^{\uprho\uptau}\bar{I}_{\bar{\uptau}\upbeta}\bar{\nabla}^{\upmu\upsigma}\bar{I}_{\bar{\upsigma}\upgamma}\right)\bm{\varphi}^{\upbeta}\bm{\varphi}^{\upgamma}\right.\\
    &\left.-\left(\bar{C}_{\bar{\updelta}\upsigma\upgamma}\bar{\nabla}^{\upsigma\upbeta}-\frac{3}{2}\bar{D}_{\bar{\updelta}\uptau\uprho}\bar{\nabla}^{\uptau\upbeta}\bar{\nabla}^{\uprho\upsigma}\bar{I}_{\bar{\upsigma}\upgamma}\right)\bm{p}_{\upbeta}\bm{\varphi}^{\upgamma} - \frac{3}{2}\bar{D}_{\bar{\updelta}\uptau\upsigma}\bar{\nabla}^{\uptau\upbeta}\bar{D}^{\upsigma\upgamma}\bm{p}_{\upbeta}\bm{p}_{\upgamma}+ \ldots\right]\,,
\end{aligned}
\end{equation}
where the ellipses denote higher-order terms, irrelevant in our case. We have denoted  $\bar{\nabla}^{\upalpha\upbeta}$ to be the inverse matrix satisfying $\bar{\nabla}^{\upalpha\upbeta}\bar{\Delta}_{\upbeta\upgamma} = \delta^{\upalpha}_{\upgamma}$. Rearranging the various terms, the Hamiltonian reads
\begin{eBox}
\begin{equation}
\label{eq:full_Hamiltonian}
\begin{aligned}
    H = \frac{1}{2}\int \mathrm{d}t\, a^3 &\left( \Delta_{\upalpha\upbeta} \bm{p}^{\upalpha}\bm{p}^{\upbeta} - M_{\upalpha\upbeta}\bm{\varphi}^\upalpha\bm{\varphi}^\upbeta - 2I_{\upalpha\upbeta}\bm{\varphi}^{\upalpha}\bm{p}^{\upbeta} \right.\\
    &\left. - A_{\upalpha\upbeta\upgamma}\bm{\varphi}^{\upalpha}\bm{\varphi}^{\upbeta}\bm{\varphi}^{\upgamma} - B_{\upalpha\upbeta\upgamma}\bm{\varphi}^{\upalpha}\bm{\varphi}^{\upbeta}\bm{p}^{\upgamma} - C_{\upalpha\upbeta\upgamma}\bm{p}^{\upalpha}\bm{p}^{\upbeta}\bm{\varphi}^{\upgamma} - D_{\upalpha\upbeta\upgamma}\bm{p}^{\upalpha}\bm{p}^{\upbeta}\bm{p}^{\upgamma}  \right)\,,
\end{aligned}
\end{equation}
\end{eBox}
where the anew defined tensors are
\begin{equation}
\begin{aligned}
\Delta_{\upalpha\upbeta} &= \bar{\nabla}^{\bar{\upalpha}\bar{\upbeta}}\,, \hspace*{0.5cm}
M_{\upalpha\upbeta} = \bar{M}_{\upalpha\upbeta} - \bar{\nabla}^{\bar{\upgamma}\bar{\updelta}}\bar{I}_{\bar{\upgamma}\upalpha}\bar{I}_{\bar{\updelta}\upbeta}\,, \hspace*{0.5cm}
I_{\upalpha\upbeta} = \bar{\nabla}^{\bar{\upbeta}\upgamma}\bar{I}_{\upalpha\upgamma}\,,\\
A_{\upalpha\upbeta\upgamma} &= \bar{A}_{\upalpha\upbeta\upgamma} - \bar{B}_{\alpha\upbeta\updelta} \bar{\nabla}^{\updelta\bar{\upsigma}}\bar{I}_{\bar{\upsigma}\upgamma} + \bar{C}_{\updelta\upsigma\upgamma} \bar{\nabla}^{\updelta\bar{\uptau}} \bar{I}_{\bar{\uptau}\upalpha} \bar{\nabla}^{\upsigma\bar{\uprho}}\bar{I}_{\bar{\uprho}\upbeta} - \bar{D}_{\updelta\upsigma\uprho} \bar{\nabla}^{\updelta \upmu} \bar{\nabla}^{\upsigma \upnu} \bar{\nabla}^{\uprho \upkappa} \bar{I}_{\bar{\upmu} \upalpha} \bar{I}_{\bar{\upnu} \upbeta} \bar{I}_{\bar{\upkappa} \upgamma}\,,\\
B_{\upalpha\upbeta\upgamma} &= \bar{B}_{\upalpha\upbeta\updelta} \bar{\nabla}^{\updelta\bar{\gamma}} - 2 \bar{C}_{\updelta\upsigma\upalpha}\bar{\nabla}^{\updelta\upgamma} \bar{\nabla}^{\upsigma\bar{\uprho}} \bar{I}_{\bar{\uprho}\upbeta} + 3\bar{D}_{\updelta\upsigma\uprho}\bar{\nabla}^{\updelta\bar{\upgamma}} \bar{\nabla}^{\upsigma\bar{\upmu}} \bar{\nabla}^{\uprho\bar{\upnu}} \bar{I}_{\bar{\upmu}\upbeta} \bar{I}_{\bar{\upnu}\upalpha}\,,\\
C_{\upalpha\upbeta\upgamma} &= \bar{C}_{\updelta\upsigma\upgamma} \bar{\nabla}^{\updelta\bar{\upalpha}} \bar{\nabla}^{\upsigma\bar{\upbeta}} - 3\bar{D}_{\updelta\upsigma\uprho}\bar{\nabla}^{\upalpha\bar{\updelta}} \bar{\nabla}^{\upalpha\bar{\upsigma}} \bar{\nabla}^{\uprho\bar{\uptau}} \bar{I}_{\bar{\uptau}\upgamma}\,,\\
D_{\upalpha\upbeta\upgamma} &= \bar{D}_{\updelta\upsigma\uprho} \bar{\nabla}^{\upalpha\bar{\updelta}} \bar{\nabla}^{\upbeta\bar{\upsigma}} \bar{\nabla}^{\upgamma\bar{\uprho}}\,.
\end{aligned}
\end{equation}
Notice that the internal target space of the fluctuations is flat.
Consequently, the tensor indices are raised and lowered using the delta function. This form of the Hamiltonian---rather than the action (\ref{eq:action})---can also be taken as a definition of the theory. Also, note that non-linear contributions to the momenta cancel in the cubic Hamiltonian. It is therefore enough to only consider linear momenta in the computation of three-point correlators.

\subsection{Flow Equations}

Given the explicit form of the Hamiltonian (\ref{eq:full_Hamiltonian}), it is an easy task to derive the $\tensor{u}{^{\sf{a}}}{_{\sf{b}}} = \epsilon^{\sf{ac}}H_{\sf{cb}}$ and $\tensor{u}{^{\sf{a}}}{_{\sf{bc}}} = \epsilon^{\sf{ad}}H_{\sf{dbc}}$ tensors entering in the flow equations (\ref{eq:flow_equation_2pt}) and (\ref{eq:flow_equation_3pt}). For the quadratic Hamiltonian, after carefully symmetrising the $H_{\sf{ab}} = H_{(\sf{ab})}$ tensor, we obtain
\begin{equation}
    H_{\mathsf{ab}} = 
    \begin{pmatrix}
    -M_{\upalpha\upbeta} & - I_{\upalpha\upbeta} \\
    - I_{\upbeta\upalpha} & \Delta_{\upalpha\upbeta}
    \end{pmatrix}\,,
\end{equation}
where the $\sf{a, b}$ indices are arranged
in such a way that the field indices are followed by momentum indices in the same order. The desired $\tensor{u}{^{\sf{a}}}{_{\sf{b}}}$ tensor is then

\begin{eBox}
\begin{equation}
    \tensor{u}{^{\sf{a}}}{_{\sf{b}}} = 
    \begin{pmatrix}
    \mathbf{0} & \mathbf{1} \\
    \mathbf{-1} & \mathbf{0}
    \end{pmatrix}
    \begin{pmatrix}
    -\tensor{M}{^{\bar{\upalpha}}}{_{\upbeta}} & - \tensor{I}{^{\bar{\upalpha}}}{_{\upbeta}} \\
    - \tensor{I}{_{\upbeta}}{^{\bar{\upalpha}}} & \tensor{\Delta}{^{\bar{\upalpha}}}{_\upbeta}
    \end{pmatrix} + 
    \begin{pmatrix}
    \mathbf{0} & \mathbf{0} \\
    \mathbf{0} & -3H\delta^{\bar{\upalpha}}_{\upbeta}
    \end{pmatrix} = 
    \begin{pmatrix}
    \tensor{I}{_{\upbeta}}{^{\bar{\upalpha}}} & \tensor{\Delta}{^{\bar{\upalpha}}}{_\upbeta} \\
    \tensor{M}{^{\bar{\upalpha}}}{_\upbeta} & \tensor{I}{^{\bar{\upalpha}}}{_\upbeta} - 3H\delta^{\bar{\upalpha}}_{\upbeta}
    \end{pmatrix}\,.
\end{equation}
\end{eBox}
The extra factor $- 3H\delta^{\bar{\upalpha}}_{\upbeta}$ comes from the momentum rescaling.\footnote{Due to the momentum rescaling $\bm{p}_{\upalpha} \rightarrow a^3 \bm{p}_{\upalpha}$, the Heisenberg equation is modified
\begin{equation}
    \frac{\mathrm{d} \bm{p}_\upalpha}{\mathrm{d}t} = i[H, \bm{p}_\upalpha] - 3H\bm{p}_\upalpha.
\end{equation}
The additional non-canonical factor can be absorbed in the $\tensor{u}{^{\mathsf{a}}}{_{\mathsf{b}}}$ tensor as done in the main text.
} Similarly, the cubic Hamiltonian reads
\begin{equation}
    H_{\sf{abc}} = 
    \begin{Bmatrix}
\begin{pmatrix}
-3A_{\upalpha\upbeta\upgamma} & -B_{\upalpha\upgamma\upbeta}\\
-B_{\upgamma\upbeta\upalpha} & -C_{\upalpha\upbeta\upgamma}
\end{pmatrix}\\
\begin{pmatrix}
-B_{\upalpha\upbeta\upgamma} & 
-C_{\upgamma\upbeta\upalpha}\\
-C_{\upalpha\upgamma\upbeta} & -3D_{\upalpha\upbeta\upgamma}
\end{pmatrix}
\end{Bmatrix}\,,
\end{equation}
where the index $\sf{c}$ labels the first (resp.~second) matrix within the braces \{$\ldots$\} if it is a field (resp.~a momentum) index, and similarly $\sf{a}$ (resp.~$\sf{b}$) labels the row (resp.~column) in each $2\times2$ block matrix. We recall that $H_{\sf{abc}}$ is constructed to be symmetric under the exchange of any pair of indices hence the multiple presence of some tensors. The $\tensor{u}{^{\sf{a}}}{_{\sf{bc}}}$ tensor is obtained by explicitly performing the matrix multiplication
\begin{eBox}
\begin{equation}
\label{eq:full_u_abc_tensor}
    \tensor{u}{^{\sf{a}}}{_{\sf{bc}}} = 
     \begin{Bmatrix}
     \begin{pmatrix}
    \mathbf{0} & \mathbf{1} \\
    \mathbf{-1} & \mathbf{0}
    \end{pmatrix}
\begin{pmatrix}
-3\tensor{A}{^{\bar{\upalpha}}}{_{\upbeta\upgamma}} & -\tensor{B}{^{\bar{\upalpha}}}{_{\upgamma\upbeta}}\\
-\tensor{B}{_{\upgamma\upbeta}}{^{\bar{\upalpha}}} & -\tensor{C}{^{\bar{\upalpha}}}{_{\upbeta\upgamma}}
\end{pmatrix}\\
\begin{pmatrix}
    \mathbf{0} & \mathbf{1} \\
    \mathbf{-1} & \mathbf{0}
    \end{pmatrix}
\begin{pmatrix}
-\tensor{B}{^{\bar{\upalpha}}}{_{\upbeta\upgamma}} & -\tensor{C}{_{\upgamma\upbeta}}{^{\bar{\upalpha}}}\\
-\tensor{C}{^{\bar{\upalpha}}}{_{\upgamma\upbeta}} & -3\tensor{D}{^{\bar{\upalpha}}}{_{\upbeta\upgamma}}
\end{pmatrix}
\end{Bmatrix}
=
\begin{Bmatrix}
\begin{pmatrix}
-\tensor{B}{_{\upgamma\upbeta}}{^{\bar{\upalpha}}} & -\tensor{C}{^{\bar{\upalpha}}}{_{\upgamma\upbeta}}\\
3\tensor{A}{^{\bar{\upalpha}}}{_{\upbeta\upgamma}} & \tensor{B}{^{\bar{\upalpha}}}{_{\upgamma\upbeta}}
\end{pmatrix}\\
\begin{pmatrix}
-\tensor{C}{^{\bar{\upalpha}}}{_{\upgamma\upbeta}} & 3\tensor{D}{^{\bar{\upalpha}}}{_{\upbeta\upgamma}}\\
\tensor{B}{^{\bar{\upalpha}}}{_{\upbeta\upgamma}} & \tensor{C}{_{\upgamma\upbeta}}{^{\bar{\upalpha}}}
\end{pmatrix}
\end{Bmatrix}.
\end{equation}
\end{eBox}
As done in~\cite{Dias:2016rjq}, we further notice that the particular arrangement of barred indices---namely the upper $\bar{\upalpha}$ index for each tensors---enables us to extract a delta function, hence defining $\bm{k}$-dependent tensor coefficients $\tensor{u}{^{a}}{_{b}}$ and $\tensor{u}{^{a}}{_{bc}}$ that we label with normal indices $a, b, \ldots$ in the following way
\begin{equation}
\label{eq:u_tensor_coefficients}
    \begin{aligned}
    \tensor{u}{^{\mathsf{a}}}{_{\mathsf{b}}} &= (2\pi)^3\delta^{(3)}(\bm{k}_a - \bm{k}_b)\,\tensor{u}{^{a}}{_{b}}(\bm{k}_a, \bm{k}_b)\,,\\
    \tensor{u}{^{\mathsf{a}}}{_{\mathsf{bc}}} &= (2\pi)^3\delta^{(3)}(\bm{k}_a - \bm{k}_b - \bm{k}_c)\,\tensor{u}{^{a}}{_{bc}}(\bm{k}_a, \bm{k}_b, \bm{k}_c)\,.
    \end{aligned}
\end{equation}
It is important to stress that in writing expressions with explicit momentum dependence, the index labelling the momentum is associated with the corresponding tensor index.
Explicitly, $\bm{k}_a$ is associated with $a$ and so forth. For example if one would want to exchange $b$ and $c$ in $\tensor{u}{^{a}}{_{bc}}$, the corresponding momenta should be exchanged too. Furthermore, due to the presence of the delta function and by isotropy, the $\tensor{u}{^{a}}{_{b}}$ tensor coefficients only depend on the magnitude of the momentum, $\tensor{u}{^{a}}{_{b}}(k)$ where $k = |\bm{k}_a| = |\bm{k}_b|$.

\paragraph{Flow equations.} Turning to the two- and three-point correlation functions, statistical isotropy and homogeneity imply that one can write 
\begin{equation}
    \begin{aligned}
    \langle \bm{X}^{\sf{a}}\bm{X}^{\sf{b}}\rangle 
    &= (2\pi)^3\delta^{(3)}(\bm{k}_a + \bm{k}_b)\,\Sigma^{ab}(k)\,, \\
    \langle \bm{X}^{\sf{a}} \bm{X}^{\sf{b}} \bm{X}^{\sf{c}}\rangle 
    &= (2\pi)^3\delta^{(3)}(\bm{k}_a + \bm{k}_b + \bm{k}_c)\, B^{abc}(k_a, k_b, k_c)\,,
    \end{aligned}
    \label{eq:def-Sigma-B}
\end{equation}
where for the two-point correlator we have set $k = |\bm{k}_a| = |\bm{k}_b|$. We do not explicitly write the time dependency of the above correlators but it should be understood that they are evaluated at the same time $t$. The three-point correlators $B^{abc}(k_a, k_b, k_c)$ are real, as we will explain below, while $\Sigma^{ab}$ is complex and can be split into real and imaginary parts
$\Sigma^{ab} = \Sigma_{\Re}^{ab} + i \Sigma_{\Im}^{ab}$. Because the flow equation (\ref{eq:flow_equation_2pt}) is linear and the $\tensor{u}{^a}{_c}$ have real elements, one can solve independently for $\Sigma_{\Re}^{ab}$ and $\Sigma_{\Im}^{ab}$, obeying
\begin{eBox}
\begin{equation}
    \frac{\mathrm{d}\Sigma_{\Re, \Im}^{ab}(k)}{\mathrm{d}t} = \tensor{u}{^a}{_c}(k)\, \Sigma_{\Re, \Im}^{cb}(k) + \tensor{u}{^b}{_c}(k)\, \Sigma_{\Re, \Im}^{ac}(k)\,.
\end{equation}
\end{eBox}
Actually, the commutation relation---valid \textit{at all times}---fully fixes the imaginary part of the two-point correlators, independently of any theory. To see this, note that the real-space operators we consider are Hermitian. In Fourier space, this implies that
\begin{equation}
\bm{X}^{\sf{a} \dagger}=\bm{X}^{\sf{\bar{a}}}\,,
\end{equation}
where $\dagger$ is the Hermitian conjugate operation, and we recall that $\sf{\bar{a}}$ indicates that the sign of the corresponding momentum is reversed. From this, one can first derive symmetry properties satisfied by the two-point correlators. Indeed, $\langle \bm{X}^{\sf{a}} \bm{X}^{\sf{b}} \rangle^* = \langle \left(\bm{X}^{\sf{a}} \bm{X}^{\sf{b}}\right)^\dagger \rangle= \langle \bm{X}^{\sf{\bar{b}}} \bm{X}^{\sf{\bar{a}}} \rangle= \langle \bm{X}^{\sf{b}} \bm{X}^{\sf{a}} \rangle $, where we used \eqref{eq:def-Sigma-B} in the last step.
This enforces the real part to be symmetric and the imaginary part to be anti-symmetric
\begin{equation}
    \Sigma_{\Re}^{ab} = \Sigma_{\Re}^{ba}\,, \hspace*{0.5cm} \Sigma_{\Im}^{ab} =- \Sigma_{\Im}^{ba}\,.
\end{equation}
Second, introducing the commutator in $\langle \bm{X}^{\sf{a}} \bm{X}^{\sf{b}}\rangle$ and taking the imaginary part, one finds $\Sigma_{\Im}^{ab} = \frac{\epsilon^{ab}}{2 a^3}$, where the tensor $\epsilon^{ab}$ is defined in (\ref{eq:commutation_relation}). As a result, the imaginary part either decays in $1/a^3$ for correlators involving a field and a conjugate momentum of the same kind (recall that the physical momenta are rescaled by $a^3$) or is strictly zero (for all other correlators). Therefore, there is no need to solve the flow equation for $\Sigma_{\Im}^{ab}$, although this equation can all the same be solved as a numerical consistency check.

\vskip 4pt
For three-point correlators, similar manipulations give
\begin{equation}
\langle \bm{X}^{\sf{a}} \bm{X}^{\sf{b}} \bm{X}^{\sf{c}} \rangle^*- \langle \bm{X}^{\sf{\bar{a}}} \bm{X}^{\sf{\bar{b}}} \bm{X}^{\sf{\bar{c}}} \rangle=- i \epsilon^{\sf{ab}} \langle X^{\sf{\bar{c}}} \rangle - i \epsilon^{\sf{ac}} \langle X^{\sf{\bar{b}}} \rangle  - i \epsilon^{\sf{bc}} \langle X^{\sf{\bar{a}}} \rangle \,.
\end{equation}
The right-hand side vanishes for connected correlators, for which none of the wavevectors individually vanishes. Defining more precisely the bispectra as the connected three-point correlators, this leads to $B^{abc}(\bm{k}_a,\bm{k}_b,\bm{k}_c)^*-B^{abc}(-\bm{k}_a,-\bm{k}_b,-\bm{k}_c)=0$. Then, statistical isotropy, implying that $B^{abc}$ in \eqref{eq:def-Sigma-B} only depends on the norms of the wavevectors, enforces that $B^{abc}$ is real. As for its time evolution, it is important to keep track of both $\Sigma_{\Re}^{ab}$ and $\Sigma_{\Im}^{ab}$ because the flow equations---being sourced by a non-linear term---requires the knowledge of the \textit{complex} two-point correlators. Taking into account that the tensor coefficients $\tensor{u}{^a}{_{bc}}$ are real, we arrive at
\begin{eBox}
\begin{gather}
    \begin{aligned}
    \frac{\mathrm{d}B^{abc}(k_a, k_b, k_c)}{\mathrm{d}t} &= \tensor{u}{^a}{_d}(k_a)B^{dbc}(k_a, k_b, k_c) + \tensor{u}{^b}{_d}(k_b)B^{adc}(k_a, k_b, k_c) + \tensor{u}{^c}{_d}(k_c)B^{abd}(k_a, k_b, k_c) \\
    &+ \tensor{u}{^a}{_{de}}(\bm{k}_a, \bm{k}_b, \bm{k}_c)\Sigma_{\Re}^{db}(k_b)\Sigma_{\Re}^{ec}(k_c) - \tensor{u}{^a}{_{de}}(\bm{k}_a, \bm{k}_b, \bm{k}_c)\Sigma_{\Im}^{db}(k_b)\Sigma_{\Im}^{ec}(k_c) \\
    &+ \tensor{u}{^b}{_{de}}(\bm{k}_b, \bm{k}_a, \bm{k}_c)\Sigma_{\Re}^{ad}(k_a)\Sigma_{\Re}^{ec}(k_c) - \tensor{u}{^b}{_{de}}(\bm{k}_b, \bm{k}_a, \bm{k}_c)\Sigma_{\Im}^{ad}(k_a)\Sigma_{\Im}^{ec}(k_c) \\
    &+ \tensor{u}{^c}{_{de}}(\bm{k}_c, \bm{k}_a, \bm{k}_b)\Sigma_{\Re}^{ad}(k_a)\Sigma_{\Re}^{be}(k_b) - \tensor{u}{^c}{_{de}}(\bm{k}_c, \bm{k}_a, \bm{k}_b)\Sigma_{\Im}^{ad}(k_a)\Sigma_{\Im}^{be}(k_b)\,.
    \end{aligned}
\raisetag{35pt}
\end{gather}
\end{eBox}
where we have deliberately written all the permutations to avoid ambiguity in the index ordering. This form of the flow equations for the three-point functions has been derived in~\cite{Dias:2016rjq}.

\vskip 4pt 
We have seen that the imaginary parts of two-point correlators are fixed from first principles, and that the ones of connected three-point correlators vanish. In this respect, it is instructive to also consider four-point correlators. Systematically inserting commutators, one finds
\begin{gather}
 \begin{aligned}
\text{Im} \langle \bm{X}^{\sf{a}} \bm{X}^{\sf{b}} \bm{X}^{\sf{c}}  \bm{X}^{\sf{d}}  \rangle &=\frac{1}{2 i} \big[  \langle \bm{X}^{\sf{a}} \bm{X}^{\sf{b}} \bm{X}^{\sf{c}}  \bm{X}^{\sf{d}}  \rangle-  \langle \bm{X}^{\sf{\bar{a}}} \bm{X}^{\sf{\bar{b}}} \bm{X}^{\sf{\bar{c}}} \bm{X}^{\sf{\bar{d}}} \rangle  \big] \\
&\hspace{-3cm} 
+\frac12 \left( \epsilon^{\sf{cd}} \langle \bm{X}^{\sf{a}} \bm{X}^{\sf{b}} \rangle+ \epsilon^{\sf{ab}} \langle \bm{X}^{\sf{d}} \bm{X}^{\sf{c}} \rangle+ \epsilon^{\sf{bd}} \langle \bm{X}^{\sf{a}} \bm{X}^{\sf{c}} \rangle+ \epsilon^{\sf{ac}} \langle \bm{X}^{\sf{d}} \bm{X}^{\sf{b}} \rangle+ \epsilon^{\sf{bc}} \langle \bm{X}^{\sf{a}} \bm{X}^{\sf{d}} \rangle+ \epsilon^{\sf{ad}} \langle \bm{X}^{\sf{c}} \bm{X}^{\sf{b}} \rangle \right)\,.
\raisetag{45pt}
\end{aligned}
\end{gather}
The second line vanishes for connected four-point functions, and simply corresponds to the imaginary part of the correlator that one would deduce from Wick theorem in terms of two-point correlators. Note, however, that we did not use a perturbative scheme to derive this. Only the first line remains for connected correlators of interest. In that case, and contrary to three-point kinematic above, $(\bm{k}_a, \ldots ,\bm{k}_d)$ and $(-\bm{k}_a, \ldots ,-\bm{k}_d)$ need not be related by a rotation, so that the imaginary parts of connected four-point correlators do not vanish in general. This happens for theories in which parity is violated.

\paragraph{Derivative interactions.} The entries of the cubic tensors $A_{\upalpha\upbeta\upgamma}, B_{\upalpha\upbeta\upgamma}, C_{\upalpha\upbeta\upgamma}$ and $D_{\upalpha\upbeta\upgamma}$ are in general time and momentum dependent.
The momentum dependency comes from considering spatial derivative interactions, that give simple factors of momenta after Fourier transform.
For instance, let us take the contact interaction $\mathcal{L}/a^3 \supset \frac{\lambda}{2}\varphi (\partial_i \varphi)^2/a^2$. In Fourier space, the corresponding cubic Hamiltonian gives the following tensor coefficient $A_{\varphi_{\bm{k}_1}\varphi_{\bm{k}_2}\varphi_{\bm{k}_3}} = \lambda\,\bm{k}_1 \cdot \bm{k}_2/a^2 + 2\text{ perms}.$
In general, these tensor elements can be more complicated, and can for example scale as inverse powers of the momenta. Such momentum dependencies arise from non-local (in space) interactions, to which the cosmological flow has already been applied in \cite{Jazayeri:2023xcj}. In this way, the cosmological flow formalism makes it convenient to implement various types of interactions without much complications.

\vskip 4pt
When computing the in-in integrals using the Schwinger-Keldysh formalism, the time derivatives appearing in a vertex should be directly applied to the attached propagators. In the cosmological flow formalism, time derivative interactions can be expressed as contact vertices between fields and momenta. For example, the interaction $\mathcal{L}/a^3 \supset -\frac{\lambda}{2}\varphi^2\dot{\varphi}$ gives the tensor coefficient 
$B_{\varphi_{\bm{k}_1} \varphi_{\bm{k}_2} p_{\bm{k}_3}} = -\lambda$, where $p$ is the (rescaled) conjugate momentum\footnote{We have assumed, for illustration purposes, that the quadratic theory is simple enough so that $p=\dot{\varphi}$.} to the field $\varphi$. Therefore, it is an easy task to treat time derivative interactions with the formalism presented in this paper.

\paragraph{Boundary terms.} When defining a theory, the action is sometimes supplemented by a boundary term. This is the case for example in the ADM formalism \cite{ADM} where the Gibbons-Hawking-York boundary term needs to be added to the Einstein-Hilbert action. Another celebrated example is Maldacena's calculation of the cubic Lagrangian in single-field inflation \cite{Maldacena:2002vr, Seery:2005wm} and its more recent generalisation to non-linear sigma models in \cite{Garcia-Saenz:2019njm, Pinol:2020kvw}. There, the cubic Lagrangian for the fluctuations is supplemented by a boundary term whose various terms come from rendering the size of interactions manifest by using the equation of motions, equivalently performing multiple integrations by parts. Although these boundary terms do not contribute to the equations of motion, they do contribute to the three-point correlators. We show here how to include these terms in correlators from first principles and in the most general case.

\vskip 4pt
A generic boundary term appears in the form of a total derivative at the level of the action $S = S_{\text{bulk}} + S_\partial$. In Fourier space, we write the boundary term $S_\partial$ as 
\begin{equation}
    S_\partial = \int \mathrm{d}t\, \mathcal{L}_\partial\,, \hspace*{0.5cm}\text{with} \hspace*{0.5cm}  \mathcal{L}_\partial = \frac{\mathrm{d}}{\mathrm{d}t}\left(\frac{1}{3!}\, \Omega_{\sf{abc}} \bm{X}^{\sf{a}} \bm{X}^{\sf{b}} \bm{X}^{\sf{c}}\right)\,,
\end{equation}
where in general the tensor $\Omega$ has time and momentum dependent elements. It is taken to be fully symmetric $\Omega_{\sf{abc}} = \Omega_{(\sf{abc})}$. With the cubic Hamiltonian, one can formally write the in-in formula for the three-point correlators
\begin{equation}
\begin{aligned}
    \langle \bm{X}^{\sf{a}} \bm{X}^{\sf{b}} \bm{X}^{\sf{c}} \rangle &\supset -i \bra{0} \int_{-\infty}^t \mathrm{d}t'\left[\frac{\mathrm{d}}{\mathrm{d}t'}\left(\frac{1}{3!}\,\Omega_{\sf{def}}\,X^{\sf{d}} X^{\sf{e}} X^{\sf{f}}\right), X^{\sf{a}} X^{\sf{b}} X^{\sf{c}} \right]\ket{0} \\
    &= -i\bra{0} \frac{1}{3!}\, \Omega_{\sf{def}}\left[X^{\sf{d}} X^{\sf{e}} X^{\sf{f}}, X^{\sf{a}} X^{\sf{b}} X^{\sf{c}} \right]\ket{0}\,,
\end{aligned}
\end{equation}
where in the last line all the interaction-picture operators are evaluated at the time $t$. As we have done previously for the $\tensor{u}{^{\sf{a}}}{_{\sf{b}}}$ and $\tensor{u}{^{\sf{a}}}{_{\sf{bc}}}$ tensors, we define the tensor elements by extracting the momentum conserving delta functions which follows from that of the correlator $\Omega_{\sf{abc}} = (2\pi)^3\delta^{(3)}(\bm{k}_a+\bm{k}_b + \bm{k}_c)\,\Omega_{abc}(\bm{k}_a, \bm{k}_b, \bm{k}_c)$. As usual, the $i\epsilon$ prescription in the far past switches off the interactions. The equation shows that a general boundary term gives a local contribution in time for the three-point correlators. The fields being free in the interaction picture, we use Wick theorem to express the contribution into explicit two-point correlators. After using the symmetry properties of $\Omega_{\sf{abc}}$ and retaining only connected contributions, one obtains
\begin{equation}
\begin{aligned}
    \langle \bm{X}^{\sf{a}} \bm{X}^{\sf{b}} \bm{X}^{\sf{c}} \rangle &\supset -i\, \Omega_{\sf{def}} \left( \langle \bm{X}^{\sf{d}} \bm{X}^{\sf{a}}\rangle\langle \bm{X}^{\sf{e}}\bm{X}^{\sf{b}}\rangle\langle \bm{X}^{\sf{f}}\bm{X}^{\sf{c}}\rangle - \langle \bm{X}^{\sf{a}} \bm{X}^{\sf{d}}\rangle\langle \bm{X}^{\sf{b}}\bm{X}^{\sf{e}}\rangle\langle \bm{X}^{\sf{c}}\bm{X}^{\sf{f}}\rangle \right)\,.
\end{aligned}
\end{equation}
We can further use the commutator to simplify the relation, obtaining
\begin{equation}
\begin{aligned}
    \langle \bm{X}^{\sf{a}}\bm{X}^{\sf{b}}\bm{X}^{\sf{c}} \rangle \supset - \Omega_{\sf{def}} &\left(\epsilon^{\sf{ad}}\langle \bm{X}^{\sf{e}} \bm{X}^{\sf{b}}\rangle\langle \bm{X}^{\sf{f}} \bm{X}^{\sf{c}}\rangle + \epsilon^{\sf{be}}\langle \bm{X}^{\sf{a}} \bm{X}^{\sf{d}}\rangle\langle \bm{X}^{\sf{c}} \bm{X}^{\sf{f}}\rangle \right.\\
    &\left.+ \epsilon^{\sf{cf}}\langle \bm{X}^{\sf{a}} \bm{X}^{\sf{d}}\rangle\langle \bm{X}^{\sf{e}} \bm{X}^{\sf{b}}\rangle \right)\,.
\end{aligned}
\end{equation}
keeping in mind that three-point correlators are real, one finds that the boundary term only needs to be added a posteriori:
\begin{equation}
B^{abc}(k_a, k_b, k_c) \rightarrow B^{abc}_{\text{bulk}}(k_a, k_b, k_c) + B^{abc}_\partial(k_a, k_b, k_c)\,,
\end{equation}
with the contribution coming from the boundary term being
\begin{eBox}
\begin{equation}
\begin{aligned}
\hspace*{-0.3cm} B^{abc}_\partial(k_a, k_b, k_c) = -\Omega_{def}(-\bm{k}_a, -\bm{k}_b, -\bm{k}_c &)\left[ \epsilon^{ad}\Sigma_{\Re}^{be}(k_b)\Sigma_{\Re}^{fc}(k_c) - \epsilon^{ad}\Sigma_{\Im}^{be}(k_b)\Sigma_{\Im}^{fc}(k_c) \right.\\
&\left.+ \epsilon^{be}\Sigma_{\Re}^{ad}(k_a)\Sigma_{\Re}^{cf}(k_c)  - \epsilon^{be}\Sigma_{\Im}^{ad}(k_a)\Sigma_{\Im}^{cf}(k_c) \right.\\
&\left.+\epsilon^{cf}\Sigma_{\Re}^{ad}(k_a)\Sigma_{\Re}^{be}(k_b)  - \epsilon^{cf}\Sigma_{\Im}^{ad}(k_a)\Sigma_{\Im}^{be}(k_b)\right]\,.
\end{aligned}
\end{equation}
\end{eBox}
Note that in general, only correlators of light degrees of freedom survive until the end of inflation. Also, terms proportional to spatial gradients are suppressed by the exponential of the time elapsed in $e$-fold since the corresponding mode exited the horizon, and are negligible for cosmological scales. In practice, a lot of the listed terms do not contribute significantly to the three-point correlators at the end of the time evolution. However for completeness we have derived the most general form for such boundary terms.

\subsection{Initial Conditions}
\label{sec:Initial_Conditions}

To render the system complete, one must provide initial conditions for $\Sigma^{ab}_{\Re}, \Sigma^{ab}_{\Im}$ and $B^{abc}$. Within the cosmological flow formalism, these initial conditions can be derived analytically provided one initialises the correlators sufficiently in the deep past. Indeed, by looking at the Hamiltonian in Eq.~(\ref{eq:full_Hamiltonian}), we see that the quadratic theory is composed of a part governing the free evolution of the fields and one encoding non-trivial quadratic mixings. These quadratic interactions constitute one reason why analytical methods to compute late-time correlators fail in the most general cases.
However, one can always choose an initial time deep enough inside the horizon for the modes of interest, so that the theory is dominated by the decoupled quadratic theory.
In this regime, all the various couplings and masses can be neglected so that the theory approaches that of a set of uncoupled degrees of freedom, and initial conditions can be computed analytically.
This is the very idea of asymptotically reaching the vacuum state.

\subsubsection*{Two-point correlation functions}

We begin with the second-order action in Fourier space,
\begin{equation}
    S^{(2)} = \frac{1}{2} \int\mathrm{d}t\, a^3\left(\bar{\Delta}_{\upalpha\upbeta}\dot{\bm{\varphi}}^{\upalpha}\dot{\bm{\varphi}}^{\upbeta} + \bar{M}_{\upalpha\upbeta}\bm{\varphi}^{\upalpha} \bm{\varphi}^{\upbeta} + 2\bar{I}_{\upalpha\upbeta}\bm{\varphi}^{\upalpha}\dot{\bm{\varphi}}^{\upbeta} \right)\,,
\end{equation}
where $\bar{M}_{\upalpha\upbeta}$ encompasses the kinetic gradient terms as well as other non-kinetic terms. We isolate the mass matrix $\bar{\mathcal{M}}_{\alpha\beta}$ by writing
\begin{equation}
    \bar{M}_{\upalpha\upbeta} = (2\pi)^3\delta^{(3)}(\bm{k}_\alpha + \bm{k}_\beta) \left(\bar{\mathcal{K}}_{\alpha\beta} + \bar{\mathcal{M}}_{\alpha\beta} \right)\,,
\end{equation}
where all gradient terms are included in the matrix $\bar{\mathcal{K}}_{\alpha\beta}$. Without loss of generality, it can be written as
\begin{equation}
    \bar{\mathcal{K}}_{\alpha\beta} = -\frac{k^2}{a^2}\, c^{(2)}_{\alpha\beta} - \frac{k^4}{a^4}\, \frac{c^{(4)}_{\alpha\beta}}{\Lambda^2} - \frac{k^6}{a^6}\, \frac{c^{(6)}_{\alpha\beta}}{\Lambda^4} + \dots \,,
\end{equation}
with the ellipses denoting terms of higher order in $k$ that have been omitted, and $\Lambda$ is some energy scale to render the elements of the $c^{(2n)}_{\alpha\beta}$ matrix dimensionless. Specifically, $c^{(2)}_{\alpha\beta}$ encapsulates all the sound speeds (squared) for the various fields.

\paragraph{Mass matrix and quadratic mixings.} In general, the mass matrix $\bar{\mathcal{M}}_{\alpha\beta}$ renders analytical treatment delicate because it models complex processes such as energy exchange between various fields at the quadratic level. It has been recently stressed that these interactions were analogous to neutrino and quark oscillations, as inflationary flavor and mass eigenstates generically mix with one other, leading to a new interesting phenomenology~\cite{Pinol:2021aun}. However, because the gradient terms in $\bar{\mathcal{K}}_{\alpha\beta}$ become increasingly important in the infinite past, they will eventually dominate the elements of $\bar{\mathcal{M}}_{\alpha\beta}$ provided that the (time-dependent) eigenvalues of the mass matrix are bounded from above. If one chooses an initial time such that contributions from the mass matrix become effectively irrelevant, each mode can be treated as massless. Therefore, we ignore the mass matrix for the remaining of this section. Similarly, one can always initialise the correlators deep enough in the past so that the gradient terms also dominate the quadratic couplings in the $\bar{I}_{\upalpha\upbeta}$ matrix. We therefore ignore the quadratic coupling $\bar{I}_{\upalpha\upbeta}$. In short, at sufficiently early times, the full quadratic theory will approach that of a set of \textit{massless} and \textit{uncoupled} degrees of freedom. Indeed, one can always diagonalise both kinetic terms with time and spatial derivatives in the same basis as they are both definite positive quadratic forms. Therefore, in the following, we will write in a non-covariant way the matrix elements $\bar{\Delta}_{\alpha\beta}=\delta^{\alpha\beta}\bar{\Delta}_{\alpha}$ and $\bar{\mathcal{K}}_{\alpha
\beta}=\delta^{\alpha\beta}\bar{\mathcal{K}}_{\alpha}$.

\paragraph{Linear dispersion relations.} For the remainder of this section, we assume that the kinetic terms for the fields only contain two spatial derivatives. We include the possibility of having different sound speeds for the fields by considering a generic diagonal matrix $c^{(2)}_{\alpha\beta}= \text{diag}(c_1^2, c_2^2, \ldots)$. We highlight that the premise of our argument of analytically deriving initial conditions is not altered by this choice. However, other quadratic and diagonal theories such as theories with higher-order derivative terms (like ghost inflation \cite{Arkani-Hamed:2003juy}) require a different treatment. The quadratic theory we consider is
\begin{equation}
    S^{(2)} = \frac{1}{2}\int \mathrm{d}t\, a^3\left(\bar{\Delta}_{\upalpha}\left(\dot{\bm{\varphi}}^{\upalpha}\right)^2 - c_{\upalpha}^2\, \frac{k^2}{a^2}\, \left(\bm{\varphi}^{\upalpha}\right)^2 \right)\,.
\end{equation}

\paragraph{Initial two-point correlators.} Referring the interested reader to the Appendix \ref{app:2pt_initial_conditions} for the derivation, the two-point correlator initial conditions are found to be
\begin{equation}
\label{eq:initial_2pt_correlators}
    \Sigma^{ab}_{\Re} = \frac{1}{2a^2c_\alpha k}\,
    \begin{pmatrix}
    \bar{\Delta}_{\alpha\beta}\left(1 + \frac{a^2 H^2}{c_\alpha^2 k^2}\right) & - H\delta^{\alpha\beta} \\
    - H\delta^{\alpha\beta} & c_\alpha^2 k^2\,\bar{\Delta}^{-1}_{\alpha\beta}
    \end{pmatrix}\,, \hspace*{0.5cm}
    \Sigma^{ab}_{\Im} = \frac{1}{2a^3} 
    \begin{pmatrix}
    0 &  \delta^{\alpha\beta} \\
    - \delta^{\alpha\beta} & 0
    \end{pmatrix}\,,
\end{equation}
where all time-dependent quantities should be evaluated at the initial time. Note that in \cite{Dias:2016rjq} the authors neglected the constant $\propto \frac{H^2}{2c_\alpha^3 k^3}$ mode in the $\braket{\bm{\varphi}^\alpha \bm{\varphi}^{\beta}}$ initial correlator. Even though this term is negligible on subhorizon scales, we include it to improve numerical stability. We also recall that $\Sigma^{ab}_\Im$ is completely fixed by the commutator, so that its initial condition in (\ref{eq:initial_2pt_correlators}) is actually valid at all times. As an aside, notice that for small sound speeds $c_\alpha \ll 1$, the amplitude of $\braket{\bm{\varphi}^\alpha \bm{\varphi}^\beta}$ is enhanced whereas the amplitude of $\braket{\bm{p}^\alpha \bm{p}^\beta}$ is damped, intuitively reflecting the fact that the fields do not propagate in the limit $c_\alpha \rightarrow 0$. 

\subsubsection*{Three-point correlation functions}

The derivation of initial three-point correlators is more involved than that of initial two-point correlators. Computing correlators in the deep past instead of in the far future requires some foresight.
Indeed, examining the Hamiltonian (\ref{eq:full_Hamiltonian}), the tensors $A_{\upalpha\upbeta\upgamma}, B_{\upalpha\upbeta\upgamma}, C_{\upalpha\upbeta\upgamma}$ and $D_{\upalpha\upbeta\upgamma}$ encode cubic interactions that can be of arbitrary order in derivatives, each gradient (squared) term giving a term of order $(k/a)^2\sim (k\tau)^2$ that grows exponentially as we approach the limit $|\tau|\rightarrow\infty$. Therefore, terms with higher-order derivative will be the dominant contributions of the various correlators. Specifically, let us extract the momentum-conserving delta function of the cubic tensors to define the associated kernels
\begin{equation}
    A_{\upalpha\upbeta\upgamma} = (2\pi)^3 \delta^{(3)}(\bm{k}_\alpha + \bm{k}_\beta + \bm{k}_\gamma) \,A_{\alpha\beta\gamma}\,,
\end{equation}
and equivalent definitions for the tensors $B_{\upalpha\upbeta\upgamma}, C_{\upalpha\upbeta\upgamma}$ and $D_{\upalpha\upbeta\upgamma}$. We further decompose the above kernels by extracting the scale factor dependency that needs to be considered in the in-in integral, according to the number of spatial derivatives it contains, as follows\footnote{Non-local interactions characterised by inverse gradient terms are completely negligible in this regime compared to usual derivative interactions so we do not consider this possibility.}
\begin{equation}
    A_{\alpha\beta\gamma} = A^{(0)}_{\alpha\beta\gamma} + \frac{1}{a^2} A^{(2)}_{\alpha\beta\gamma} + \frac{1}{a^4} A^{(4)}_{\alpha\beta\gamma}+ \ldots\,,
\end{equation}
and similarly for the other kernels. We have absorbed all momentum dependency in the newly defined kernels, and indicate with the superscript $2n$ ($n\geq 0$) the number of spatial derivatives that appears in the cubic interaction. For a given cubic interaction, the higher the number of spatial derivatives, the (exponentially) faster these terms grow on subhorizon scales. Consequently, it is essential to keep track of every order in powers of $\tau$ to avoid creating a large hierarchy between cubic interactions.
By explicitly computing the in-in integrals, we generalise the results of \cite{Dias:2016rjq, Butchers:2018hds} to include all possible cubic interactions to all order in (spatial) derivatives.
The full derivation is a bit lengthy, so we jump directly to the final results without loss of continuity, leaving all the details in Appendix \ref{app:3pt_initial_conditions}.

\paragraph{Three fields initial correlator:}

\begin{gather}
\label{eq:3pt_fff_initial_conditions}
    \begin{aligned}
    &\braket{\bm{\varphi}_{\bm{k}_1}^\alpha \bm{\varphi}_{\bm{k}_2}^\beta \bm{\varphi}_{\bm{k}_3}^\gamma}_0 = \frac{(2\pi)^3\delta^{(3)}(\Sigma_i \bm{k}_i)}{4 \textcolor{pyblue}{a^4} e_3 k_t}
    \left\{ \frac{A_{\alpha\beta\gamma}^{(0)}}{2}\, \textcolor{pyblue}{a^2} + \frac{A_{\alpha\beta\gamma}^{(2)}}{2}\left[1 + \frac{\textcolor{pyblue}{a^2}H^2}{e_3^2}\left(e_2\left(e_2+\frac{e_3}{k_t}\right)\right) - k_t e_3\right] \right.\\
    &\left. + \frac{A_{\alpha\beta\gamma}^{(4)}}{2}\left[\frac{1}{\textcolor{pyblue}{a^2}} + \frac{H^2}{e_3^2}\left(e_2\left(e_2+\frac{3e_3}{k_t}\right)\right) - k_t e_3\right] + \ldots - \frac{B_{\alpha\beta\gamma}^{(0)}}{2}\, \frac{\textcolor{pyblue}{a^2}H}{k_1k_2} \left[e_2 - k_3(k_1+k_2)\right] \right.\\
    &\left. - \frac{B_{\alpha\beta\gamma}^{(2)}}{2}\,  \frac{H}{k_1k_2}\left[ e_2 - \frac{1}{k_t}(2e_3 + k_t(k_1+k_2)k_3)\right] - \frac{B_{\alpha\beta\gamma}^{(4)}}{2}\,  \frac{H}{\textcolor{pyblue}{a^2}\,k_1k_2}\left[ e_2 - \frac{1}{k_t}(4e_3 + k_t(k_1+k_2)k_3)\right] \right.\\
    &\left. - \frac{C_{\alpha\beta\gamma}^{(0)}}{2}\left[k_1 k_2 + \textcolor{pyblue}{a^2} H^2 \left(\frac{e_2}{k_3^2}\left(1 + \frac{k_3}{k_t}\right)\right) - \frac{k_t}{k_3}\right] - \frac{C_{\alpha\beta\gamma}^{(2)}}{2}\left[\frac{k_1k_2}{\textcolor{pyblue}{a^2}} + H^2 \left(\frac{e_2}{k_3^2}\left(1 + \frac{3k_3}{k_t}\right)\right) - \frac{k_t}{k_3}\right] \right.\\
    &\left. - \frac{C_{\alpha\beta\gamma}^{(4)}}{2}\left[\frac{k_1k_2}{\textcolor{pyblue}{a^4}} + \frac{H^2}{\textcolor{pyblue}{a^2}} \left(\frac{e_2}{k_3^2}\left(1 + \frac{5k_3}{k_t}\right)\right) - \frac{k_t}{k_3}\right] + \ldots + \frac{D_{\alpha\beta\gamma}^{(0)}}{2} H \left[e_2 - \frac{2e_3}{k_t}\right] \right.\\
    &\left. + \frac{D_{\alpha\beta\gamma}^{(2)}}{2} \frac{H}{\textcolor{pyblue}{a^2}} \left[e_2 - \frac{4e_3}{k_t}\right] + \frac{D_{\alpha\beta\gamma}^{(4)}}{2} \frac{H}{\textcolor{pyblue}{a^4}} \left[e_2 - \frac{6e_3}{k_t}\right] + \dots + \text{ 5 perms} \right\}\,.
    \end{aligned}
    \raisetag{50pt}
\end{gather}

\paragraph{Two fields and one momentum initial correlator:}

\begingroup
\allowdisplaybreaks
\begin{align}
\label{eq:3pt_pff_initial_conditions}
    &\braket{\bm{p}_{\bm{k}_1}^\alpha \bm{\varphi}_{\bm{k}_2}^\beta \bm{\varphi}_{\bm{k}_3}^\gamma}_0 = \frac{-(2\pi)^3\delta^{(3)}(\Sigma_i \bm{k}_i)\, H}{4 \textcolor{pyblue}{a^3} e_3^2 k_t}
    \left\{ -k_1 \left[-\frac{A^{(0)}_{\alpha\beta\gamma}}{2}\, \textcolor{pyblue}{a^2}\left(e_2 - \frac{e_3}{k_t}\right) -\frac{A^{(2)}_{\alpha\beta\gamma}}{2}\, \left(e_2 + \frac{e_3}{k_t}\right) \right.\right.\nonumber\\
    &\left.\left. -\frac{A^{(4)}_{\alpha\beta\gamma}}{2}\,\frac{1}{\textcolor{pyblue}{a^2}}\, \left(e_2 + \frac{3e_3}{k_t}\right) + \dots +  \frac{B^{(0)}_{\alpha\beta\gamma}}{2H}\, e_3k_3 + \frac{B^{(2)}_{\alpha\beta\gamma}}{2H}\, \frac{e_3k_3}{\textcolor{pyblue}{a^2}} + \frac{B^{(4)}_{\alpha\beta\gamma}}{2H}\, \frac{e_3k_3}{\textcolor{pyblue}{a^4}} + \dots \right.\right.\nonumber\\
    &\left.\left. + \frac{C^{(0)}_{\alpha\beta\gamma}}{2}\, k_1^2 k_2^2\left(1+\frac{k_3}{k_t}\right) + \frac{C^{(2)}_{\alpha\beta\gamma}}{2}\, \frac{k_1^2 k_2^2}{\textcolor{pyblue}{a^2}}\left(1+\frac{3k_3}{k_t}\right) + \frac{C^{(4)}_{\alpha\beta\gamma}}{2}\, \frac{k_1^2 k_2^2}{\textcolor{pyblue}{a^4}}\left(1+\frac{5k_3}{k_t}\right) + \dots \right.\right.\\
    &\left.\left. - \frac{D^{(0)}_{\alpha\beta\gamma}}{2H} \frac{e_3^2}{\textcolor{pyblue}{a^2}} - \frac{D^{(2)}_{\alpha\beta\gamma}}{2H} \frac{e_3^2}{\textcolor{pyblue}{a^4}} - \frac{D^{(4)}_{\alpha\beta\gamma}}{2H} \frac{e_3^2}{\textcolor{pyblue}{a^6}} + \dots + \text{ 5 perms} \right] - k_1^2(k_2+k_3)\left[ \frac{A^{(0)}_{\alpha\beta\gamma}}{2}\, \textcolor{pyblue}{a^2} \right.\right.\nonumber\\
    &\left.\left.+
    \frac{A^{(2)}_{\alpha\beta\gamma}}{2} + \frac{A^{(4)}_{\alpha\beta\gamma}}{2}\, \frac{1}{\textcolor{pyblue}{a^2}} + \dots + \frac{B^{(0)}_{\alpha\beta\gamma}}{2}\, \textcolor{pyblue}{a^2}H\,\frac{(k_1+k_2)k_3^2}{e_3} + \frac{B^{(2)}_{\alpha\beta\gamma}}{2}\,\frac{Hk_3}{e_3k_t}\left(2e_3 + k_t(k_1+k_2)k_3\right) \right.\right.\nonumber\\
    &\left.\left. + \frac{B^{(4)}_{\alpha\beta\gamma}}{2}\,\frac{H}{\textcolor{pyblue}{a^2}}\,\frac{k_3}{e_3k_t}\left(4e_3 + k_t(k_1+k_2)k_3\right) + \dots - \frac{C^{(0)}_{\alpha\beta\gamma}}{2}\,k_1k_2 - \frac{C^{(2)}_{\alpha\beta\gamma}}{2}\,\frac{k_1k_2}{\textcolor{pyblue}{a^2}}- \frac{C^{(4)}_{\alpha\beta\gamma}}{2}\,\frac{k_1k_2}{\textcolor{pyblue}{a^4}} + \dots \right.\right.\nonumber\\
    &\left.\left. - D^{(0)}_{\alpha\beta\gamma}\, \frac{He_3}{k_t} - 2D^{(0)}_{\alpha\beta\gamma}\,\frac{H}{\textcolor{pyblue}{a^2}}\, \frac{e_3}{k_t} - 3D^{(0)}_{\alpha\beta\gamma}\,\frac{H}{\textcolor{pyblue}{a^4}}\, \frac{e_3}{k_t} + \text{ 5 perms}\right] \right\}\,.\nonumber
\end{align}
\endgroup

\paragraph{Two momenta and one field initial correlator:}

\begin{gather}
\label{eq:3pt_ppf_initial_conditions}
    \begin{aligned}
    &\braket{\bm{p}_{\bm{k}_1}^\alpha \bm{p}_{\bm{k}_2}^\beta \bm{\varphi}_{\bm{k}_3}^\gamma}_0 = \frac{-(2\pi)^3\delta^{(3)}(\Sigma_i \bm{k}_i)}{4 \textcolor{pyblue}{a^4} e_3 k_t}
    \left\{ k_1k_2 \left[ \frac{A^{(0)}_{\alpha\beta\gamma}}{2}\,\textcolor{pyblue}{a^2} + \frac{A^{(2)}_{\alpha\beta\gamma}}{2} + \frac{A^{(4)}_{\alpha\beta\gamma}}{2}\,\frac{1}{\textcolor{pyblue}{a^2}} + \dots \right.\right.\\
    &\left.\left. + \frac{B^{(0)}_{\alpha\beta\gamma}}{2}\,\textcolor{pyblue}{a^2}H\, \frac{(k_1+k_2)k_3^2}{e_3} + \frac{B^{(2)}_{\alpha\beta\gamma}}{2}\,H\, \frac{k_3}{e_3k_t}\left(2e_3 + k_t(k_1+k_2)k_3\right) + \right.\right.\\
    &\left.\left. + \frac{B^{(4)}_{\alpha\beta\gamma}}{2}\,\frac{H}{\textcolor{pyblue}{a^2}}\, \frac{k_3}{e_3k_t}\left(4e_3 + k_t(k_1+k_2)k_3\right) + \dots - \frac{C^{(0)}_{\alpha\beta\gamma}}{2}\,k_1k_2 - \frac{C^{(2)}_{\alpha\beta\gamma}}{2}\,\frac{k_1k_2}{\textcolor{pyblue}{a^2}} - \frac{C^{(4)}_{\alpha\beta\gamma}}{2}\,\frac{k_1k_2}{\textcolor{pyblue}{a^4}}+ \dots \right.\right.\\
    &\left.\left. - D^{(0)}_{\alpha\beta\gamma}\,\frac{He_3}{k_t} - 2D^{(2)}_{\alpha\beta\gamma}\,\frac{H}{\textcolor{pyblue}{a^2}}\,\frac{e_3}{k_t} - 3D^{(4)}_{\alpha\beta\gamma}\,\frac{H}{\textcolor{pyblue}{a^4}}\,\frac{e_3}{k_t} + \dots + \text{ 5 perms}\right] \right.\\
    &\left.+\, k_1^2k_2^2\left[\frac{A^{(0)}_{\alpha\beta\gamma}}{2}\,\frac{\textcolor{pyblue}{a^4}H^2}{e_3}\left(e_2 - \frac{e_3}{k_t}\right) + \frac{A^{(2)}_{\alpha\beta\gamma}}{2}\,\frac{\textcolor{pyblue}{a^2}H^2}{e_3}\left(e_2 + \frac{e_3}{k_t}\right) + \frac{A^{(4)}_{\alpha\beta\gamma}}{2}\,\frac{H^2}{e_3}\left(e_2 + \frac{3e_3}{k_t}\right) + \dots \right.\right.\\
    &\left.\left. - \frac{B^{(0)}_{\alpha\beta\gamma}}{2}\, \textcolor{pyblue}{a^2}Hk_3 - \frac{B^{(2)}_{\alpha\beta\gamma}}{2}\, Hk_3 - \frac{B^{(4)}_{\alpha\beta\gamma}}{2}\, \frac{H}{\textcolor{pyblue}{a^2} k_3} + \dots - \frac{C^{(0)}_{\alpha\beta\gamma}}{2}\, \textcolor{pyblue}{a^2} H^2\, \frac{k_1^2k_2^2}{e_3^2}\left(1+\frac{k_3}{k_t}\right) \right.\right.\\
    &\left.\left. - \frac{C^{(2)}_{\alpha\beta\gamma}}{2}\,H^2\, \frac{k_1^2k_2^2}{e_3^2}\left(1+\frac{3k_3}{k_t}\right) - \frac{C^{(4)}_{\alpha\beta\gamma}}{2}\,\frac{H^2}{\textcolor{pyblue}{a^2}}\, \frac{k_1^2k_2^2}{e_3^2}\left(1+\frac{5k_3}{k_t}\right) + \frac{D^{(0)}_{\alpha\beta\gamma}}{2}\,H + \frac{D^{(2)}_{\alpha\beta\gamma}}{2}\,\frac{H}{\textcolor{pyblue}{a^2}} \right.\right.\\
    &\left.\left. + \frac{D^{(4)}_{\alpha\beta\gamma}}{2}\,\frac{H^2}{\textcolor{pyblue}{a^4}} + \dots + \text{ 5 perms}\right]\right\}\,.
    \end{aligned}
    \raisetag{100pt}
\end{gather}

\paragraph{Three momenta initial correlator:}

\begin{equation}
\label{eq:3pt_ppp_initial_conditions}
    \begin{aligned}
    &\braket{\bm{p}_{\bm{k}_1}^\alpha \bm{p}_{\bm{k}_2}^\beta \bm{p}_{\bm{k}_3}^\gamma}_0 = \frac{(2\pi)^3\delta^{(3)}(\Sigma_i \bm{k}_i)\,H}{4 \textcolor{pyblue}{a^3} e_3 k_t}
    \left\{\frac{A^{(0)}_{\alpha\beta\gamma}}{2}\, \textcolor{pyblue}{a^2}\left(e_2 - \frac{e_3}{k_t}\right) + \frac{A^{(2)}_{\alpha\beta\gamma}}{2}\, \left(e_2 + \frac{e_3}{k_t}\right) \right.\\
    &\left.+ \frac{A^{(4)}_{\alpha\beta\gamma}}{2}\, \frac{1}{\textcolor{pyblue}{a^2}}\left(e_2 + \frac{3e_3}{k_t}\right) + \dots - \frac{B^{(0)}_{\alpha\beta\gamma}}{2H}\,e_3k_3 - \frac{B^{(2)}_{\alpha\beta\gamma}}{2H}\,\frac{e_3k_3}{\textcolor{pyblue}{a^2}} - \frac{B^{(4)}_{\alpha\beta\gamma}}{2H}\,\frac{e_3k_3}{\textcolor{pyblue}{a^4}} + \dots \right.\\
    &\left. - \frac{C^{(0)}_{\alpha\beta\gamma}}{2}\, k_1^2 k_2^2 \left(1 + \frac{k_3}{k_t}\right) - \frac{C^{(2)}_{\alpha\beta\gamma}}{2}\, \frac{k_1^2 k_2^2}{\textcolor{pyblue}{a^2}} \left(1 + \frac{3k_3}{k_t}\right) - \frac{C^{(4)}_{\alpha\beta\gamma}}{2}\, \frac{k_1^2 k_2^2}{\textcolor{pyblue}{a^4}} \left(1 + \frac{5k_3}{k_t}\right) + \dots \right.\\
    &\left. + \frac{D^{(0)}_{\alpha\beta\gamma}}{2H}\frac{e_3^2}{\textcolor{pyblue}{a^2}} + \frac{D^{(2)}_{\alpha\beta\gamma}}{2H}\frac{e_3^2}{\textcolor{pyblue}{a^4}} + \frac{D^{(4)}_{\alpha\beta\gamma}}{2H}\frac{e_3^2}{\textcolor{pyblue}{a^6}} + \dots + \text{ 5 perms} \right\}\,,
    \end{aligned}
\end{equation}
where we have introduced the symmetric polynomials $e_1 \equiv k_t = k_1+k_2+k_3$, $e_2 \equiv k_1k_2 + k_1k_3 + k_2k_3$, and $e_3 \equiv k_1k_2k_3$. We have coloured in \textcolor{pyblue}{blue} the powers of the scale factor to render the size of the various terms in the subhorizon regime manifest. The momentum dependencies of the various kernels are not written explicitly but these need to be regarded as functions of $\bm{k}_1, \bm{k}_2$ and $\bm{k}_3$ associated with each index $\alpha, \beta$ and $\gamma$ respectively. When writing the permutations for the kernels $A^{(2n)}_{\alpha\beta\gamma}, B^{(2n)}_{\alpha\beta\gamma}, C^{(2n)}_{\alpha\beta\gamma}$ and $D^{(2n)}_{\alpha\beta\gamma}$, one should also exchange the associated momenta. Additional manipulations are needed when the fields are not canonically normalised. We leave this detail to Appendix~\ref{app:3pt_initial_conditions}. In the end, note that these initial conditions are completely fixed by the tensor elements, or equivalently by the Hamiltonian.

\vskip 4pt
An alternative valid procedure consists in initialising the three-point correlators to zero and letting them build up as the cubic interactions are switched on adiabatically. This method essentially encapsulates the essence and \textit{defines} the $i\epsilon$ prescription. Such prescription is simple and useful when dealing with other interactions and higher-order correlators.
We have checked numerically for several examples that it is equivalent to using the initial conditions~(\ref{eq:3pt_fff_initial_conditions})--(\ref{eq:3pt_ppp_initial_conditions}).

\newpage
\part*{Applications}
\addcontentsline{toc}{section}{Applications}

\section{Inflationary Fluctuations}
\label{sec:Application_EFT_multifield}

With the cosmological flow method at hand, we begin our exploration with the study of primordial fluctuations in the line of the EFT of inflationary fluctuations \cite{Cheung:2007st}. We will present an exhaustive study of the quadratic theory of the Goldstone boson of broken time translations coupled to an additional massive field. We will also provide new insight on regimes that are not accessible by analytical means, which includes the strong mixing regime.

\subsection{Goldstone Description}
\label{subsec:EFT}

First, we will construct and study a multi-field theory at the level of the fluctuations built upon the EFT of (multi-field) inflation. We refer the reader to the original papers~\cite{Creminelli:2006xe, Cheung:2007st, Senatore:2010wk} or review \cite{Piazza_2013} for more details. This framework provides a unified way of studying inflationary correlators because it is independent of the microscopic details of the theory that gives rise to the quasi-de Sitter background. The unavoidable degree of freedom describing inflationary fluctuations is the Goldstone boson associated with the spontaneous breaking of time-translation invariance. 

\paragraph{Goldstone boson sector.} In the unitary gauge, where the spatial part of the metric reads $g_{ij} = a^2(t)e^{2\zeta(t, \bm{x})}\delta_{ij}$ after disregarding tensor modes, the most general local action is
\begin{equation}
\label{eq:unitary_gauge_action}
    \begin{aligned}
    S_\pi = \int \mathrm{d}^4x \sqrt{-g} &\left[
    \frac{1}{2}M_{\text{pl}}^2 R + M_{\text{pl}}^2 \dot{H} g^{00} - M_{\text{pl}}^2(3H^2 + \dot{H}) \right.\\
    &\left.+ \frac{1}{2!}M_2^4(t) \left(\delta g^{00}\right)^2 + \frac{1}{3!} M_3^4(t)\left(\delta g^{00}\right)^3 + \ldots \right.\\
    &\left. - \frac{1}{2}\bar{M}_1^3(t)\delta g^{00} \delta K_\mu^\mu - \frac{1}{2}\bar{M}_2^2(t) \left(\delta K_\mu^\mu\right)^2 - \frac{1}{2}\bar{M}_3^2(t) \delta K^{\mu\nu} \delta K_{\mu\nu} + \ldots
    \right]\,,
    \end{aligned}
\end{equation}
where $\delta g^{00} = g^{00} + 1$, $\delta K_{\mu\nu} = K_{\mu\nu} - a^2 H h_{\mu\nu}$ with $h_{\mu\nu}$ the induced spatial metric, and $M_{2, 3}$ and $\bar{M}_{1, 2, 3}$ are time-dependent mass scales. The dots stand for higher-order terms in the fluctuations or with more spatial derivatives.
This effective action encapsulates a large class of single-field models of inflation. In the following, we will be interested in the case where $\dot{H}$ and the mass scales $M_{2, 3}$ and $\bar{M}_{1, 2, 3}$ are slowly time-varying,\footnote{We will explore the possibility of strongly time-varying mass scales in Section \ref{sec:features}, leading to features.} so that our treatment captures the behaviour of primordial fluctuations in the limit of approximate scale invariance.

\vskip 4pt
To render the dynamics of this theory more explicit, we introduce the Goldstone boson $\pi$ associated with the spontaneous breaking of time-translation invariance, i.e.~$\pi$ restores full diffeomorphism invariance by transforming as $\pi\rightarrow \pi - \xi(t, \bm{x})$ under a time reparameterisation $t \rightarrow t + \xi(t, \bm{x})$ such that $t+\pi$ is invariant. This is known as the Stückelberg trick. In this way, the field $\pi$ parametrises adiabatic perturbations corresponding to a local shift in time for the matter fields. A precise definition of the gauge in which $\pi$ appears is often not mentioned, as this becomes irrelevant when considering the decoupling limit. In the following, we will also concentrate on theories and observables for which the mixing with gravity is negligible. For definiteness, however, we define $\pi$ in the spatially flat gauge, with $g_{ij}=a^2(t) \delta_{ij}$, so that the link between $\pi$ and $\zeta$ is unambiguous. The decoupling limit $M_{\text{pl}}\rightarrow \infty$ and $\dot{H}\rightarrow 0$ is taken while keeping the product $M_{\text{pl}}^2\dot{H}$ fixed. In this way, the inflationary correlators determined by the decoupled dynamics of $\pi$ are accurate up to slow-roll corrections $\varepsilon = -\dot{H}/H^2$. After introducing the Goldstone boson and in the decoupling limit, $\delta g^{00}$ becomes $\delta g^{00} \rightarrow -2 \dot{\pi} - \dot{\pi}^2 + (\partial_i \pi)^2/a^2$. Implementing this procedure in the action (\ref{eq:unitary_gauge_action}), neglecting terms that involve the extrinsic curvature\footnote{Operators that involve the extrinsic curvature---being terms with higher derivatives---are suppressed by extra powers of the cutoff of the theory and hence are irrelevant at low energy. Even though these terms can become important in some cases---e.g. in ghost inflation \cite{Arkani-Hamed:2003juy}---we do not treat them in this paper. Nonetheless, they can be easily implemented in the cosmological flow formalism.} and only keeping the leading deformation of the slow-roll action, one obtains the following single-field Lagrangian for the $\pi$-sector
\begin{equation}
\label{eq: Goldstone boson Lagrangian with pi}
    \mathcal{L}_{\pi}/a^3 = \frac{M_{\text{pl}}^2|\dot{H}|}{c_s^2}\left[\dot{\pi}^2 - c_s^2\frac{(\partial_i\pi)^2}{a^2} + (1 - c_s^2)\left(\dot{\pi}^3 - \dot{\pi}\frac{(\partial_i\pi)^2}{a^2}\right) - \frac{4}{3}M_3^4 \frac{c_s^2}{M_{\text{pl}}^2|\dot{H}|}\dot{\pi}^3\right]\,,
\end{equation}
where $c_s^{-2} = 1 - 2M_2^4/M_{\text{pl}}^2|\dot{H}|$ is the intrinsic speed of sound for the propagation of $\pi$, generated by the lowest-order Lorentz-breaking operator proportional to $M_2$. This effective theory makes the non-linearly realised symmetry manifest, i.e. a small sound speed in the quadratic theory necessarily engenders a large $\dot{\pi}(\partial_i \pi)^2$ cubic self-interaction.

\paragraph{Additional field sector.} We are interested in coupling the adiabatic mode $\pi$ to an additional massive scalar degree of freedom $\sigma$ which may have significant self-interactions. We consider the Lagrangian up to dimension-4 operators for the $\sigma$-sector
\begin{equation}
\label{eq:Lagrangian-sigma}
    \mathcal{L}_{\sigma}/a^3 = -\frac{1}{2}(\partial_\mu \sigma)^2 - \frac{1}{2}m^2\sigma^2 - \mu \sigma^3\,.
\end{equation}
At cubic order, the $\sigma$ self-interactions are completely characterised by $\mu$.\footnote{The dimension-4 operator $\dot{\sigma}\sigma^2$ is redundant with $\sigma^3$ upon integration by parts.} 

\paragraph{Mixing sector.} Our interest is the coupling between the two sectors, both converting intrinsic non-linearities of $\sigma$ to the $\pi$ sector via a quadratic mixing, and generating non-linear interactions amongst them.
In the unitary gauge, we consider the following mixing action
\begin{equation}
\label{eq:mixing_action_unitary_gauge}
    S_{\pi-\sigma} = -\int\mathrm{d}^4x\sqrt{-g}\left[ \tilde{M}_1^3(t)\delta g^{00} \sigma + \tilde{M}_2^2(t)\delta g^{00} \sigma^2 + \tilde{M}_3^3(t)\left(\delta g^{00}\right)^2\sigma + \dots\right]\,,
\end{equation}
where $\tilde{M}_{1, 2, 3}$ are some---in general time-dependent---mass scales.\footnote{The most general mixing operators are of the form $\left(\delta g^{00}\right)^m\mathcal{O}_n$ where $\mathcal{O}_n$ is an operator of dimension $n$ function of the following combinations: $\sigma$, $g^{0\mu}\partial_\mu\sigma$ and $g^{\mu\nu}\partial_\mu\partial_\nu\sigma$. These operators have been fully classified in \cite{Baumann:2011nk} up to dimension-6 operators. We consider an example of such operators in Section~\ref{subsec: Sound Speed Collider}. Note that we do not impose additional symmetries on the $\sigma$ field to make it naturally light because we will be mostly interested in the cases where $\sigma$ is (effectively) heavier than the Hubble scale. Theories with additional symmetries imposed on the $\sigma$-sector have been treated in \cite{Senatore:2010wk}.} Taking the decoupling limit and introducing $\pi$, the theory up to cubic interactions reads
\begin{equation}
\label{eq:pi_sigma_decoupling_limit}
    \mathcal{L}_{\pi-\sigma}/a^3 = 2\tilde{M}_1^3 \dot{\pi}\sigma + \tilde{M}_1^3(\partial_\mu \pi)^2\sigma + 2\tilde{M}_2^2\dot{\pi}\sigma^2 - 4\tilde{M}_3^3\dot{\pi}^2\sigma\,.
\end{equation}
Similarly to the $\pi$-sector treated previously, we note that the Wilson coefficient of the $(\partial_i \pi)^2\sigma$ operator is completely fixed by the $\dot{\pi}\sigma$ operator that appears in the quadratic Lagrangian, again a consequence of the fact that $\pi$ non-linearly realises time diffeomorphisms.

\paragraph{Full theory.} Up to cubic order including self-interactions in the $\pi$-sector, the complete Lagrangian for the coupled $\pi$-$\sigma$ system that we consider is then
\begin{equation}
\label{eq:Full_pi_sigma_theory}
\begin{aligned}
    \mathcal{L}/a^3 = &-\frac{1}{2}\left[-\dot{\pi}_c^2 + c_s^2 \frac{(\partial_i \pi_c)^2}{a^2}\right] - \frac{1}{2}\left[ (\partial_\mu \sigma)^2 + m^2\sigma^2\right] + \rho \dot{\pi}_c\sigma \\
    & -\lambda_1\dot{\pi}_c\frac{(\partial_i \pi_c)^2}{a^2} - \lambda_2 \dot{\pi}_c^3
    -\mu\sigma^3 - \frac{1}{2}\alpha\dot{\pi}_c \sigma^2 - \frac{1}{2\Lambda_1} \frac{(\partial_i \pi_c)^2}{a^2}\sigma - \frac{1}{2\Lambda_2}\dot{\pi}_c^2\sigma\,,
\end{aligned}
\end{equation}
where we have defined the canonically normalised field $\pi_c^2 = c_s^{-3}f_{\pi}^4\,\pi^2$ with $f_\pi^4 \equiv 2M_{\text{pl}}^2|\dot{H}|c_s$ being the symmetry breaking scale,\footnote{For energies above $f_\pi$, one should consider an infinite number of operators in the Lagrangian because higher-order terms---that we truncated---become relevant. At energies below $f_\pi$, a description in terms of Goldstone boson is appropriate.} and redefined the coupling constants
\begin{equation}
\label{eq:definition_couplings}
\begin{aligned}
    &\rho \equiv 2c_s^{3/2} \frac{\tilde{M}_1^3}{f_\pi^2}, \hspace*{0.5cm} \lambda_1 \equiv c_s^{3/2} \frac{c_s^2-1}{2f_\pi^2}, \hspace*{0.5cm} \lambda_2 \equiv -\frac{c_s^{3/2}}{2  f_{\pi}^2}\left[1-c_s^2 - \frac{8}{3}\frac{ M_3^4}{f_{\pi}^4}c_s^3\right],\\
    &\alpha \equiv -4c_s^{3/2}\frac{H^2}{f_\pi^2}\,\tilde{\alpha}^2, \hspace*{0.5cm} \Lambda_1^{-1} \equiv c_s^{3/2}\frac{\rho}{f_\pi^2}, \hspace*{0.5cm} \Lambda_2^{-1} \equiv c_s^{3/2}\frac{\tilde{\rho}}{f_\pi^2}\,,
\end{aligned}
\end{equation}
with $\tilde{\alpha} = \frac{\tilde{M}_2}{H}$ and $\tilde{\rho} = 8c_s^{3/2}\frac{\tilde{M}_3^3}{f_\pi^2}-\rho$. These couplings have mass dimensions $[\Lambda_{1,2}] = [\mu] = [\rho] = 1, [\alpha] = 0$, and $[\lambda_{1, 2}] = -2$. As we will see in Section \ref{sec:NG_pheno}, the couplings $\rho, \tilde{\rho}$ and $\tilde{\alpha}$ enter in the genuine typical size of each contribution to the dimensionless shape function of the bispectrum in the case $c_s=1$. Both $\rho$ and $\Lambda_1$ are correlated since they are determined by the same parameter $\tilde{M}_1$, which is a consequence of the non-linearly realised time-diffeomorphism invariance. Similarly, $M_2$ governs both the Goldstone boson sound speed $c_s$ and $\lambda_1$. Note that there is a priori no model-building requirement on the size of the quadratic coupling $\rho/H$, and we crucially allow this mixing parameter to be order one or larger, which from now on we call the strong mixing regime.

\paragraph{Cosmological flow implementation.} In order to implement the flow equations for the fields $\bm{\varphi}^\alpha = (\pi_c, \sigma)$, one needs to write the theory in a Hamiltonian form. The rescaled linear conjugate momenta associated with the fields are
\begin{equation}
    p_{\pi_c} \equiv \frac{1}{a^3}\frac{\delta S}{\delta \dot{\pi}_c} = \dot{\pi}_c + \rho \sigma, \hspace*{0.5cm} p_{\sigma} \equiv \frac{1}{a^3}\frac{\delta S}{\delta \dot{\sigma}} = \dot{\sigma}.
\label{eq:linear-conjugate-momenta}    
\end{equation}
After performing the Legendre transform, the Hamiltonian can be arranged in the form (\ref{eq:full_Hamiltonian}) and the identification of the various tensors yields
\begingroup
\allowdisplaybreaks
\begin{align}
\label{eq:tensor_primordial_fluctuations_interactions}
    \Delta_{\alpha\beta} &= \delta_{\alpha\beta}\,, \hspace*{0.8cm} M_{\alpha\beta} = 
    \begin{pmatrix}
- c_s^2\frac{k^2}{a^2} & 0 \\
    0 & -\left(\frac{k^2}{a^2} + m^2+\rho^2\right)
    \end{pmatrix}\,, \hspace*{0.8cm} I_{\alpha\beta} = 
    \begin{pmatrix}
    0 & \rho \\
    0 & 0
    \end{pmatrix}\,, \nonumber\\
    A_{\sigma\sigma\sigma} &= -2\mu + \alpha\rho - \frac{\rho^2}{\Lambda_2} + 2\lambda_2\rho^3\,, \hspace*{0.2cm} A_{\pi_c \pi_c \sigma} = \left(\frac{1}{\Lambda_1}-2\lambda_1\rho\right) \frac{\bm{k}_{\pi_c}\cdot\bm{k}_{\pi_c}}{a^2}\,,\\
    B_{\sigma\sigma p_{\pi_c}} &= -\alpha + 2\frac{\rho}{\Lambda_2} - 6\lambda_2\rho^2\,, \hspace*{0.85cm} B_{\pi_c \pi_c p_{\pi_c}} = 2\lambda_1 \frac{\bm{k}_{\pi_c}\cdot\bm{k}_{\pi_c}}{a^2}\,,\nonumber\\
    C_{p_{\pi_c} p_{\pi_c}\sigma} &= -\frac{1}{\Lambda_2} + 6\lambda_2\rho\,, \hspace*{1.65cm} D_{p_{\pi_c} p_{\pi_c} p_{\pi_c}} = -2\lambda_2\,.\nonumber
\end{align}
\endgroup
\noindent Deriving these tensors fully fixes the flow equations and the initial conditions for the two- and three-point functions.

\paragraph{From $\pi$ to $\zeta$.} One needs to relate the various correlators to primordial observables, namely correlators of the curvature perturbation $\zeta$. Actually, this link is only geometrical and can be expressed simply in terms of the Goldstone boson $\pi$, independently of the specific theory that one considers. Note that the link between $\pi$ and the canonically normalised field $\pi_c$ is theory-dependent. Our aim is only to compute the primordial power spectrum and bispectrum, i.e.~the late-time two- and three-point correlators of $\zeta$ at the end of inflation, when all modes are well outside the horizon.
In this regime, the relation between $\pi$ and $\zeta$ reads (see e.g.~\cite{Behbahani:2011it})
\begin{equation}
\label{eq:change-gauge}
    \zeta(t, \bm{x}) =
    \log \left(\frac{ a(t+\xi(t, \bm{x}))}{a(t)} \right)
    \quad \text{with} \quad \pi(t+\xi(t, \bm{x}), \bm{x}) + \xi(t, \bm{x}) = 0\,,
\end{equation}
where the implicit equation defining $\xi$ is found by working out the time-diffeomorphism mapping the spatially flat gauge to the unitary one. Beyond linear order, a spatial diffeomorphism is also required for the mapping, but its effect is negligible when all modes are well outside the horizon. Using Eq.~\eqref{eq:change-gauge}, one obtains 
\begin{equation}
    \zeta = -H \pi + H \pi \dot{\pi}+ \frac{1}{2}\dot{H}\pi^2 + \mathcal{O}(\pi^3)\,,
\end{equation}
up to quadratic order, which is sufficient to work out the power spectrum and bispectrum. Explicitly, one finds
\begin{equation}
\label{eq:zeta-correlators}
    \begin{aligned}
    \braket{\zeta_{\bm{k}} \zeta_{-\bm{k}}}' &= H^2 \braket{\pi_{\bm{k}} \pi_{-\bm{k}}}'\,, \\
    \braket{\zeta_{\bm{k}_1} \zeta_{\bm{k}_2} \zeta_{\bm{k}_3}}' &= -H^3 \braket{\pi_{\bm{k}_1} \pi_{\bm{k}_2} \pi_{\bm{k}_3}}' + H^3 \big[ \Sigma^{\pi \pi}(k_1) \Sigma_{\Re}^{\pi \dot{\pi}}(k_2) + \text{5 perms}   \big] \\
    &+ H^2 \dot{H} \big[  \Sigma^{\pi \pi}(k_1) \Sigma^{\pi \pi}(k_2) + \text{ 2 perms}\big]\,,
    \end{aligned}
\end{equation}
where all correlators should be evaluated at the end of inflation.
A few comments are in order. First, the result~\eqref{eq:zeta-correlators} is valid in any theory, be it single-clock or with multiple degrees of freedom. Second, no decoupling limit or slow-roll approximation has been considered here, only that modes are well outside the horizon. Note also that even in single-clock inflation, $\pi$ is strictly speaking not massless and is time-dependent even outside the horizon. However, the time dependence of $H, \pi$ and $\dot{\pi}$ are such that the correlators of $\zeta$ themselves become time-independent in that case.

\vskip 4pt
Eventually, in the applications considered in this work, where any time-dependence of $H$ is neglected, only the first line is relevant in the expression of the bispectrum. Notice also that the surviving correction to the bispectrum of $\pi$ can be rewritten in terms of correlators of fields and conjugate momenta, see Eq.~\eqref{eq:linear-conjugate-momenta}. This shift of perspective is useful as even when multi-field interactions keep building up on super-horizon scales, the cross power spectrum involving the momentum of $\pi$ quickly decays and only remains the contribution from the mixed $\pi$-$\sigma$ power spectrum, see~\cite{Garcia-Saenz:2019njm} for a related discussion in the context of the contribution of boundary terms in multi-field inflation. This situation occurs only when the field $\sigma$ is almost massless. However, in this situation, $\zeta$ keeps evolving until the end of inflation, which effectively breaks scale invariance and makes the theory non-predictive without a complete description of reheating. We discuss this particular case in Appendix~\ref{app:LightField}, see also~\cite{Wang:2022eop} for a bootstrap perspective on this parameter space. When $\sigma$ is effectively heavy, it quickly decays on super-horizon scales, an adiabatic limit is then reached by the end of inflation, and the bispectrum of $\zeta$ is directly proportional to the one of $\pi$. The large majority of our applications lie in this regime.

\vskip 4pt
In the single-field limit $\rho=0$, the amplitude of curvature perturbations $\zeta$ is fixed by the ratio of $H$ and $f_\pi$
\begin{equation}
\label{eq:single_field_power_spectrum}
    \Delta_{\zeta, 0}^2 = \frac{k^3}{2\pi^2} \braket{\zeta_{\bm{k}} \zeta_{-\bm{k}}}' = \frac{1}{4\pi^2}\left(\frac{H}{f_\pi}\right)^4\,.
\end{equation}
The measured amplitude of the dimensionless power spectrum on CMB scales is $\Delta_\zeta^2 = 2.2\times 10^{-9}$ \cite{Planck:2018jri}.

\subsection{Quadratic Theory}
\label{subsec:quadratic_theory}

We begin the study of this theory by examining the time evolution of the various two-point functions. In this section, we start by introducing the notion of effective mass that emerges from a sizeable quadratic mixing. Then, we provide exact results---derived for the first time using the cosmological flow---for the power spectrum in all regimes of interest including the strong mixing regime. This provides a comprehensive analysis of the entire quadratic phase space.

\subsubsection{Effective mass and resonance at strong mixing} 
\label{subsub: Effective mass and resonance at strong mixing}

Writing the theory in the Hamiltonian form makes it explicit that the quadratic mixing induces an effective mass for the $\sigma$ field
\begin{equation}
    m^2_{\text{eff}} = m^2 + \rho^2\,,
\label{eq:def-effective-mass}   
\end{equation}
that can be appreciated from the form of the $M$ matrix in Eq.~(\ref{eq:tensor_primordial_fluctuations_interactions}). This is a consequence of considering the full linear momenta, i.e.~the conjugate momentum of $\pi_c$ is corrected due to the quadratic mixing.

\paragraph{Late-time leading behaviour.} To understand the origin of the effective mass at energy scales $\omega\ll H$, it is sufficient to look at the leading late-time behaviour of the $\pi_c-\sigma$ system. Looking for solutions of the form (see Appendix~\ref{app:LightField})
\begin{equation}
    \lim\limits_{\tau \to 0} 
    \begin{pmatrix}
    \pi_c(\tau, \bm{x}) \\
    \sigma(\tau, \bm{x})
    \end{pmatrix} = 
    \begin{pmatrix}
    \pi_c^+(\bm{x}) \\
    \sigma^+(\bm{x})
    \end{pmatrix} \tau^{\Delta_+} + 
    \begin{pmatrix}
    \pi_c^-(\bm{x}) \\
    \sigma^-(\bm{x})
    \end{pmatrix} \tau^{\Delta_-}\,,
\end{equation}
where $\tau$ is the conformal time, one finds four modes with scaling dimensions
\begin{equation}
    \Delta_-^{\pi_c} = 0, \hspace*{0.5cm} \Delta_+^{\pi_c} = 3, \hspace*{0.5cm} \Delta_{\pm}^{\sigma} = \frac{3}{2} \pm i \mu_{\text{eff}}\,,
\end{equation}
with $\mu_{\text{eff}}^2 = m_{\text{eff}}^2/H^2  - 9/4$, which makes it manifest that the system is composed of two modes corresponding to a massless field ($\Delta^{\pi_c}_{\pm}$) and a decaying massive field ($\Delta^{\sigma}_{\pm}$) with effective mass $m_{\text{eff}}$. 

\paragraph{Flat-space leading behaviour.} At sufficient high energies $\omega\gg H$,\footnote{In the strong mixing regime $\rho\gg H$, the flat-space limit is reached for energies such that $\omega\gg \rho$.} the dynamics of the theory can be approximated by the flat-space limit $H\rightarrow 0$, in which case the system can be solved analytically up to Hubble friction corrections.\footnote{By looking at plane-wave solutions of the form $\pi_c \propto e^{i\omega t}$ and $\sigma \propto e^{i\omega t}$, the Feynman propagator for the fields $\varphi^\alpha=(\pi_c, \sigma)$ is found by inverting the quadratic operator
\begin{equation}
\label{eq:Feynman flat-space full propagators}
    \braket{\text{T}(\varphi_{\bm{k}}^T \varphi_{-\bm{k}})} = \frac{i}{\left(\omega^2 - c_s^2k^2\right)\left(\omega^2 - k^2 - m^2\right) - \rho^2\omega^2 + i\epsilon} 
    \begin{pmatrix}
    \omega^2 - k^2 - m^2 & -i\rho\omega \\
    i\rho\omega & \omega^2 - c_s^2k^2
    \end{pmatrix}\,.
\end{equation}
} Here, we explicitly show that, even if the quadratic mixing cannot a priori be treated perturbatively, the analytical solution of the quadratic theory follows from resumming an infinite number of quadratic mixing insertions in the propagators of the two degrees of freedom. Focusing on the field $\sigma$, its dynamics is affected by the surrounding $\pi_c$ medium that interacts with it, leading to a self-energy correction. At tree-level as we only consider quadratic interactions, it reads
\begin{equation}
    \Pi(\omega, \bm{k}) = 
\raisebox{0pt}{
\begin{tikzpicture}[line width=1. pt, scale=2]
\draw[pyblue] (0.1, 0) -- (0.25, 0);
\draw[pyred] (0.25, 0) -- (0.85, 0);
\draw[pyblue] (0.85, 0) -- (1, 0);
\draw[fill=black] (0.25, 0) circle (.03cm);
\draw[fill=black] (0.85, 0) circle (.03cm);
\end{tikzpicture} 
}
= \frac{-i\rho^2\omega^2}{\omega^2 - c_s^2 k^2 + i\epsilon}\,,
\end{equation}
that is found from the following elementary Feynman rules
\begin{equation}
\begin{aligned}
    G_{\pi_c}^0(\omega, \bm{k}) = 
    \raisebox{2pt}{
\begin{tikzpicture}[line width=1. pt, scale=2]
\draw[pyred] (0.1, 0) -- (0.9, 0);
\end{tikzpicture} 
} &= \frac{i}{\omega^2 - c_s^2k^2 + i\epsilon},\hspace*{1cm}
\begin{tikzpicture}[line width=1. pt, scale=2]
\draw[pyblue] (0.1, 0) -- (0.5, 0);
\draw[pyred] (0.5, 0) -- (0.9, 0);
\draw[fill=black] (0.5, 0) circle (.03cm);
\end{tikzpicture} = -i\rho\omega\,,\\
G_{\sigma}^0(\omega, \bm{k}) = 
    \raisebox{2pt}{
\begin{tikzpicture}[line width=1. pt, scale=2]
\draw[pyblue] (0.1, 0) -- (0.9, 0);
\end{tikzpicture} 
} &= \frac{i}{\omega^2 - k^2 - m^2 + i\epsilon}\,,
\end{aligned}
\end{equation}
where $G_{\pi_c}^0$ (resp. $G_\sigma^0$) is the free $\pi_c$ (resp. $\sigma$) Green two-point function. From the bare propagator, the dressed propagator $G_\sigma(\omega, \bm{k})$ is found by computing the associated Dyson series summing over all one-particle irreducible diagrams---in our case the tree-level contribution. Summing the geometrical series, we have
\begin{equation}
    \begin{aligned}
    G_\sigma(\omega, \bm{k}) &= 
    \raisebox{2pt}{
\begin{tikzpicture}[line width=1. pt, scale=2]
\draw[pyblue] (0.1, 0) -- (0.9, 0);
\end{tikzpicture} 
}+
    \raisebox{0pt}{
\begin{tikzpicture}[line width=1. pt, scale=2]
\draw[pyblue] (0.1, 0) -- (0.35, 0);
\draw[pyred] (0.35, 0) -- (0.85, 0);
\draw[pyblue] (0.85, 0) -- (1.1, 0);
\draw[fill=black] (0.35, 0) circle (.03cm);
\draw[fill=black] (0.85, 0) circle (.03cm);
\end{tikzpicture} 
}+
\raisebox{0pt}{
\begin{tikzpicture}[line width=1. pt, scale=2]
\draw[pyblue] (0.1, 0) -- (0.35, 0);
\draw[pyred] (0.35, 0) -- (0.85, 0);
\draw[pyblue] (0.85, 0) -- (1.1, 0);
\draw[fill=black] (0.35, 0) circle (.03cm);
\draw[fill=black] (0.85, 0) circle (.03cm);
\draw[pyblue] (1.1, 0) -- (1.35, 0);
\draw[pyred] (1.35, 0) -- (1.85, 0);
\draw[pyblue] (1.85, 0) -- (2.1, 0);
\draw[fill=black] (1.35, 0) circle (.03cm);
\draw[fill=black] (1.85, 0) circle (.03cm);
\end{tikzpicture} 
}+\dots \\
&=G_\sigma^0(\omega, \bm{k})\sum_{n=0}^{\infty}\left[\Pi(\omega, \bm{k}) G_\sigma^0(\omega, \bm{k})\right]^n = \frac{i}{\omega^2 - k^2 - m^2 - \rho^2\left[\frac{\omega^2}{\omega^2-c_s^2k^2}\right] + i\epsilon}\,,
    \end{aligned}
\end{equation}
which agrees with~(\ref{eq:Feynman flat-space full propagators}). Non-trivial plane-wave solutions for the propagating modes are found at the poles of the dressed propagator
\begin{equation}
    \omega_\pm^2 = \frac{1+c_s^2}{2}\,k^2 + \frac{m_{\text{eff}}^2}{2} \pm \frac{1}{2}\sqrt{m_{\text{eff}}^4 + \left(1-c_s^2\right)^2k^4 + 2(1+c_s^2)\, k^2\, \left[\rho^2 + \frac{1-c_s^2}{1+c_s^2}\,m^2\right]}.
\end{equation}
We wish to highlight several features of this result by examining various limits. At very high momenta $k\gg\rho,m$, the dispersion relation boils down to $\omega_-^2\approx c_s^2 k^2$ and $\omega_+\approx k^2$ which describes two independent massless free fields, i.e.~the vacuum state in the asymptotic past limit of the inflationary spacetime. In the strong mixing regime $\rho \gg k, m$, one finds
\begin{equation}
    \omega_-^2 \approx \frac{k^2}{\rho^2}\left(c_s^2k^2 + m^2\right), \hspace*{0.5cm} \omega_+^2 \approx m_{\text{eff}}^2\approx \rho^2,
\end{equation}
which describes a massless degree of freedom and a massive degree of freedom with mass $m_{\text{eff}}^2 = m^2+\rho^2$. We note that in the UV limit $k\gg m$, the massless degree of freedom has a modified dispersion relation $\omega\propto k^2$. In the rest of this section, we set $c_s = 1$ to avoid clutter.

\begin{figure}[t!]
   \centering
   \hspace*{0.5cm}
    \includegraphics[width=0.9\textwidth]{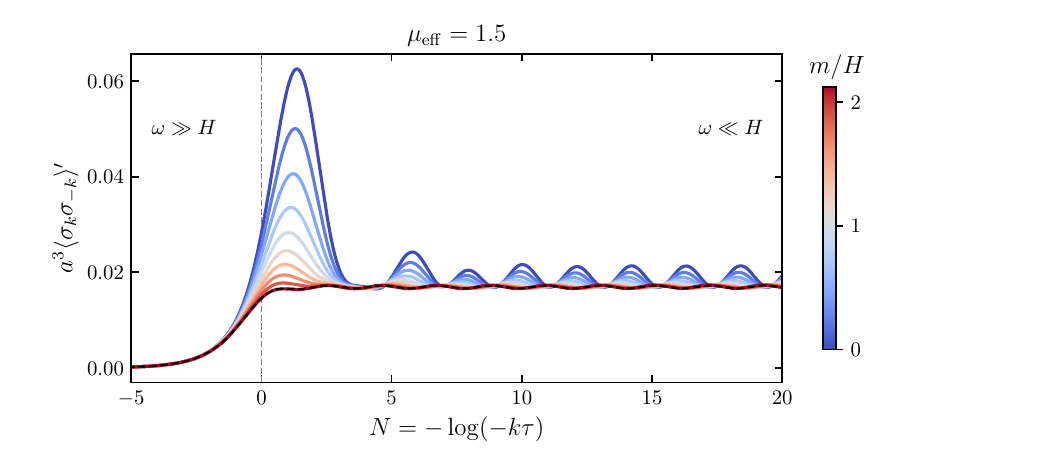}
   \caption{Time evolution of the rescaled $\sigma$ two-point correlation function in Hubble units for a fixed \textit{effective} mass such that $\mu_{\text{eff}} = 1.5$ and $c_s=1$, varying the \textit{bare} mass $m$. The dotted black line corresponds to the analytical prediction in the decoupled regime $\rho/H=0$. The vertical dotted line shows the energy scale at which the correlators cross the horizon $-k\tau=1$.}
  \label{fig:sigma_2pt}
\end{figure}

\paragraph{Resonance at strong mixing.} 
The appearance of the effective mass both in the deep UV and in the late IR suggests that a similar resummation in the intermediate regime---around horizon crossing---is compelling. 
Figure~\ref{fig:sigma_2pt} shows the time evolution of the $\sigma$ two-point correlation function rescaled by $a^3$ for a fixed \textit{effective} mass $\mu_{\text{eff}}=1.5$ while varying the \textit{bare} mass (hence also adjusting the quadratic mixing). At late times $\omega\ll H$, as explained by the previous late-time analysis, one can appreciate that the correlators oscillate at the same frequency set by $\mu_{\text{eff}}$. The amplitude of the oscillations and the mean value of the two-point function grow as the \textit{bare} mass $m$ decreases. Furthermore, another notable feature is that the oscillations present a \textit{phase shift} as the quadratic mixing constant $\rho/H$ increases. This is due to a noticeable phenomenon, namely the amplification of the massive correlator soon after Hubble crossing, which induces a time delay before the late-time asymptotic regime is reached. This effect can be explained by examining the strong mixing regime of the $\pi_c - \sigma$ dynamics:\footnote{We recall that we have set $c_s=1$ so that the following arguments are valid for a unity intrinsic sound speed only.} 

\begin{itemize}
    \item[($i$)] At the \textit{effective mass-shell horizon} $|k\tau|\sim \mu_{\text{eff}}$, the massive field starts to decay as $a^{-3/2}$,
    \item[($ii$)] At the $\rho$\textit{-horizon crossing} $|k\tau|\sim \sqrt{\rho/H}$, the Goldstone boson freezes, as we will see in the next section.
\end{itemize}
At strong quadratic mixing, the event $(i)$ occurs before $(ii)$.\footnote{The reverse case, i.e. $\sqrt{\rho/H}>\mu_{\text{eff}}$, is reached only in a very limited region of the parameter space.} Therefore, when the massive field starts to decay, it transfers its power to the Goldstone boson, creating the $\sqrt{\rho/H}$ amplification of the power spectrum (see next section). Because the dynamics is dominated by the quadratic mixing, the Goldstone boson backreacts on the massive field after horizon crossing $\omega = H$, creating the visible boost in Figure~\ref{fig:sigma_2pt}. The precise dynamics can only be accessed numerically. This interesting phenomenon is also visible in the cosmological collider signal, as we will see in Section~\ref{sec:CC_phycis}.

\subsubsection{Phase diagram} 
\label{subsubsec: phase diagram}

\begin{figure}[h!]
   \centering
    \includegraphics[width=0.5\textwidth]{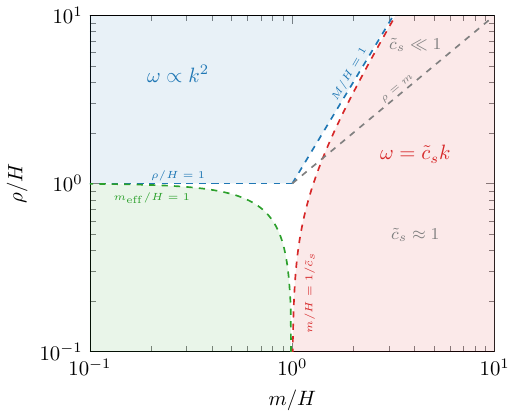}
   \caption{Phase diagram of the quadratic theory in the $(m/H, \rho/H)$ plane, for $c_s=1$, that delimits three distinct regions: (i) the \textcolor{pyblue}{modified dispersion relation} regime ($M\ll H$ and $\rho \gg H$ where $M=m^2/\rho$), (ii) the \textcolor{pyred}{reduced speed of sound} regime ($m\gg H/\tilde{c}_s$ with $\tilde{c}_s^{-2} = 1 + \rho^2/m^2$), and (iii) the \textcolor{pygreen}{light field} regime ($m_{\text{eff}}\ll H$). One should notice that the regime with a significantly reduced speed of sound, $\tilde{c}_s\ll 1$, is rather limited as it requires $\rho\gg m$ in region (ii). This is shown by the grey dashed line $\rho=m$. Of course, the borderlines of these regions should not be taken as exact but rather as faint boundaries: the deeper in the regions, the better the approximations. More details on each regime are given below.}
  \label{fig:Analytic_Phase_Diagram}
\end{figure}

\begin{figure}[t!]
  \centering
  \hspace*{-2.2cm}
  \subfloat{\includegraphics[width=0.6\textwidth]{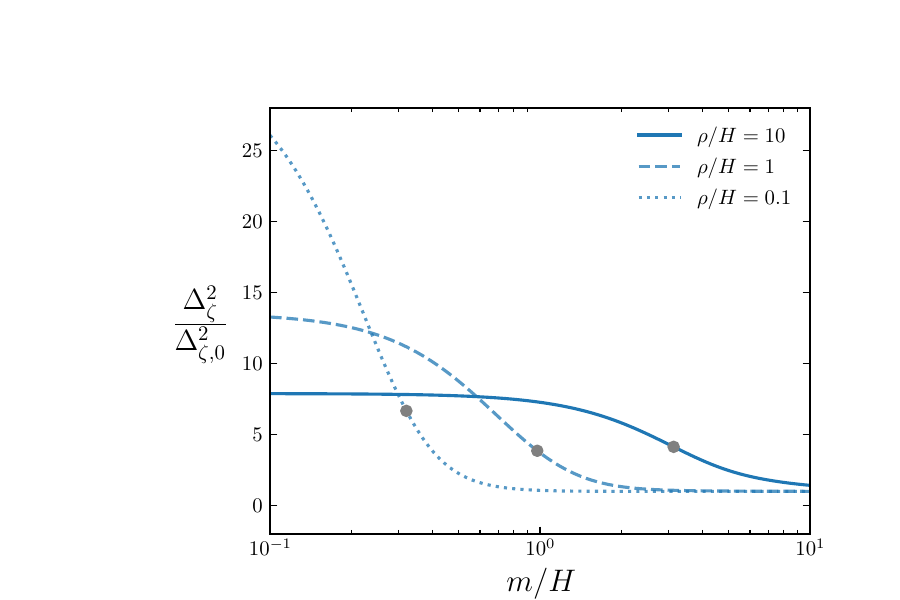}}
  \hspace*{-0.5cm}
  \subfloat{\includegraphics[width=0.6\textwidth]{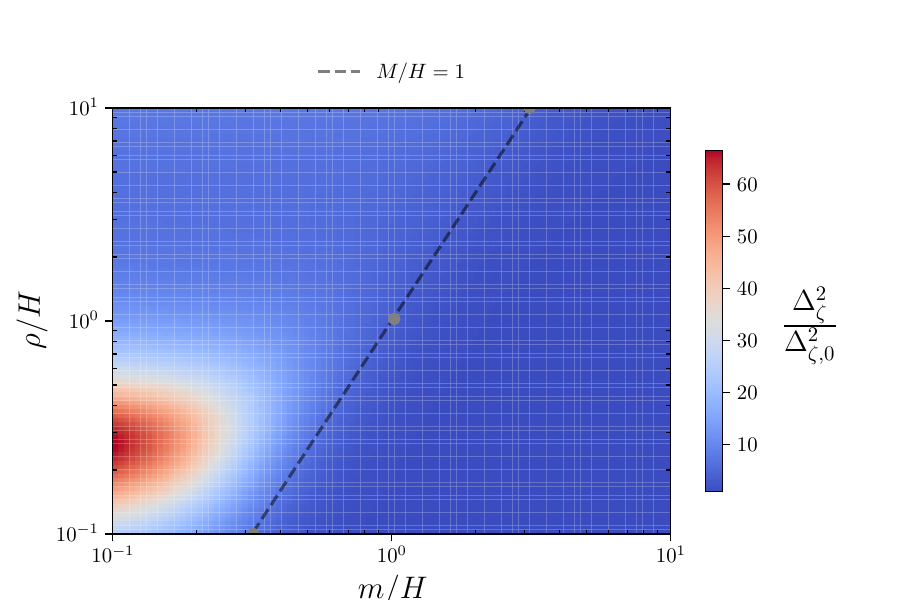}}
  \caption{\textit{Right panel} : Phase diagram of the dimensionless power spectrum $\Delta_\zeta^2$ (in units of $\Delta_{\zeta, 0}^2$). We have set $N=60$ $e$-folds of super-horizon evolution. The dashed line corresponds to the $M=m^2/\rho = H$ delimitation, isolating the modified dispersion relation regime ($M\ll H$ and $\rho \gg H$) and the reduced speed of sound regime ($M\gg H$) (see Figure~\ref{fig:Analytic_Phase_Diagram} for more details). \textit{Left panel} : Slices of the phase diagram for $\rho/H=10$, $\rho/H=1$, and $\rho/H=0.1$ as a function of the bare mass $m/H$ in Hubble units. The grey dots mark the $M/H=1$ delimitation. Note that the the quadratic interaction between $\pi_c$ and $\sigma$ alters the symmetry breaking scale $f_\pi \approx 58H\times (\Delta_\zeta^2/\Delta_{\zeta, 0}^2)^{1/4}$.}
  \label{fig:Gaussian_phase_diagram}
\end{figure}

Fixing $c_s=1$, the quadratic theory depends on two variables $m$ and $\rho$. The entire phase space is not accessible analytically. In Figure~\ref{fig:Gaussian_phase_diagram}, we show the dimensionless power spectrum as a function of $\rho/H$ and $m/H$, computed by solving the flow equations. This result is exact and is consistent with previous studies~\cite{2011Cremoni,Bravo:2020wdr}. To get some physical intuition, one can understand the qualitative behaviour of the quadratic theory in some asymptotic limits that we summarise in Figure~\ref{fig:Analytic_Phase_Diagram}. In the following, we give insights on these regimes.

\paragraph{Modified dispersion relation.} When the mixing is sufficiently strong, the quadratic interaction dominates the dynamics, and one can drop the kinetic terms $\dot{\pi}_c^2$ and $\dot{\sigma}^2$ in~(\ref{eq:Full_pi_sigma_theory}). The equation of motion for the massive field in Fourier space becomes $(k^2/a^2 + m^2)\sigma = \rho \dot{\pi}_c$, which, if we also neglect the mass term $m^2 \ll k^2/a^2$ at relevant times, simplifies to 
\begin{equation}
\label{eq: MDR constrain equation}
    \frac{k^2}{a^2}\,\sigma = \rho\dot{\pi}_c\,.
\end{equation}
In this case, the equation of motion for the Goldstone boson is found to be
\begin{equation}
    \label{eq: EOM in MDR regime}
    \ddot{\pi}_c + 5H\dot{\pi}_c +  \frac{k^4}{\rho^2 a^4} \pi_c = 0\,. 
\end{equation}
Therefore, the theory can be described by an effective single-field theory---because the field $\sigma$ is non-propagating i.e.~its equation of motion is a pure constraint---with a modified (quadratic) dispersion relation $\omega = k^2/\rho$, inferred from~(\ref{eq: EOM in MDR regime}). Let us now derive the precise conditions of validity of this regime.
First, we have considered that the quadratic mixing $\rho \dot{\pi}_c \sigma$ dominates over the kinetic term $\dot{\sigma}^2$. Using the constraint~(\ref{eq: MDR constrain equation}) and the dispersion relation, this condition translates into the following bound
\begin{equation}
    \frac{\rho}{H} \gtrsim 1\,.
\end{equation}
Imposing that the mixing term also dominates the kinetic term $\dot{\pi}_c^2$ leads to the same bound.\footnote{We work here in the limit $c_s=1$. However, in the general case, the most stringent bound is instead found to be $\rho/H \gtrsim 1/c_s$~\cite{Jazayeri:2023xcj}.} Similarly, neglecting the mass term $m^2\sigma^2$ leads to\footnote{In the general case with $c_s<1$, the bound is found to be $c_s m^2/\rho \lesssim H$~\cite{Jazayeri:2023xcj}.}
\begin{equation}
    M \equiv \frac{m^2}{\rho} \lesssim H\,,
\end{equation}
where we see the appearance of a new energy scale $M$. These two bounds define the modified dispersion relation regime, as depicted in Figure~\ref{fig:Analytic_Phase_Diagram}. In this regime, the field $\sigma$ plays the role of the conjugate momentum of the massless field $p_{\pi_c} \approx \rho\sigma$, and the effective single-field quadratic Lagrangian becomes
\begin{equation}
\label{eq:MDR_mode_function}
    \mathcal{L}^{(2)}/a^3 \approx 
    \rho \dot{\pi}_c\sigma-\frac{1}{2}  \frac{(\partial_i \pi_c)^2}{a^2} - \frac{1}{2}\frac{(\partial_i \sigma)^2}{a^2}\,,
\end{equation}
which leads to the equation of motion~(\ref{eq: EOM in MDR regime}) after using~(\ref{eq: MDR constrain equation}). This can be solved exactly, yielding the following mode function for the light field after imposing the Bunch-Davies vacuum~\cite{Baumann:2011su}
\begin{equation}
    \pi_{c, k}(\tau) = \sqrt{\frac{\pi}{4}}\frac{H}{\rho}\frac{H}{\sqrt{2k^3}}(-k\tau)^{5/2} H_{5/4}^{(1)}\left(\frac{1}{2}\frac{H}{\rho}(k\tau)^2\right)\xrightarrow{k\tau\rightarrow0}e^{-i\frac{\pi}{2}}\,\frac{2\Gamma(5/4)}{\sqrt{\pi}} \frac{H}{k^{3/2}}\left(\frac{\rho}{H}\right)^{1/4}\,.
\end{equation}
From the analytical expression of the mode function, or simply evaluating the non-linear dispersion relation at $\omega\sim H$, one observes the emergence of a new time scale in this regime, given by $|k\tau|\sim \sqrt{\rho/H} \gtrsim 1$, hence the $\pi_c$ mode freezes before the usual horizon crossing $|k\tau|\sim 1$.
This regime leads to the amplification of the power spectrum
\begin{equation}
\label{eq:MDR_power_spectrum}
    \Delta_\zeta^2 = \frac{2\Gamma(5/4)^2}{\pi^3} \left(\frac{H}{f_{\pi}}\right)^4 \left(\frac{\rho}{H}\right)^{1/2}\,.
\end{equation}
We numerically confirm this expected asymptotic behaviour as $M\ll H$, and the fact that the power spectrum becomes independent of $m/H$, hence reaching a plateau in the left panel of Figure~\ref{fig:Gaussian_phase_diagram} for $\rho/H=10$.

\paragraph{Reduced speed of sound.} When the bare mass of $\sigma$ becomes large $m\gg H$ while maintaining $M\gg H$, the field $\sigma$ can be integrated out by setting it on-shell\footnote{We will come back to the regime of validity of this regime and give more details in Section \ref{subsec:features_setup}.} $(-\Box+m^2)\sigma = \rho\dot{\pi}_c$ \cite{Tolley:2009fg, Achucarro:2012yr}. The quadratic Lagrangian then inherits the imprint of this heavy mode in an infinite number of irrelevant operators which---at leading order---induces the well-known speed of sound\footnote{The Goldstone boson can in principal have an additional speed of sound $c_s$ governed by the coefficient $M_2$ in (\ref{eq:unitary_gauge_action}). For simplicity, this \textit{intrinsic} speed of sound has been set to unity in this section. However, note that the total speed of sound can be further reduced when the heavy field is integrated out: $\tilde{c}_s^{-2} = c_s^{-2}(1 + \rho^2/m^2)$.} for $\pi_c$
\begin{equation}
\label{eq:induced_sound_speed}
    \tilde{c}_s^{-2} = 1 + \frac{\rho^2}{m^2}\,.
\end{equation}
In this effective theory, the massless mode crosses the sound horizon $|\tilde{c}_s k\tau|\sim 1$ and freezes while its dispersion relation is linear $\omega=\tilde{c}_sk$. The power spectrum becomes
\begin{equation}
\label{eq:power_spectrum_reduced_sound_speed}
    \Delta_{\zeta}^2 =  \frac{1}{4\pi^2\tilde{c}_s}\left(\frac{H}{f_\pi}\right)^4\,.
\end{equation}
We will see in Section \ref{subsec:features_setup} that the actual regime of validity of this regime is $m\gg H/\tilde{c}_s$. However, as depicted in Figure~\ref{fig:Analytic_Phase_Diagram}, the actual region in the phase diagram where $\tilde{c}_s\ll 1$, which typically leads to enhanced equilateral non-Gaussianities, is limited.

\vskip 4pt
Although one might naively think that $\rho\lesssim H$ is the correct condition on the quadratic coupling to have weak mixing, it has been shown that this condition can be extended to $\rho \lesssim m$ for $m\gtrsim H$, see~\cite{Jazayeri:2022kjy,Jazayeri:2023xcj}. In this regime, $\rho<m$, that we will from now on refer to as the weak mixing, one finds that the correction to the power spectrum scales as $\rho^2/m^2$. This correction has also been computed in \cite{Chen:2009zp, Pi:2012gf} in the full two-field theory for $c_s = 1$, see also~\cite{Lee:2016vti} for numerical and~\cite{Jazayeri:2022kjy} for analytical computations in the case of an \textit{intrinsic} reduced speed of sound $c_s <1$. At strong mixing $\rho\gtrsim m$, the speed of sound is significantly reduced and the power spectrum scales as $\rho/m$.

\vskip 4pt
The mass dependency of the correction, scaling as $(H/m)^2$, reflects the imprint of the heavy field $\sigma$. Note that it also induces higher-dimensional operators suppressed by additional powers of $(H/m)^2$. We numerically confirm this mass dependence and that this scaling extends to would-be non-perturbative values of the quadratic mixing $\rho/H>1$. In fact, we find that the perturbative regime holds up to $\rho \sim m > H$, as can be understood by the emergence of the effective mass \eqref{eq:def-effective-mass} when going beyond perturbative treatment of the quadratic mixing, and as also shown by the regime of validity of the expansion in~(\ref{eq:power_spectrum_reduced_sound_speed}).  

\vskip 4pt
We will show in Section~\ref{sec:NG_pheno} that these two regimes, of a modified dispersion relation and of a reduced speed of sound, can be captured by a unified effective theory where the heavy field is integrated out in a (spatially) non-local manner.

\paragraph{Light additional field.} Our numerical analysis is also applicable in the light $\sigma$ field regime where $m_{\text{eff}}\lesssim H$.
In this regime, the field $\sigma$ decays at a very slow pace on super-horizon scales. The massless field $\pi_c$ is therefore sourced by $\sigma$ on super-horizon scales so that the value of $\Delta_\zeta^2$ is sensitive to the amount of $e$-folds elapsed from the moment at which the mode crosses the horizon and the end of inflation. The corresponding enhancement of the power spectrum is well visible in the lower left part of the phase diagram in Figure~\ref{fig:Gaussian_phase_diagram}. In Appendix \ref{app:LightField}, we derive the following analytical approximation that reproduces this amplification
\begin{equation}
    \frac{\Delta_\zeta^2}{\Delta_{\zeta, 0}^2} -1 \propto \left[\frac{3\rho H}{m_{\text{eff}}^2}\left(1 - e^{-\frac{m_{\text{eff}}^2}{3H^2} \Delta N}\right)\right]^2\,,
\end{equation}
where $\Delta N$ is the number of $e$-folds elapsed since horizon crossing.

\section{Strong Coupling Scales and Naturalness}
\label{sec:strong_coupling_scales}

It is important to ensure that the effective description of fluctuations that we study is under theoretical control and identify the natural hierarchy of scales of the system. Because the theory contains non-renormalisable interactions, the theory will become strongly coupled at a certain energy scale.\footnote{This scale can be derived by finding the maximum energy at which the tree-level scattering of the fields is unitary, through partial wave expansion \cite{Grall:2020tqc}. Here, we give an heuristic derivation of these strong coupling scales based on power counting which, of course, misses order-one factors.} In this section, we derive the strong coupling scale, or unitarity bound, in the weak and strong quadratic mixing regimes. We gather all the found bounds in Appendix \ref{app:strong_coupling_scales_tabulars}. These bounds translate into bounds on the size of non-Gaussianities, see Section~\ref{sec:NG_pheno}, but only give a parametric estimation, possibly missing large dimensionless numbers.

\paragraph{Goldstone boson sector.} The effective description of primordial fluctuations in (\ref{eq:Full_pi_sigma_theory}) is valid up to some UV cutoff scale $\Lambda_\star$, called the strong coupling scale, beyond which the perturbative description breaks down. When $c_s = 1$, analogous to the chiral Lagrangian describing the dynamics of pions, the strong coupling scale of the effective theory is $\Lambda_\star = 4\pi f_\pi$.\footnote{After reintroducing the Goldstone boson, the action (\ref{eq: Goldstone boson Lagrangian with pi}) without truncating to a certain order in fields is an expansion in $\pi_c/f_\pi$ with $\pi_c\sim \omega$, and higher dimensional operators become relevant if the effective dimensionless coupling constant $\omega/f_\pi$ is of order one.} However, when the sound speed $c_s$ is sufficiently small, it is well-known that $\Lambda_\star$ is not too far from the Hubble scale. Indeed, it is given by~\cite{Cheung:2007st, Baumann:2011su}\footnote{This strong coupling scale is associated with the operator $(\partial_i \pi_c)^4$.} 
\begin{equation}
\label{eq:Goldstone_boson_strong_coupling_scale}
\Lambda_\star^4 = 2\pi f_\pi^4 \,\frac{c_s^4}{1 - c_s^2}\,,
\end{equation}
which, in the limit $c_s\ll 1$, gives $\Lambda_\star\sim c_s f_\pi$. Imposing $\Lambda_\star>H$ for the EFT to be valid---otherwise fluctuations cannot be initialised in the proper vacuum inside the horizon---and assuming the single-field limit (i.e.~$\rho/H=0$) gives a theoretical lower bound on the sound speed $c_s\geq 0.01$.\footnote{Assuming pure single-field $c_s$-theory, Planck constraints on primordial non-Gaussianities give a more stringent lower bound on the speed of sound: $c_s\geq 0.021$ ($95\%$ CL) \cite{Planck:2019kim}.} The additional degree of freedom $\sigma$ changes the physical description, and interactions between $\pi_c$ and $\sigma$ will modify $\Lambda_\star$. Yet, the quadratic mixing $\dot{\pi}_c\sigma$ makes the analysis more subtle. Therefore, we examine separately the weak mixing regime and the modified dispersion relation regime---in the following loosely denoted as strong mixing---corresponding to the \textcolor{pyblue}{blue} region in the phase diagram \ref{fig:Analytic_Phase_Diagram}.

\subsection{Weak Mixing}
\label{subsec: weak mixing}

In the weak mixing regime, the quadratic theory is dominated by the standard kinetic terms. The free fields can then be considered decoupled as a first approximation. Usually, the strong coupling scale $\Lambda_\star$ is derived from operators constructed out of canonically-normalised fields with a Lorentz invariant kinetic term. Here, the non-relativistic dispersion relation of $\pi_c$ makes it impossible to put time and space on the same footing, so that one needs to consider energy and momentum separately.\footnote{In single-field theories, up to a redefinition of spacetime coordinates, fake Lorentz invariance can be restored and standard dimensional analysis can then be applied, see~\cite{Baumann:2011su}. However, this trick does not work in our situation with several degrees of freedom propagating at different sound speeds.}

\vskip 4pt
Let us first consider the cubic interaction coming from $\delta g^{00}\sigma$ in the unitary gauge, so that the action we examine---together with the quadratic part but neglecting the quadratic mixing---reads
\begin{equation}
    S \supset \frac{1}{2}\int \mathrm{d}t\,\mathrm{d}^3x \,a^3 \left[\dot{\pi}_c^2 - c_s^2 \frac{(\partial_i \pi_c)^2}{a^2} + \dot{\sigma}^2 - \frac{(\partial_i \sigma)^2}{a^2} - m^2\sigma^2 - c_s^{3/2} \frac{\rho}{f_\pi^2}\frac{(\partial_i \pi_c)^2}{a^2} \sigma\right]\,.
\end{equation}
In order to estimate how various terms in this action scale with energy and momentum, we need to chose a reference. Let us fix the momentum $k$ carried by the propagating modes, whose energies should be understood as derived concepts depending on their dispersion relations. The energy of $\pi_c$ then reads $\omega_\pi = c_s k$, whereas the energy of $\sigma$ is $\omega_\sigma = k$, from which we deduce that $\omega_\pi=c_s \omega_\sigma$. In the following, we will express all quantities in terms of the energy of the Goldstone boson $\omega\equiv\omega_\pi$ because it is the energy probed experimentally at the Hubble scale $\omega\sim H$. Therefore, the kinetic term $\dot{\pi}_c^2$ scales as $\sim \omega^2\pi_c^2$---so does the gradient term---, and the kinetic term $\dot{\sigma}^2$ scales as $\sim \omega_\sigma^2\sigma^2 \sim c_s^{-2}\omega^2\sigma^2$, and likewise for the gradient term. Yet, a question needs to be answered. What is the typical size of the fluctuations? By dimensional analysis, $\pi_c$ scales as $\omega^{-1/2} k^{3/2}\sim c_s^{-3/2}\omega$. Similarly, $\sigma$ scales as $\omega_\sigma \sim c_s^{-1}\omega$. Combining all scalings, we obtain that the Goldstone boson kinetic term scales as $\sim c_s^{-3}\omega^4$, and that of $\sigma$ scales as $c_s^{-4}\omega^4$. We notice that, at small sound speed, the kinetic term of $\sigma$ is the dominant one in the quadratic action.\footnote{Note also that this hierarchy is reversed when the massive fluctuations propagate slower than the Goldstone boson fluctuations, $c_\sigma<c_\pi$, a situation discussed in Section~\ref{subsec: Sound Speed Collider}.} No strong coupling scale is associated with the quadratic mixing operator $\dot{\pi}_c\sigma$ because it is of dimension 3. However, the theory is under control in the weak mixing regime if this interaction can be treated perturbatively. Requiring that this interaction---that scales as $c_s^{-5/2}\rho\,\omega^3$---is smaller than the kinetic term of the Goldstone boson gives
\begin{equation}
\label{eq: bound rho weak mixing}
    \frac{\rho}{H}\lesssim c_s^{-1/2}\,.
\end{equation}
Now, the cubic interaction scales as $c_s^{-9/2}\rho/f_\pi^2\,\omega^5$. Requiring that this term should be smaller than the kinetic term of the Goldstone boson---giving the more stringent bound---sets the strong coupling scale to\footnote{Let us comment on the mass term $m^2\sigma^2$, that we neglected so far. This term changes the dispersion relation of $\sigma$ to $\omega_\sigma^2 = k^2 + m^2$, which leads to $\omega_\sigma^2 = \omega^2/c_s^2 + m^2$. The kinetic term of $\sigma$ including this mass term therefore scales as $\sim (\omega^2/c_s^2 + m^2)\sigma^2$. The typical size of $\sigma$ is found to be $\sigma \sim \omega/c_s\, \left[1 + (c_s m/\omega)^2\right]^{-1/4}$. In the end, the kinetic term of $\sigma$ scales as $\sim c_s^{-4}\omega^4\, \left[1 + (c_s m/\omega)^2\right]^{1/2}$. Note that we recover the scaling presented in the main text in the limit $\omega \gg c_s m$. Fixing the energy scale to the Hubble scale, we deduce that the mass term is negligible when $m/H\ll c_s^{-1}$, corresponding to the low-speed collider regime~\cite{Jazayeri:2022kjy,Jazayeri:2023xcj}. Still, we note that properly taking into account the mass term increases the size of the kinetic term of $\sigma$. For $c_s < 1$ and including the mass of $\sigma$, the smallest contribution of the quadratic theory is still the kinetic term of the Goldstone boson. Therefore, we can safely discard the mass term in our analysis because it would play no role in the derivation of the strong coupling scales.}
\begin{equation}
    \Lambda_\star = c_s^{3/2}\, \frac{f_\pi^2}{\rho}\,.
\end{equation}
Imposing $H \lesssim \Lambda_\star$ leads to $\rho/H\lesssim c_s^{3/2}/2\pi\Delta_\zeta$. This bound is less stringent that the previous one, and therefore~(\ref{eq: bound rho weak mixing}) defines the weak mixing regime. Additionally, it has been shown that $\rho \lesssim m$ is also required in order to treat the quadratic mixing perturbatively, see Section~\ref{subsubsec: phase diagram}. Depending on the value of $c_s$, this bound can be stronger. 

\vskip 4pt
Next, we consider the interactions coming from $\delta g^{00}\sigma^2$. After introducing the Goldstone boson, the interactions we examine are
\begin{equation}
    S\supset \frac{1}{2}\int \mathrm{d}t\,\mathrm{d}^3x\, a^3\left[-\alpha \dot{\pi}_c\sigma^2 + \frac{c_s^{3/2}}{2}\frac{\alpha}{f_{\pi}^2} (\partial_\mu \pi_c)^2\sigma^2\right]\,.
\end{equation}
The interaction $\dot{\pi}_c\sigma^2$ is of dimension 4 so no strong coupling scale is associated with it. However, requiring that the size of this term in the action be smaller than the kinetic term of $\pi_c$ to be in the perturbative regime leads\footnote{In the standard Lorentz invariant case, the perturbative criteria on marginal operators is that the dimensionless couplings have to be less than unity. Here, the non-relativistic dispersion relation makes things more subtle, introducing a sound speed dependence in the perturbative criterion.} to 
\begin{equation}
    \alpha\lesssim c_s^{1/2}\,,\hspace*{0.5cm} \text{i.e.} \hspace*{0.5cm} \tilde{\alpha}^2\lesssim c_s^{-1}/4\times(2\pi \Delta_\zeta)^{-1}\,.
\end{equation}
Repeating the power counting for the dimension-6 operator gives the following strong coupling scale\footnote{Note that, for $c_s\ll 1$, the gradient term in $(\partial_\mu \pi_c)^2$ dominates over the $\dot{\pi}_c^2$ term, hence giving the strong coupling scale by requiring that it should be smaller than the kinetic term of $\pi_c$. However, if $c_s>1$, the hierarchy is reversed and the strong coupling scale is found by comparing the time derivative interaction term to the kinetic term of $\sigma$. This gives $\Lambda_\star = c_s^{-1/4}f_\pi / \sqrt{\alpha}$.}
\begin{equation}
    \Lambda_\star = c_s^{5/4} \frac{f_\pi}{\sqrt{\alpha}}\,.
\end{equation}
Requiring that primordial fluctuations are weakly coupled at the Hubble scale $\Lambda_\star>H$ gives the constraint $\alpha \lesssim \frac{c_s^{5/4}}{2\pi\Delta_\zeta}$, equivalently $\tilde{\alpha}^2 \lesssim c_s^{-1/4}\,(4\pi \Delta_\zeta)^{-2}$, on the coupling constant, which is less stringent than the previous one. 

\vskip 4pt
Lastly, we consider the interactions coming from $\left(\delta g^{00}\right)^2\sigma$, which in terms of the Goldstone boson gives the following interactions
\begin{equation}
\label{eq:interaction_deltag002_sigma}
\begin{aligned}
    S\supset \frac{1}{2}\int \mathrm{d}t\,\mathrm{d}^3x\, a^3&\left[-\frac{1}{\Lambda_2}\dot{\pi}_c^2\sigma + \frac{c_s^{3/2}}{f_\pi^2}\dot{\pi}_c\frac{(\partial_\mu \pi_c)^2}{\Lambda_2}\sigma - \frac{c_s^3}{4f_\pi^4}\frac{1}{\Lambda_2}\dot{\pi}_c^4\sigma \right.\\
    &\left.+\frac{c_s^3}{2f_\pi^4}\frac{1}{\Lambda_2}\dot{\pi}_c^2 \frac{(\partial_i \pi_c)^2}{a^2}\sigma - \frac{c_s^3}{4f_\pi^4}\frac{1}{\Lambda_2}\frac{(\partial_i \pi_c)^4}{a^4}\sigma\right]\,,
\end{aligned}
\end{equation}
where we have defined $\Lambda_2^{-1} = 8c_s^3 \tilde{M}_3^3/f_\pi^4$. Note that this definition differs from (\ref{eq:definition_couplings}). The reason is that the interaction $\dot{\pi}_c^2\sigma$ is also generated by $\delta g^{00}\sigma$, so that its coupling constant depends on both $\tilde{M}_1$ and $\tilde{M}_3$. Here, we isolate interactions coming from $\left(\delta g^{00} \right)^2\sigma$ only, effectively setting $\tilde{M}_1 = 0$. Estimating the size of all the interactions in (\ref{eq:interaction_deltag002_sigma}) gives several strong coupling scales. The smallest one is associated with the interaction $\dot{\pi}_c^2\sigma$ and reads
\begin{equation}
    \Lambda_\star = c_s \Lambda_2\,.
\end{equation}
Requiring $H<\Lambda_\star$ gives the bound $H/\Lambda_2 \lesssim c_s$, equivalently $\tilde{\rho}/H\lesssim c_s^{-1/2}(2\pi \Delta_\zeta)^{-1}$. 

\vskip 4pt
Finally, the operator $\sigma^3$ being relevant and the field $\sigma$ having a relativistic dispersion relation, the criteria for a perturbative treatment of this interaction is simply $\mu/H\lesssim 1$.

\subsection{Strong Mixing} 

At large quadratic mixing, as discussed in Section \ref{subsec:quadratic_theory}, the mixing term $\dot{\pi}_c\sigma$ dominates over the kinetic terms $\dot{\pi}_c^2$ and $\dot{\sigma}^2$. We recall that, in the general case with $c_s\leq1$, the condition of validity for this regime is \cite{Jazayeri:2023xcj}
\begin{equation}
    \frac{\rho}{H} \gtrsim c_s^{-1} \,, \hspace*{0.5cm} \text{and} \hspace*{0.5cm} c_s\, \frac{m^2}{\rho} \lesssim H\,.
\end{equation}
The theory being effectively described by a single degree of freedom, it is possible to rescale the spatial coordinates and the fields to canonically normalise the propagating degree of freedom. This is completely equivalent to the power counting method used previously. By performing the rescaling $x\rightarrow \tilde{x}=\rho^{1/2}c_s^{-1/2}x$, $\pi_c\rightarrow \tilde{\pi}_c=\rho^{-1/4} c_s^{5/4}\pi_c$, and $\sigma\rightarrow \tilde{\sigma} = \rho^{-1/4} c_s^{1/4}\sigma$, we obtain for the interactions coming from $\delta g^{00}\sigma$---and the quadratic part but neglecting the time derivative terms---the following action
\begin{equation}
    S \supset \frac{1}{2}\int \mathrm{d}t\,\mathrm{d}^3\tilde{x} \,a^3 \left[\dot{\tilde{\pi}}_c\tilde{\sigma} - \frac{(\tilde{\partial}_i \tilde{\pi}_c)^2}{a^2} - \frac{(\tilde{\partial}_i \tilde{\sigma})^2}{a^2} - M^2\tilde{\sigma}^2 - c_s^{-3/4} \frac{\rho^{5/4}}{f_\pi^2}\frac{(\tilde{\partial}_i \tilde{\pi}_c)^2}{a^2} \tilde{\sigma}\right]\,,
\end{equation}
where $M \equiv c_s \tfrac{m^2}{\rho}$. By dimensional analysis, we have $[\tilde{\partial}] = 1/2$ and $[\tilde{\pi}_c] = [\tilde{\sigma}] = 3/4$ in mass units. This implies that the strong coupling scale is
\begin{equation}
    \Lambda_\star = c_s\left(\frac{f_\pi^8}{\rho^5}\right)^{1/3}\,,
\end{equation}
which generalises the strong coupling scale found in \cite{Baumann:2011su} for the $c_s=1$ case. Using (\ref{eq:MDR_power_spectrum}) and imposing $H \lesssim \Lambda_\star$, we obtain the following constraint\footnote{Note that imposing the validity of the decoupling limit---that is neglecting the mixing with gravity through terms like $\delta g^{00} \dot{\pi}$---also gives a lower bound on $\rho/H$.}
\begin{equation}
    \frac{\rho}{H}\lesssim c_s^{3/4}\frac{\kappa^{1/2}}{\Delta_\zeta}\,,
\end{equation}
where $\kappa = 2\Gamma(5/4)^2/\pi^3\approx 0.053$. Importantly, this bound shows that the theory can be strongly mixed without being strongly coupled.

\vskip 4pt
Let us now consider the interactions coming from $\delta g^{00}\sigma^2$. After performing the same rescaling, the action we examine is
\begin{equation}
    S\supset -\frac{1}{2}\int \mathrm{d}t\,\mathrm{d}^3\tilde{x}\,a^3\left[c_s^{-1/4} \alpha \rho^{-3/4} \dot{\tilde{\pi}}_c \tilde{\sigma}^2 + \frac{1}{2}\frac{\alpha}{f_\pi^2}\left(\rho^{-1/2}\dot{\tilde{\pi}}_c^2 \tilde{\sigma}^2 - c_s^{-1}\rho^{1/2} \frac{(\tilde{\partial}_i \tilde{\pi}_c)^2}{a^2}\tilde{\sigma}^2\right)\right]\,.
\end{equation}
We notice that the would-be Lorentz invariant interaction $(\partial_\mu \pi_c)^2\sigma^2$ is broken due to the rescaling of spatial (but not time) coordinates. Therefore, we expect that different strong coupling scales will be associated with $\dot{\tilde{\pi}}_c^2 \tilde{\sigma}^2$ and $(\tilde{\partial}_i \tilde{\pi}_c)^2 \tilde{\sigma}^2$. The interaction $\dot{\tilde{\pi}}_c\tilde{\sigma}^2$ can be treated as a perturbative correction to the quadratic action as long as 
\begin{equation}
\label{eq:perturbativity_alpha_MDR}
    \alpha \lesssim c_s^{1/4} \left(\frac{\rho}{H}\right)^{3/4}\,, \hspace*{0.5cm} \text{i.e.} \hspace*{0.5cm} \tilde{\alpha}^2\lesssim \frac{c_s^{-1/2}}{4} \frac{\kappa}{\Delta_\zeta^2} \,.
\end{equation}
The smallest strong coupling scale---associated with the interaction $(\tilde{\partial}_i \tilde{\pi}_c)^2 \tilde{\sigma}^2$---is
\begin{equation}
    \Lambda_\star = \left(\frac{c_s}{\alpha}\right)^{2/3} \left(\frac{f_\pi^4}{\rho}\right)^{1/3}\,,
\end{equation}
which gives the following constraint to be in the perturbative regime at energies $\omega\sim H$
\begin{equation}
\label{eq:perturbativity_alpha2_MDR}
    \alpha\lesssim c_s \left(\frac{\rho}{H}\right)^{-1/2} \left(\frac{f_\pi}{H}\right)^2\,, \hspace*{0.5cm} \text{i.e.} \hspace*{0.5cm} \tilde{\alpha}^2 \lesssim \frac{c_s^{-1/2}}{4} \frac{\kappa}{\Delta_\zeta^2}\,.
\end{equation}
Interestingly, both conditions~(\ref{eq:perturbativity_alpha_MDR}) and~(\ref{eq:perturbativity_alpha2_MDR}) give the same perturbativity bounds. Therefore, it is taken to be the criteria for a consistent perturbative treatment of all interactions coming from $\delta g^{00}\sigma^2$. 

\vskip 4pt
Next, we consider the interactions coming from $(\delta g^{00})^2\sigma$. Following the same strategy as we did previously, we find that the smallest strong coupling scale is the one associated with the interaction $\dot{\tilde{\pi}}_c^2\sigma$, which is
\begin{equation}
    \Lambda_\star = \left(c_s^5 \rho^3 \Lambda_2^4\right)^{1/7}\,.
\end{equation}
Requiring $H\lesssim \Lambda_\star$ gives us the following condition 
\begin{equation}
    \frac{H}{\Lambda_2}\lesssim c_s^{5/4} \left(\frac{\rho}{H}\right)^{3/4}\,, \hspace*{0.5cm} \text{i.e.} \hspace*{0.5cm} \frac{\tilde{\rho}}{H}\lesssim c_s^{1/2}\frac{\kappa}{\Delta_\zeta^2}\,.
\end{equation}
Lastly, the condition for perturbativity of the interaction $\sigma^3$ is found to be $\mu/H \lesssim c_s^{-3/4} \left(\rho/H\right)^{3/4}$.\footnote{We have checked that the requirement that no group velocity exceeds the speed of light gives no further constraints on the parameters.}

\subsection{Radiative Corrections}

Finally, we derive bounds on the size of the coupling constants based on the requirement that the theory is stable under radiative corrections, hence requiring the theory to be \textit{technically natural}. These bounds are not strict, and rely on several assumptions: (i) we only require that the mass of the field $\sigma$ does not receive large loop corrections\footnote{Similar bounds were derived in \cite{Baumann:2011nk, Lee:2016vti}. Of course, all couplings receive corrections under renormalisation but the entire analysis is beyond the scope of this paper.} $\delta m^2\lesssim m^2\sim H^2$, (ii) we restrict ourselves to one-loop order, therefore neglecting self-interactions of $\pi_c$. Because the UV-divergent diagrams are dominated by the UV dynamics of the system $\omega>H$, the fields---even at strong mixing provided that the loops mostly receive contributions for energies $\omega>\rho$---are considered decoupled. Moreover, this enables us to use flat-space propagators already encountered in Section~\ref{subsub: Effective mass and resonance at strong mixing}. Divergent diagrams are cut off (in energy) at the scale where the theory breaks down, namely the strong coupling scale associated with the Goldstone sector $\Lambda_\star\sim c_s f_\pi$.\footnote{In principle, the UV-divergent integrals should be cut off at the most stringent strong coupling scale, see Section~\ref{subsec: weak mixing}. However, these scales are strongly parameter dependent, and we consider to set the UV-cutoff to $\Lambda_\star\sim c_s f_\pi$ for definiteness~\cite{Baumann:2011nk}. Note that in the case of a reduced sound speed $c_s\ll 1$, it is the most stringent bound. Let us reiterate that the naturalness conditions we derive here are contingent upon several assumptions, and therefore, they should not be regarded as absolute.}

\vskip 4pt
$\bullet$ The interaction $\sigma^3$ generates the following mass shift
\begin{equation}
    \delta m^2 = 
\vcenter{\hbox{
\begin{tikzpicture}[line width=1. pt, scale=2]
\draw[pyblue] (0.1, 0) -- (0.5, 0);
\draw[fill=none, pyblue] (0.7, 0) circle (.2cm);
\draw[pyblue] (0.9, 0) -- (1.3, 0);
\draw[fill=black] (0.5, 0) circle (.03cm);
\draw[fill=black] (0.9, 0) circle (.03cm);
\end{tikzpicture}}} \sim \mu^2 \int \frac{\mathrm{d}\omega\mathrm{d}^3k}{(\omega^2 - k^2 - m^2)^2} \sim \mu^2\,,
\end{equation}
hence giving a finite loop correction. Naturalness of the mass of the field $\sigma$ therefore requires $\mu\lesssim H$.

\vskip 4pt
$\bullet$ Let us now consider the mass correction arising from the interaction fixed by the non-linearly realised symmetry $(\partial_i \pi_c)^2\sigma$. It reads
\begin{equation}
\label{eq: mass correction from rho}
    \delta m^2 = 
\vcenter{\hbox{
\begin{tikzpicture}[line width=1. pt, scale=2]
\draw[pyblue] (0.1, 0) -- (0.5, 0);
\draw[fill=none, pyred] (0.7, 0) circle (.2cm);
\draw[pyblue] (0.9, 0) -- (1.3, 0);
\draw[fill=black] (0.5, 0) circle (.03cm);
\draw[fill=black] (0.9, 0) circle (.03cm);
\end{tikzpicture}}} \sim c_s^{3}\,\frac{\rho^2}{f_\pi^4} \int\mathrm{d}\omega\mathrm{d}^3k\, \frac{k^4}{(\omega^2 - c_s^2k^2)^2} \sim \rho^2\,,
\end{equation}
where we have set the cutoff in energy to $\Lambda_\omega = c_s f_\pi$ and the cutoff in momentum to $\Lambda_k = c_s^{-1}\Lambda_\omega$ to take into account the non-relativistic dispersion relation of $\pi_c$. 
Hence, the naturalness criterion we consider requires the quadratic mixing to be weak $\rho/H\lesssim 1$. We see that a large quadratic mixing \textit{and} a small mass---and consequently the modified dispersion relation regime---is \textit{not} technically natural, and can only be reached with fine-tuning.

\vskip 4pt
$\bullet$ Next, we consider the interaction $\dot{\pi}_c^2\sigma$ which gives the following mass correction
\begin{equation}
    \delta m^2 = 
\vcenter{\hbox{
\begin{tikzpicture}[line width=1. pt, scale=2]
\draw[pyblue] (0.1, 0) -- (0.5, 0);
\draw[fill=none, pyred] (0.7, 0) circle (.2cm);
\draw[pyblue] (0.9, 0) -- (1.3, 0);
\draw[fill=black] (0.5, 0) circle (.03cm);
\draw[fill=black] (0.9, 0) circle (.03cm);
\end{tikzpicture}}} \sim \frac{1}{\Lambda_2^2} \int\mathrm{d}\omega\mathrm{d}^3k\, \frac{\omega^4}{(\omega^2 - c_s^2k^2)^2} \sim c_s \frac{f_\pi^4}{\Lambda_2^2}\,.
\end{equation}
In the weak mixing regime, we use (\ref{eq:single_field_power_spectrum}) to write the condition for radiative stability in term of the dimensionless power spectrum. It reads
\begin{equation}
    \frac{H}{\Lambda_2}\lesssim \frac{2\pi\Delta_\zeta}{c_s^{1/2}}\,, \hspace*{0.5cm} \text{i.e.} \hspace*{0.5cm} \frac{\tilde{\rho}}{H} \lesssim c_s^{-2}\,.
\end{equation}

\vskip 4pt
$\bullet$ Finally, we consider the radiative correction induced by the interaction $\dot{\pi}_c\sigma^2$. This reads
\begin{equation}
    \delta m^2 = 
\vcenter{\hbox{
\begin{tikzpicture}[line width=1. pt, scale=2]
\vspace*{-2cm}
\draw[pyblue] (0.1, 0) -- (0.5, 0);
\draw [pyblue, xshift=0.7cm, domain=180:360] plot(\x:0.2);
\draw [pyred, xshift=0.7cm, domain=180:360] plot(-\x:0.2);
\draw[pyblue] (0.9, 0) -- (1.3, 0);
\draw[fill=black] (0.5, 0) circle (.03cm);
\draw[fill=black] (0.9, 0) circle (.03cm);
\end{tikzpicture}}} \sim \alpha^2 \int\mathrm{d}\omega\mathrm{d}^3k\, \frac{\omega^2}{(\omega^2 - c_s^2k^2)(\omega^2 - k^2 - m^2)} \sim \alpha^2 \frac{f_\pi^2}{c_s}\,.
\end{equation}
In the weak mixing regime, the naturalness condition reads
\begin{equation}
    \alpha\lesssim c_s^{1/2}\, (2\pi\Delta_\zeta)^{1/2}\,, \hspace*{0.5cm} \text{i.e.} \hspace*{0.5cm} \tilde{\alpha}^2 \lesssim c_s^{-1}\, (2\pi \Delta_\zeta)^{-1/2}\,.
\end{equation}

\section{Size of non-Gaussianities beyond Weak Mixing}
\label{sec:NG_pheno}

The cosmological flow provides a unique opportunity to systematically study non-Gaussian signals in the strong mixing regime by fully treating the quadratic coupling. Most previous studies have examined the bispectrum in particular regimes. In this section, we numerically solve the flow equations for the bispectrum and inspect their phenomenological implications in the entire parameter space. In particular, we provide results in regions of the parameter space that are not easily accessible by analytical means.

\vskip 4pt
Following standard conventions, we define the dimensionless shape function of the bispectrum $S$ such that
\begin{equation}
\braket{\zeta_{\bm{k}_1} \zeta_{\bm{k}_2} \zeta_{\bm{k}_3}}' \equiv (2\pi)^4 \, \frac{S(k_1, k_2, k_3)}{(k_1k_2k_3)^2}\Delta_\zeta^4\,.
\end{equation}
At weak quadratic mixing $\rho\lesssim m$, the $\pi_c-\sigma$ conversion process can be treated perturbatively. At leading order, the cubic interactions we consider lead to single-, double-, and triple-exchange diagrams, respectively for $(\partial_i \pi_c)^2\sigma$ and $\dot{\pi}_c^2\sigma$, $\dot{\pi}_c\sigma^2$, and $\sigma^3$. In the strong mixing regime $\rho\gtrsim m$, the quadratic coupling needs to be treated non-perturbatively. As a consequence, the notion of diagram becomes meaningless as one needs to sum up all contributions with any number of quadratic coupling insertions. In this sense, we present ``non-perturbative" results. The shape function we compute should be understood as the \textit{resummed} one. In the following, we quantify the size of non-Gaussianities by the usual parameter 
\begin{equation}
    \fnl \equiv \frac{10}{9}\, S(k,k,k)\,,
\end{equation}
where the shape function is evaluated in the equilateral configuration $k_1=k_2=k_3=k$. The shapes generated here typically resemble the equilateral template, so that it makes sense to compare $\fnl$ to the corresponding constraint $\fnl^{\text{eq}} = -26 \pm 47$ ($68\%$ CL, statistical) \cite{Planck:2019kim}.

\subsection{Non-local EFT}

Before exposing exact numerical results and for later convenience, it is instructive to realise that one can derive an effective theory that provides a unified picture covering a large part of the phase diagram in Figure~\ref{fig:Analytic_Phase_Diagram}.
In fact, as long as $m_{\text{eff}} \gg H$, the phase diagram can be described by the following non-local single-field theory \cite{Castillo:2013sfa, Gwyn:2012mw, Jazayeri:2022kjy, Jazayeri:2023xcj}
\begin{equation}
\label{eq:non-local_Lagrangian}
\begin{aligned}
    \mathcal{L}/a^3 &= \frac{1}{2}\dot{\pi}_c\left[1 + \frac{\rho^2}{-\partial_i^2/a^2 + m^2}\right] \dot{\pi}_c - \frac{1}{2}\frac{(\partial_i \pi_c)^2}{a^2} - \mu \rho^3 \left[\left(-\partial_i^2/a^2 + m^2\right)^{-1}\dot{\pi}_c\right]^3 \\
    &- \frac{1}{2}\alpha\rho^2 \dot{\pi}_c \left[\left(-\partial_i^2/a^2 + m^2\right)^{-1}\dot{\pi}_c\right]^2 - \frac{\rho}{2}\left[\frac{1}{\Lambda_1} \frac{(\partial_i \pi_c)^2}{a^2} + \frac{1}{\Lambda_2} \dot{\pi}_c^2 \right]\left(-\partial_i^2/a^2 + m^2\right)^{-1}\dot{\pi}_c\,,
\end{aligned}
\end{equation}
where we have omitted self-interactions of $\pi_c$ parameterised by $\lambda_{1, 2}$ in (\ref{eq:Full_pi_sigma_theory}).
We have also fixed the intrinsic speed of sound of  $\pi_c$ to unity for this Section; $c_s=1$.
This Lagrangian results from integrating out the massive field $\sigma = \left(-\Box + m^2\right)^{-1}\rho\dot{\pi}_c$ and neglecting the time derivatives in the d’Alembert operator (see Eq.~(\ref{eq: EOM pi-sigma dynamics})). As such, Eq.~(\ref{eq:non-local_Lagrangian}) should be understood as the leading order contribution in a time derivative expansion, higher-order terms being encoded in an infinite number of operators coming from
\begin{equation}
    \frac{1}{-\Box + m^2} = \frac{1}{-\partial_i^2/a^2 + m^2} \sum_{n=0}^{\infty} (-1)^n \left[(\partial_t^2 + 3H\partial_t)(-\partial_i^2/a^2 + m^2)^{-1}\right]^n\,.
\end{equation}
Let us now take the flat-space limit to develop our intuition. One immediately sees that the above expansion is valid whenever the low-energy modes obey a non-relativistic dispersion relation such that $\omega^2\ll k^2 + m^2$. The quadratic part of the Lagrangian (\ref{eq:non-local_Lagrangian}) gives the following dispersion relation
\begin{equation}
    \omega^2 = \frac{k^2}{1 + \frac{\rho^2}{k^2 + m^2}}\,. 
\end{equation}
In the limit $k\ll m$, this boils down to $\omega = \tilde{c}_s k$ with $\tilde{c}_s$ defined in (\ref{eq:induced_sound_speed}). At strong mixing such that $\rho\gg k \gg m$, one recovers $\omega = k^2/\rho$. Therefore, the non-local effective theory interpolates between the reduced speed of sound and the modified dispersion relation regimes. However, being intrinsically single-field, this theory misses the particle production and therefore does not describe the cosmological collider signal in the squeezed limit of the three-point correlators (see Section \ref{sec:CC_phycis}). We will see later that (\ref{eq:non-local_Lagrangian}) is very convenient to estimate the amplitude of the bispectrum using dimensional analysis.

\subsection{Phenomenology} 
\label{subsec: NG size pheno}

\textit{Conversion from $\pi$ to $\zeta$.} Independently of cubic interactions in the Lagrangian~\eqref{eq:Full_pi_sigma_theory}, the non-linear relation between $\pi$ and $\zeta$ generates a contribution to the bispectrum, see the last two terms in Eq.~\eqref{eq:zeta-correlators} and the discussion below. It can be computed solely from the linear dynamics, and it is exponentially suppressed in all parameter space of the $m$-$\rho$ phase diagram, except in the effectively light field regime $m_\textrm{eff} \ll H$. Explicit results, see Figure~\ref{fig:nonlinearCorrectionfNL} in Appendix~\ref{app:LightField}, show that this conversion contribution to $\fnl$ is at most of order $10^{-2}$ in the parameter space we display, for $N=60$ \textit{e}-folds of super-horizon evolution. Such a small value is negligible for today’s cosmology, but should be taken into account in general. 

\begin{figure}[t!]
  \centering
  \hspace*{-0.8cm}
  \subfloat{\includegraphics[width=0.6\textwidth]{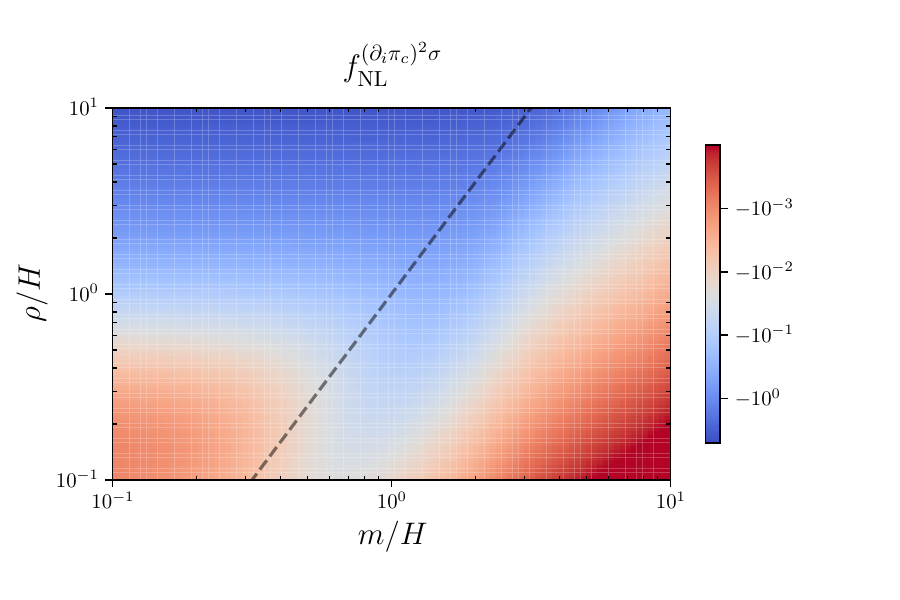}}
  \hspace*{-1cm}
  \subfloat{\includegraphics[width=0.6\textwidth]{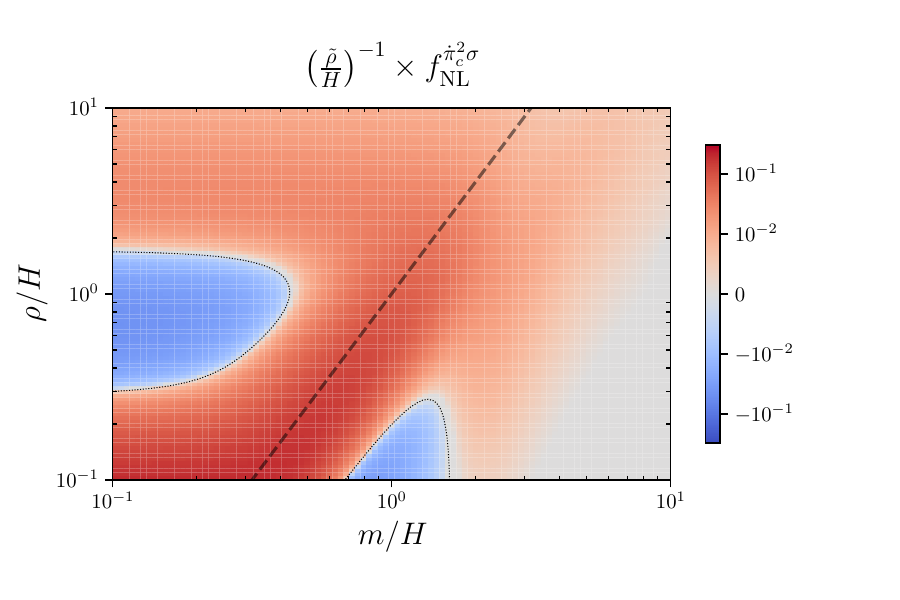}}
  \hfill
  \vspace*{-1cm}
  \centering
  \hspace*{-0.8cm}
  \subfloat{\includegraphics[width=0.6\textwidth]{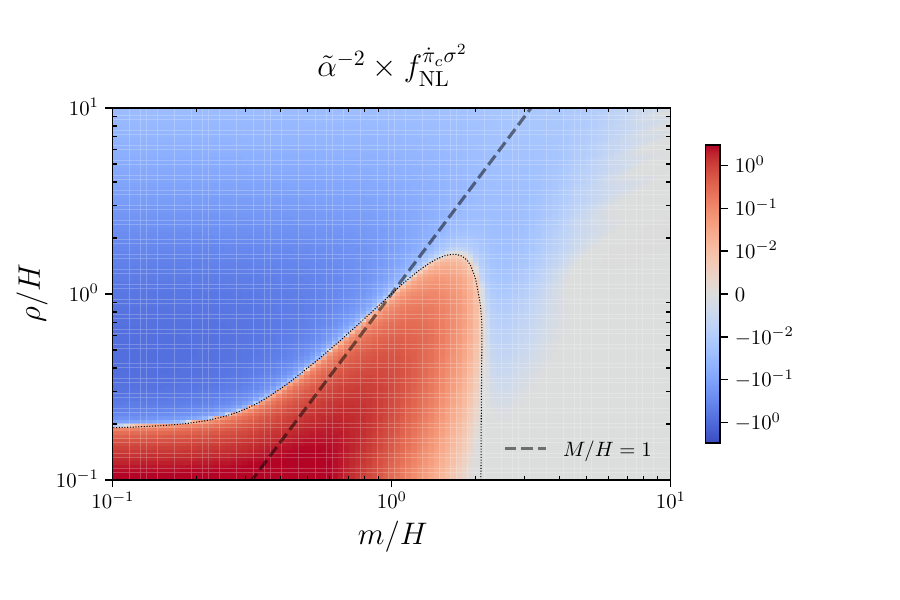}}
  \hspace*{-1cm}
  \subfloat{\includegraphics[width=0.6\textwidth]{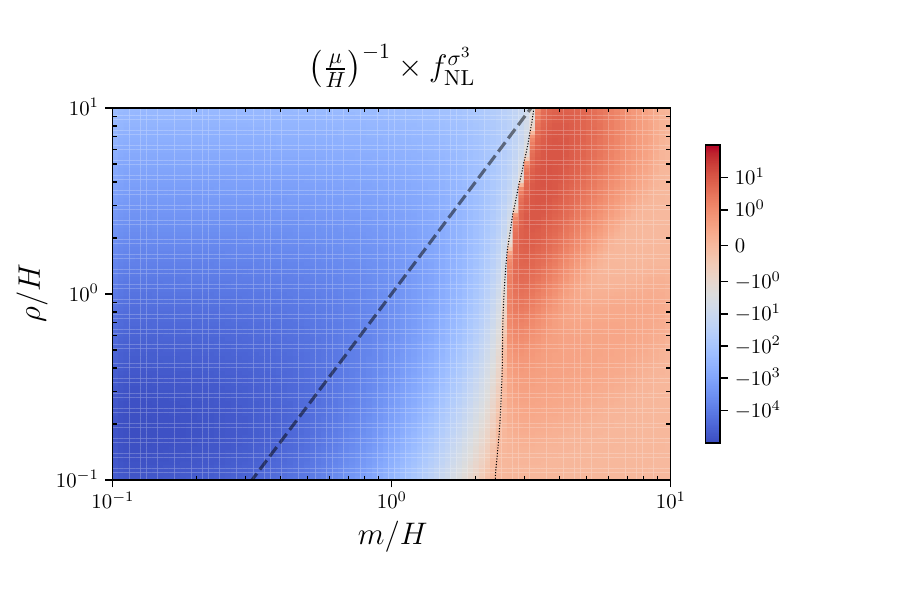}}
  \caption{Shape function phase diagram for the $(\partial_i \pi_c)^2\sigma$ (\textit{top left}), $\dot{\pi}_c^2\sigma$ (\textit{top right}), $\dot{\pi}_c\sigma^2$ (\textit{bottom left}), and $\sigma^3$ (\textit{bottom right}) interactions, in the equilateral configuration $k_1=k_2=k_3=k$. The dashed line corresponds to the $M=H$ limit, isolating the modified dispersion relation regime $M/H\ll 1$ and the reduced speed of sound regime $M/H\gg 1$. The dotted thin line corresponds to $S(k, k, k) = 0$  where the shape function changes sign. We have set $N=60$ $e$-folds of super-horizon evolution. Slices of the phase diagram for various interactions can be found in Figure~\ref{fig:slices_bispectrum}.
  }
  \label{fig:phase_diagrams_bispectrum}
\end{figure}

\begin{figure}[h!]
    \centering
    \hspace*{-1.8cm}
    \subfloat{\includegraphics[width=1.2\textwidth]{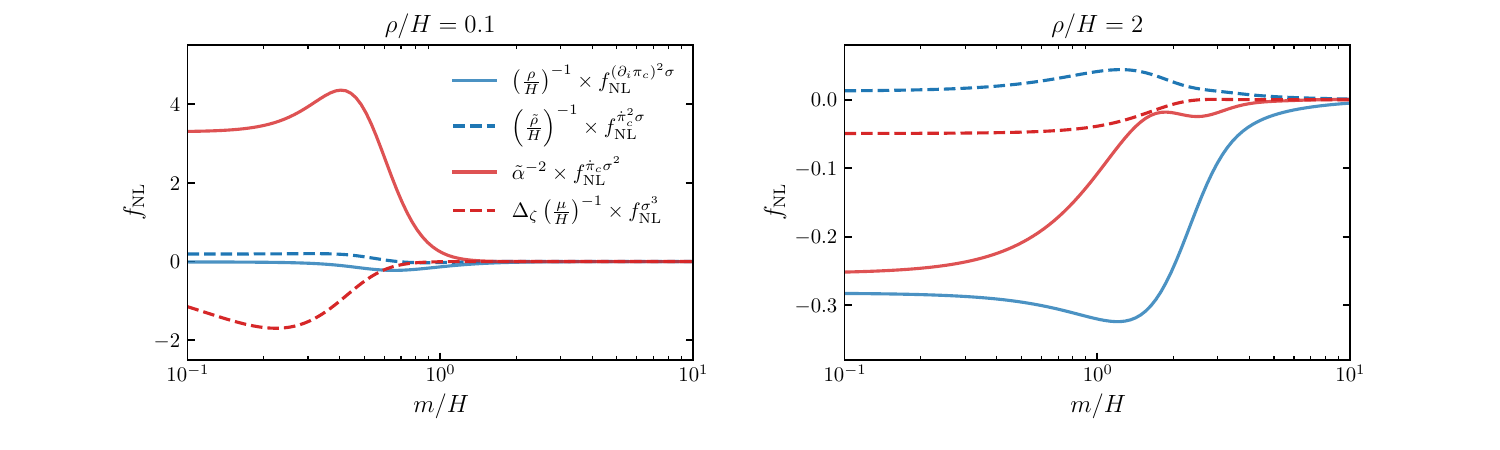}}
    \hfill
    \centering
    \hspace*{-1.8cm}
    \subfloat{\includegraphics[width=1.2\textwidth]{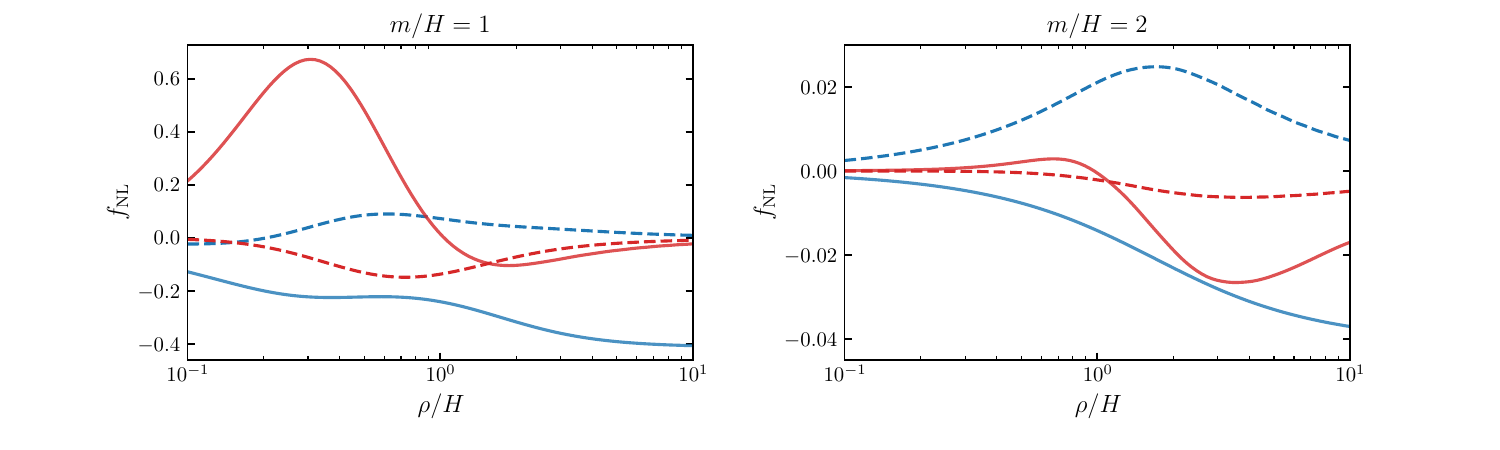}}
    \caption{Slices of the shape function phase diagrams as function of the bare mass $m/H$ for $\rho/H=0.1$ (\textit{top left}) and $\rho/H=2$ (\textit{top right}), and as function of the quadratic mixing $\rho/H$ for $m/H=1$ (\textit{bottom left}) and $m/H=2$ (\textit{bottom right}), in the equilateral configuration $k_1=k_2=k_3=k$. For better visualisation, the shape generated by the interaction $\sigma^3$ has been multiplied by $\Delta_\zeta$.}
    \label{fig:slices_bispectrum}
\end{figure}

\vskip 4pt
In the following, we concentrate on the ``intrinsic’’ bispectrum generated by cubic interactions in the Lagrangian~\eqref{eq:Full_pi_sigma_theory}, and which can give sizeable contributions. Figure~\ref{fig:phase_diagrams_bispectrum} shows the $\fnl$ parameter in the $\left(m/H, \rho/H\right)$ plane for the various interactions. For easy comparison and better visualisation, we also present slices of the phase diagrams in Figure~\ref{fig:slices_bispectrum}. For each interaction, we consider the three regions of the phase diagram in turn.

\vskip 4pt
\textit{$(\partial_i \pi_c)^2\sigma$ interaction.} First, we consider the bispectrum generated by the interaction $(\partial_i \pi_c)^2\sigma$. In the weak mixing regime $\rho \lesssim m$, the size of the single-exchange diagram can be estimated (see insert below), yielding\footnote{Note that the non-linearly realised symmetry fixes this cubic interaction coupling constant to $f_{\pi}/\Lambda_1 = \rho/f_{\pi}$.}
\begin{equation}
\label{eq:scaling_fNL_weak_mixing_1}
    \fnl^{(\partial_i \pi_c)^2\sigma} \sim \left(\frac{\rho}{H}\right)^2\,.
\end{equation}
It is found that this scaling indeed extends up to values of the quadratic coupling $\rho\sim m$ as we increase the mass. Hence the leading perturbative result still gives a correct estimation for $\rho/H\gtrsim 1$ when the field $\sigma$ is heavy enough. The associated non-Gaussian signal cannot reach order unity.

\vskip 4pt
As function of the bare mass $m/H$, the bispectrum signal presents a suppression whose leading contribution scales as $(H/m)^2$ for large mass.\footnote{The signal also contains other terms suppressed by the exponential Boltzmann factor $e^{-\pi\mu}$ where $\mu^2 = (m/H)^2 - 9/4$. However for large mass, the $(H/m)^2$ suppression dominates in the equilateral configuration.} This comes from the leading imprint of higher-dimension operators when the heavy field is integrated out, as can be explicitly seen in (\ref{eq:non-local_Lagrangian}) in the limit $\partial_i^2/a^2 \ll m^2$. The transition to this EFT regime is found to occur at $m/H\gtrsim 1$, as can be observed in the upper panel of Figure~\ref{fig:slices_bispectrum}. The hilltop visible around $m/H\sim1$ is due to the fact that the intrinsic bispectrum decreases slower than the power spectrum squared as function of $m/H$, resulting in a boosted shape function.

\vskip 4pt
In the strong mixing regime $\rho \gtrsim m$, the relevant mass scale to describe the dynamics is $M=m^2/\rho$ as explained in Section \ref{subsec:quadratic_theory}. Therefore, the transition to the EFT regime is shifted to higher values of the bare mass $m/H \gtrsim \sqrt{\rho/H}$. In the modified dispersion relation regime, one can use (\ref{eq:non-local_Lagrangian}) and estimate the size of the non-Gaussianity, obtaining
\begin{equation}
\label{eq:scaling_fNL_strong_mixing_1}
    \fnl^{(\partial_i \pi_c)^2 \sigma} \sim \frac{\rho}{H}\,.
\end{equation}
We show in the insert below how this scaling is found using dimensional analysis. The found scaling is confirmed numerically. For the same reason as explained previously, this regime is reached for higher values of the quadratic mixing as the heavy field mass increases. Importantly, note that the size of this non-Gaussian signal monotonically increases as a function of the quadratic mixing, and can be large.

\begin{framed}
{\small \noindent {\it Estimating the size of the bispectrum.}---In this insert, we show how one can easily recover the scalings~(\ref{eq:scaling_fNL_weak_mixing_1}) and (\ref{eq:scaling_fNL_strong_mixing_1}) using purely dimensional analysis arguments and the non-local effective Lagrangian (\ref{eq:non-local_Lagrangian}). This theory being single-field, we can estimate the size of the bispectrum using 
\begin{equation}
    \fnl \sim \frac{\mathcal{L}_3}{\mathcal{L}_2} \, \Delta_{\zeta}^{-1}\,,
\end{equation}
evaluated at the energy scale that we probe during inflation i.e. $\omega\sim H$. In the weak mixing regime, using the linear dispersion relation (we recall that we have fixed $c_s=1$ for simplicity), we obtain
\begin{equation}
    \fnl^{(\partial_i \pi_c)^2 \sigma} \sim \left(\frac{\rho}{H}\right)^2\,,
\end{equation}
where we have evaluated the gradient term in $(-\partial_i^2/a^2 + m^2)^{-1}$ at the Goldstone boson horizon crossing $|k\tau|\sim 1$ and assumed $m \sim H$. 
In the strong mixing regime, using $\dot{\pi}_c \sim H \pi_c$ and the modified dispersion relation $\omega=k^2/\rho$, one obtains
\begin{equation}
    \fnl^{(\partial_i \pi_c)^2 \sigma} \sim \frac{\rho}{H} \left(\frac{H}{f_\pi}\right)^2 \frac{\pi_c}{H}\, \Delta_{\zeta}^{-1}\,.
\end{equation}
We finally use $\Delta_\zeta \sim (H/f_\pi)^2 (\rho/H)^{1/4}$ (see e.g. (\ref{eq:MDR_power_spectrum})) to obtain $\pi_c/H \sim (\rho/H)^{1/4}$ which, in the end, gives the desired scaling. A similar analysis can be performed for the other interactions to find the scalings (\ref{eq:scaling_fNL_strong_mixing_2}), (\ref{eq:scaling_fNL_strong_mixing_4}) and (\ref{eq:scaling_fNL_strong_mixing_3}). In these cases, the ``propagator" $(-\partial_i^2/a^2 + m^2)^{-1}$ can be estimated by evaluating the gradient term at the Goldstone boson $\rho$-horizon $|k\tau| \sim \sqrt{\rho/H}$, leading to 
\begin{equation}
    (k^2/a^2 + m^2)/H^2 \sim \rho/H\,. 
\end{equation}

 }
\end{framed}

\vskip 4pt
When the field is effectively light $m/H\ll 1$ and $\rho/H< 1$, we see that the $\fnl$ parameter does not present an amplification due to super-horizon sourcing, visible in the phase diagram of the power spectrum in Figure~\ref{fig:Gaussian_phase_diagram}. We recall that this enhancement depends on the number of super-horizon $e$-folds until the end of inflation. In fact, the bispectrum is indeed enhanced but this boost is balanced by the amplification of the power spectrum to give no visible effect in the $\fnl$ parameter.

\vskip 4pt
\textit{$\dot{\pi}_c^2\sigma$ interaction.} At weak mixing $\rho\lesssim m$, the size of the non-Gaussianity arising from the interaction $\dot{\pi}_c^2\sigma$ is\footnote{Due to spontaneously broken de Sitter boosts, the covariance of the cubic interaction $(\partial_\mu\pi_c)^2\sigma$ is violated. The non-linearly realised symmetry only fixes the spatial derivative interaction, see Eq.~(\ref{eq:pi_sigma_decoupling_limit}).}
\begin{equation}
    \fnl^{\dot{\pi}_c^2\sigma} \sim \frac{\rho}{H}\times\frac{\tilde{\rho}}{H}\,,
\end{equation}
which we confirm numerically up to $\rho\sim m$, similarly to the previous interaction. At strong mixing $\rho \gtrsim m$, the EFT regime is reached for higher values of the bare mass $m/H\gtrsim \sqrt{\rho/H}$ for the same reason as the interaction $(\partial_i \pi_c)^2\sigma$. Accounting for perturbativity bounds on the couplings, we can accommodate for large non-Gaussianities in this regime. In the modified dispersion relation regime, the size of the non-Gaussianity is
\begin{equation}
\label{eq:scaling_fNL_strong_mixing_2}
    \fnl^{\dot{\pi}_c^2\sigma} \sim \frac{\tilde{\rho}}{H} \left(\frac{\rho}{H} \right)^{-1}\,,
\end{equation}
which, for $\rho/H\gg 1$, gives a lower signal compared to that in the weakly coupled regime. Therefore, for this interaction, a large quadratic mixing does not necessarily imply large non-Gaussianities. In the lower panel of Figure~\ref{fig:slices_bispectrum}, one can appreciate that this interaction is damped at large quadratic mixing, as opposed to the spatial derivative interaction. In fact, from the non-local Lagrangian (\ref{eq:non-local_Lagrangian}) and the modified dispersion relation $\omega=k^2/\rho$, it is easy to realise that interactions with strictly negative powers of spatial gradients---including the ``propagators" $(k^2/a^2 + m^2)^{-1}$---are suppressed at strong quadratic mixing.

\vskip 4pt
In the effectively light field case and as can be better seen in Figure~\ref{fig:slices_bispectrum}, the shape function reaches a plateau as the boost in the bispectrum is compensated by that of the power spectrum. The transition to this regime, either from the $\rho$ or the $m$ directions, is delimited by the $\fnl$ parameter changing sign, as represented by the dotted line in Figure~\ref{fig:phase_diagrams_bispectrum}.

\vskip 4pt
Recall that because we consider interactions that break de Sitter boosts, we have split $(\partial_\mu \pi_c)^2\sigma$ into $(\partial_i \pi_c)^2\sigma$ and $\dot{\pi}_c^2\sigma$, the coefficient of the first one being fixed in terms of $\rho/H$. We have shown that the time derivative interaction leads to a completely different phenomenology than the spatial derivative interaction, a fact that was not enough appreciated before. 

\vskip 4pt
\textit{$\dot{\pi}_c\sigma^2$ interaction.} At weak mixing $\rho\lesssim m$, the non-Gaussian signal arising from the interaction $\dot{\pi}_c\sigma^2$ scales as
\begin{equation}
    \fnl^{\dot{\pi}_c\sigma^2} \sim \tilde{\alpha}^2 \left(\frac{\rho}{H}\right)^2\,.
\end{equation}
We observe that this scaling is valid up to $\rho\sim m$. In the modified dispersion relation regime, the amplitudes scales as
\begin{equation}
\label{eq:scaling_fNL_strong_mixing_4}
    \fnl^{\dot{\pi}_c\sigma^2} \sim \tilde{\alpha}^2 \left(\frac{\rho}{H}\right)^{-1}\,,
\end{equation}
which again gives a comparable amplitude compared to that in the weakly coupled regime. As shown in Figure~\ref{fig:slices_bispectrum}, the cosmological flow approach enables us to effortlessly probe the intermediate regime where analytical computations fail. A large non-Gaussian signal in the strong mixing regime is possible while remaining under perturbative control. 

\vskip 4pt 
In the light field regime, one can appreciate once more that the shape function is almost constant, for the same reason as for the previous interaction.

\vskip 4pt
\textit{$\sigma^3$ interaction.} At leading order in the weak mixing regime, the interaction $\sigma^3$ leads to the triple-exchange diagram whose amplitude is 
\begin{equation}
\label{eq:sigma3_weak_coupling_scaling}
    \fnl^{\sigma^3} \sim \frac{\mu}{H}\left(\frac{\rho}{H}\right)^3\Delta_\zeta^{-1}\,.
\end{equation}
It is well-known that this amplitude of non-Gaussianity is boosted due to the factor $\Delta_\zeta^{-1}\sim 10^5$, which leads to the largest non-Gaussian signal. This is explained by the fact that, for $\mu\sim H$ and $\rho\sim H$, the entire non-Gaussian signal in the $\sigma$-sector should be transferred to the $\zeta$-sector.
The actual size of non-Gaussianities in the bispectrum being $\fnl\Delta_\zeta$, one should have $\fnl\Delta_\zeta\sim (\mu/H)\left(\rho/H\right)^3$. As function of $m/H$, the intrinsic bispectrum scales as $(H/m)^{6}$ \cite{Gong:2013sma}---it drops faster than $(H/m)^{2}$---, leading to no characteristic enhancement around $m/H\sim1$ in the shape function, as one can observe in the upper panel of Figure~\ref{fig:slices_bispectrum}. When the bare mass increases, the modified dispersion relation regime is found for higher values of the quadratic mixing. We indeed confirm numerically that the scaling (\ref{eq:sigma3_weak_coupling_scaling}) extends up to $\rho/H\sim 1$ when increasing the bare mass. For $M/H\ll 1$ and $\rho/H\gg 1$, the non-Gaussian signal scales as
\begin{equation}
\label{eq:scaling_fNL_strong_mixing_3}
    \fnl^{\sigma^3} \sim \frac{\mu}{H}\left(\frac{\rho}{H}\right)^{-3/4}\Delta_\zeta^{-1}\,,
\end{equation}
which---although a priori suppressed by $\rho/H$---gives a larger contribution than in the weak mixing regime for typical values of the quadratic coupling.
Here, as well as for other interactions, the change of sign of the shape function for $m/H\sim 2$ can be attributed to a cancellation between two contributions resembling equilateral shapes and giving rise to an orthogonal shape.

\vskip 4pt
In the effectively light field regime the $\fnl$ parameter is decreasing as the bispectrum is not enhanced like the power spectrum.

\paragraph{Summary.} Accounting for perturbativity bounds derived in Section~\ref{sec:strong_coupling_scales} and the exact size of non-Gaussianities presented in Figure~\ref{fig:phase_diagrams_bispectrum}, one can obtain sizable non-Gaussianities---typically $\fnl\gg 1$---in the strong mixing regime for all interactions. At weak mixing, all interactions except $(\partial_i\pi_c)^2\sigma$ can lead to $\fnl\sim\mathcal{O}(1)$ signals.
Typically, the signal is largest for a mass of $\sigma$ of order $H$ or less. We have checked that, for each of these four cubic interactions, both in the weak and the strong mixing regimes, saturating the perturbativity bounds universally gives $\fnl \Delta_\zeta \lesssim 1$, as expected.

\paragraph{Self-regularised bispectrum.} It is worth insisting on the convenience of the cosmological flow approach when computing the bispectrum. When using analytical mode functions in calculations treating the quadratic mixing as a perturbative interaction---therefore necessarily narrowing down the range of applications to the weak mixing regime---the integrals appearing in the in-in formula are very difficult to compute for at least two reasons: (i) they involve several nested integrals, and (ii) they usually do not converge in the IR or UV, and need to be regularised. 

\vskip 4pt
Traditionally, the convergence in the UV is achieved by considering the $i\epsilon$-prescription---slightly tilting the integration contour in the imaginary direction in the complex plane---that turns highly oscillatory mode functions into damped exponentials.\footnote{Other methods can be used to speed up the numerical convergence in the UV, see e.g.~\cite{Chen:2006xjb, Chen:2008wn, Junaid:2015hga, Tran:2022euk}. Note also that such a difficulty is bypassed as well in the modal approach \cite{Funakoshi:2012ms,Clarke:2020znk}.} This regularisation requires the use of the factorised form of the in-in formula. In the IR, the factorised form of the in-in formula leads to individual integrals each containing spurious divergences.
The convergence is instead more easily achieved by using the commutator form of the in-in formula. As a consequence, one needs to split the in-in formula into two contributions, one converging explicitly in the UV and the other one in the IR. This mixed form requires introducing a fiducial intermediate time as a regulator (see \cite{Chen:2009zp} for more details).

\vskip 4pt
Going beyond a perturbative treatment of the mixing by numerically solving the matrix mode functions of the full quadratic theory removes the need to perform nested integrals to compute the bispectrum. As for the issue of the UV convergence, it can then be addressed by Wick-rotating the theory, i.e.~solving the mode functions in Euclidean time before evaluating the corresponding time integral defining the bispectrum \cite{Assassi:2013gxa, An:2017hlx}. However, this method has the drawback that if one wants to determine the time-evolution of correlators during inflation, one needs to evaluate \textit{different} integrals for each time considered, whereas this is automatically obtained from solving in one go the set of cosmological flow equations in our approach. Furthermore, going beyond the bispectrum, nested integrals describing exchange processes necessarily reappear.

\vskip 4pt
The cosmological flow completely bypasses all these complications because: (i) this method does not require the use of neither analytical nor numerical mode functions, (ii) it relies on evolving the correlators on the real time axis, hence there is no need to Wick rotate the integrand, and (iii) it systematically works for all tree-level $n$-point functions.

\section{Cosmological Collider Phenomenology}
\label{sec:CC_phycis}

The squeezed limit of the bispectrum is believed to be a robust probe of the particle content during inflation. One major advantage of the cosmological flow is that it provides a generic framework to study the squeezed limit of the bispectrum in a systematic and exact manner.
In this section, we present several cosmological collider signals, including the strong mixing regime for which it is important to take into account the notion of \textit{effective} mass $m_{\text{eff}}^2 = m^2+\rho^2$ instead of the bare mass. Moreover, we show that a phase shift and a non-trivial amplitude of the oscillations can be seen as a signature of primordial fluctuations undergoing strong mixing. We also give novel insight on the role of the sound  speeds on cosmological collider signals. We finish by exposing the cosmological collider flow---i.e. the squeezed limit of the bispectrum as function of time---that enables us to shed light on the characteristic time scales of the dynamics.

\vskip 4pt
In this section, for definiteness, we will focus on the bispectrum arising from the $(\partial_i \pi_c)^2\sigma$ and $\dot{\pi}_c\sigma^2$ interactions (except in \ref{subsec: Sound Speed Collider}), and we will be interested in the case where the field $\sigma$ is heavy; $\mu_{\text{eff}}>0$.

\subsection{Strongly Mixed Colliders}
\label{subsec: CC quadratic coupling}

\begin{figure}[b!]
    \centering
    \hspace*{-1.8cm}
    \subfloat{\includegraphics[width=1.2\textwidth]{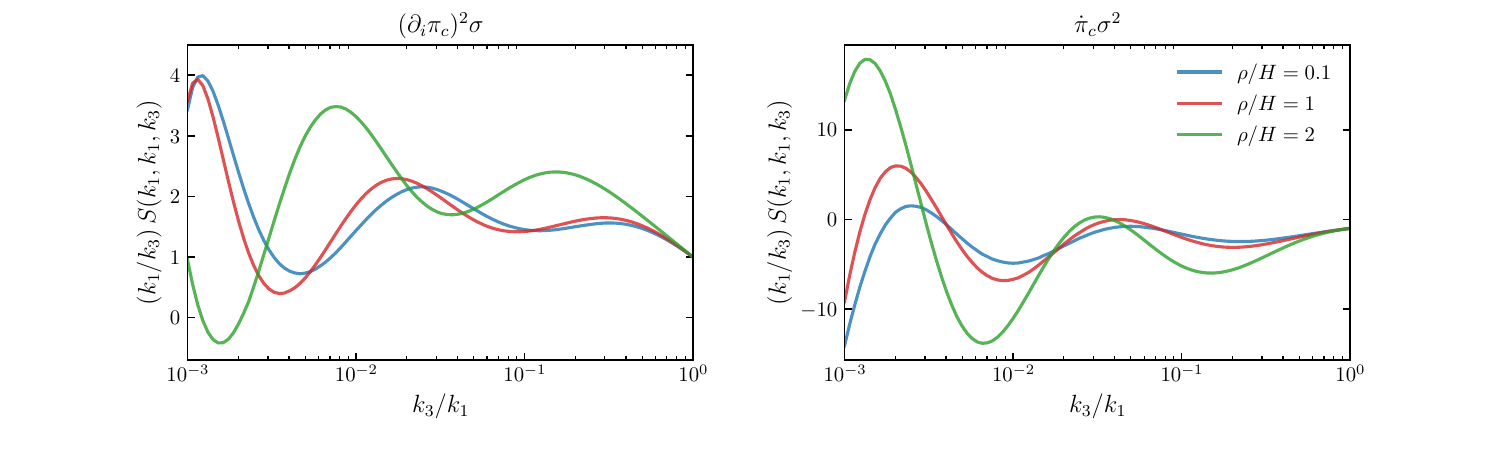}}
    \hfill
    \centering
    \hspace*{-1.8cm}
    \subfloat{\includegraphics[width=1.2\textwidth]{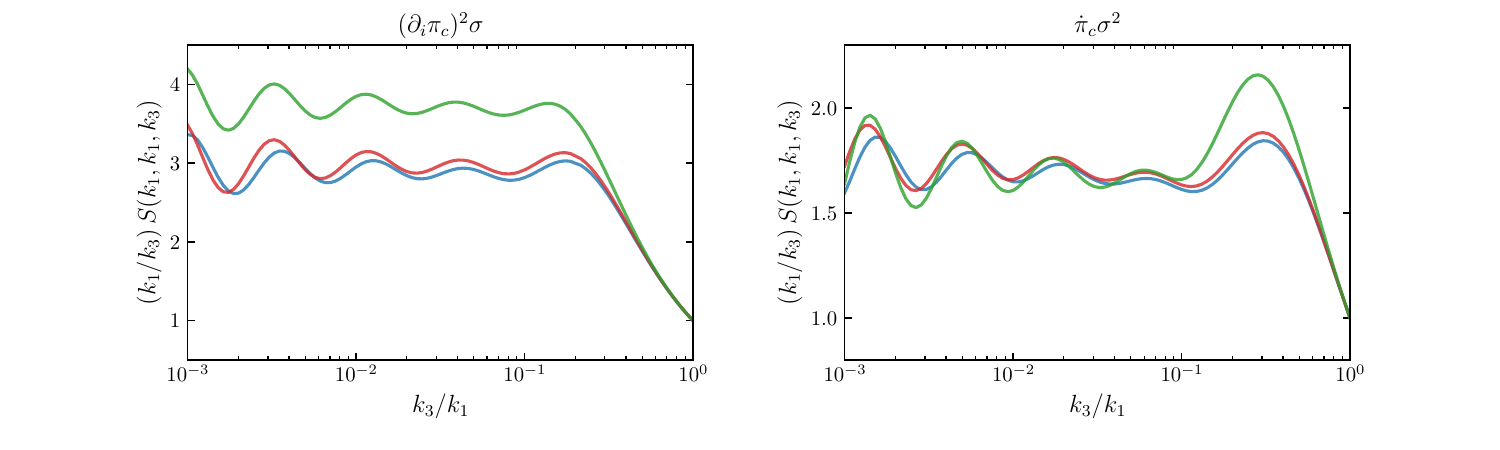}}
    \vspace*{-0.2cm}
    \caption{Shape functions for the $(\partial_i \pi_c)^2\sigma$ interaction (\textit{left panel}) and $\dot{\pi}_c\sigma^2$ interaction (\textit{right panel}) in the isosceles-triangle configuration $k_1=k_2$, for $\mu_{\text{eff}} = 2$ (\textit{top}) and $\mu_{\text{eff}} = 5$ (\textit{bottom}) varying the quadratic mixing $\rho/H$. For illustration purposes, we have normalised the amplitudes of the signals to be the same in the equilateral configuration. For $\mu_{\text{eff}} = 5$, we also have set $c_s=0.1$ to amplify the oscillations of the cosmological collider signal (signals for different values for the sound speed---including $c_s=1$---are presented in figure \ref{fig:Sound_Speed_Collider}).
    }
    \label{fig:CC_signal}
\end{figure}

We first look at how the bispectrum shape in the squeezed limit depends on the quadratic mixing. In this limit, the bispectrum is sensitive to the heavy field mass, as it governs the frequency of the oscillations.
Figure~\ref{fig:CC_signal} shows the bispectrum shapes as functions of the momentum ratio in the isosceles-triangle configuration, for fixed $\mu_{\text{eff}}$ but varying the quadratic mixing (hence adjusting the bare mass). We observe that the cosmological collider signals oscillate at the same frequency. Indeed as we have seen in Section \ref{subsec:quadratic_theory}, the relevant notion of mass for this system is the effective mass $m_{\text{eff}}$ and not the bare mass $m$. As a consequence, a light but strongly mixed field and a heavy but weakly mixed field cannot be distinguished based only on the signal frequency. For illustration purposes and to focus on the oscillation frequencies only, the signals have been normalised to unity in the equilateral configuration (see Section~\ref{sec:NG_pheno}).
A notable feature is the phase shift of the signal to smaller momentum ratios for larger quadratic mixings. This constitutes a striking example that having complete predictions for the signal---amplitude, frequency, phase, as well as contaminations from equilateral shapes---is important to disentangle the bare mass from the effective one, and will be needed to correctly infer the model parameters from the cosmological data. As discussed in Section \ref{subsec:quadratic_theory}, this phase shift is due to the resonance of the heavy field before it starts oscillating. 

\begin{figure}[t!]
    \centering
    \hspace*{-1.8cm}
    \subfloat{\includegraphics[width=1.2\textwidth]{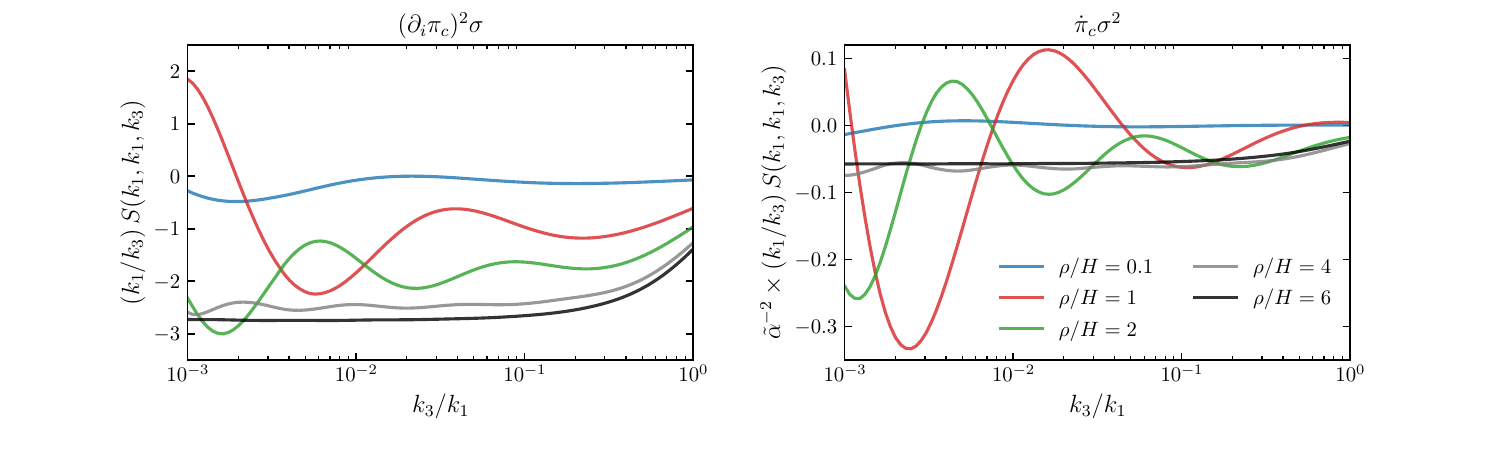}}
    \hfill
    \vspace*{-1cm}
    \caption{Shape functions for the $(\partial_i \pi_c)^2\sigma$ interaction (\textit{left panel}), and $\dot{\pi}_c\sigma^2$ interaction (\textit{right panel}) in the isosceles-triangle configuration $k_1=k_2$, for a fixed bare mass $m/H = 2$ and varying the quadratic mixing $\rho/H$.}
    \label{fig:CCsignal_amplitudes}
\end{figure}

\vskip 4pt
Because of the arbitrary normalisation of the shape function, Figure~\ref{fig:CC_signal} does not exhibit the size of non-Gaussianities nor the amplitude of the cosmological collider signals. To shed light on these two aspects, we show in Figure~\ref{fig:CCsignal_amplitudes} the shape function in the isosceles-triangle configuration for different values of the quadratic mixing $\rho/H$ while fixing the bare mass $m/H$. In the equilateral configuration $k_3/k_1=1$, the parametric dependence of $\fnl$ on $\rho/H$ has been discussed in Section~\ref{sec:NG_pheno}. Yet, a noticeable feature of the cosmological collider signals is the strong dependence of its amplitude on the quadratic coupling $\rho/H$ that is not monotonic. In the weak mixing regime $\rho\lesssim m$, the amplitude grows with $\rho$, after which it decreases in the strong mixing regime. At strong mixing for the $(\partial_i \pi_c)^2\sigma$ interaction, this indicates that although the size of the total signal is amplified, signatures of the cosmological collider signal are harder to detect, as it is exponentially suppressed by the effective mass $e^{-\pi\mu_{\text{eff}}}$.

\vskip 4pt
To sum up, the cosmological collider (denoted CC in the following equation) signal in the squeezed limit can be written as
\begin{equation}
    \lim\limits_{k_3/k_1 \to 0}S^{\text{CC}}(k_1, k_1, k_3) \sim \left(\frac{k_3}{k_1}\right)^{1/2} \mathcal{A}\left(\frac{m}{H}, \frac{\rho}{H}\right)\,\cos\left[ \mu_{\text{eff}} \log\left(\frac{k_3}{k_1}\right) + \delta\left(\frac{m}{H}, \frac{\rho}{H}\right)\right]\,.
\end{equation}
It qualitatively describes the squeezed limit even beyond the analytical lamppost of the weakly mixed regime, with an amplitude and phase information that can be efficiently obtained from the cosmological flow approach. The $m/H$-dependence of the amplitude $\mathcal{A}$ and the phase $\delta$ has been computed in \cite{Arkani-Hamed:2015bza, Lee:2016vti, Pinol:2021aun, Qin:2022lva} in the weakly mixed regime and only for the single-exchange diagram. It should be noted that for realistically squeezed configurations like the ones presented in the figures, the shape function does not oscillate around zero. This means that it always contains a contamination part---analytic in momenta---which can be described by the (possibly non-local) single-field EFT.

\vskip 4pt
Eventually, let us highlight that if we ever detect the cosmological collider signal, it is more likely to be attributed to a substantial mixing $\rho\sim m$ for which the non-analytic signal is the largest. This illustrates one of the main aspects motivating the development of the cosmological flow approach: interesting physics lying beyond the lamppost of analytical calculations can be missed, therefore biasing the interpretation of data.

\subsection{Sound-Speed Colliders} 
\label{subsec: Sound Speed Collider}

Next, we present the effect of varying the speed of sound $c_s$ on the cosmological collider signal. More precisely, throughout the paper, we have used $c_s$ to denote the speed of sound of the Goldstone boson $\pi_c$. However, in this section exclusively, we allow the field $\sigma$ to also have a non-unity speed of sound $c_\sigma$, and $c_s$ will actually denote the ratio of the sound speed of $\pi_c$ and the massive field: $c_s \equiv c_\pi/c_\sigma$. Therefore, $c_s<1$ means that the Goldstone boson propagates slower than the field $\sigma$, and the opposite for $c_s>1$, which by no means implies superluminal propagation. Crucially, as a consequence of the non-linearly realised symmetry, allowing $c_\sigma\neq 1$ induces additional mixings between $\pi_c$ and $\sigma$ that one is forced to consider.

\begin{figure}[h!]
    \centering
    \hspace*{-1.8cm}
    \subfloat{\includegraphics[width=1.2\textwidth]{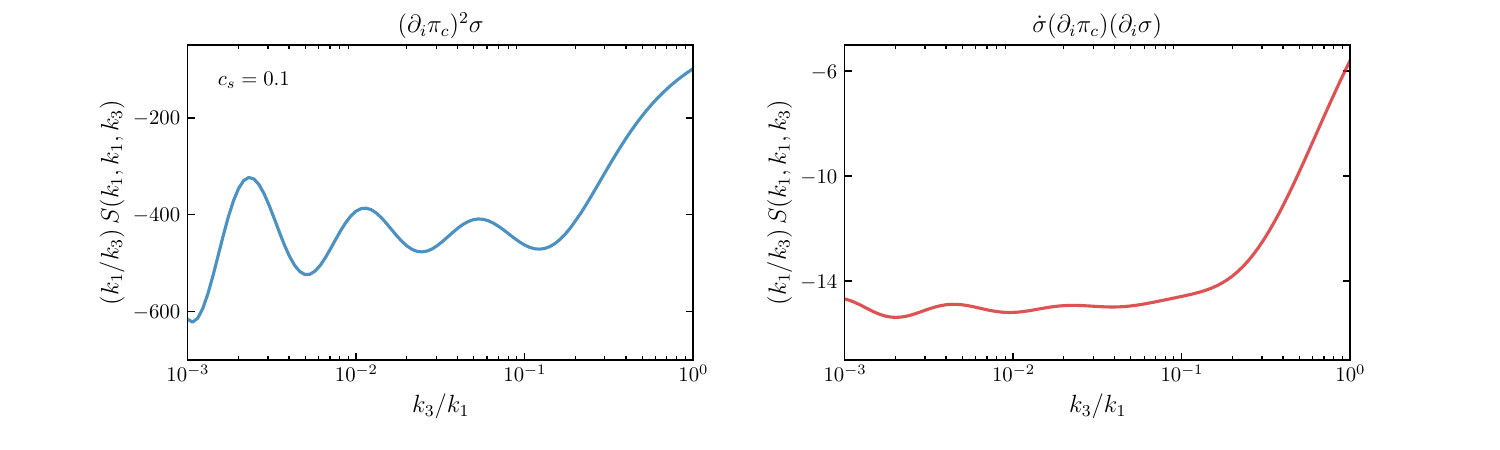}}
    \hfill
    \vspace*{-0.5cm}
    \centering
    \hspace*{-1.8cm}
    \subfloat{\includegraphics[width=1.2\textwidth]{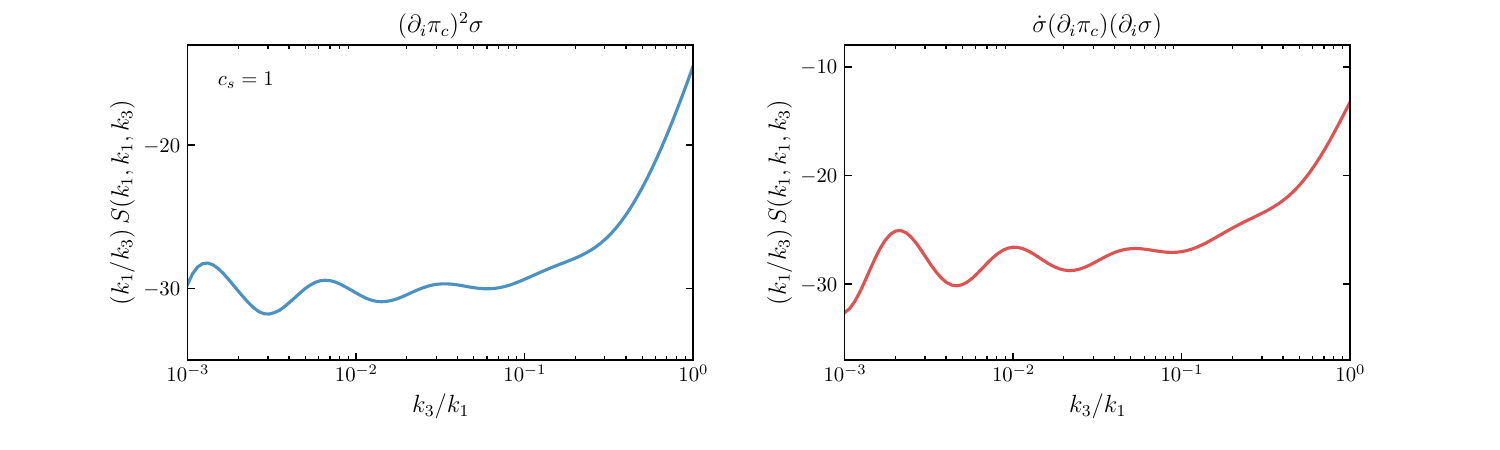}}
    \hfill
    \vspace*{-0.5cm}
    \centering
    \hspace*{-1.8cm}
    \subfloat{\includegraphics[width=1.2\textwidth]{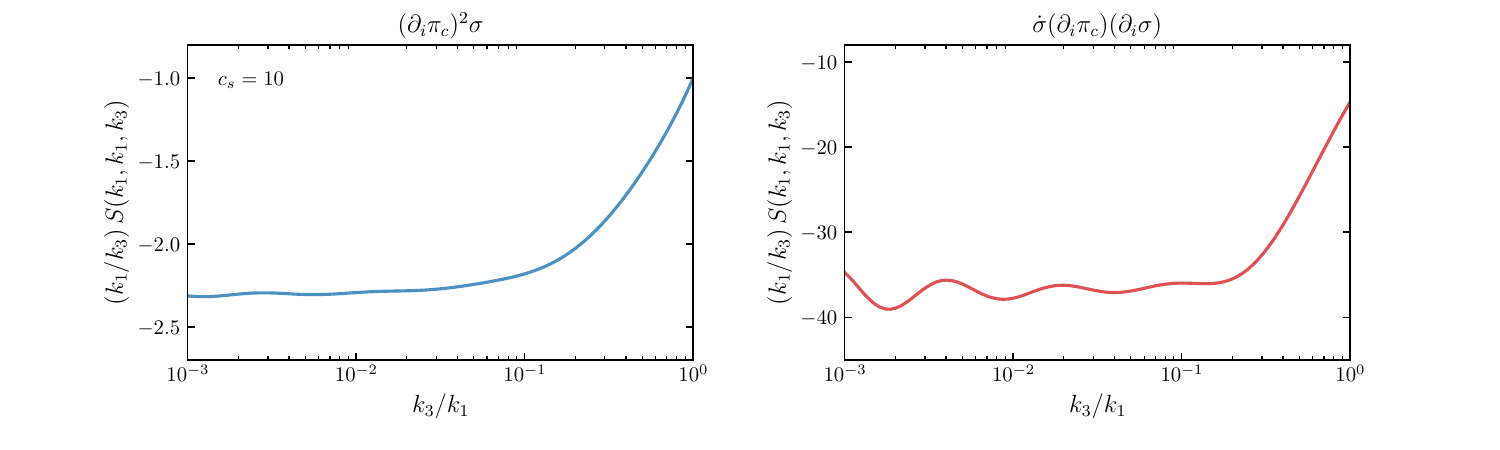}}
    \vspace*{-0.2cm}
    \caption{Shape functions for the $(\partial_i \pi_c)^2\sigma$ interaction (\textit{left panel}) and $\dot{\sigma} (\partial_i \pi_c) (\partial_i \sigma)$ interaction (\textit{right panel}) in the isosceles-triangle configuration $k_1=k_2$, for $c_\sigma=0.1$, $\mu_{\text{eff}} = 4$ and $\rho/H=4$, and with $c_s=0.1$ (\textit{top}), $c_s=1$ (\textit{middle}), and $c_s=10$ (\textit{bottom}). We recall that $c_s = c_\pi/c_\sigma$ and that both interactions are fully fixed by the non-linearly realised symmetry (either by $\rho$ or $c_\sigma$).}
    \label{fig:Sound_Speed_Collider}
\end{figure}

\vskip 4pt
Indeed, a non trivial propagation speed for $\sigma$ can only be generated by the unitary-gauge operator $(g^{0\mu} \partial_\mu \sigma)^2$, invariant under time-dependent spatial diffeomorphism \cite{Senatore:2010wk}. Therefore, as long as $c_\sigma\neq 1$, one should consider it in the theory, i.e.
\begin{equation}
    S \supset - \int \mathrm{d}^4x\sqrt{-g}\, \tilde{e} \,(g^{0\mu} \partial_\mu \sigma)^2\,,
    \label{eq:csigma-operator}
\end{equation}
where $\tilde{e}$ is a dimensionless coupling that we assume to be constant. Reintroducing the Goldstone boson in the decoupling limit via $g^{0\mu} \partial_\mu \sigma \rightarrow -(1+\dot{\pi}) \dot \sigma +\frac{1}{a^2} \partial_i\pi \partial_i \sigma$, and discarding quartic terms in the fields---not relevant for the bispectrum---, one obtains\footnote{Under a time diffeomorphism $t \rightarrow \tilde{t}=t+\pi(t, \bm{x})$, the four-dimensional partial derivative (inversely) transforms as $\tilde{\partial}_\mu \rightarrow \partial_\mu = \partial_\mu \tilde{x}^\nu \tilde{\partial}_\nu = \partial_\mu(t+\pi)\tilde{\partial}_0 + \partial_\mu \tilde{x}^i\tilde{\partial}_i$. The temporal component therefore transforms as $\tilde{\partial}_0 \rightarrow (1+\dot{\pi})\tilde{\partial}_0$ which implies $\partial_0 \rightarrow \tilde{\partial}_0 = \tfrac{\partial_0}{1+\dot{\pi}}$. The spatial ones transform as $\tilde{\partial}_i \rightarrow \partial_i = \tilde{\partial}_i + (\partial_i \pi)\tilde{\partial}_0$ which implies $\partial_i \rightarrow \tilde{\partial}_i = \partial_i - \tfrac{\partial_i\pi}{1+\dot{\pi}}\partial_0$. The off-diagonal metric components transform as $g^{0i}(t, \bm{x}) \rightarrow \tilde{g}^{0i}(\tilde{t}, \tilde{\bm{x}}) = \partial_\mu(t+\pi) \partial_\nu \tilde{x}^i g^{\mu\nu}(t, \bm{x}) = \tfrac{1}{a^2}\partial_i\pi$ where we have ultimately evaluated the metric on the background. Collecting everything together, and noting that $g^{00}(t,\bm{x}) \rightarrow \tilde{g}^{00}(\tilde{t},\tilde{\bm{x}}) = -(1+\dot{\pi})^2 + \tfrac{1}{a^2}(\partial_i\pi)^2$ (see Section~\ref{subsec:EFT}), we obtain the following exact transformation $g^{0\mu}\partial_\mu\sigma \rightarrow -(1+\dot{\pi})\dot{\sigma} + \tfrac{1}{a^2}(\partial_i\pi)(\partial_i\sigma)$, valid to all orders in the fields.}
\begin{equation}
    \mathcal{L}/a^3 \supset -\tilde{e} \left[\dot{\sigma}^2 + 2\dot{\pi}\dot{\sigma}^2 - 2\dot{\sigma}\frac{(\partial_i \pi)(\partial_i \sigma)}{a^2} + \dots \right]\,.
\end{equation}
Because the second term is also generated by the operator $\delta g^{00}(g^{0\mu} \partial_\mu \sigma)^2$, its coupling constant is not fully fixed by $\tilde{e}$.
We therefore do not consider it in the following.
By contrast, the interaction $\dot{\sigma} (\partial_i \pi)(\partial_i \sigma)$ can only be generated by the operator~\eqref{eq:csigma-operator}. Hence, in the same way as a non-trivial sound speed for $\pi$ universally comes with the interaction $\dot{\pi} (\partial_i \pi)^2$ whose size is controlled by $c_\pi$, a propagation speed for $\sigma$ different from unity universally comes with the interaction in $\dot{\sigma} (\partial_i \pi)(\partial_i \sigma)$ whose size is fixed by $c_\sigma$. The only difference is that the observable non-Gaussian signal it generates also depends on the $\sigma$ to $\pi$ transfer, i.e.~on the quadratic mixing between the two fields. 

\vskip 4pt
Explicitly, when adding the effects of \eqref{eq:csigma-operator} to the action considered in Section~\ref{subsec:EFT}, one obtains the following Lagrangian
\begin{equation}
\label{eq:Full_pi_sigma_theory-csigma}
\begin{aligned}
    \mathcal{L}/a^3 &= \frac{1}{2}\left[\dot{\pi}_c^2 - c_s^2 \frac{(\partial_i \pi_c)^2}{a^2}\right] +\frac{1}{2}\left[ \dot{\sigma}_c^2 - \frac{(\partial_i \sigma_c)^2}{a^2} - m^2\sigma_c^2\right] + \rho \dot{\pi}_c\sigma_c \\
    &- \frac{c_s^{3/2}}{2 f_\pi^2}\frac{\rho}{c_\sigma^2} \sigma_c \frac{(\partial_i \pi_c)^2}{a^2}
 - \frac{c_s^{3/2}}{f_\pi^2}\left(\frac{1}{c_\sigma^2}-1 \right) \dot{\sigma}_c \frac{\partial_i \pi_c \partial_i \sigma_c}{a^2}\,,
\end{aligned}
\end{equation}
where we concentrated on the new effects coming from $c_\sigma \neq 1$, with 
\begin{equation}
\begin{aligned}
 c_\sigma^{-2} &= 1 - 2\tilde{e}\,, \quad  c_s=c_\pi/c_\sigma\,,\quad  f_\pi^4= 2M_{\text{pl}}^2|\dot{H}|c_\pi\,, \quad \rho=2c_\pi^{3/2} c_\sigma \frac{\tilde{M}_1^3}{f_\pi^2}\,,\\ 
 \pi_c&=c_s^{-3/2} f_\pi^2 \pi\,, \quad \sigma_c=c_\sigma^{1/2} \sigma\,, \quad m \to c_\sigma m\,.
 \end{aligned}
\end{equation}
Here, we rescaled spatial coordinates $\tilde{\bm{x}}=\bm{x}/c_\sigma$. This does not affect the physics in our scale-invariant theories, hence we omit the tilde for simplicity. In this way, the quadratic Lagrangian in terms of canonically normalised fields takes the same form as the one studied in the rest of the paper, simply allowing for $c_s >1$, and where the mass $m$ in Eq.~\eqref{eq:Full_pi_sigma_theory-csigma} has been rescaled by a factor $c_\sigma$ compared to the bare mass in \eqref{eq:Lagrangian-sigma}.
We leave for future work the study of the full phase diagram and non-Gaussianities, varying $m, \rho$ and $c_s$. Nonetheless, an important aspect made transparent in the Lagrangian \eqref{eq:Full_pi_sigma_theory-csigma} is worth highlighting: a small sound speed $c_\sigma$ for the additional field boosts the overall amplitude of non-Gaussianities by a factor $\sim 1/c_\sigma^2$. Indeed, at fixed $m, \rho$ and $c_s$, the sound speed $c_\sigma$ only appears in the overall multiplicative factors, $1/c_\sigma^2$ and $1/c_\sigma^2-1$  respectively,  setting the sizes of the cubic interactions in $(\partial_i \pi_c)^2\sigma$ and $\dot{\sigma}(\partial_i \pi_c)(\partial_i \sigma)$. The corresponding rather large non-Gaussian signals are well visible in Figure~\ref{fig:Sound_Speed_Collider}, where we show the shape functions as a function of the squeezing parameter $k_3/k_1$ at strong mixing $\rho/H=4$, for $c_\sigma=0.1$ and for $c_s\in \{0.1, 1, 10\}$.
 
\vskip 4pt
For the $(\partial_i \pi_c)^2\sigma$ interaction on the left panel, the amplitude of the non-analytic part of the signal, i.e.~the cosmological collider signal, increases when $c_s$ decreases.
In the weak mixing regime, the scaling of the amplitude with $c_s$ is known analytically. For $c_s=1$, it is known that the amplitude of the non-analytic part of the signal is suppressed by $e^{-\pi\mu_{\text{eff}}}$.\footnote{These scalings were derived in the weak mixing regime, and for single-exchange diagrams only. In Section \ref{subsec:quadratic_theory}, we have given evidence that these scalings extend to the strong mixing regime albeit with the replacement $\mu\rightarrow \mu_{\text{eff}}$. It should also be mentioned that this Boltzmann suppression is only the leading behaviour of a more complicated scaling of the amplitude.}
However, for $c_s\ll 1$, this suppression takes the form $e^{-(\pi/2 + c_s)\mu_{\text{eff}}}$ \cite{Jazayeri:2022kjy}, hence leading to an amplification of the cosmological collider signal.
We find indeed that this tendency extends to the strong mixing regime, and that for $c_s>1$ the signal is further damped.
On the contrary, the amplitude of the cosmological collider signal arising from the $\dot{\sigma}(\partial_i \pi_c)(\partial_i \sigma)$ interaction is mildly dependent on $c_s$.

\vskip 4pt
For $c_s > 1$ and in the weak mixing regime, the cosmological collider signal is shifted towards larger momentum ratios, closer to the equilateral configuration~\cite{Pimentel:2022fsc}, as shown explicitly for single-exchange diagrams (see~\cite{Jazayeri:2022kjy} for an explanation of this phenomenon related to the various time scales involved). This phenomenon is not apparent at strong mixing for our parameter choices. However, more importantly, in this regime, the total observable signal is dominated by the (would-be) double-exchange diagram, coming from the interaction unavoidably generated by the non-linearly realised symmetry in the presence of a non-trivial sound speed for the additional field, a point that was not appreciated before.

\subsection{Cosmological Collider Flow}
\label{subsec: cosmo flow collider}

\begin{figure}[h!]
   \centering
   \hspace*{0.5cm}
        \includegraphics[width=0.9\textwidth]{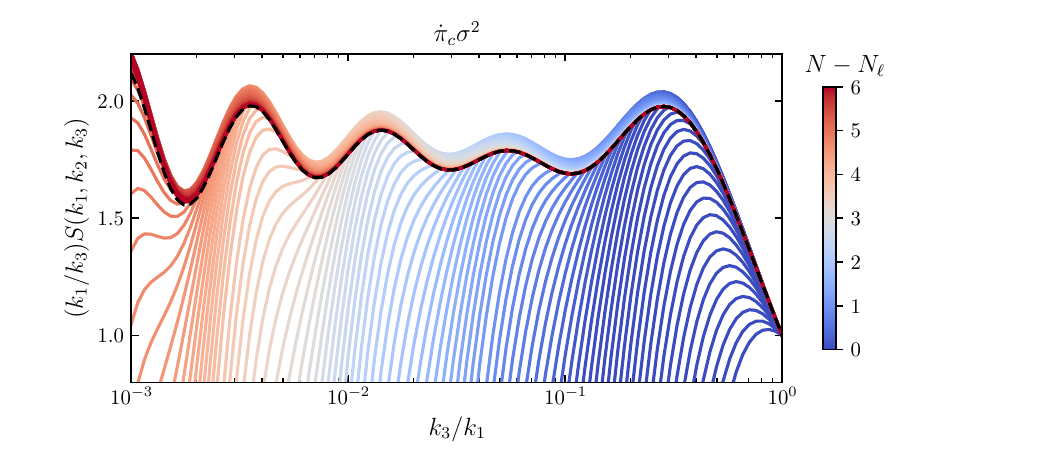}
   \caption{Time evolution of the shape function for the $\dot{\pi}_c\sigma^2$ interaction in the isosceles-triangle configuration $k_1=k_2$, for $\mu_{\text{eff}} = 5, \rho/H=0.1$ and $c_s=0.1$. The measure of time $N-\Nl$ is defined to be the time in $e$-folds elapsed since sound-horizon crossing of the long mode $\kl = k_3$. We have normalised the amplitude of the signal to unity in the equilateral configuration for illustration purposes. The dotted black line corresponds to the shape function at $N - \Nl=20$, a time by which the signal has become stationary.}
  \label{fig:CoCoFlow}
\end{figure}

Following the time evolution of the three-point correlators for a fixed momentum configuration allows us to reconstruct the flow of the cosmological collider signal. This is shown in Figure~\ref{fig:CoCoFlow} where we depict the shape function in the isosceles-triangle configuration as function of time. The measure of time is taken to be the number of $e$-folds elapsed since sound horizon crossing of the long mode: $N-\Nl$, with $\Nl$ such that $\kl=k_3=a (\Nl)H(\Nl)$. One can notice that the cosmological collider signal builds up progressively as function of time, starting with configurations close to the equilateral one. This representation makes it clear that the squeezed limit of the three-point correlator is a probe of the super-horizon evolution of the additional massive field $\sigma$ between $\Nl$ and $N$. This example showcases the utility of the cosmological flow approach as it enables us to reveal the physics of various processes by analysing the characteristic time scales at play, allowing for a direct access to the bulk time evolution of correlators. A video of the cosmological collider flow can be found \href{https://github.com/deniswerth/Cosmological-Collider-Flow}{here}.\footnote{\href{https://github.com/deniswerth/Cosmological-Collider-Flow}{github.com/deniswerth/Cosmological-Collider-Flow}} It is instructing to notice the different behaviours between the weakly and strongly mixed cases. At weak mixing, the system is underdamped and the signal exhibits temporal oscillations about its late-time asymptotic form before converging. At strong mixing, the two fields are so bound that the late-time signal is asymptotically reached with overdamped oscillations.

\section{Primordial Features}
\label{sec:features}

We have illustrated that the cosmological flow is a generic framework that enables one to study non-Gaussian phenomenology in scale-invariant cases that are not easily accessible by analytical means. However, realistic models of inflation generically lead to features in the scale dependence of primordial correlation functions. In this section, we will show that features can be accounted for systematically, both in the scale-dependence and the shape-dependence of correlators. As a proof of concept to illustrate the utility of our approach, we will first perform an exact calculation of the power spectrum and the bispectrum scale-dependence in the case of a time-varying quadratic mixing, and compare it with the predictions given by an effective single-field theory with a time-varying speed of sound. We will then show how features are encoded in the bispectrum squeezed limit, giving a new way of probing the inflationary landscape through soft limits of cosmological correlators, and therefore extending the reach of cosmological collider phenomenology.

\subsection{Effective Description of Features}
\label{subsec:features_setup}

Let us first discuss how time-varying background quantities arise in the effective description of fluctuations presented in Section \ref{subsec:EFT}. We refer the reader to this section for details.

\paragraph{Non-linearly realised symmetry.} The mixing action in the unitary gauge (\ref{eq:mixing_action_unitary_gauge}) is parameterised by a set of mass scales $\tilde{M}_i(t)$. After reintroducing the Goldstone boson $\pi$, expanding the mass scales to first order around the inflationary attractor gives
\begin{equation}
    \tilde{M}_i(t) \rightarrow \tilde{M}_i(t + \pi) \approx \tilde{M}_i(t) + \dot{\tilde{M}}_i(t)\pi + \ldots\,.
\label{eq:expansion-feature}    
\end{equation}
If the time scales of variation of the $\tilde{M}_i(t)$ are too small, i.e.~if the features that are considered are too sharp, the expansion~\eqref{eq:expansion-feature} may break down. This corresponds to a violation of perturbative unitarity, see e.g~\cite{Behbahani:2011it,Bartolo:2013exa,Adshead:2014sga,Cannone:2014qna}. Concentrating for simplicity on $c_s=1$ and at weak mixing, the (minimum) time scale of variation $\Delta t$ of any time-dependent quantity should obey $\Delta t> 1/f_\pi$. This comes from requiring that the Taylor expansion in \eqref{eq:expansion-feature} is valid, i.e.~$\pi < \Delta t$, with $\pi=\pi_c/f_\pi^2$ and $\pi_c \sim \omega$, at the energy $\omega \sim 1/\Delta t$ excited by the feature. Only a time-dependence in $\tilde{M}_1$ generates an additional cubic interaction compared to the ones present in (\ref{eq:pi_sigma_decoupling_limit}), that reads
\begin{equation}
    \mathcal{L}_{\pi-\sigma}/a^3 \supset 6 \tilde{M}_1^2(t) \dot{\tilde{M}}_1(t)\, \pi \dot{\pi}\sigma\,.
\end{equation}
For this reason, we will only consider a time dependence for $\tilde{M}_1$ throughout this section, setting $\tilde{M}_2 = \tilde{M}_3 = 0$. We also set the intrinsic sound speed of $\pi$ to unity $c_s=1$. In terms of the canonically normalised field $\pi_c$, the mixing Lagrangian we study is
\begin{equation}
\label{eq:mixing_lagrangian_time_dependent_rho}
    \mathcal{L}_{\pi-\sigma}/a^3 = \rho(t) \dot{\pi}_c \sigma - \frac{\rho(t)}{2f_\pi^2} (\partial_\mu \pi_c)^2 \sigma + \frac{\dot{\rho}(t)}{f_\pi^2} \pi_c \dot{\pi}_c \sigma\,,
\end{equation}
where we have defined the quadratic mixing $\rho(t) = 2 \tilde{M}_1^3(t)/f_\pi^2$. Notice that operators weighted by higher derivatives of $\rho$ appear if $\tilde{M}_1(t+\pi)$ is expanded to higher order around $\tilde{M}_1(t)$. We discard such terms as they appear at quartic order (and higher) in the fields. Also note that we have neglected any time variation of the Hubble parameter $H$ and the mass of the field $\sigma$ in this effective description of fluctuations. We will come back to this point in the next section.

\paragraph{Effective single-field theory.} When the field $\sigma$ is sufficiently massive, one can integrate it out and obtain an effective theory for $\pi$. Indeed, solving the linear equation of motion for $\sigma$ to leading order in powers of $\Box/m^2$, that is $\sigma = \rho\dot{\pi}_c/m^2$, and plugging the solution back in the original Lagrangian leads to
\begin{equation}
\label{eq:single-field_Lagrangian_features}
    \mathcal{L}/a^3 = \frac{1}{2\tilde{c}_s^2(t)}\dot{\pi}_c^2 - \frac{1}{2}\frac{(\partial_i \pi_c)^2}{a^2} - \frac{1}{2f_\pi^2} \left(\frac{1}{\tilde{c}_s^2(t)} - 1\right)\dot{\pi}_c (\partial_\mu \pi_c)^2 - \frac{\dot{\tilde{c}}_s(t)}{f_\pi^2 \tilde{c}_s(t)} \pi_c \dot{\pi}_c^2\,,
\end{equation}
where the time-dependent sound speed $\tilde{c}_s(t)$ is defined in (\ref{eq:induced_sound_speed}). Let us now discuss the regime of validity of this effective theory. Clearly, neglecting the gradient term compared to the mass term $\partial_i^2/a^2 \ll m^2$ means that (\ref{eq:single-field_Lagrangian_features}) describes the dynamics of low-energy modes with physical momenta $k/a$ such that $k^2/a^2\ll m^2$. In fact, it can be shown that requiring that the derivative expansion in powers of $\Box/m^2$ be under control implies a set of \textit{generalised adiabaticity conditions}\footnote{The first two conditions for $n=1, 2$ can be easily found by considering the next-to-leading order term $\Box\dot{\pi}_c/m^2$ in the derivative expansion. Using the quadratic equation of motion for $\pi_c$, one obtains
\begin{equation}
    \Box\dot{\pi}_c = \left(1-\tilde{c}_s^2\right)\frac{\partial_i^2\dot{\pi}_c}{a^2} - 4\tilde{c}_s^2 \, \frac{\dot{\tilde{c}}_s}{\tilde{c}_s} \frac{\partial_i^2\pi_c}{a^2} + H^2 \left[6\frac{\dot{\tilde{c}}_s}{\tilde{c}_s} - 2 \left(\frac{\dot{\tilde{c}}_s}{H\tilde{c}_s}\right)^2 + 3\frac{\dot{H}}{H^2} - 2\frac{\ddot{\tilde{c}}_s}{H^2 \tilde{c}_s}\right]\dot{\pi}_c\,.
\end{equation}
Requiring $\Box\dot{\pi}_c \ll m^2\dot{\pi}_c$, so that the derivative expansion is under perturbative control, imposes some restrictions. Looking at the third term gives the following conditions
\begin{equation}
    \frac{\dot{\tilde{c}}_s}{\tilde{c}_s} \ll m, \hspace*{0.5cm} \epsilon \ll \left(\frac{m}{H}\right)^2, \hspace*{0.5cm} \frac{\ddot{\tilde{c}}_s}{\tilde{c}_s} \ll m^2\,,
\end{equation}
whereas the first and the second terms are negligible as long as $k^2/a^2\ll m^2$ (see e.g. \cite{Garcia-Saenz:2019njm} for more details).
} \cite{Cespedes:2012hu, Achucarro:2012sm, Ach_carro_2014, Garcia-Saenz:2019njm}
\begin{equation}
\label{eq:generalised_adiabaticity_conditions}
    \frac{1}{\tilde{c}_s}\frac{\mathrm{d}^n \tilde{c}_s}{\mathrm{d}t^n} \ll m^n\,,
\end{equation}
for $n\geq 1$. These conditions indicate that the time scale of variation $\Delta t$ of \textit{all} time-varying quantities are bounded from below by the mass scale of the heavy field $\Delta t \gg m^{-1}$. In particular, this leaves room for the speed of sound to vary significantly on a time scale smaller than $H^{-1}$ but still larger than $m^{-1}$ if the field $\sigma$ is heavy enough. 

\vskip 4pt
In addition to the previous adiabaticity conditions, since the dispersion relation of the low-energy modes is $\omega^2 = \tilde{c}_s^2 k^2$, the derivative expansion is reliable if \cite{2011Cremoni, Baumann:2011su}
\begin{equation}
\label{eq:mass_condition}
    \omega \ll \omega_{\text{new}} = \tilde{c}_s m\,,
\end{equation}
where $\omega_{\text{new}}$ marks the energy scale at which higher-order derivative terms should be taken into account. Now, in order for the effective theory to have predictable power, one must require that $\omega_{\text{new}}$ be well above the Hubble scale. Indeed, if $\omega_{\text{new}} \ll H$, the effective theory would describe modes that are outside the sound horizon $k/a\ll m\ll H/\tilde{c}_s$. Such modes could not be initialised in the proper vacuum without the knowledge of the underlying UV completion. Therefore, for the effective theory to be valid, we also require
\begin{equation}
\label{eq:mass_condition2}
    m \gg H/\tilde{c}_s\,.
\end{equation}

\subsection{Correlated Sharp Features}
\label{subsec:correlated_features}

Due to the non-linearly realised symmetry, features in the scale-dependence of the power spectrum are also imprinted in the bispectrum scale-dependence. This can be explicitly seen from the mixing Lagrangian in the full theory~(\ref{eq:mixing_lagrangian_time_dependent_rho}) or in the effective theory~(\ref{eq:single-field_Lagrangian_features}). Such correlated features have been extensively studied in the context of the single-field effective theory with a time-varying sound speed \cite{Achucarro:2012fd, Gong:2014spa, Achucarro:2014msa,Palma:2014hra}. Here, we will perform exact calculations in both full and effective theories.

\paragraph{Time dependence.} Although the mechanisms leading to features can be very model-dependent, we focus on a single form of the quadratic mixing. We choose its time dependence to be parameterised by
\begin{equation}
\label{eq: rho time dependence}
    \rho(t) = \rho_0 \left(a/a_0\right)^{-n} \sin[\omega_c(t-t_0)] \times \frac{1}{2}\left[1+\tanh\left(\frac{t-t_0}{\Delta t}\right)\right] \,,
\end{equation}
where $\mu_c = \omega_c/H$ determines the background oscillation frequency in Hubble units, and $\rho_0$ is the amplitude of the feature. In the following, we will choose $\rho_0$ such that the deviation of the curvature power spectrum from scale invariance satisfies current observational bounds on cosmological scales.
The feature is turned on continuously at $t = t_0$ with a step of width $\Delta t$ that dictates its sharpness. We also consider that the feature is diluted by $(a/a_0)^{-n}$ where $a_0=a(t_0)$. This overall time dependence of the mixing describes well multi-field inflationary scenarios in which the background trajectory undergoes a turn in field space. If the turn is sharp, it will excite negative-frequency modes and produce correlation functions with oscillations linear in $k$~\cite{Achucarro:2010da,Chen:2011zf}. Those oscillations are characteristic of any sharp feature, be it single field or multi-field, and they are independent of $\omega_c$. They only probe a very short time range located at the sharp feature. Depending on the values of the different parameters, the turn may also transiently excite a background massive field, leading to damped oscillations until the trajectory settles down along the light direction, corresponding to a dilution parameter $n=3/2$.  
For $\omega_c \gtrsim H$, these background oscillations resonate with the quantum oscillations of fluctuations inside the Hubble radius, leading to oscillations linear in $\log k$ and of frequency $\omega_c$~\cite{Chen:2008wn}. Resonant features are particularly interesting as they can tell apart the inflationary scenario from alternative ones, as exemplified by the primordial standard clocks program~\cite{Chen:2011zf,Chen:2011tu, Noumi:2013cfa, Saito:2012pd, Chen:2014joa, Chen:2014cwa, Chen:2015lza, Braglia:2021rej, Braglia:2022ftm}. Both sharp and resonant features may also be encountered in single-field scenarios with either transient (bump, dip), or ever-present (tiny oscillatory component) deviations from the slow-roll dynamics,
see for instance~\cite{Chen:2006xjb, Chen:2008wn,Chluba:2015bqa, Slosar:2019gvt, Achucarro:2022qrl}.

\vskip 4pt
Before presenting numerical results, we briefly digress to discuss the purpose and limitations of our approach. From an EFT point of view, all couplings of the theory are generic functions of time that need to be inferred from data. However, notice that specific background models result in general in relationships between the time-dependencies of these couplings, whose knowledge can be used to parameterise the EFT couplings. In the following, for simplicity, we adopt a modelling where only the time dependence of the quadratic mixing is taken into account. First, we assume that the time dependence of the quadratic mixing is not reflected in the evolution of the background spacetime, that is we assume that there is no variation of the Hubble parameter. Moreover, we do not assume any time dependence for the mass of the fluctuation $\sigma$, although in concrete multi-field realisations it is known to contain---in addition to a contribution from the second derivative of the potential---an explicit dependence on the quadratic mixing and the background field-space geometry~\cite{Gordon:2000hv,GrootNibbelink:2001qt,Renaux-Petel:2015mga}.\footnote{From $m^2(t)\sigma^2 \rightarrow m^2(t+\pi) \sigma^2 \approx \left(m + \dot{m}\,\pi\right)^2\sigma^2$, a non-trivial time variation of the mass of the field $\sigma$ would generate the cubic interaction $\mathcal{L}_{\pi-\sigma}/a^3 \supset \frac{\dot{(m^2)}}{2f_\pi^2}\, \pi_c \sigma^2$ (see~\cite{Garcia-Saenz:2019njm} for the complete cubic action for inflationary fluctuations written in comoving gauge in non-linear two-field sigma model, and~\cite{Pinol:2020kvw} for the generalisation to any number of fields, where it is explicitly shown that this operator is present.).
}
Therefore, our approach disregards effects such as resonances from the time variation of the mass of the heavy field fluctuation with the same frequency $\omega_c$. Of course, it would be possible to solve the background dynamics and feed \textsf{CosmoFlow} with the actual time evolution of all varying quantities. This is precisely the purpose of the already existing tools \textsf{PyTransport} and \textsf{CppTransport} \cite{Mulryne:2016mzv, Seery:2016lko, Ronayne:2017qzn}. We instead embrace an effective description of fluctuations, having the advantage of rendering the symmetries manifest, and focusing directly on observables rather than the underlying model building.

\begin{figure}[t!]
    \centering
    \hspace*{-1.8cm}
    \subfloat{\includegraphics[width=1.2\textwidth]{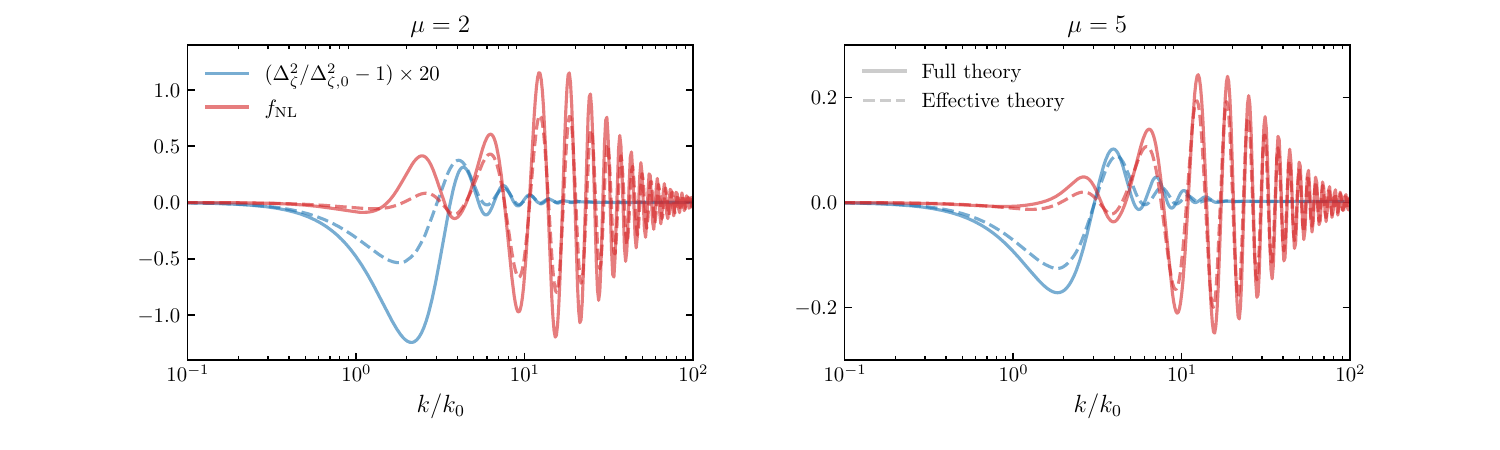}}
    \hfill
    \vspace*{-1cm}
    \caption{The relative power spectrum to the featureless one $\Delta_{\zeta,0}^2$ and the bispectrum $\fnl$ in the equilateral configuration $k_1=k_2=k_3=k$ as functions of $k/k_0$, computed in the full theory (\textit{solid lines}) and in the effective single-field theory (\textit{dashed lines}) both for $\mu=2$ (\textit{left panel}) and $\mu=5$ (\textit{right panel}). The feature, parameterised by $\rho_0/H=20, \mu_c = 0.1, n=3/2$, and $\Delta t = 0.1H^{-1}$, is localised at $t = t_0$ which corresponds to the scale $k=k_0$.}
    \label{fig:Scale-Dependence}
\end{figure}

\paragraph{Numerical results.} For $(\omega_c, 1/\Delta t )\ll m$, the adiabaticity conditions~(\ref{eq:generalised_adiabaticity_conditions}) are satisfied and we expect that the effects coming from the heavy physics are well captured by a reduced speed of sound, provided that~\eqref{eq:mass_condition2} holds. Whenever $\Delta t \ll m^{-1}$ or $\omega_c \gg m$, the effective theory is expected to break down. Here, we focus on the particularly interesting regime $\Delta t \lesssim m^{-1}$ while maintaining $\mu_c \ll \mu \equiv \sqrt{m^2/H^2-9/4}$. Figure~\ref{fig:Scale-Dependence} shows the numerical results for $\rho_0/H=20, \mu_c=0.1$, $\Delta t = 0.1 H^{-1}$ and $n=3/2$, both for $\mu=2$ and $\mu=5$. These parameters describe a sharp feature where the resonant effects from the oscillations, that occur only for $\mu_c \gtrsim 1$, are absent. We display the power spectrum (\textcolor{pyblue}{blue} line) and the bispectrum in the equilateral configuration (\textcolor{pyred}{red} line), computed both in the full (solid line) and the effectively single-field (dashed line) theories. For illustration purposes, we have normalised the power spectrum to the featureless one and mutiplied it by a factor of 20. We have chosen the overall amplitude $\rho_0$ so that the power spectrum presents features of order a few percents for $\mu=2$ and is compatible with current bounds from CMB observations. This figure exhibits the presence of correlated oscillations in $2k/k_0$ for the power spectrum and $3k/k_0$ for the bispectrum, characteristic of a sharp feature that eventually decays at smaller scales. It is interesting to observe that the bispectrum envelop reaches a maximum at $k/k_0 \sim 20$. These features are compatible with previous findings in the literature, including general statements about the universal running and model-dependent envelop in~\cite{Chen:2011zf}. With $\Delta t = 0.1 H^{-1}$ and our choice of masses, it is clear that the adiabaticity conditions~(\ref{eq:generalised_adiabaticity_conditions}) are not satisfied. Nonetheless, it is interesting to see that the effective theory already provides a reasonable description, modestly reproducing the correct amplitude of the feature but replicating the oscillatory profile.
For larger masses $m \gg \Delta t^{-1} = 10H$, we expect the power spectrum and the bispectrum in equilateral configurations obtained with the effective theory to be indistinguishable from the ones obtained with the full theory.

\subsection{Probing Inflationary Landscapes with Cosmological Colliders}
\label{subsec:feature_CC_signal}

We now study the bispectrum shape dependence at a given overall scale. Importantly, the explicit breaking of scale invariance by the time dependence of the quadratic mixing, and therefore of the cubic one by symmetry, induces unobservable contributions that should be subtracted from any bispectrum calculation. Here, we prove that after subtracting these gauge artefacts, the observable bispectrum encodes non-standard cosmological collider signals whose frequency is partially dictated by the background frequency. 

\paragraph{Consistency relation and observable bispectrum.} In its squeezed limit, the observable primordial bispectrum measured in cosmological data, is found by carefully subtracting the leading-order contribution coming from an effective rescaling of the background experienced by short-wavelength fluctuations, in presence of a long-wavelength $\zeta$ mode~\cite{Tanaka:2011aj, Baldauf:2011bh, Pajer:2013ana}. This gauge artefact---since a rescaling of the background does not constitute an observable quantity---which is present in soft limits of all primordial correlation functions, is determined by so-called consistency relations.
For the bispectrum, squeezed configurations are famously dominated by the power spectrum scale-dependence~\cite{Maldacena:2002vr, Creminelli:2004yq, Cheung:2007sv}
(see also e.g.~\cite{Creminelli:2011rh,Creminelli:2012ed, Senatore:2012wy, Assassi:2012zq,Hinterbichler:2013dpa,Kundu:2014gxa, Kundu:2015xta, Cabass:2016cgp, Suyama:2020akr} for related works)
\begin{equation}
\label{eq: bispectrum consistency relation}
    \underset{k_3 \ll k_1}{\text{lim}}\frac{\braket{\zeta_{\bm{k}_1} \zeta_{\bm{k}_2}\zeta_{\bm{k}_3} }' }{P_\zeta (k_3) }=-P_\zeta(\ks)\,\frac{\partial \log[\ks^3 P_\zeta(\ks)]}{\partial\log \ks} + \mathcal{O}((\kl/\ks)^2)\,,
\end{equation}
where $P_\zeta(k) = \braket{\zeta_{\bm{k}} \zeta_{-\bm{k}}}'$, $\bm{k}_{\ell} = \bm{k}_3$ and $\bm{k}_{\text{s}} = \bm{k}_1+\bm{k}_{\ell}/2 = (\bm{k}_1  - \bm{k}_2)/2$. When computing the bispectrum in the pure scale-invariant limit, i.e.~$\partial \log[k^3 P_\zeta(k)]/\partial\log k = n_s-1 = 0$, the gauge artefact contribution to the bispectrum vanishes in the squeezed limit, and observable effects, which normally show up at next-to-leading order, become automatically the dominant contribution. However, in the presence of features, $n_s(k)\neq 1$ and the consistency relation~(\ref{eq: bispectrum consistency relation}) should be subtracted to obtain theoretical predictions that can be readily compared to actual cosmological data. As we will see with a concrete example supported by both an exact numerical calculation and analytical insights, it is crucial to properly subtract the consistency relations in cosmological collider signals when features are present.
Indeed, one could wrongly conclude that the observable signal in the squeezed limit is hidden below a dominant and universal contribution, which is nothing but the aforementioned gauge artefact.

\vskip 4pt
When the quadratic mixing strength is weak, some analytical insight can be gained by evaluating the bispectrum shape using perturbative techniques. Using the standard in-in formalism~\cite{Weinberg:2005vy} (see also~\cite{Chen:2010xka, Wang:2013zva} for reviews), the bispectrum shape is composed of two contributions $S(k_1, k_2, k_3) = S_1(k_1, k_2, k_3) + S_2(k_1, k_2, k_3)$, where $S_1$ corresponds to a quadratic mixing insertion prior to the cubic interaction and $S_2$ the opposite. Specifying to isosceles configurations $k_1=k_2 \simeq \ks$ and $k_3=\kl$ where $\ks$ is the short (i.e.~hard) momentum and $\kl$ is the long (i.e.~soft) momentum, each of the two shape contributions have two fundamental permutations, $S_i = S_i^{\mathtt{a}} + S_i^{\mathtt{b}}$ for $i=\{1, 2\}$, corresponding to having either $\ks$ (denoted in the following permutation ``$\mathtt{a}$") or $\kl$ (denoted in the following permutation ``$\mathtt{b}$") carried by a $\zeta$ mode attached to the quadratic mixing. Note that each contribution ``$\mathtt{a}$" includes two additional internal permutations accounting for the exchange of $\kl$ and $\ks$ for both $\zeta$ modes attached to the cubic interaction. Diagrammatically and using the above notations, the total shape is given by the four following contributions\footnote{Time runs from bottom to top until the end of inflation denoted by the \textcolor{gray}{grey} line, \textcolor{pyred}{red} lines represent propagators of $\zeta$ that are independent---in this perturbative scheme---of those of $\sigma$ in \textcolor{pyblue}{blue}, black dots are quadratic and cubic vertices.}
\begin{equation}
\label{eq: feature shape four contributions}
    \begin{aligned}
    \hspace*{-0.2cm}
    \raisebox{-\height/2}{
    \begin{tikzpicture}[line width=1. pt, scale=2.5]
\draw[pyred] (0.2, 0) -- (0.4, -0.3);
\draw[pyred] (0.6, 0) -- (0.4, -0.3);
\draw[pyred] (1.2, 0) -- (1, -0.6);
\draw[pyblue] (0.4, -0.3) -- (1, -0.6);
\draw[fill=black] (0.4, -0.3) circle (.03cm);
\draw[fill=black] (1, -0.6) circle (.03cm);
\draw[lightgray, line width=0.8mm] (0.1, 0) -- (1.3, 0);
\node at (0.2, 0.15) {$\ks$};
\node at (0.6, 0.15) {$\kl$};
\node at (1.2, 0.15) {$\ks$};
\node at (0.7, -0.8) {$S_1^{\mathtt{a}}$};
    \end{tikzpicture}
    }
    +
    \raisebox{-\height/2}{
    \begin{tikzpicture}[line width=1. pt, scale=2.5]
\draw[pyred] (0.2, 0) -- (0.4, -0.3);
\draw[pyred] (0.6, 0) -- (0.4, -0.3);
\draw[pyred] (1.2, 0) -- (1, -0.6);
\draw[pyblue] (0.4, -0.3) -- (1, -0.6);
\draw[fill=black] (0.4, -0.3) circle (.03cm);
\draw[fill=black] (1, -0.6) circle (.03cm);
\draw[lightgray, line width=0.8mm] (0.1, 0) -- (1.3, 0);
\node at (0.2, 0.15) {$\ks$};
\node at (0.6, 0.15) {$\ks$};
\node at (1.2, 0.15) {$\kl$};
\node at (0.7, -0.8) {$S_1^{\mathtt{b}}$};
    \end{tikzpicture} 
    }
    +
    \raisebox{-\height/2}{
    \begin{tikzpicture}[line width=1. pt, scale=2.5]
\draw[pyred] (0.2, 0) -- (0.4, -0.6);
\draw[pyred] (0.6, 0) -- (0.4, -0.6);
\draw[pyred] (1.2, 0) -- (1, -0.3);
\draw[pyblue] (0.4, -0.6) -- (1, -0.3);
\draw[fill=black] (0.4, -0.6) circle (.03cm);
\draw[fill=black] (1, -0.3) circle (.03cm);
\draw[lightgray, line width=0.8mm] (0.1, 0) -- (1.3, 0);
\node at (0.2, 0.15) {$\ks$};
\node at (0.6, 0.15) {$\kl$};
\node at (1.2, 0.15) {$\ks$};
\node at (0.7, -0.8) {$S_2^{\mathtt{a}}$};
    \end{tikzpicture} 
    }
    +
    \raisebox{-\height/2}{
    \begin{tikzpicture}[line width=1. pt, scale=2.5]
\draw[pyred] (0.2, 0) -- (0.4, -0.6);
\draw[pyred] (0.6, 0) -- (0.4, -0.6);
\draw[pyred] (1.2, 0) -- (1, -0.3);
\draw[pyblue] (0.4, -0.6) -- (1, -0.3);
\draw[fill=black] (0.4, -0.6) circle (.03cm);
\draw[fill=black] (1, -0.3) circle (.03cm);
\draw[lightgray, line width=0.8mm] (0.1, 0) -- (1.3, 0);
\node at (0.2, 0.15) {$\ks$};
\node at (0.6, 0.15) {$\ks$};
\node at (1.2, 0.15) {$\kl$};
\node at (0.7, -0.8) {$S_2^{\mathtt{b}}$};
    \end{tikzpicture} 
    }
    \end{aligned}\,.
\end{equation}
The contributions $S_1^{\mathtt{b}}$ and $S_2^{\mathtt{b}}$ contain the cosmological collider signals as these channels effectively probe the super-horizon decay of the massive field for a large hierarchy of scales $\kl \ll \ks$. As we will see in what follows and for the quadratic mixing given in Eq.~(\ref{eq: rho time dependence}), the contribution $S_1^{\mathtt{b}}$ contains the dominant cosmological collider signal. On the other hand, the contributions $S_1^{\mathtt{a}}$ and $S_2^{\mathtt{a}}$, scaling as $\sim (\kl/\ks)^{-1+2n}$ (see insert below), become large and dominant in the squeezed limit. Note that for an ever-present feature $n=0$, the latter shape contributions are of the local type. One might expect that the cosmological collider signal is therefore completely out of reach. However, these large contaminating signals are not observable. Indeed, translating the bispectrum consistency relation to the dimensionless shape function for isosceles kinematic configurations, we obtain the remarkable identity
\begin{eBox}
\begin{equation}
\label{eq: subtracting consistency relation}
    \begin{aligned}
    \hspace*{-0.4cm}
    \raisebox{-\height/2}{
    \begin{tikzpicture}[line width=1. pt, scale=2.5]
\draw[pyred] (0.2, 0) -- (0.4, -0.3);
\draw[pyred] (0.6, 0) -- (0.4, -0.3);
\draw[pyred] (1.2, 0) -- (1, -0.6);
\draw[pyblue] (0.4, -0.3) -- (1, -0.6);
\draw[fill=black] (0.4, -0.3) circle (.03cm);
\draw[fill=black] (1, -0.6) circle (.03cm);
\draw[lightgray, line width=0.8mm] (0.1, 0) -- (1.3, 0);
\node at (0.2, 0.15) {$\ks$};
\node at (0.6, 0.15) {$\kl$};
\node at (1.2, 0.15) {$\ks$};
\node at (0.7, -0.8) {$S_1^{\mathtt{a}}$};
    \end{tikzpicture}
    }
    \hspace*{-0.2cm}
    +
    \hspace*{-0.2cm}
    \raisebox{-\height/2}{
    \begin{tikzpicture}[line width=1. pt, scale=2.5]
\draw[pyred] (0.2, 0) -- (0.4, -0.6);
\draw[pyred] (0.6, 0) -- (0.4, -0.6);
\draw[pyred] (1.2, 0) -- (1, -0.3);
\draw[pyblue] (0.4, -0.6) -- (1, -0.3);
\draw[fill=black] (0.4, -0.6) circle (.03cm);
\draw[fill=black] (1, -0.3) circle (.03cm);
\draw[lightgray, line width=0.8mm] (0.1, 0) -- (1.3, 0);
\node at (0.2, 0.15) {$\ks$};
\node at (0.6, 0.15) {$\kl$};
\node at (1.2, 0.15) {$\ks$};
\node at (0.7, -0.8) {$S_2^{\mathtt{a}}$};
    \end{tikzpicture} 
    }
    \hspace*{-0.4cm}
    \underset{\kl\ll\ks}{=}
    - \frac{\ks}{4\kl} \frac{\partial\log k^3}{\partial\log k}\left[ 
    \raisebox{-\height/3}{
    \begin{tikzpicture}[line width=1. pt, scale=2.5]
\draw[pyred] (0.2, 0) -- (0.4, -0.6);
\draw[pyred] (1, 0) -- (0.8, -0.6);
\draw[pyblue] (0.4, -0.6) -- (0.8, -0.6);
\draw[fill=black] (0.4, -0.6) circle (.03cm);
\draw[fill=black] (0.8, -0.6) circle (.03cm);
\draw[lightgray, line width=0.8mm] (0, 0) -- (1.2, 0);
\node at (0.2, 0.15) {$k$};
\node at (1, 0.15) {$k$};
    \end{tikzpicture} 
    }
    \right]_{k=\ks} \hspace*{-0.2cm} + \mathcal{O}\left(\kl/\ks\right)
    \end{aligned}
\end{equation}
\end{eBox}
We leave the details of the derivation in the insert below. This property holds for any time dependence of the quadratic mixing. Importantly, this means that the gauge artefact encoded by the consistency relation is included in the two contributions ``$\mathtt{a}$'', and therefore that the observable squeezed bispectrum is composed only of the two contributions of the type ``$\mathtt{b}$", as well as the remaining ``$\mathtt{a}$" contributions. Whenever the various contributions are calculated separately, as in the analytical calculation in the perturbative regime, one can simply focus on the type ``$\mathtt{b}$" contributions and discard the leading-order ``$\mathtt{a}$" ones. On the contrary, whenever the bispectrum shape is computed all at once, as in the cosmological flow approach, one has to explicitly subtract the contribution proportional to the tilt of the power spectrum. The effect from rescaling the background by a long mode also partially constrains higher-order corrections, see e.g.~\cite{Creminelli:2011rh,Creminelli_2012, Assassi:2012zq, Hinterbichler:2013dpa, Berezhiani:2013ewa}, especially terms of order $(\kl/\ks)^2$ in Eq.~\eqref{eq: bispectrum consistency relation} which are physical. Still, the $S_1^{\mathtt{b}}$ channel remains the dominant contribution in the squeezed limit, as we will see.

\begin{framed}
{\small \noindent {\it Derivation.}---In this insert, we derive~(\ref{eq: subtracting consistency relation}) for a quadratic mixing $\rho(t)$ with arbitrary time dependence. The leading-order power spectrum correction due to the massive field exchange can be recast as
\begin{equation}
    (\Delta_\zeta^2 - \Delta_{\zeta, 0}^2)/\Delta_{\zeta, 0}^2 = -\pi \left(\frac{\rho_0}{H}\right)^2 \int_0^\infty\d x \int_{x}^\infty \d y\, \mathcal{I}(x, y; x_0)\,,
\end{equation}
where the integrand reads
\begin{equation}
    \mathcal{I}(x, y; x_0) = (xy)^{-\frac{1}{2}}\tilde{\rho}(x) \tilde{\rho}(y) \sin(x)\text{Im}\left[H_{i\mu}^{(1)}(x) H_{i\mu}^{(2)}(y) e^{-iy}\right]\,,
\end{equation}
with $\tilde{\rho}=\rho/\rho_0$ denoting the dimensionless linear coupling. For the specific choice~(\ref{eq: rho time dependence}) and neglecting the window function for simplicity, one has $\tilde{\rho}(x) = (x/x_0)^n \sin[\mu_c\log(x/x_0)]$ with $x_0=-k\tau_0$, although the derivation is applicable to any time dependence of the linear coupling. In the isosceles kinematic configuration, the dimensionless shape function generated by the single exchange of the massive field is a function of $\kappa \equiv \kl/\ks$, encoding the shape dependence, and $\lambda \equiv \kl/k_0 = \kappa x_0$ which encodes the overall scale dependence due to the feature. We focus on the cubic interaction $\dot{\rho}(t)\pi_c \dot{\pi}_c\sigma$ as the other interactions---shift symmetric in $\pi_c$---do not enter the consistency relation. The two contributions to the shape function for which a long mode is attached to the cubic interaction read
\begin{equation}
    S_i^{\mathtt{a}}(\kappa, \lambda) = -\frac{\pi}{4} \left(\frac{\rho_0}{H}\right)^2 \kappa^{-1} \int_0^\infty\d x \int_{x}^\infty \d y\, \mathcal{J}_i^{\mathtt{a}}(x, y; \kappa, \lambda)\,,
\end{equation}
where $i=\{1, 2\}$ refers to either having the quadratic or the cubic interaction inserted at earlier times, as depicted in~(\ref{eq: feature shape four contributions}), and the integrands read
\begin{gather}
    \begin{aligned}
        \mathcal{J}_1^{\mathtt{a}}(x, y; \kappa, \lambda) &= \tfrac{\partial_t\tilde{\rho}}{H}(x)\tilde{\rho}(y) (xy)^{-\frac{1}{2}} \text{Im}\left[(1+i\kappa x + \kappa^2(1+ix))e^{-i(1+\kappa)x}\right]\text{Im}\left[H_{i\mu}^{(1)}(x) H_{i\mu}^{(2)}(y) e^{-iy}\right] \,,\\
        \mathcal{J}_2^{\mathtt{a}}(x, y; \kappa, \lambda) &= \tilde{\rho}(x)\tfrac{\partial_t\tilde{\rho}}{H}(y) (xy)^{-\frac{1}{2}} \sin(x)\text{Im}\left[H_{i\mu}^{(1)}(x) H_{i\mu}^{(2)}( y) (1+i\kappa y + \kappa^2(1+iy))e^{-i(1+\kappa)y}\right]\,.
    \end{aligned}
    \raisetag{17pt}
\end{gather}
Let us now estimate the scalings of these contributions in the squeezed limit $\kappa\ll 1$. The exponential $e^{-iy}$ in the integrand $\mathcal{J}_1^{\mathtt{a}}$ effectively renders all contributions from the $y\geq 1$ region negligible, so that the integral is dominated by the saddle point around $y\sim 1$. As a result, one can expand in $\kappa y \ll 1$ inside the integral for $\kappa\ll 1$. Similarly inside the second nested integral for $x\leq y$, one can take the $\kappa x\ll 1$ limit, which could also be seen from the factor $e^{-i(1+\kappa)x}$ killing all contributions from $(1+\kappa)x\approx x\geq 1$. At leading order in $\kappa$, the integrand $\mathcal{J}_1^{\mathtt{a}}$ simplifies to
\begin{equation}
    \mathcal{J}_1^{\mathtt{a}}(x, y; \kappa, \lambda) = -\tfrac{\partial_t\tilde{\rho}}{H}(x)\tilde{\rho}(y) (xy)^{-\frac{1}{2}} \sin(x)\text{Im}\left[H_{i\mu}^{(1)}(x) H_{i\mu}^{(2)}(y) e^{-iy}\right]\,.
\end{equation}
For the quadratic mixing~(\ref{eq: rho time dependence}), and taking into account the dependence of $\tilde{\rho}$ on $x_0=\kappa/\lambda$, the overall scaling in the squeezed limit is found to be
\begin{equation}
    S_1^{\mathtt{a}}(\kappa, \lambda) \sim \kappa^{-1+2n} \lambda^{2n} \,\text{Im}\left[\alpha_{1, +}^{\mathtt{a}} \kappa^{2i\mu_c} + \alpha_{1, -}^{\mathtt{a}} \kappa^{-2i\mu_c} + \alpha_{1, 0}^{\mathtt{a}} \right]\,,
\end{equation}
where the coefficients $\alpha_{\pm}$ and $\alpha_0$ depend on $\mu, n$ and $\lambda$. Note that in the presence of a slowly-decaying feature in the mixing such that $n\leq 1/2$, this shape contribution dominates in the squeezed limit as it grows as function of $\kappa$, even for a process involving the exchange of a heavy field. Notably, the shape oscillates in $\log \kappa$ at a frequency solely set by the background frequency $\mu_c$. Proceeding in the same way, the contribution $S_2^{\mathtt{a}}$ has the same scaling in the squeezed limit. Note that these scalings and the oscillations with frequency $2\mu_c$ can be recovered using the effective single-field Lagrangian~(\ref{eq:single-field_Lagrangian_features}) in the corresponding regime of validity.

\vskip 4pt
Additionally, from
\begin{equation}
    \frac{\partial \tilde{\rho}(x)}{\partial k} = \frac{x}{k^2} \left.\frac{\partial \tilde{\rho}(\tau)}{\partial \tau}\right|_{\tau=-\tfrac{x}{k}} = \frac{1}{k} \frac{\partial_t\tilde{\rho}}{H}(x)\,,
\end{equation}
we obtain the remarkable identity
\begin{equation}
    \frac{\partial\mathcal{I}}{\partial k} = \frac{1}{k} \left(\mathcal{J}_1^{\mathtt{a}} + \mathcal{J}_2^{\mathtt{a}}\right)\left(1+\mathcal{O}(\kappa^2 )\right)\,.
\end{equation}
Lastly, one needs to pull the derivative out of the time integrals. Defining $I(x_0) = \int_0^\infty \d x f(x, x_0) \int_x^\infty \d y g(y, x_0)$ where $\mathcal{I}(x, y; x_0) = f(x, x_0)g(y, x_0)$, and differentiating with respect to $x_0$, one straightforwardly obtains
\begin{equation}
    \frac{\partial I}{\partial x_0} = \int_0^\infty \d x \,\frac{\partial f}{\partial x_0}(x, x_0)\int_x^\infty \d y  \,g(y, x_0) + \int_0^\infty \d x \,f(x, x_0)\int_x^\infty \d y  \, \frac{\partial g}{\partial x_0}(y, x_0) \,.
\end{equation}
Altogether, we obtain the identity
\begin{equation}
    \sum_{i=1, 2} S_i^{\mathtt{a}}(\kappa, \lambda) = -\frac{1}{4\kappa} \left.\frac{\partial\log (k^3 \Delta_\zeta^2)}{\partial\log k}\right|_{k = \ks}+\mathcal{O}(\kappa)\,,
\end{equation}
which is precisely the consistency relation at the level of the shape function.
 }
\end{framed}

\begin{figure}[t!]
    \centering
    \hspace*{-1.8cm}
    \subfloat{\includegraphics[width=1.2\textwidth]{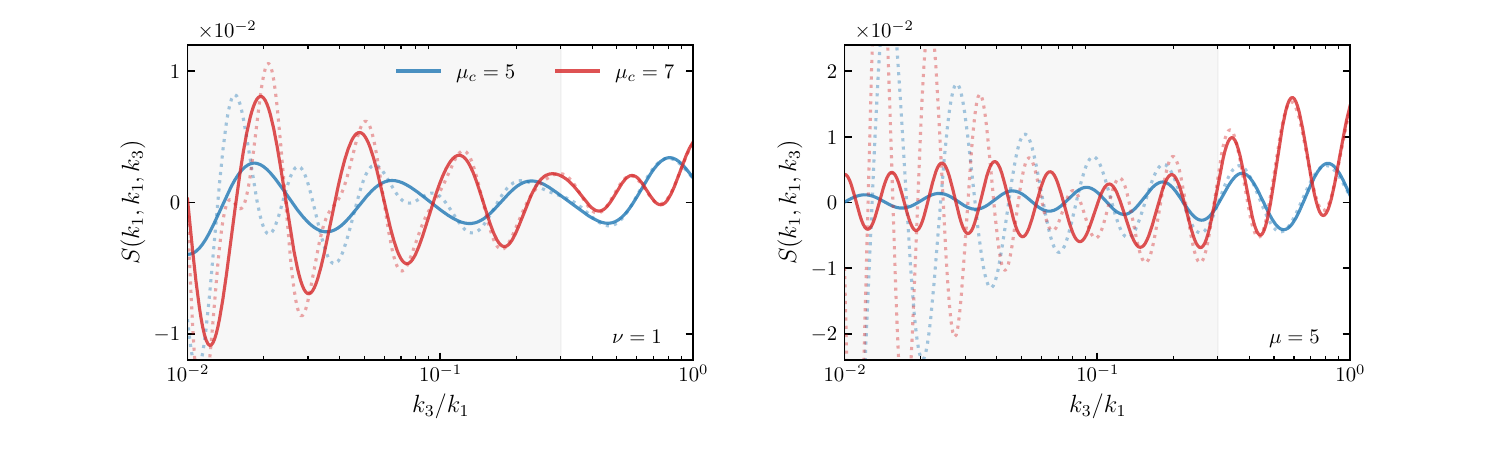}}
    \hfill
    \vspace*{-1cm}
    \caption{Observable shape functions (\textit{solid lines}) dictated by the mixing Lagrangian~(\ref{eq:mixing_lagrangian_time_dependent_rho}) with the time-dependent mixing $\rho(t)$ given in Eq.~(\ref{eq: rho time dependence}), in the isosceles kinematic configuration $k_1=k_2$ at a fixed overall scale $\log(k_3/k_0)=2$ where $k_0=(aH)(t_0)$ and $t_0$ is the characteristic time of the feature, for a light field $\nu=1$ (\textit{left panel}) and a heavy field $\mu=5$ (\textit{right panel}), varying the background frequency $\mu_c = \textcolor{pyblue}{5}, \textcolor{pyred}{7}$, and setting $n=0$. The observable signal is reconstructed from the computed one (\textit{dotted line}) upon numerically subtracting the consistency relation~(\ref{eq: subtracting consistency relation}). The shaded region corresponds to the squeezed limit where the templates~(\ref{eq: feature shape templates}) are valid. We have chosen $\rho_0/H=0.1$ to respect the current bounds on the power spectrum~\cite{Planck:2018jri}.}
    \label{fig:Shape Features}
\end{figure}

\paragraph{Cosmological collider signals.} We focus on the bispectrum arising from the cubic interaction $\dot{\rho}(t) \pi_c \dot{\pi}_c\sigma$ because it is the dominant one for rapid oscillations $\mu_c>1$, with the quadratic mixing time dependence of Eq.~(\ref{eq: rho time dependence}). We show in Figure~\ref{fig:Shape Features} the corresponding shape functions for a light and heavy field $\sigma$, for $n=0$ and different background frequencies $\mu_c$. The dashed line corresponds to the cosmological flow results---hence containing all contributions, even the unobservable ones corresponding to $S_{1,2}^\mathtt{a}$ in the perturbative analysis---and the solid lines correspond to the shapes that are directly measurable in cosmological data as they are found after subtracting the consistency relation given solely in terms of the power spectrum. Note that in the cosmological flow framework, both the two- and three-point correlators are computed simultaneously as the flow equations form a closed system. It therefore requires no additional computations to directly obtain the observable signals out of a single computation. Note also that our numerical approach is precise enough (both at the level of the power spectrum and of the bispectrum) to allow for a clean subtraction of dominant unobservable contributions in order to recover a subdominant scaling behaviour. This is readily visible, for example, in the light-field case for $\mu_c=5$ where the raw signal (\textit{dashed line}) presents a distorted oscillatory pattern due to the interference of various signals with different frequencies, as well as a quickly growing envelope (as $\ks/\kl$) in the squeezed limit. The observable signal after subtracting gauge artefacts (\textit{solid line}), on the other hand, features a single frequency and is growing at a slower pace (as $(\kl/\ks)^{-1/2}$).

\vskip 4pt
A notable feature of these signals is that their frequencies are partially set by the background frequency $\mu_c$. This offers the possibility to probe inflationary landscapes with cosmological collider signals.\footnote{Instead, for cubic interactions of constant strengths, cosmological collider signals are conventional with frequencies only set by the mass of the additional field, despite the time-dependence of the quadratic mixing \cite{Chen:2022vzh}. This can also be seen in our proof in the insert below.
}
Indeed, in the perturbative analysis and considering only the shape contributions $S_1^{\mathtt{b}}$ and $S_2^{\mathtt{b}}$, as the other two are removed by the consistency relation, the non-analytical signal lying in the squeezed limit ($k_3\ll k_1$) of the shape function both for the exchange of a \textit{light} ($m/H\leq 3/2$) and a \textit{heavy} ($m/H\geq 3/2$) field is found to be
\begin{eBox}
\begin{equation}
\label{eq: feature shape templates}
    S(k_1, k_2, k_3) =
    \begin{cases}
      \left(\frac{k_3}{k_1}\right)^{1/2+n - \nu} \mathcal{A}\cos\left( \mu_c \log\left(\frac{k_3}{k_1}\right) + \varphi\right)\,, & \text{(light)}\\
      \left(\frac{k_3}{k_1}\right)^{1/2+n}\sum_{\pm}\mathcal{A}_\pm \cos\left((\mu\pm\mu_c)\log\left(\frac{k_3}{k_1}\right) + \varphi_\pm\right)\,, & \text{(heavy)}
    \end{cases} 
\end{equation}
\end{eBox}
with $\mu=\sqrt{(m/H)^2-9/4}$ and $\nu=i\mu$. Details on the derivation can be found in the insert below. A generic feature of these signals is that the power-law scaling acquires an additional $(k_3/k_1)^n$ suppression compared to the standard one~\cite{Chen:2009zp}. This is a direct consequence of the dilution of the feature in $(a/a_0)^{-n}$. For the exchange of a light field, in addition to the power-law scaling that depends on the mass, the shape presents oscillations in $\log(k_3/k_1)$ whose frequency is set by the background frequency $\mu_c$. The light-field template is not valid for the massless case ($\nu=3/2$) and an ever-present feature $n=0$ as curvature modes never freeze on super-horizon scales. In the case of a heavy-field exchange process, the power-law scaling is saturated and the cosmological collider signals present modulated frequencies in $\mu\pm\mu_c$. These beats are characteristic of interference patterns between the usual oscillations of the heavy field on super-horizon scales and an additional oscillating signal, in this case generated by the background. Cross-correlating the analytic template~(\ref{eq: feature shape templates}) with features in the power spectrum, namely their amplitudes, and the scale-dependence of the bispectrum, can remove the degeneracy between the mass of the additional field and the oscillating background frequency. Remarkably, we find numerically that the templates~(\ref{eq: feature shape templates}) hold in the strong mixing regime. Altogether, these results provide a new way of probing the landscape of inflationary backgrounds through soft limits of cosmological correlators. In particular, for light fields with $\nu>1/2+n$, the oscillations that modulate a growing shape envelope in the squeezed limit are expected to generate distinctive oscillations in the scale-dependent galaxy bias. It would also be interesting to find a concrete background model that reproduces these features, as our analysis entirely relies on symmetries at the level of inflationary fluctuations. We leave these open questions for an upcoming work.

\begin{framed}
{\small \noindent {\it Derivation.}---In this insert, we estimate the scaling of the \textit{observable} shape contributions $S_1^{\mathtt{b}}$ and $S_2^{\mathtt{b}}$ in the squeezed limit and derive the analytical templates~(\ref{eq: feature shape templates}). After subtracting the consistency relation at leading order in $\kl/\ks\ll1$, the physical shape is given by
\begin{equation}
    S(\kappa, \lambda) = -\frac{\pi}{4} \left(\frac{\rho_0}{H}\right)^2 \kappa^{-1} \int_0^\infty\d x \int_x^\infty \d y \sum_{i=1, 2} \mathcal{J}_i^{\mathtt{b}}(x, y; \kappa, \lambda)\,,
\end{equation}
where $\kappa \equiv \kl/\ks$ and $\lambda \equiv \kl/k_0$. The integrands read
\begin{equation}
    \begin{aligned}
        \mathcal{J}_1^{\mathtt{b}}(x, y; \kappa, \lambda) &= \tfrac{\partial_t\tilde{\rho}}{H}(x)\tilde{\rho}(y) \kappa^2(xy)^{-\frac{1}{2}} \text{Im}\left[(1+ix)e^{-2ix}\right]\text{Im}\left[H_{i\mu}^{(1)}(\kappa x) H_{i\mu}^{(2)}(\kappa y) e^{-i\kappa y}\right]\,,\\
        \mathcal{J}_2^{\mathtt{b}}(x, y; \kappa, \lambda) &= \tfrac{\partial_t\tilde{\rho}}{H}(x)\tilde{\rho}(y) \kappa^2(xy)^{-\frac{1}{2}} \sin(\kappa x)\text{Im}\left[H_{i\mu}^{(1)}(\kappa x) H_{i\mu}^{(2)}(\kappa y) (1+iy)e^{-2i y}\right]\,.
    \end{aligned}
\end{equation}
We now estimate the scalings and behaviours of these contributions in the squeezed limit $\kappa\ll1$. The exponential $e^{-2ix}$ in the integrand $\mathcal{J}_1^{\mathtt{b}}$ selects the integration region $x<1$, discarding possible resonances between this exponential and the oscillating feature. Note that we have neglected the window function, which corresponds to considering $\kl/k_0\gg 1$, i.e.~that both the long and the short modes exit the horizon well after the characteristic scale of the feature. We can then expand in $\kappa x\ll 1$ inside the integral and use the late-time behaviour of the Hankel function $H_{i\mu}^{(1)}(z) \approx A_\mu^+ z^{i\mu} + A_\mu^- z^{-i\mu}$ for $z\ll1$. After changing variables $y\rightarrow \tilde{y} = \kappa y$, the lower bound of the inner integral can be extended up to $\kappa x\rightarrow 0$, effectively decoupling the two integrals. Altogether, the overall scaling in the squeezed limit for the shape contribution $S_1^{\mathtt{b}}$ is found to be
\begin{equation}
    \begin{aligned}
    S_{1, \text{light}}^{\mathtt{b}}(k_1, k_2, k_3) &\sim \kappa^{\tfrac{1}{2}+n-\nu} \lambda^{-2n} \text{Im}\left[\alpha_{1, +}^{\mathtt{b}} \kappa^{i \mu_c} + \alpha_{1, -}^{\mathtt{b}} \kappa^{-i \mu_c}\right]\,,\\
    S_{1, \text{heavy}}^{\mathtt{b}}(k_1, k_2, k_3) &\sim \kappa^{\tfrac{1}{2}+n} \lambda^{-2n} \text{Im}\left[\alpha_{1, ++}^{\mathtt{b}} \kappa^{ i(\mu + \mu_c)} + \alpha_{1, +-}^{\mathtt{b}} \kappa^{ i(\mu - \mu_c)} \right.\\
    &\left.\hspace*{2.5cm}+ \alpha_{1, -+}^{\mathtt{b}} \kappa^{ -i(\mu - \mu_c)} + \alpha_{1, --}^{\mathtt{b}} \kappa^{ -i(\mu + \mu_c)}\right]\,,
    \end{aligned}
\end{equation}
for the light and the heavy case, respectively. Similarly, for the integrand $\mathcal{J}_2^{\mathtt{b}}$, the exponential $e^{-2iy}$ selects the region $y<1$ and we can take the $\kappa y\ll 1$ limit inside the integrand. As $x<y$, we can also take the limit $\kappa x\ll 1$. Using the Hankel expansion and at leading order in $\kappa$, the scalings are found to be
\begin{equation}
    \begin{aligned}
    S_{2, \text{light}}^{\mathtt{b}}(k_1, k_2, k_3) &\sim \kappa^{2+2n-2\nu} \lambda^{-2n} \text{Im}\left[\alpha_{2, +}^{\mathtt{b}} \kappa^{2 i \mu_c} + \alpha_{2, 0}^{\mathtt{b}} + \alpha_{2, -}^{\mathtt{b}} \kappa^{-2 i \mu_c} \right]\,,\\
    S_{2, \text{heavy}}^{\mathtt{b}}(k_1, k_2, k_3) &\sim \kappa^{2+2n} \lambda^{-2n} \text{Im}\left[\alpha_{2, ++}^{\mathtt{b}} \kappa^{2i(\mu+\mu_c)} + \alpha_{2, --}^{\mathtt{b}} \kappa^{-2i(\mu+\mu)} \right.\\
    &\left.+ \alpha_{2, +-}^{\mathtt{b}} \kappa^{2i(\mu-\mu_c)} + \alpha_{2,-+}^{\mathtt{b}} \kappa^{-2i(\mu-\mu_c)} + \alpha_{2, 0+}^{\mathtt{b}} \kappa^{2i\mu_c} \right.\\
    &\left.+ \alpha_{2, 0-}^{\mathtt{b}} \kappa^{-2i\mu_c} + \alpha_{2, +0}^{\mathtt{b}} \kappa^{2i\mu} + \alpha_{2, -0}^{\mathtt{b}} \kappa^{-2i\mu} + \alpha_{2, 00}^{\mathtt{b}}\right]\,.
    \end{aligned}
\end{equation}
The contribution $S_1^{\mathtt{b}}$ is therefore the dominant contribution. Defining $\mathcal{A}e^{i\varphi} = (\alpha_{1, +}^{\mathtt{b}} - \alpha_{1, -}^{\mathtt{b}*})/i$, and $\mathcal{A}_\pm e^{i\varphi_\pm} = (\alpha_{1, +\pm}^{\mathtt{b}} - \alpha_{1, -\mp}^{\mathtt{b}*})/i$, we obtain the templates~(\ref{eq: feature shape templates}).
Setting aside intricacies related to IR ($x, y \rightarrow 0$) divergences of the corresponding integrals that cancel when adding all contributions, the amplitudes and the phases of these signals can be computed analytically. 
}
\end{framed}

\newpage
\section{Conclusions and Outlook}
\label{sec:Conclusion}

In this paper, following a shorter letter~\cite{Werth:2023pfl}, we have presented a systematic framework---the cosmological flow---to compute and study primordial correlators. This numerical approach consists in solving differential equations \textit{in time} satisfied by the correlators, and enables us to obtain exact results in regimes that are challenging to probe analytically. We have exemplified this method with an extensive study of correlators coming from the exchange of a massive scalar particle, both at strong mixing---adopting a non-perturbative treatment of quadratic interactions---and in non scale-invariant theories. This development provides an efficient way to study the rich phenomenology of primordial non-Gaussianities as it makes predictions straightforward. In order for the community to progress faster in these challenging directions, we have decided to publicly release the numerical implementation of our new approach as a \href{https://github.com/deniswerth/CosmoFlow}{freely available resource}.\footnote{In this repository, we have also gathered numerical results for power spectra and non-Gaussianities computed with our code, for different theories and parameter values. We plan to turn this into a bank of theoretical data to be nurtured by the work of the community.}

\vskip 4pt
In more details, we have reformulated the computation of tree-level in-in correlators by deriving a closed set of differential equations in time, providing direct insights into the bulk dynamics of equal-time correlators. While the structure of these flow equations is universal, the specific form of the (time-dependent) coefficients is theory dependent. These theories, formulated at the level of fluctuations, can include any number of scalar degrees of freedom with arbitrary dispersion relations and masses, coupled through any type of time-dependent interactions. Our approach is convenient as it enables one to decouple from a specific background and to focus directly on  observables. We have established the already-existing flow equations for the two- and three-point correlators by introducing handy diagrammatic rules. This has the advantage of following the usual Leibniz product rule of differentiation and making the contributions coming from an exchange or a contact diagram manifest. Using the diagrammatic representation of the flow equations, we have also derived the flow equations for all tree-level $n$-point correlators.

\vskip 4pt
In a second phase, we have applied our formalism to the study of correlators emerging from a theory where the Goldstone boson of broken time translations is coupled to an additional massive scalar field. After an extensive analysis of the quadratic theory, we have provided an exhaustive dictionary of strong coupling scales, perturbativity bounds and naturalness criteria for the theory under scrutiny. Such glossary is crucial for theoretical consistency and observability of primordial signals. We have provided exact results for the size of the non-Gaussian signal in the entire parameter space, including the strong mixing regime. Then, we have studied the shape dependence, varying the quadratic mixing strength and the speed of sound. We found that the resulting shapes have interesting new phenomenology compared to the known previous cases, namely the shapes coming from a single-exchange correlator at weak mixing. In particular, we have highlighted two features that had not been appreciated enough. First, as the quadratic mixing becomes stronger, the additional field acquires an \textit{effective mass} that sets the frequency of the cosmological collider signal. We have also presented exact results for the amplitude of the non-analytic signal, pointing out that the maximal amplitude is achieved when the mixing strength is comparable to the mass of the exchanged particle. Second, a large quadratic mixing results in a slight boost for the heavy fluctuation upon crossing the horizon, due to backreaction from the curvature fluctuation. This interesting effect is intrinsic to the strong mixing regime, and creates a phase shift in the cosmological collider signal. Differences between the weak and the strong mixing regimes can be appreciated by examining videos of the building of the cosmological collider signal as time passes, a unique opportunity offered by our approach. We have also unveiled a large cosmological collider signal, built from a non-linear interaction automatically generated by symmetry as soon as the \textit{additional field} has a reduced intrinsic speed of sound. Finally, we have applied the cosmological flow to non scale-invariant scenarios, where the quadratic mixing picks up a time dependence. We have provided exact results for the power spectrum and bispectrum---both its scale and shape dependence---in the regime where a single-field effective description breaks down. With minimal assumptions---a non-linearly realised symmetry and an oscillating background---, we have shown that the observable cosmological collider signal encodes information about the background, offering a novel approach to probe the inflationary landscape. This generic phenomenon, beyond the standard lore, breaks the direct link between the frequency of the cosmological collider signal and the mass of an additional field. 

\vskip 4pt
Heading towards a universal program to generate theoretical data, the cosmological flow has a number of benefits. First, it requires no analytical nor numerical mode functions as it focuses directly on correlators. Second, it effortlessly deals with the early-time $i\epsilon$ prescription as the correlators are evolved on the real time axis. Third, the procedure of setting the flow equations given a certain Lagrangian is straightforward. Finally, the cosmological flow systematically works for all $n$-point correlators. With this formalism at hand, we have paved the way towards a systematic investigation of the rich and fascinating subject of primordial correlators. Since observational progresses are expected to be made in the near future, the coming years will lead to a wealth of observational data and constraints on the statistical properties of primordial fluctuations. It is therefore important to provide an adequate amount of theoretical benchmarks, and establish the capability for an efficient exploration of the vast landscape of possible signals.\footnote{More ambitiously, we could include the cosmological flow routine in the already-existing chain of late-time cosmological tools. This would automate the generation of theoretical primordial data that can be directly used for CMB or LSS observables, hence extending the cosmological flow to the late-time Universe.} In this respect, the cosmological flow can both extend the reach of primordial correlators phenomenology and complement analytical computations by providing useful checks and generating intuition.

\vskip 4pt
Finally, yet importantly, our work suggests a number of directions for future exploration:

\begin{itemize}
    \item \textbf{Small-scale phenomenology.} Primordial density fluctuations on cosmological scales are probed by observations of the CMB and LSS. Density fluctuations on smaller scales that are forged in the ``dark era'' of inflation are not directly observable by themselves. Yet, they shape many properties of new observables like gravitational wave backgrounds and primordial black holes. In this respect, we highlight that the cosmological flow approach is not restricted to the largest cosmological scales but is also perfectly applicable to these questions. As the inflationary dynamics involved in this context often relies on a strong breaking of scale invariance and substantial non-linear effects which are difficult to handle analytically, a systematic approach like the cosmological flow provides an essential tool to illuminate the dark era of inflation.

    \item \textbf{Spinning fields.} It would be interesting to extend the cosmological flow to include spinning particles. On the one hand, in conventional setups of weak mixing and under the lamppost of weakly broken scale-invariance, it is known that the cosmological collider signal arising from the exchange of a spinning particle displays a characteristic angular dependence.
    On the other hand, computing correlators with external spin (for example gravitons, hence accounting for gravitational waves) is notoriously difficult because it is challenging to track the detailed time evolution of the physics in the bulk. The cosmological flow would encapsulate both cases as the flow equations couple \textit{all} correlators, that need to be solved at once. More practically, implementing particles with spin within the cosmological flow requires a careful treatment of the tensorial structures of the spinning fields, and hence necessitates additional indices that would account for polarisation states. A possible direction for future work would be to relax the assumptions of weak mixing and/or scale-invariance, and to extend the phenomenology of spinning fields to regimes that have not yet been explored.
    
    \item \textbf{Path integral formulation.} It would also be preferable to derive the flow equations using the path integral formulation instead of the canonical operator formalism. The reasons for this a priori cumbersome change of perspective are at least twofold. First, most of the theories at the level of the fluctuations are defined by a Lagrangian (or equivalently an action), whereas the current formalism required switching to the Hamiltonian, expanding the Dyson series, and doing field contractions by hand order by order. Second, from a particle physics perspective, a path integral formulation of the flow approach would allow us to include gauge theories.
    Indeed, when working with spinning fields, one needs to deal with problems such as gauge redundancies. In flat space, these complications are avoided by using Feynman diagrams to express the $S$-matrix elements order by order in perturbation theory. Using a path integral approach and importing standard techniques from flat-space QFT, one could expect to render the implementation of fields with spin within the cosmological flow more limpid.
    
    \item \textbf{Loop-level.} Mostly for technical reasons, the large and rich phenomenology of inflationary correlators mainly comes from explicit calculations at tree level. \textit{Per contra}, pushing the available technology further in perturbation theory and developing new techniques to compute loops is of primary importance. Practically, it would extend the reach of the cosmological collider phenomenology by including particles within and beyond the Standard Model (it is known that spin-$1/2$ fermions or the Higgs field contribute to primordial non-Gaussianities at loop level). Yet, aside from isolated examples, there is no systematic treatment of loops in cosmological spacetimes. That is why it would be interesting to derive the flow equations at one-loop level. One can already notice that extending the cosmological flow to loop-level would require way more computational capability. For example, computing the two-point correlators at one-loop level requires to solve the flow equations for the three-point functions (working at loop-level precisely means that we do not neglect anymore non-linearities of the three-point correlators on the two-point correlators), which requires first the solutions of the two-point correlators flow equations at tree-level (this is equivalent to solving the quadratic theory). Therefore, one is not able to solve all equations at once---as we have done so far---but rather order by order.
    
    \item \textbf{Unitarity.} More conceptually, the cosmological flow---by evolving correlators in time---brings the question of unitarity. Indeed, a unitary time evolution ensures that the total probability is conserved over time, and therefore the theory is complete in the sense that it makes self-contained predictions for observables. In flat-space QFT, important properties follow from unitarity. For example, correlation functions must \textit{factorize} into products of lower-order ones. Within the cosmological flow, we have seen that the time evolution of correlators is \textit{sourced} by products of lower-order ones. This unforeseen concordance implores a better understanding of the status of the flow equations and their flat-space analogues. Finally, it would be interesting to see how non-unitary effects arising from the interference between the two branches of the in-in contour---such as decoherence and dissipation---can be incorporated in the flow equations.
\end{itemize}

\paragraph{Acknowledgements.}

We are grateful to Angelo Caravano, Sebasti\'an C\'espedes, Xingang Chen, Paolo Creminelli, Tanguy Grall, Sadra Jazayeri, Liam McAllister, Scott Melville, David Mulryne, Toshifumi Noumi, Enrico Pajer, Gui Pimentel, David Seery, David Stefanyszyn, Xi Tong, Dong-Gang Wang, Lukas Witkowski and Yuhang Zhu for helpful discussions. We would like to especially thank David Mulryne and David Seery for enlightening conversations when this work was initiated. DW and SRP are supported by the European Research Council under the European Union’s Horizon 2020 research and innovation programme (grant agreement No 758792, Starting Grant project GEODESI). L.P. acknowledges funding support from the Initiative Physique des Infinis (IPI), a research training program of the Idex SUPER at Sorbonne Université. This article is distributed under the Creative Commons Attribution International Licence (\href{https://creativecommons.org/licenses/by/4.0/}{CC-BY 4.0}).

\newpage
\appendix

\section{Details on the Cosmological Flow}
\label{app:CosmoFlow_Approach}

In this appendix, we provide additional materials on the cosmological flow that have been used in this work. In Section \ref{app:Fourier_summation}, we give details about the Fourier summation. We then derive the initial conditions for the flow equations of the two- and three-point correlators in Section \ref{app:2pt_initial_conditions} and \ref{app:3pt_initial_conditions} respectively, generalising the results of \cite{Dias:2016rjq} to arbitrary cubic interactions.

\subsection{Extended Fourier Summation}
\label{app:Fourier_summation}

We deal in full generality with $N$ scalar degrees of freedom. Latin letters ($a, b, \ldots$) run over phase-space coordinates from $1$ to $2N$. Greek letters ($\alpha, \beta, \ldots$) run over all fields or their conjugate momenta from $1$ to $N$. 

\paragraph{Fourier indices.} We have seen in the main text that it is helpful to condense the notations and the Fourier integrals in the summation convention. We indicate that this extended interpretation is used by typesetting the labels to which it applies in a \textsf{sans serif} face. An index contraction such as $A_{\sf{a}}B^{\sf{a}}$ reads
\begin{equation}
A_{\sf{a}}B^{\sf{a}} = \sum_{a} \int \frac{\mathrm{d}^3k_a}{(2\pi)^3} A_a(\bm{k}_a)B^{a}(\bm{k}_a)\,.
\end{equation}
Note that we position indices to respect the normal rules for covariant expressions. There is an extra complexity with $\bm{k}$-labels included because Fourier-space expressions
sometimes produce the $\delta$-function $\delta_{\sf{ab}} = (2\pi)^3 \delta_{ab}\,\delta^{(3)}(\bm{k}_a + \bm{k}_b)$. Within a Fourier integral, this reverses the sign of a $\bm{k}$-label, which we indicate by decorating the label with a bar as in $B^{\bar{\sf{a}}}$. Hence, 
\begin{equation}
A_{\sf{a}}B^{\bar{\sf{a}}} =  \sum_{a} \int \frac{\mathrm{d}^3k_a}{(2\pi)^3} A_a(\bm{k}_a)B^{a}(-\bm{k}_a)\,.
\end{equation}
A contraction with $\tensor{\delta}{^{\sf{a}}_{\sf{b}}}$ will bar an index $A^{\bar{\sf{a}}} = \delta{^{\sf{a}\sf{b}}}A_{\sf{b}}$. When dealing with field Greek indices $(\alpha, \beta)$ running from $1$ to $N$, we also indicate the existence of an extended Fourier summation by typesetting the labels in \textsf{sans serif} face $(\upalpha, \upbeta)$.

\paragraph{Commutator in Fourier space.} With $\bm{X}^a = (\bm{\varphi}^\alpha, \bm{p}^\beta)$ being the phase-space vector in the Heisenberg picture, the canonical commutation algebra has a compact expression
\begin{equation}
[\bm{X}^{\sf{a}}, \bm{X}^{\sf{b}}] = i \epsilon^{\sf{a}\sf{b}}\,, 
\end{equation}
where $\epsilon^{\sf{a}\sf{b}} \equiv (2\pi)^3\delta^{(3)}(\bm{k}_a + \bm{k}_b)\,\epsilon^{ab}$ and the $2N\times2N$ matrix can be written in block form
\begin{equation}
\epsilon^{ab} = 
\begin{pmatrix}
\bf{0} & \bf{1} \\
\bf{-1} & \bf{0}
\end{pmatrix}\,.
\end{equation}
We also assume that phase-space indices are organised so that a block of field labels are followed by a block of momentum labels, in the same order.

\subsection{Two-point Correlator Initial Conditions}
\label{app:2pt_initial_conditions}

The derivation of initial conditions for the two-point correlators is done using the path integral formalism~\cite{Chen:2017ryl} instead of the canonical in-in formalism, phrased in operator language.\footnote{We will closely follow the derivation in \cite{Butchers:2018hds}.} In line with the point of view of the ``cosmological flow", this approach circumvents the use of mode functions. 

\paragraph{Path integral formalism.} The quadratic theory that we consider (see Section \ref{sec:Initial_Conditions}) is
\begin{equation}
    S^{(2)} = \frac{1}{2}\int \mathrm{d}t\, a^3\left(\bar{\Delta}_{\upalpha}\left(\dot{\bm{\varphi}}^{\upalpha}\right)^2 - c_{\upalpha}^2\, \frac{k^2}{a^2}\, \left(\bm{\varphi}^{\upalpha}\right)^2 \right)\,.
\end{equation}
After integrating by parts while assuming the boundary terms vanish at infinity and introducing conformal time $\mathrm{d}\tau = \mathrm{d}t/a$, the above action in real space reads
\begin{equation}
    S = -\frac{1}{2}\int \mathrm{d}\tau\, \mathrm{d}^3x\, a^2\, \bm{\varphi}^{\alpha} \bm{\mathcal{O}}_{\alpha} \bm{\varphi}^{\alpha}\,, \hspace*{0.5cm} \text{with} \hspace*{0.5cm} \bm{\mathcal{O}}_{\alpha} = \bar{\Delta}_{\alpha} \left(\partial_\tau^2 + 2 \frac{a'}{a}\partial_\tau\right) - c_{\alpha}^2\, \partial_i^2\,,
\end{equation}
where a prime denotes derivative with respect to conformal time. It can be shown that the partition function can be recast into two copies of path integrals, one going forward in time for the time-ordered factor in Eq.~(\ref{eq:in-in}), and the other going backward in time for the anti-time ordered factor. As a consequence, two sets of field configurations $\bm{\varphi}_+^{\alpha}$ and $\bm{\varphi}_-^{\alpha}$ are introduced\footnote{One also needs to introduce two sets of conjugate momenta $\bm{p}_+^\alpha$ and $\bm{p}_-^{\alpha}$ but the resulting generating functional is quadratic in the momenta so that the path integral is Gaussian and can be evaluated exactly. In this way one can trade a Hamiltonian formulation for a Lagrangian one.}, and the in-in generating functional can be written in the following way
\begin{equation}
\label{eq:generating_functional}
\begin{aligned}
    Z = &\int \left[\mathcal{D}\bm{\varphi}_+^{\alpha} \mathcal{D}\bm{\varphi}_-^{\alpha}\right] \exp\left\{-\frac{i}{2} \int_{\tau_0}^{\tau}\mathrm{d}\tau\mathrm{d}^3x\, a^2\, \bar{\bm{\varphi}}^{\alpha}
    \begin{pmatrix}
    \bm{\mathcal{O}} &  \\
       & -\bm{\mathcal{O}}
    \end{pmatrix}_{\alpha} \bm{\varphi}^{\alpha}
    \right\} \\
    &\times \prod_{\alpha, \bm{x}}\delta\left(\bm{\varphi}_+^{\alpha}(\tau, \bm{x}) - \bm{\varphi}_-^{\alpha}(\tau, \bm{x})\right)\,,
\end{aligned}
\end{equation}
where, in this formula, $\bm{\varphi}^\alpha=(\bm{\varphi}_+^\alpha, \bm{\varphi}_-^\alpha)$ and $\bar{\bm{\varphi}}^\alpha$ denotes the transpose vector. The time $\tau_0$ should be set in the infinite past and $\tau$ is the time at which we want to initialise the two-point correlation functions. The $\delta$-function enforces that the fields coincide at the upper bound $\bm{\varphi}_+^{\alpha}(\tau, \bm{x}) = \bm{\varphi}_-^{\alpha}(\tau, \bm{x})$. We define the time-ordered two-point function of fields as
\begin{equation}
    \mathsf{\Delta}_{++}^{\alpha\beta}(\tau_1, \bm{x}; \tau_2, \bm{y}) = \braket{\text{T}\,\bm{\varphi}^{\alpha}_+(\tau_1, \bm{x})\bm{\varphi}^{\beta}_+(\tau_2, \bm{y})}\,,
\end{equation}
with similar definitions for $\mathsf{\Delta}_{+-}, \mathsf{\Delta}_{-+}$ and $\mathsf{\Delta}_{--}$. The theory being quadratic and diagonal, it is an easy task to compute the $\mathsf{\Delta}_{\pm\pm}$. They are obtained by formally inverting the quadratic operator $\bm{\mathcal{O}}$,
\begin{equation}
    \begin{pmatrix}
    \bm{\mathcal{O}} &  \\
       & -\bm{\mathcal{O}} 
    \end{pmatrix}
    \begin{pmatrix}
    \mathsf{\Delta}_{++} & \mathsf{\Delta}_{+-} \\
    \mathsf{\Delta}_{-+} & \mathsf{\Delta}_{--} 
    \end{pmatrix}
    =
    -\frac{i}{a^2}\, 
    \begin{pmatrix}
    1 & 0 \\
    0 & 1 
    \end{pmatrix}\,
    \delta(\tau_1-\tau_2)\delta^{(3)}(\bm{x} - \bm{y})\,.
\end{equation}
These inhomogeneous equations can be solved exactly in Fourier space where the spatial dependence becomes diagonal. With the proper Bunch-Davies vacuum boundary conditions and imposing that $\mathsf{\Delta}_{++}$ (in Fourier space) oscillates with positive frequency in the far past, the solution is given by
\begin{equation}
\label{eq:time-ordered_2pt_functions}
    \mathsf{\Delta}_{++}^{\alpha\beta} = (2\pi)^3\delta^{\alpha\beta}\delta^{(3)}(\bm{k}_1 + \bm{k}_2) \, \bar{\Delta}_{\alpha} \, \frac{H^2}{2c_\alpha^3k^3}\, (1 + i c_{\alpha} k\tau_1) (1 - i c_{\alpha} k\tau_2) e^{ic_\alpha k(\tau_2 - \tau_1)}\,,
\end{equation}
assuming $\tau_2<\tau_1$. Here, $k = |\bm{k}_1| = |\bm{k}_2|$, and there is no sum on the repeated $\alpha$ index. We also recall that the index position is meaningless because the field space for fluctuations is flat i.e. field-space indices are lowered and raised with the flat Euclidean metric $\delta_{\alpha\beta}$. The remaining time-ordered two-point correlators can then be found. The $\delta$-function in (\ref{eq:generating_functional}) enforces that $\mathsf{\Delta}_{-+}(\tau, \tau_2) = \mathsf{\Delta}_{++}(\tau, \tau_2)$ for all $\tau_2$ when $\tau_1=\tau$ is the external time.
From the properties of the path integral, $\mathsf{\Delta}_{--}$ and $\mathsf{\Delta}_{+-}$ are Hermitian conjugates of $\mathsf{\Delta}_{++}$ and $\mathsf{\Delta}_{-+}$, respectively.

\paragraph{Equal-time correlators.} From the above solutions, we can get the equal-time two-point correlators of fields by setting $\tau_1 = \tau_2 = \tau$. Trading the time-dependence for a scale factor dependence using $\tau = -1/aH$, we obtain
\begin{equation}
\braket{\bm{\varphi}^{\alpha}_{\bm{k}}(\tau) \bm{\varphi}^{\beta}_{-\bm{k}}(\tau)}' = \delta^{\alpha\beta} \bar{\Delta}_{\alpha} \, \frac{H^2}{2c_\alpha^3k^3} \,\left(1 + \frac{c_\alpha^2 k^2}{a^2 H^2}\right)\,,
\end{equation}
where the prime denotes that we have extracted the momentum conservation $\delta$-function. Similarly one can derive the equal-time (rescaled) momentum-field and momentum-momentum two-point correlators.
Eventually, one finds the two-point correlator initial conditions in Eq.~(\ref{eq:initial_2pt_correlators}), expressed in terms of the diagonal matrix $\Delta_{\alpha\beta}=\delta^{\alpha\beta} \bar{\Delta}_{\alpha}$ (no sum).

\subsection{Three-point Correlator Initial Conditions}
\label{app:3pt_initial_conditions}

Let us now use the results of the previous section to derive the initial conditions for the three-point correlators. We are interested in computing the three-point correlators at some initial conformal time $\tau_0$ in the deep past with the Hamiltonian (\ref{eq:full_Hamiltonian}). Expanding the exponentials to leading-order in Eq.~(\ref{eq:in-in}) leads to
\begin{equation}
\hspace*{-0.2cm}
    \braket{\bm{X}^a(\tau_0, \bm{k}_1)\bm{X}^b(\tau_0, \bm{k}_2)\bm{X}^c(\tau_0, \bm{k}_3)} = i\, \Bigl< \int_{-\infty}^{\tau_0} \mathrm{d}\tau \left[ H_{\text{I}}(\tau), X^a(\tau_0, \bm{k}_1)X^b(\tau_0, \bm{k}_2)X^c(\tau_0, \bm{k}_3)\right]\Bigr>\,,
\end{equation}
where the expectation value is evaluated with respect to the usual vacuum of the free theory and $H_\text{I}$ is the full cubic Hamiltonian that can be read of Eq.~(\ref{eq:full_Hamiltonian}). To condensate notations, we will now on write the momentum dependency of the operators as subscripts. Assuming it is clear at which time the operators are evaluated, we will also omit the time dependency of the operators. By Fourier transforming the operators in $H_\text{I}$, the first term in the commutator becomes
\begin{equation}
    \braket{\bm{X}^a_{\bm{k}_1}\bm{X}^b_{\bm{k}_2}\bm{X}^c_{\bm{k}_3}} = i\, \int_{-\infty}^{\tau_0} \mathrm{d}\tau \,H_{\mathsf{def}}\int \frac{\Pi_{i}\,\mathrm{d}^3q_i}{(2\pi)^9}\, (2\pi)^3 \delta^{(3)}\left(\Sigma_i \bm{q}_i\right)\Bigl<X^{\mathsf{d}}_{\bm{q}_1}X^{\mathsf{e}}_{\bm{q}_2}X^{\mathsf{f}}_{\bm{q}_3}X^a_{\bm{k}_1}X^b_{\bm{k}_2}X^c_{\bm{k}_3} \Bigr>\,.
\end{equation}
We have written the integrals over Fourier modes to be explicit---this time not using the Fourier summation convention---to be clear about momentum dependencies. We can now perform Wick contractions between the different operators to rewrite the expectation value as a combination of two-point functions. Discarding disconnected contributions, we obtain
\begin{gather}
\begin{aligned}
    \braket{\bm{X}^a_{\bm{k}_1}\bm{X}^b_{\bm{k}_2}\bm{X}^c_{\bm{k}_3}} = i\, \int_{-\infty}^{\tau_0} \mathrm{d}\tau \,H_{\mathsf{def}}\int \frac{\Pi_{i}\,\mathrm{d}^3q_i}{(2\pi)^9}\, (2\pi)^3 \delta^{(3)}\left(\Sigma_i \bm{q}_i\right)&\left\{\langle X^{\mathsf{d}}_{\bm{q}_1}X^a_{\bm{k}_1}\rangle \langle X^{\mathsf{e}}_{\bm{q}_2}X^b_{\bm{k}_2}\rangle \langle X^{\mathsf{f}}_{\bm{q}_3}X^c_{\bm{k}_3}\rangle \right.\\
    &\left.+ \text{ 5 perms}\right\}\,.
\end{aligned}
\raisetag{18pt}
\end{gather}
We now need to specify the form of $H_{\sf{def}}$. For pedagogical reasons before treating the general case, we start with a simple example to illustrate how the machinery works.

\paragraph{Preliminaries.} Let us treat the concrete case of a cubic field-field-field interaction with no derivative for simplicity. Using the notations in Eq. (\ref{eq:full_Hamiltonian}), the cubic Hamiltonian reads
\begin{equation}
    H_{\mathsf{def}}\bm{X}^{\mathsf{d}}\bm{X}^{\mathsf{e}}\bm{X}^{\mathsf{f}} = A_{\upalpha\upbeta\upgamma} \bm{\varphi}^{\upalpha}\bm{\varphi}^{\upbeta}\bm{\varphi}^{\upgamma}\,,
\end{equation}
where we recall that the Latin indices run over the \textit{phase}-space variables and the Greek indices run over \textit{field}-space variables. Taking into account the possible non-canonical normalisation of the fields in the $\Delta_{\upalpha\upbeta}$ tensor and using the time-ordered two-point function (\ref{eq:time-ordered_2pt_functions}) derived in the previous section, the three-point initial correlator reads
\begin{gather}
\label{eq:preliminaries_3pt_initial_conditions}
\begin{aligned}
    &\braket{\bm{\varphi}^{\alpha}_{\bm{k}_1}\bm{\varphi}^{\beta}_{\bm{k}_2}\bm{\varphi}^{\gamma}_{\bm{k}_3}}  = 6i\, (2\pi)^3 \delta^{(3)}(\Sigma_i \bm{k}_i) \,\frac{H^6}{8e_3^3}\,\Delta_{\alpha\alpha} \Delta_{\beta\beta} \Delta_{\gamma\gamma}\times(1+ik_1\tau_0)(1+ik_2\tau_0)(1+ik_3\tau_0)e^{-ik_t\tau_0} \\
    &\times\left\{\int_{-\infty}^{\tau_0} \mathrm{d}\tau\,\Delta_{\alpha\alpha}(\tau) \Delta_{\beta\beta}(\tau) \Delta_{\gamma\gamma}(\tau)\, A_{\alpha\beta\gamma}(\tau)\, (1-ik_1\tau)(1-ik_2\tau)(1-ik_3\tau)e^{ik_t\tau}\right\} + \text{c.c.}\,,
\end{aligned}
\raisetag{22pt}
\end{gather}
where $k_t = k_1+k_2+k_3$. For convenience in later calculations, we also introduce the other symmetric polynomials  
\begin{equation}
    e_2 \equiv k_1 k_2 + k_1 k_3 + k_2 k_3\,, \hspace*{0.5cm}
    e_3 \equiv k_1 k_2 k_3\,.
\end{equation}
Note that one needs to carry the $\Delta$ dependency both for external (at time $\tau_0$) and internal (at time $\tau$) operators. There is no summation for the repeated indices $\alpha, \beta$ and $\gamma$. If we assume that the background parameters are slowly varying near $\tau=\tau_0$, then we can Taylor expand the third-order vertex function to first order $\left[\Delta_{\alpha\alpha} \Delta_{\beta\beta} \Delta_{\gamma\gamma}\,A_{\alpha\beta\gamma}\right](\tau) \approx \left[\Delta_{\alpha\alpha} \Delta_{\beta\beta} \Delta_{\gamma\gamma}\,A_{\alpha\beta\gamma}\right](\tau_0)$ and remove it from the time integral. In the end, the three-point correlator is given by
\begin{equation}
    \begin{aligned}
    \braket{\bm{\varphi}^{\alpha}_{\bm{k}_1}\bm{\varphi}^{\beta}_{\bm{k}_2}\bm{\varphi}^{\gamma}_{\bm{k}_3}} & =6\,\Delta_{\alpha\alpha} \Delta_{\beta\beta} \Delta_{\gamma\gamma}\, i\, (2\pi)^3 \delta^{(3)}(\Sigma_i \bm{k}_i) \,\frac{H^6}{8e_3^3} \, \Delta_{\alpha\alpha} \Delta_{\beta\beta} \Delta_{\gamma\gamma}\,A_{\alpha\beta\gamma}\\
    &\times (1+ik_1\tau_0)(1+ik_2\tau_0)(1+ik_3\tau_0)e^{-ik_t\tau_0} \\
    &\times \int_{-\infty}^{\tau_0} \mathrm{d}\tau\left\{ (1-ik_1\tau)(1-ik_2\tau)(1-ik_3\tau)e^{ik_t\tau}\right\} + \text{c.c.}\,.
    \end{aligned}
\end{equation}
In the first line, we have collected all ``constant" terms. These terms include the normalisation of the external and internal operators encoded in the $\Delta_{\upalpha\upbeta}$ tensor and the third-order vertex kernel from the considered cubic interaction. The remaining factor appears in all three-point correlators. The second line contains the external polynomial and the third line the internal polynomial. This simple example can now be generalised to the computation of all three-point correlators including all cubic interactions.

\paragraph{External polynomials.} The external polynomial depends on the correlator considered. There are four different types of such polynomials. Formally setting $\tau_1=\tau_0$ and $\tau_2=0$ in the time-ordered two-point function (\ref{eq:time-ordered_2pt_functions}), we see that the operators $\bm{\varphi}^\alpha$ and $\bm{p}^\alpha$ contribute the following polynomials
\begin{equation}
\label{eq:free_massless_mode_function}
    \bm{\varphi}^\alpha \rightarrow  (1 + ik \tau_0)e^{-ik\tau_0}\,, \hspace*{0.5cm} \bm{p}^\alpha \rightarrow  k^2 \tau_0 e^{-ik\tau_0}\,.
\end{equation}
Using these expressions, it is an easy task to derive the external polynomials
\begin{equation}
\label{eq:external_polynomials}
    \begin{aligned}
    \braket{\bm{\varphi}^{\alpha}_{\bm{k}_1}\bm{\varphi}^{\beta}_{\bm{k}_2}\bm{\varphi}^{\gamma}_{\bm{k}_3}} &\rightarrow \left(1 + ik_t \tau_0 - e_2\tau_0^2 - ie_3\tau_0^3\right)e^{-ik_t \tau_0}\,, \\
    \braket{\bm{p}^{\alpha}_{\bm{k}_1}\bm{\varphi}^{\beta}_{\bm{k}_2}\bm{\varphi}^{\gamma}_{\bm{k}_3}} &\rightarrow \left(k_1^2 \tau_0 + i k_1^2(k_2 + k_3)\tau_0^2 - e_3 k_1 \tau_0^3\right)e^{-ik_t \tau_0}\,, \\
    \braket{\bm{p}^{\alpha}_{\bm{k}_1}\bm{p}^{\beta}_{\bm{k}_2}\bm{\varphi}^{\gamma}_{\bm{k}_3}} &\rightarrow \left(k_1^2 k_2^2\tau_0 + i e_3 k_1 k_2 \tau_0^2 \right)e^{-ik_t \tau_0}\,, \\
    \braket{\bm{p}^{\alpha}_{\bm{k}_1}\bm{p}^{\beta}_{\bm{k}_2}\bm{p}^{\gamma}_{\bm{k}_3}} &\rightarrow e_3^2 \tau_0^3 e^{-ik_t \tau_0}\,.
    \end{aligned}
\end{equation}

\paragraph{Internal polynomials.} The internal polynomials depend on the cubic interaction we consider. From Eq.~(\ref{eq:full_Hamiltonian}), we see that we have four different cubic vertex integrals to perform, each of them depending on the third-order kernels
\begin{equation}
\label{eq:kernels}
    \begin{aligned}
    A_{\upalpha\upbeta\upgamma} &= (2\pi)^3 \delta^{(3)}(\bm{k}_1 + \bm{k}_2 + \bm{k}_3)\,A_{\alpha\beta\gamma}\,, \\
    B_{\upalpha\upbeta\upgamma} &= (2\pi)^3 \delta^{(3)}(\bm{k}_1 + \bm{k}_2 + \bm{k}_3)\,B_{\alpha\beta\gamma}\,, \\
    C_{\upalpha\upbeta\upgamma} &= (2\pi)^3 \delta^{(3)}(\bm{k}_1 + \bm{k}_2 + \bm{k}_3)\,C_{\alpha\beta\gamma}\,, \\
    D_{\upalpha\upbeta\upgamma} &= (2\pi)^3 \delta^{(3)}(\bm{k}_1 + \bm{k}_2 + \bm{k}_3)\,D_{\alpha\beta\gamma}\,.
    \end{aligned}
\end{equation}
When choosing an initial time well inside the horizon, we want to keep only the highest-order terms in $\tau$. Although the computation of the internal polynomials is straightforward, we still need to deal with higher-order derivative interactions. Thus in Fourier space, the third-order kernels in Eq.~(\ref{eq:kernels}) above may contain explicit $\left(k^2/a^2\right)^n$ dependencies for some integer $n$ that needs to be tracked when evaluating the integrals. In order to take into account the scale factor dependency in such terms, we split the kernels according to the number of spatial derivatives they contain in the following way
\begin{equation}
    \begin{aligned}
    A_{\alpha\beta\gamma} &= A^{(0)}_{\alpha\beta\gamma} + \frac{1}{a^2} A^{(2)}_{\alpha\beta\gamma} + \frac{1}{a^4} A^{(4)}_{\alpha\beta\gamma} + \ldots\,, \hspace*{0.5cm}
    B_{\alpha\beta\gamma} = B^{(0)}_{\alpha\beta\gamma} + \frac{1}{a^2} B^{(2)}_{\alpha\beta\gamma} + \frac{1}{a^4} B^{(4)}_{\alpha\beta\gamma} + \ldots\,,\\
    C_{\alpha\beta\gamma} &= C^{(0)}_{\alpha\beta\gamma} + \frac{1}{a^2} C^{(2)}_{\alpha\beta\gamma} + \frac{1}{a^4} C^{(4)}_{\alpha\beta\gamma} + \ldots\,,\hspace*{0.5cm}
    D_{\alpha\beta\gamma} = D^{(0)}_{\alpha\beta\gamma} + \frac{1}{a^2} D^{(2)}_{\alpha\beta\gamma} + \frac{1}{a^4} D^{(4)}_{\alpha\beta\gamma} + \ldots\,,
    \end{aligned}
\end{equation}
where we have absorbed all momentum dependency in the newly defined kernels. We indicate with the superscript $2n$ ($n\geq 0$) the number of spatial derivatives that appears in the cubic interaction. Let us start with the internal polynomial involving the $A_{\alpha\beta\gamma}$ tensor elements. For $n\geq0$, it reads
\begin{gather}
\label{eq:A_internal_polynomial_Gamma}
    \begin{aligned}
    \underline{A}^{(2n)}_{\alpha\beta\gamma} &= \frac{1}{2} \int_{-\infty}^{\tau_0} \mathrm{d}\tau\, a^4 \,\frac{A^{(2n)}_{\alpha\beta\gamma}}{a^{2n}}\, (1 - ik_1\tau)(1 - ik_2\tau)(1 - ik_3\tau)\, e^{ik_t \tau} + \text{ 5 perms} \\
    &= \frac{-A^{(2n)}_{\alpha\beta\gamma}}{2i^{3-2n}H^{4-2n}} \left\{ \frac{e_3}{k_t^{2n}}\,\Gamma(2n, -ik_t\tau_0) + \frac{e_2}{k_t^{2n-1}}\,\Gamma(2n-1, -ik_t\tau_0) \right.\\
    &\left.\hspace*{3cm}+ \frac{1}{k_t^{2n-2}}\,\Gamma(2n-2, -ik_t\tau_0) + \frac{1}{k_t^{2n-3}}\,\Gamma(2n-3, -ik_t\tau_0) \right\} + \text{ 5 perms}\,,
    \end{aligned}
    \raisetag{55pt}
\end{gather}
where $\Gamma(m, z)$ for some integer $m$ and a complex entry $z$ is the upper incomplete $\Gamma-$function. Details about this function can be found in the following insert.

\begin{framed}
{\small \noindent {\it Incomplete $\Gamma-$function.}---The incomplete $\Gamma-$function is defined as
\begin{equation}
    \Gamma(s, z) = \int_{z}^{+\infty} t^{s-1}\,e^{-t}\,\mathrm{d}t\,,
\end{equation}
with $s$ being a complex parameter with $\Re(s)>0$. This function is defined without restrictions on the integration paths in the complex plane. In our case, we are interested in the case where $s$ is an integer (positive or negative) and $z$ is purely imaginary. For $s = n \geq 1$, the incomplete $\Gamma-$function can be expressed as a \textit{finite} sum 
\begin{equation}
    \Gamma(n, z) = (n-1)!\, \sum_{j=0}^{n-1}\frac{z^j}{j!}\, e^{-z}\,,
\end{equation}
with the special value $\Gamma(1, z) = e^{-z}$. For a fixed $n\leq 0$, $\Gamma(n, z)$ has an asymptotic expansion -- with \textit{infinite} terms -- at large $z$ of the form
\begin{equation}
    \Gamma(n, z) = z^{n-1}e^{-z}\, \left(\sum_{j=0}^{m-1} \frac{u_j}{z^j} + R_m(n, z)\right)\,, \hspace*{0.5cm} m = 1, 2, \ldots\,,
\end{equation}
where the series coefficients $u_j$ are defined as
\begin{equation}
    u_j = (-1)^j(1-n)_j = (n-1)(n-2)...(n-j)\,,
\end{equation}
and $R_m(n, z)$ is the remainder that decays as $\mathcal{O}\left(1/z^{m}\right)$.
 }
\end{framed}

\noindent From inspecting Eq.~(\ref{eq:A_internal_polynomial_Gamma}), we see that $\underline{A}^{(0)}$ and $\underline{A}^{(2)}$ have an infinite number of terms when expanding at large $|\tau_0|$. Because one cannot implement an infinite number of terms numerically, we choose to keep only the two dominant powers of $\tau_0$. Bearing in mind that this internal polynomial will eventually be multiplied with the external polynomial, this series truncation is an approximation. Indeed, the corrections to the leading behaviour of higher-order derivative interactions will not be of the same order as the leading behaviour of lower-order in derivative interactions. Keeping only the two leading-order terms in $\tau$, the $\underline{A}$ internal polynomials read
\begin{equation}
\label{eq:internal_A}
    \begin{aligned}
    \underline{A}^{(2n)}_{\alpha\beta\gamma} &= \frac{A^{(2n)}_{\alpha\beta\gamma}}{2H^{4-2n} k_t} \left\{e_3\tau_0^{2n-1} + i\tau_0^{2n-2}\left[e_2 + \frac{(2n-1)e_3}{k_t}\right] + \mathcal{O}\left(\tau_0^{2n-3}\right) \right\}e^{ik_t \tau_0} \\
    &+ \text{ 5 perms}\,.
    \end{aligned} 
\end{equation}
Note that it is important to keep track of factors of $i$ because the three-point correlator is real. The integral for the internal polynomial involving the $B$ tensor for $n\geq 0$ reads
\begin{equation}
\label{eq:B_internal_polynomial_Gamma}
    \begin{aligned}
    \underline{B}^{(2n)}_{\alpha\beta\gamma} &= \frac{1}{2} \int_{-\infty}^{\tau_0} \mathrm{d}\tau\, a^4 \,\frac{B^{(2n)}_{\alpha\beta\gamma}}{a^{2n}}\, \frac{1}{a}\,(1 - ik_1\tau)(1 - ik_2\tau)k_3^2\tau\, e^{ik_t \tau} + \text{ 5 perms} \\
    &= \frac{-B^{(2n)}_{\alpha\beta\gamma}}{2i^{3-2n}H^{3-2n}} \left\{ \frac{e_3 k_3}{k_t^{2n+1}}\,\Gamma(2n+1, -ik_t\tau_0) + \frac{(k_1+k_2)k_3^2}{k_t^{2n}}\,\Gamma(2n, -ik_t\tau_0) \right.\\
    &\left.\hspace*{3cm}+ \frac{k_3^2}{k_t^{2n-1}}\,\Gamma(2n-1, -ik_t\tau_0) \right\} + \text{ 5 perms}\,.
    \end{aligned}
\end{equation}
When expanding the $\Gamma$-function, only $\underline{B}^{(0)}$ has an infinite number of terms. Keeping only the two leading-order terms, one obtains
\begin{equation}
\label{eq:internal_B}
    \begin{aligned}
    \underline{B}^{(2n)}_{\alpha\beta\gamma} &= \frac{-B^{(2n)}_{\alpha\beta\gamma}}{2H^{3-2n} k_t} \left\{ie_3\tau_0^{2n}  -\frac{k_3}{k_t}\left[2ne_3 + k_t(k_1+k_2)k_3\right]\tau_0^{2n-1} + \mathcal{O}\left(\tau_0^{2n-2}\right) \right\}e^{ik_t \tau_0} \\
    &+ \text{ 5 perms}\,.
    \end{aligned} 
\end{equation}
Similarly, the integral for the internal polynomial involving the $C$ tensor for $n\geq 0$ is
\begin{equation}
\label{eq:C_internal_polynomial_Gamma}
    \begin{aligned}
    \underline{C}^{(2n)}_{\alpha\beta\gamma} &= \frac{1}{2} \int_{-\infty}^{\tau_0} \mathrm{d}\tau\, a^4 \,\frac{C^{(2n)}_{\alpha\beta\gamma}}{a^{2n}}\, \frac{1}{a^2}\,k_1^2\tau k_2^2\tau(1-ik_3\tau)\, e^{ik_t \tau} + \text{ 5 perms} \\
    &= \frac{C^{(2n)}_{\alpha\beta\gamma}i^{2n+1}}{2H^{2-2n}} \left\{ -\frac{e_3k_1k_2}{k_t^{2n+2}}\,\Gamma(2n+2, -ik_t\tau_0) - \frac{k_1^2k_2^2}{k_t^{2n+1}}\, \Gamma(2n+1, -ik_t\tau)\right\} \\
    &+ \text{ 5 perms}\,.
    \end{aligned}
\end{equation}
We note that $\underline{C}$ has a finite number of terms for all $n\geq0$. However for consistency we still keep only the first two leading-order terms, giving
\begin{gather}
\label{eq:internal_C}
    \begin{aligned}
    \underline{C}^{(2n)}_{\alpha\beta\gamma} &= \frac{-C^{(2n)}_{\alpha\beta\gamma}}{2H^{2-2n} k_t} \left\{e_3k_1 k_2\tau_0^{2n+1}  + ik_1^2 k_2^2 \left( 1 + \frac{(2n+1)k_3}{k_t}\right)\tau_0^{2n} + \mathcal{O}\left(\tau_0^{2n-1}\right) \right\}e^{ik_t \tau_0} \\
    &+ \text{ 5 perms}\,,
    \end{aligned} 
    \raisetag{18pt}
\end{gather}
where the first line is an exact result. Finally for the internal polynomial involving the $D$ tensor for $n\geq 0$, we obtain
\begin{equation}
\label{eq:D_internal_polynomial_Gamma}
    \begin{aligned}
    \underline{D}^{(2n)}_{\alpha\beta\gamma} &= \frac{1}{2} \int_{-\infty}^{\tau_0} \mathrm{d}\tau\, a^4 \,\frac{D^{(2n)}_{\alpha\beta\gamma}}{a^{2n}}\, \frac{1}{a^3}\,e_3^2\tau^3\, e^{ik_t \tau} + \text{ 5 perms} \\
    &= \frac{D^{(2n)}_{\alpha\beta\gamma}}{2H^{1-2n}}\,\frac{e_3^2\,i^{2n+3}}{k_t^{2n+3}}\, \Gamma(2n+3, -ik_t\tau_0) + \text{ 5 perms}\,,
    \end{aligned}
\end{equation}
from which we get
\begin{equation}
\label{eq:internal_D}
    \begin{aligned}
    \underline{D}^{(2n)}_{\alpha\beta\gamma} &= \frac{D^{(2n)}_{\alpha\beta\gamma}}{2H^{1-2n} k_t}\, e_3^2\, \left\{i \tau_0^{2n+2} - \frac{(2n+2)\tau_0^{2n+1}}{k_t} + \mathcal{O}\left(\tau_0^{2n}\right) \right\}e^{ik_t \tau_0} + \text{ 5 perms}\,.
    \end{aligned} 
\end{equation}
Like the previous case, $\underline{D}$ has a finite number of terms for all $n\geq 0$. We have now all the necessary ingredients to obtain the initial conditions for the three-point correlators. Combining the external polynomials in Eq.~(\ref{eq:external_polynomials}), the truncated internal polynomials in Eqs.~(\ref{eq:internal_A})-(\ref{eq:internal_B})-(\ref{eq:internal_C}) and (\ref{eq:internal_D}), and taking into account the common overall constant, the final results are the ones given in (\ref{eq:3pt_fff_initial_conditions}), (\ref{eq:3pt_pff_initial_conditions}), (\ref{eq:3pt_ppf_initial_conditions}) and (\ref{eq:3pt_ppp_initial_conditions}).

\paragraph{Tensorial structure.} We have omitted the tensorial structure coming from the field normalisations in (\ref{eq:3pt_fff_initial_conditions}), (\ref{eq:3pt_pff_initial_conditions}), (\ref{eq:3pt_ppf_initial_conditions}) and (\ref{eq:3pt_ppp_initial_conditions}) to cleanse the expressions. Combinations of $\Delta$ tensors can be included a posteriori by performing the following replacement on the correlators (no summation on repeated indices)
\begin{equation}
    \begin{aligned}
    \braket{\bm{\varphi}_{\bm{k}_1}^\alpha \bm{\varphi}_{\bm{k}_2}^\beta \bm{\varphi}_{\bm{k}_3}^\gamma}_0 &\rightarrow \Delta_{\alpha\alpha}\Delta_{\beta\beta}\Delta_{\gamma\gamma}\,\braket{\bm{\varphi}_{\bm{k}_1}^\alpha \bm{\varphi}_{\bm{k}_2}^\beta \bm{\varphi}_{\bm{k}_3}^\gamma}_0\,,\\
    \braket{\bm{p}_{\bm{k}_1}^\alpha \bm{\varphi}_{\bm{k}_2}^\beta \bm{\varphi}_{\bm{k}_3}^\gamma}_0 &\rightarrow \Delta^{-1}_{\alpha\alpha}\Delta_{\beta\beta}\Delta_{\gamma\gamma}\,\braket{\bm{p}_{\bm{k}_1}^\alpha \bm{\varphi}_{\bm{k}_2}^\beta \bm{\varphi}_{\bm{k}_3}^\gamma}_0\,,\\
    \braket{\bm{p}_{\bm{k}_1}^\alpha \bm{p}_{\bm{k}_2}^\beta \bm{\varphi}_{\bm{k}_3}^\gamma}_0 &\rightarrow \Delta^{-1}_{\alpha\alpha}\Delta_{\beta\beta}\Delta^{-1}_{\gamma\gamma}\,\braket{\bm{p}_{\bm{k}_1}^\alpha \bm{p}_{\bm{k}_2}^\beta \bm{\varphi}_{\bm{k}_3}^\gamma}_0\,,\\
    \braket{\bm{p}_{\bm{k}_1}^\alpha \bm{p}_{\bm{k}_2}^\beta \bm{p}_{\bm{k}_3}^\gamma}_0 &\rightarrow \Delta^{-1}_{\alpha\alpha}\Delta^{-1}_{\beta\beta}\Delta^{-1}_{\gamma\gamma}\,\braket{\bm{p}_{\bm{k}_1}^\alpha \bm{p}_{\bm{k}_2}^\beta \bm{p}_{\bm{k}_3}^\gamma}_0\,,
    \end{aligned}
\end{equation}
and by redefining the third-order vertex functions as follows
\begin{equation}
    \begin{aligned}
    A^{(2n)}_{\alpha\beta\gamma} &\rightarrow \Delta_{\alpha\alpha}\Delta_{\beta\beta}\Delta_{\gamma\gamma}\,A^{(2n)}_{\alpha\beta\gamma}\,,\\
    B^{(2n)}_{\alpha\beta\gamma} &\rightarrow \Delta_{\alpha\alpha} \Delta_{\beta\beta}\Delta^{-1}_{\gamma\gamma} \, B^{(2n)}_{\alpha\beta\gamma}\,,\\
    C^{(2n)}_{\alpha\beta\gamma} &\rightarrow \Delta^{-1}_{\alpha\alpha} \Delta^{-1}_{\beta\beta}\Delta_{\gamma\gamma} \, C^{(2n)}_{\alpha\beta\gamma}\,,\\
    D^{(2n)}_{\alpha\beta\gamma} &\rightarrow \Delta^{-1}_{\alpha\alpha} \Delta^{-1}_{\beta\beta}\Delta^{-1}_{\gamma\gamma} \, D^{(2n)}_{\alpha\beta\gamma}\,.
    \end{aligned}
\end{equation}
for all $n\geq 0$. These manipulations are necessary if one or several fields are not canonically normalised.

\newpage
\section{Light Limit of the Dynamics}
\label{app:LightField}

In this appendix, we first derive a simple formula for the power spectrum in the light field and weak mixing regime of the phase diagram. This reproduces the observed enhancement in Figure~\ref{fig:Gaussian_phase_diagram}. We then comment on the effect of the non-linear conversion from $\pi$ to $\zeta$ on the bispectrum. 

\paragraph{Linear Dynamics.} In Section \ref{subsec:quadratic_theory}, we obtained exact results for the Goldstone boson power spectrum in the entire phase space $(m/H, \rho/H)$. Our result shows a characteristic enhancement of $\Delta_\zeta^2$ in the light limit $m/H\ll 1$ and at weak mixing $\rho/H\ll 1$. Here, we explicitly show that this enhancement is a consequence of the light $\sigma$ field not decaying fast enough on super-horizon scales, therefore leaving an imprint in $\Delta_\zeta^2$ which strongly depends on the number of super-horizon $e$-folds. A very similar analysis has been applied to the spinning case in \cite{Bordin:2018pca} so that we will closely follow their derivation. We set $c_s=1$ for simplicity throughout this appendix.

\vskip 4pt
Our starting point is the quadratic Lagrangian in (\ref{eq:Full_pi_sigma_theory}) that leads to the following coupled equations of motion
\begin{equation}
\label{eq: EOM pi-sigma dynamics}
    \begin{aligned}
    &\ddot{\pi}_c + 3H\dot{\pi}_c + \frac{k^2}{a^2}\pi_c = -\rho[3H\sigma + \dot{\sigma}]\,, \\
    &\ddot{\sigma} + 3H\dot{\sigma} + \left(\frac{k^2}{a^2}+m^2\right)\sigma = \rho \dot{\pi}_c\,.
    \end{aligned}
\end{equation}
We are interested in obtaining the power spectrum at late times so we want to find the late-time behaviour of the mode functions. Hence, we look for a solution of the form $\pi_c = \pi_0 \tau^\Delta$ and $\sigma = \sigma_0 \tau^\Delta$.
Plugging this ansatz in the system and neglecting the gradient term leads to
\begin{equation}
    \begin{pmatrix}
    \Delta(\Delta-3) & \frac{\rho}{H}(3-\Delta) \\
    \frac{\rho}{H}\Delta & \Delta(\Delta - 3) + \frac{m^2}{H^2}
    \end{pmatrix}
    \begin{pmatrix}
    \pi_0 \\
    \sigma_0
    \end{pmatrix}
    =0\,.
\end{equation}
The requirement of a vanishing determinant gives us the following scaling dimensions
\begin{equation}
    \Delta_-^{\pi_c} = 0\,, \hspace*{0.5cm} \Delta_+^{\pi_c} = 3\,, \hspace*{0.5cm} \Delta_{\pm}^{\sigma} = \frac{3}{2} \pm \sqrt{\frac{9}{4} - \frac{m_{\text{eff}}^2}{H^2} }\,,
\end{equation}
with $m_{\text{eff}}^2 = m^2 + \rho^2$. Plugging back in the matrix equation the found solutions and trying to solve for $\pi_0$ and $\sigma_0$ gives us constraints satisfied by the mode amplitudes:

\begin{itemize}
    \item $\Delta = 0$ enforces $\sigma_0=0$ but leaves $\pi_0$ undertermined. In fact, the precise amplitude of this massless mode is fixed by the quantisation condition in the far past.
    \item $\Delta = \frac{3}{2}\pm \sqrt{9/4 - m_{\text{eff}}^2/H^2}$ gives $\frac{\pi_0}{\sigma_0} = \frac{\rho}{H}\frac{\Delta_{\mp}^\sigma}{\Delta_-^\sigma\Delta_+^\sigma}$ and $\frac{\sigma_0}{\pi_0} = \frac{H}{\rho}\Delta^\sigma_{\pm}$. Both solutions are equivalent. This solution is the manifestation of the exponential dilution of $p_{\pi_c}$ with time as the Universe inflates. Indeed at late times neglecting gradient terms, one can show that the conjugate momentum to the field $\pi_c$ satisfies
    \begin{equation}
        \dot{p}_{\pi_c} + 3Hp_{\pi_c} = 0\,,
    \end{equation}
    so that it decays as $p_{\pi_c}\propto 1/a^3$. Close to the infinite future boundary, we then have $\dot{\pi}_c \approx -\rho \sigma$. If the leading behaviour of the massive field is $\sigma\sim \tau^{\Delta_\pm^\sigma}$, then after integration one finds $\pi_c \sim  \text{cst} + \frac{\rho}{H}\frac{\sigma}{\Delta_{\pm}^\sigma}$ which gives us the found constraint on the mode amplitudes. The constant is found by imposing the quantisation condition.
    
    \item $\Delta=3$ enforces $\frac{\rho}{H}\pi_0 + \frac{m^2}{H^2}\sigma_0 = 0$. In the $M = \frac{m^2}{\rho}\gg H$ limit, we obtain $\sigma_0=0$ if the field is massive. This is expected because a massive field in de-Sitter does not contain any $\Delta=3$ mode. In the $M = \frac{m^2}{\rho}\ll H$, we obtain $\pi_0=0$. This is not surprising as the $\sigma$ field can be integrated out, leading to a modified dispersion relation for $\pi_c$. One can explicitly check that the exact form of the $\pi_c$ mode function\footnote{The leading term of $z^{5/4}H^{(1)}_{5/4}(z^2)$ as $z\rightarrow0$ is $\propto z^4$.} in this regime does not contain a $\Delta=3$ mode.
\end{itemize}

Imposing the previous constraints on the mode amplitudes, the general solution for the system at late times is
\begin{equation}
    \begin{pmatrix}
    \pi_c \\
    \sigma
    \end{pmatrix} = \pi_0^-
    \begin{pmatrix}
    1 \\
    0
    \end{pmatrix} + \pi_0^+ (-k\tau)^3
    \begin{pmatrix}
    \frac{m^2}{H^2} \\
    -\frac{\rho}{H}
    \end{pmatrix} + \sigma_0^- (-k\tau)^{\Delta^\sigma_-}
    \begin{pmatrix}
    \frac{\rho}{H}\Delta_+^\sigma \\
    \frac{m_{\text{eff}}^2}{H^2}
    \end{pmatrix} + \sigma_0^+ (-k\tau)^{\Delta_+^\sigma}
    \begin{pmatrix}
    \frac{\rho}{H} \\
    \Delta_+^\sigma
    \end{pmatrix}\,,
\end{equation}
where $\pi_0^-, \pi_0^+, \sigma_0^-$ and $\sigma_0^+$ are integration constants. When $\sigma$ is \textit{effectively light} $m_{\text{eff}}\ll H$ but not strictly speaking massless, its $\sigma_0^-$ mode decays very slowly on super-horizon scales with scaling dimension $\Delta_-^\sigma \approx \frac{m_{\text{eff}}^2}{3H^2}$. Only considering the two lightest modes, the solution for the system is
\begin{equation}
    \begin{pmatrix}
    \pi_c \\
    \sigma
    \end{pmatrix} \approx \pi_0^-
    \begin{pmatrix}
    1 \\
    0
    \end{pmatrix} + \sigma_0^- (-k\tau)^{\frac{m_{\text{eff}}^2}{3H^2}}
    \begin{pmatrix}
    \frac{3\rho}{H} \\
    \frac{m_{\text{eff}}^2}{H^2}
    \end{pmatrix}\,.
\end{equation}
The $\sigma_0^-$ mode has a component in the $\sigma$ direction that is parametrically smaller than its component in the $\pi_c$ direction in the regime of interest. We now impose that at horizon crossing, the $\sigma_0^-$ mode is dominated by the component in the $\sigma$ direction
\begin{equation}
    \pi_c |_\star = 0, \hspace*{0.5cm} \sigma |_\star = A_0\, (-k\tau)^{\frac{m_{\text{eff}}^2}{3H^2}}\,,
\end{equation}
where $\star$ means that the quantity is evaluated at horizon crossing $k=aH$. This condition enforces that we chose $A_0 = \frac{m_{\text{eff}}^2}{H^2}\sigma_0^-$ and $\pi_0^- = -\frac{3\rho}{H}\sigma_0^-$ in order to have an exact cancellation between the large $\pi_c$ modes at horizon crossing, i.e.~this amounts to only study the sourcing by $\sigma$ on $\pi_c$ on super-horizon scales. The solution for $\pi_c$ is then determined up to an overall constant
\begin{equation}
    \pi_c = -A_0\, \frac{3\rho H}{m_{\text{eff}}^2}\left(1 - e^{-\frac{m_{\text{eff}}^2}{3H^2} \Delta N}\right),
\end{equation}
where $\Delta N = N - N_\star$ is the number of $e$-folds elapsed since horizon crossing. The power spectrum correction is found to be
\begin{equation}
\label{eq:LightField}
    \frac{\Delta_\zeta^2}{\Delta_{\zeta, 0}^2} - 1 \propto \left[\frac{3\rho H}{m_{\text{eff}}^2}\left(1 - e^{-\frac{m_{\text{eff}}^2}{3H^2} \Delta N}\right)\right]^2\,.
\end{equation}

\begin{figure}[t!]
   \centering
   \hspace*{0.5cm}
         \includegraphics[width=0.8\textwidth]{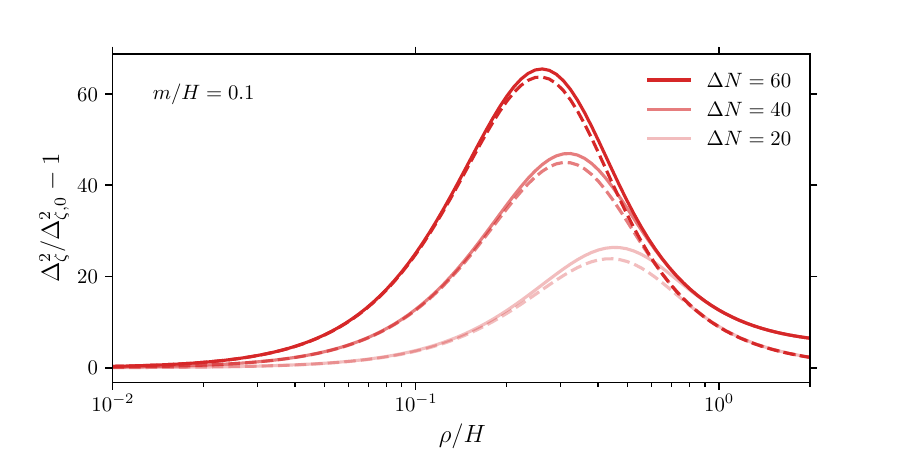}
   \caption{Correction to the power spectrum (normalised to its single-field value for $\rho=0$) in the effectively light regime $m_{\text{eff}}\ll H$ for $\Delta N = 60$, $\Delta N = 40$, and $\Delta N = 20$, where $\Delta N$ is the number of super-horizon $e$-folds. The solid and dashed lines correspond to the exact numerical results and the analytical approximation in (\ref{eq:LightField}), respectively. The amplitude of the approximation~(\ref{eq:LightField}) was determined to match the exact result at $\rho/H = 10^{-2}$. The mismatch between the solid and dashed lines for $\rho/H \gtrsim 0.5$ indicates that the field $\sigma$ is effectively massive enough so that the approximation breaks down.}
  \label{fig:LightField}
\end{figure}

At weak coupling $\rho\ll H$ when the light mode does not have enough time to decay, this formula reproduces well the leading correction found in perturbative calculations \cite{Achucarro:2016fby}
\begin{equation}
    \frac{\Delta_{\zeta}^2}{\Delta_{\zeta, 0}^2} - 1 \propto \frac{\rho^2}{H^2} \Delta N^2\,,
\end{equation}
where the precise coefficient has been computed in \cite{Chen:2012ge, Pi:2012gf}. However when the effective mass becomes of order $m_{\text{eff}}\approx H$, this analytical approximation breaks down and does not capture the modified dispersion relation behaviour. This is illustrated in Figure~\ref{fig:LightField} where we show a slice of the phase diagram at fixed bare mass $m/H=0.1$ with the exact and approximated power spectrum for different values of $\Delta N$. A notable feature is that the amplitude and the location of the peak is well reproduced by the analytical approximation (\ref{eq:LightField}). This qualitatively explains the boost of the power spectrum in the case of a light $\sigma$ field. Note that the finite duration of inflation induces an intrinsic violation of scale-invariance, and that a complete prediction even just for the power spectrum would require to describe the reheating epoch.

\begin{figure}[t!]
   \centering
   \hspace*{0.5cm}
         \includegraphics[width=0.6\textwidth]{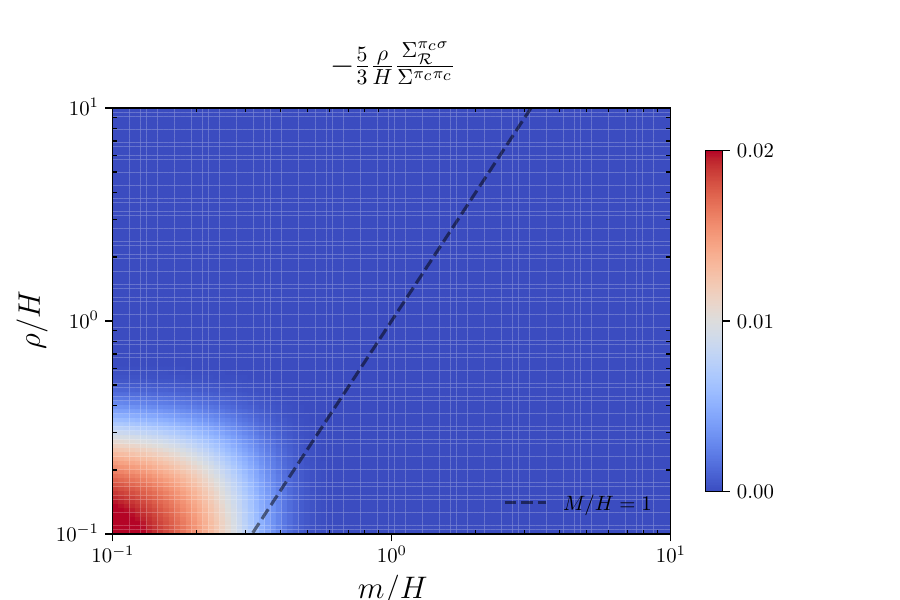}
   \caption{Correction to the ``intrinsic" $\fnl$ parameter coming from the non-linear conversion from $\pi$ to $\zeta$, see Eq.~(\ref{eq:zeta-correlators}), in the limit $\dot{H}\rightarrow 0$. We have set $N=60$ $e$-folds of super-horizon evolution. The dashed line corresponds to $M = m^2/\rho=H$, see Section~\ref{subsubsec: phase diagram} for more details.}
  \label{fig:nonlinearCorrectionfNL}
\end{figure}

\paragraph{Non-linear conversion and the bispectrum.} The non-linear conversion from $\pi$ to $\zeta$ engenders additional terms in the bispectrum that should be added to the ``intrinsic" bispectrum generated by cubic interactions. These terms are displayed in Eq.~(\ref{eq:zeta-correlators}) for any theory in the super-horizon limit. In the decoupling limit, only the second term in the first line contributes. Specifying to equilateral configurations $k_1 = k_2 = k_3$ and in the theory~(\ref{eq:Full_pi_sigma_theory}), the size of non-Gaussianities needs to be corrected by
\begin{equation}
    \fnl \rightarrow \fnl -\frac{5}{3} \frac{\rho}{H} \frac{\Sigma^{\pi_c\sigma}_\Re}{\Sigma^{\pi_c\pi_c}}\,.
\end{equation}
The exact correction is displayed in Figure~\ref{fig:nonlinearCorrectionfNL} in the phase space $(m/H, \rho/H)$. As anticipated, this contribution only affects the light-field regime, where $\sigma$ decays slowly on super-horizon scales. In the parameter space under display, its amplitude is comparable to the expected gravitational floor ($\sim 10^{-2}$), and therefore negligible for current and near-future cosmological surveys.
In the effectively light field limit, one expects the bispectrum to resemble a local shape in its squeezed limit.
However, predicting the observable part of it would require subtracting the artefact coming from a rescaling of the background by a long wavelength perturbation, including the usual consistency relation that is proportional to the tilt of the power spectrum.
All this depends on the arbitrary choice of the duration of inflation, and is also degenerate with a description of reheating.
A full study of these effects is beyond the scope of this paper.

\newpage
\section{Perturbativity Bounds}
\label{app:strong_coupling_scales_tabulars}

In this appendix, for the reader's convenience, we collect all the perturbativity bounds on the coupling constants of the theory (\ref{eq:Full_pi_sigma_theory}) found in Section \ref{sec:strong_coupling_scales}, both in the weak and strong mixing regimes.\footnote{We define $\kappa = 2\Gamma(5/4)^2/\pi^3 \approx 0.053$, see e.g.~(\ref{eq:MDR_power_spectrum}).} We recall that these bounds follow from imposing that the strong coupling scales of irrelevant operators be higher than the Hubble scale, or imposing standard perturbativity criteria on relevant/marginal operators. We have coloured in \textcolor{pyred}{red} the most stringent bounds.

\begin{figure}[h!]
   \centering
\begin{tabular}{ M{2.5cm} M{3cm} M{3cm} M{3cm} M{2.5cm}  }
 \hline
 \\[0em]
 \multicolumn{5}{c}{\textbf{Weak Mixing Regime}} \\[10pt]
 \\[-0.5em]
 \footnotesize{\textbf{Operators (unitary gauge)}} & $\delta g^{00}\sigma$ & $\delta g^{00}\sigma^2$ & $\left(\delta g^{00}\right)^2\sigma$ & $\sigma^3$ \\[10pt]
 \\[-0.5em]
 \footnotesize{\textbf{Interactions}}   & $\dot{\pi}_c\sigma, (\partial_i \pi_c)^2\sigma$ & $\dot{\pi}_c\sigma^2, \dot{\pi}_c^2\sigma^2$, $(\partial_i \pi_c)^2\sigma^2$ & $\dot{\pi}_c^2\sigma, \dot{\pi}_c (\partial_\mu \pi_c)^2\sigma, \dot{\pi}_c^4\sigma$, $\dot{\pi}_c^2(\partial_i \pi_c)^2\sigma, (\partial_i \pi_c)^4\sigma$ & $\sigma^3$\\
 \\[-0.5em]
 $\Lambda_\star$ & $c_s^{3/2}\frac{f_\pi^2}{\rho}$ & $c_s^{3/4}\frac{f_\pi}{\sqrt{\alpha}}$ & $c_s^{-1/2}\frac{f_\pi^2}{\tilde{\rho}}$ & \crossmark \\
 \\[-0.5em]
 $H\lesssim \Lambda_\star$ & $\frac{\rho}{H}\lesssim \frac{c_s^{3/2}}{2\pi\Delta_\zeta}$ & $\tilde{\alpha}\lesssim \frac{c_s^{-1/8}}{4\pi\Delta_\zeta}$ & $\textcolor{pyred}{\frac{\tilde{\rho}}{H} \lesssim \frac{c_s^{-1/2}}{2\pi\Delta_\zeta}}$ & \crossmark \\
 \\[-0.5em]
 \footnotesize{\textbf{Perturbativity of relevant or marginal operators}} & $\textcolor{pyred}{\frac{\rho}{H}\lesssim c_s^{-1/2}}$ & $\textcolor{pyred}{\tilde{\alpha} \lesssim \frac{c_s^{-1/2}}{2}\frac{1}{(2\pi \Delta_\zeta)^{1/2}}}$ & \crossmark & $\textcolor{pyred}{\frac{\mu}{H}\lesssim 1}$\\
 \\[-0.5em]
 \hline
\end{tabular}
\end{figure}

\vspace*{-0.7cm}

\begin{figure}[h!]
   \centering
\begin{tabular}{ M{2.5cm} M{2cm} M{4cm} M{3cm} M{2.5cm}  }
 \\[0em]
 \multicolumn{5}{c}{\textbf{Strong Mixing Regime}} \\[10pt]
 \footnotesize{\textbf{Operators (unitary gauge)}} & $\delta g^{00}\sigma$ & $\delta g^{00}\sigma^2$ & $\left(\delta g^{00}\right)^2\sigma$ & $\sigma^3$ \\[10pt]
 \\[-0.5em]
 \footnotesize{\textbf{Interactions}}   & $\dot{\tilde{\pi}}_c\sigma, (\tilde{\partial}_i \tilde{\pi}_c)^2\sigma$ & $\dot{\tilde{\pi}}_c\sigma^2, \dot{\tilde{\pi}}_c^2\sigma^2$, $(\tilde{\partial}_i \tilde{\pi}_c)^2\sigma^2$ & $\dot{\tilde{\pi}}_c^2\sigma, \dot{\tilde{\pi}}_c (\partial_\mu \tilde{\pi}_c)^2\sigma, \dot{\tilde{\pi}}_c^4\sigma$, $\tilde{\pi}_c^2(\tilde{\partial}_i \tilde{\pi}_c)^2\sigma, (\tilde{\partial}_i \tilde{\pi}_c)^4\sigma$ & $\sigma^3$\\
 \\[-0.5em]
 $\Lambda_\star$ & $c_s\left(\frac{f_\pi^8}{\rho^5}\right)^{1/3}$ & $\left( \frac{\rho f_\pi^4}{\alpha^2}\right)^{1/5}, \left(\frac{c_s^2f_\pi^4}{\alpha^2\rho}\right)^{1/3}$ & $\left(\frac{\rho^3 f_\pi^8}{c_s\tilde{\rho}^4} \right)^{1/7}$ & \crossmark \\
 \\[-0.5em]
 $H\lesssim \Lambda_\star$ & $\textcolor{pyred}{\frac{\rho}{H}\lesssim c_s^{3/4}\frac{\kappa^{1/2}}{\Delta_\zeta}}$ &  $\tilde{\alpha}\lesssim \frac{\kappa^{3/4}}{2c_s^{3/8}\Delta_\zeta^{3/2}}$, $\textcolor{pyred}{\tilde{\alpha}\lesssim \frac{\kappa^{1/2}}{2c_s^{1/4}\Delta_\zeta}}$ & $\textcolor{pyred}{\frac{\tilde{\rho}}{H} \lesssim c_s^{1/2}\frac{\kappa}{\Delta_\zeta^2}}$ & \crossmark \\
 \\[-0.5em]
 \footnotesize{\textbf{Perturbativity of relevant or marginal operators}} & \crossmark & $\textcolor{pyred}{\tilde{\alpha}\lesssim \frac{\kappa^{1/2}}{2c_s^{1/4}\Delta_\zeta}}$ & \crossmark & $\textcolor{pyred}{\frac{\mu}{H}\lesssim c_s^{-3/4}\left(\frac{\rho}{H}\right)^{3/4}}$\\
 \\[-0.5em]
 \hline
\end{tabular}
\end{figure}

\newpage
\phantomsection
\addcontentsline{toc}{section}{References}
\small
\bibliographystyle{utphys}
\bibliography{CosmologicalFlow}

\end{document}